\documentclass{article}
\usepackage{comment}
\usepackage{graphicx}
\usepackage{amssymb}
\usepackage{amsmath}
\usepackage{url}
\usepackage{eurosym}
\usepackage{wrapfig}
\usepackage{epsfig}
\usepackage{bbm}
\usepackage{multirow}
\usepackage{authblk}
\usepackage{caption}
\usepackage{subcaption}

\usepackage[breaklinks=true,colorlinks=true]{hyperref}
\begin{document}
\begin{center}
{\Huge Hall A Annual Report \\ 2012}
\end{center}
\vspace{4cm}
\begin{center}
\includegraphics[width=0.8\textwidth, angle = 0.]{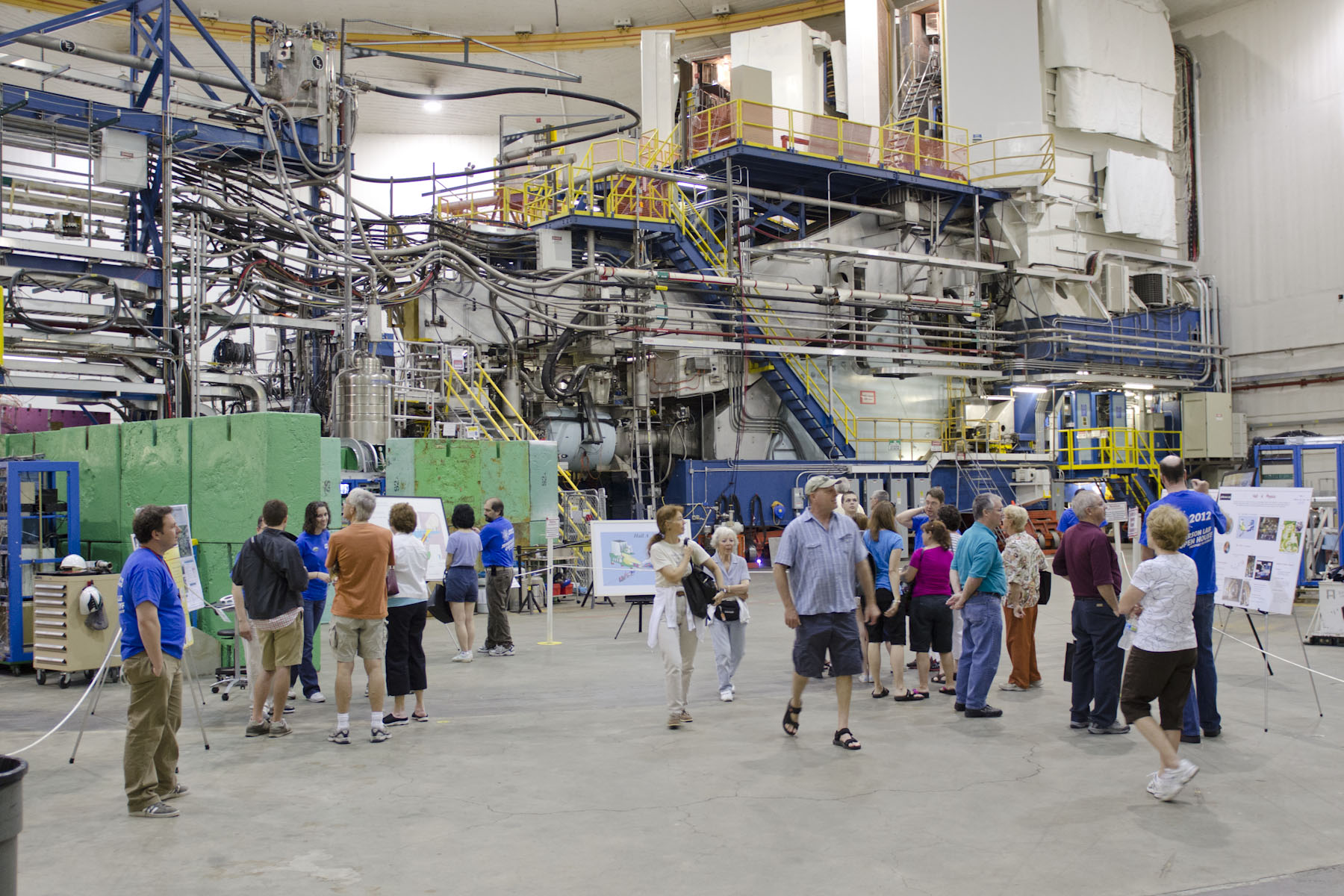}
\end{center}
\vspace{6cm}
\begin{center}
{\Large Edited by Seamus Riordan \& Cynthia Keppel}
\end{center}
\title{}  
\date{}   

\author[1]{S.~Riordan}
\author[2]{C.~Keppel}
\author[3]{K.~Aniol}
\author[4]{J.~Annand}
\author[5]{J.~Arrington}

\author[6]{T.~Averett}

\author[6]{C.~Ayerbe~Gayoso}
\author[7]{E.~Brash}
\author[8]{G.D.~Cates}
\author[2]{J.-P.~Chen}
\author[2]{E.~Chudakov}
\author[9]{D.~Flay}
\author[10]{G.B.~Franklin}
\author[11]{M.~Friedman}
\author[12]{O.~Glamazdin}
\author[2]{J.~Gomez}
\author[8]{C.~Hanretty}
\author[2]{J.-O.~Hansen}
\author[13]{C.~Hyde}
\author[2]{M.K.~Jones}
\author[14]{I.~Korover}
\author[2]{J.J.~LeRose}
\author[8]{R.A.~Lindgren}
\author[8]{N.~Liyanage}
\author[15]{E.~Long}
\author[10]{V.~Mamyan}
\author[16]{M.~Mihovilovic}
\author[17]{N.~Muangma}
\author[2]{S.~Nanda}
\author[18]{D.~Parno}
\author[6]{C.F.~Perdrisat}
\author[12]{R.~Pomatsalyuk}
\author[9]{M.~Posik}
\author[19]{V.~Punjabi}
\author[4]{G.~Rosner}
\author[4]{J.~Sj\"{o}gren}
\author[16]{S.~\v{S}irca}
\author[8]{L.C.~Smith}
\author[2]{P.~Solvignon}
\author[9]{N.F.~Sparveris}
\author[17]{V.~Sulkosky}
\author[12]{V.~Vereshchaka}
\author[2]{B.~Wojtsekhowski}
\author[8]{Z.~Ye}
\author[2]{J.~Zhang}
\author[20]{Y.W.~Zhang}
\author[8]{X.~Zheng}
\author[21]{R.~Zielinski}
\author[ ]{the~Hall~A~Collaboration}

\affil[1]{\small University of Massachusetts, Amherst, MA 01003, USA}
\affil[2]{Thomas Jefferson National Accelerator Facility, Newport News, VA 23606, USA}

\affil[3]{California State University Los Angeles, Los Angeles, CA 90032, USA}

\affil[4]{University of Glasgow, Glasgow G12 8QQ, Scotland, UK}

\affil[5]{Physics Division, Argonne National Laboratory, Argonne, IL, 60439, USA}

\affil[6]{College of William and Mary, Williamsburg, VA 23187, USA}
\affil[7]{Department of Physics, Christopher Newport University, Newport News, Virginia 23606, USA}
\affil[8]{University of Virginia, Charlottesville, VA 22901, USA}
\affil[9]{Temple University, Philadelphia, PA 19122, USA}
\affil[10]{Carnegie Mellon University, Pittsburgh, PA 15217, USA}
\affil[11]{Racah Institute of Physics, Hebrew University of Jerusalem, Givat Ram 91904, Israel}
\affil[12]{National Science Center Kharkov Institute of Physics and Technology, Kharkov 61108, Ukraine}
\affil[13]{Old Dominion University, Norfolk, VA 23529, USA}
\affil[14]{Tel Aviv University, Tel Aviv, 69978 Israel}
\affil[15]{Kent State University, Kent, OH 44242, USA}
\affil[16]{University of Ljubljana, SI-1000 Ljubljana, Slovenia}
\affil[17]{Massachusetts Institute of Technology, Cambridge, MA 02139, USA}
\affil[18]{University of Washington, Center for Experimental Nuclear Physics and Astrophysics and Department of Physics, Seattle, WA 98195, USA}
\affil[19]{Norfolk State University, Norfoik, VA 23504, USA}
\affil[20]{Rutgers, The State University of New Jersey, Piscataway, NJ 08854, USA}
\affil[21]{University of New Hampshire, Durham, NH 03824, USA}

\maketitle
\newpage
\tableofcontents
\newpage
\listoffigures
\newpage

\section{Introduction}
\begin{center}
contributed by Cynthia Keppel
\end{center}

The year 2012 marked the end of the 6~GeV era at Jefferson Lab, and the beginning of the next phase in making the 12~GeV upgrade to the Continuous Electron Beam Accelerator Facility a reality. The final months of the 6~GeV scientific program in Hall A facilitated completion of the $g_2^p$ and $G_E^p$ experiments. A long process of removing, upgrading and re-installing existing components and systems, and installing new ones, has now begun in preparation for 12~GeV operations, with the expectation to see the first beam back in Hall A in 2014.

The 12~GeV upgrade plans for Hall A, as a subset of the overall laboratory project, are modest, composed largely of requisite beam line upgrades to the beam transport, polarimetry and arc energy measurements. However, the 12~GeV scientific plans for the hall are ambitious, and include multiple new experiment installations such as the Super Bigbite Spectrometer (SBS) program to measure high precision nucleon form factors, the MOLLER experiment to measure the parity-violating asymmetry in electron-electron (Moller) scattering, and the SoLID (Solenoidal Large Intensity Device) program to provide a facility for parity violating and semi-inclusive deep inelastic scattering experiments. In addition to these large-scale efforts, many other compelling experiments will utilize the standard Hall A equipment, some with slight modifications, in conjunction with the higher energy beam. 

The last part of the year has been dedicated to preparing for two of the latter, experiments E12-07-108, a measurement of the proton magnetic form factor G$_M^p$~\cite{ref:e1207108}, and E12-06-114, a measurement of deeply virtual Compton scattering~\cite{ref:e1206114}, which will be the first experiments to receive beam in the 12~GeV era. It is interesting to note that these first experiments both harken back to the early days of Hall A, where first round experiments then also measured virtual Compton scattering and form factors, albeit at lower Q$^2$ and with somewhat different physics foci. It seems perhaps the proverb is true, ``The more things change the more they remain the same.'' Preparations for E12-07-108 and E12-06-114 have included plans for a largely combined run period, employing both HRS’s, where significant detector upgrades are underway, a hydrogen target, and other complimentary equipment. 

This year also brought DOE approval to begin the Super Bigbite Spectrometer (SBS) project. This project consists of a set of three form factor experiments centered around somewhat common equipment and new experimental capabilities. First activities to begin this program include re-design of a magnet from the Brookhaven National Laboratory, pre-research and development of GEM tracking detectors, and a host of scientific development activities including detector construction projects, data acquisition upgrades, and refined physics projections. 

Work has continued effectively as well on many other fronts, including infrastructure improvements in data acquisition, offline analysis, and core hall capabilities. Technical preparations have begun for the $^3$H target experiments anticipated for 2015. Ideas to improve and upgrade also the polarized $^3$He target, required for instance for the measurement of A$^n_1$, are being implemented. Efforts continue also for 2015 and beyond planned experiments such as PREX-II, and APEX.  Moreover, there has been active engagement in analyses of past experiments. Here, ten new publications related to Hall A experiments were authored by members of the Hall A collaboration, and seven new Hall A related doctoral theses were successfully defended.

In all, this has been a year of transition - for the laboratory, for Hall A, and also for me as the new Hall Leader. It is an exciting challenge to facilitate the progression of the characteristically excellent standards established here by the Hall A staff and user community into the 12~GeV era. Please accept my many, many thanks to you all for your shared wisdom, valuable advice, and patient support. I look forward to welcoming the higher energy beam into Hall A with you!

\clearpage
\newpage
\section{General Hall Developments}

\subsection{M{\o}ller Polarimeter}
\label{sec:moller_status}

\begin{center}
\bf Status of the Hall A M{\o}ller Polarimeter
\end{center}

\begin{center}
contributed by $^1$O.~Glamazdin, $^2$E.~Chudakov, $^2$J.~Gomez, $^1$R.~Pomatsalyuk, $^1$V.~Vereshchaka, $^2$J. Zhang  \\
$^1$National Science Center Kharkov Institute of Physics and Technology, Kharkov 61108, Ukraine  \\
$^2$Thomas Jefferson National Accelerator Facility, Newport News, VA23606, USA  .\\
\end{center}

\subsubsection{Introduction}
\label{sec:moll_intro}

The Hall A M{\o}ller polarimeter \cite{FizikaB} had been built in 1997. It was successfully used to measure a beam polarization for all Hall A experiments with polarized electron beam. 

\subsubsection{General description}
\label{sec:setup}

\begin{figure}[htb]
   \begin{center}
        \includegraphics[width=0.7\linewidth]{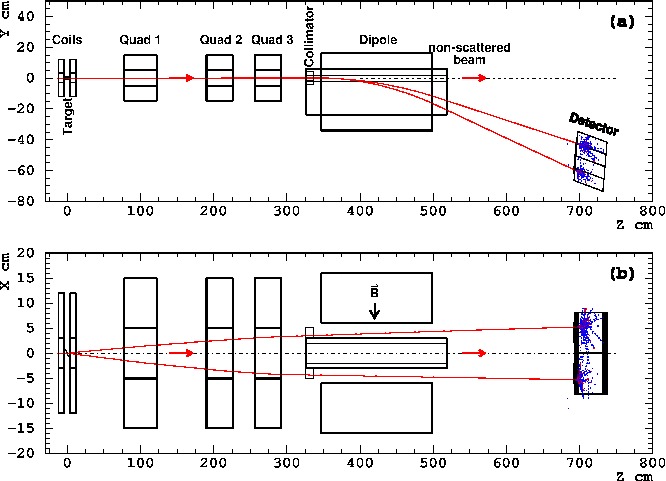}
   \end{center}
   \caption{\label{fig:layout} Layout of the M{\o}ller polarimeter before 11~GeV upgrade, 
          (a) presents the side view while (b) presents the top view. The trajectories
          displayed belong to a simulated event of M{\o}ller scattering 
          at $\theta{}_\mathrm{CM}=80^{\circ}$ and $\phi{}_\mathrm{CM}=0^{\circ}$, 
           at a beam energy of 4~GeV. }
\end{figure}

The M{\o}ller scattering events are detected with a magnetic spectrometer 
(see Fig.\ref{fig:layout}) consisting of a sequence of three quadrupole magnets and a 
dipole magnet. The electrons scattered in a plane close to the horizontal plane are transported by 
the quadrupole magnets to the entrance of the dipole which deflects the electrons down, toward
the detector. The optics of the spectrometer is optimized in order to
maximize the acceptance for pairs scattered at about 90$^\circ$ in CM.
The acceptance depends on the beam energy. The typical range for 
the accepted polar and azimuthal angles in CM is $75^{\circ} < \theta{}_\mathrm{CM}
< 105^{\circ}$ and $-6^{\circ} < \phi{}_\mathrm{CM} < 6^{\circ}$. The non-scattered
electron beam passes through a 4~cm diameter hole in a vertical steel
plate 6~cm thick, which is positioned at the central plane of the
dipole and provides a magnetic shielding for the beam area. The plate,
combined with the magnet's poles, make two 4~cm wide gaps, which
serve as two $\theta{}_\mathrm{CM}$ angle collimators for the
scattered electrons. Two additional lead collimators restrict the
$\phi{}_\mathrm{CM}$ angle range. The polarimeter can be used at beam
energies from 0.8 to 6~GeV, by setting the appropriate fields in the
magnets. The lower limit is defined by a drop of the acceptance at lower
energies, while the upper limit depends mainly on the magnetic shielding of the beam area
inside the dipole.

The detector consists of total absorption calorimeter modules, split
into two arms in order to detect two scattered electrons in
coincidence. There are two aperture plastic scintillator detectors at 
the face of the calorimeter. The beam helicity driven asymmetry of the coincidence
counting rate (typically about 10$^5$~Hz) is used to derive the beam
polarization.  Additionally to detecting the counting rates, about
300~Hz of ``minimum bias'' events containing the amplitudes and
timings of all the signals involved are recorded with a soft trigger
from one of the arms. These data are used for various checks and tuning, and
also for studying of the non--M{\o}ller background. The estimated
background level of the coincidence rate is below 1~\%.

\subsubsection{12~GeV Upgrade Status}
\label{subsec:goal}

The Hall A M\o{}ller polarimeter originally was designed for an electron beam energy range of  1~--~6 GeV.  Two factors limit the useful energy range of the polarimeter:
\begin{itemize}
  \item {} the spectrometer acceptance, defined by the positions of the magnets and 
         the available field strength, and also the positions and of the collimators;
  \item {} the beam deflection in the M\o{}ller dipole caused by the residual
         field in the shielding insertion.
\end{itemize}

In order to operate the polarimeter at 11~GeV a considerable upgrade of the polarimeter was required. 
In order to minimize the interference of such an upgrade with the rest of the beam line we did not 
consider moving the M\o{}ller target or the M\o{}ller dipole magnet and the M\o{}ller detector, as 
well as replacing the shielding insertion in the dipole magnet.  

A few items have to be considered for the higher energy polarimeter design:

\begin{enumerate}
  \item the positions and settings of the quadrupole magnets;
  \item the dipole magnet bending angle;
  \item the dipole shielding insertion;
  \item the detector position;
  \item the beam line downstream of the M\o{}ller dipole.
\end{enumerate}

\paragraph{Quadrupole magnets position} 
\label{sec:quads}

The acceptance of a M\o{}ller polarimeter  is defined as the accepted range of the scattering 
angles in CM, around 90$^{\circ}$. In Hall A polarimeter a collimator, consisting of two 
vertical slits between the poles of the dipole magnet and the shielding insertion in the dipole 
gap plays the most important role in limiting the acceptance. The goal of the quadrupole magnets is 
to direct the scattered electrons into the slits. With the old (6~GeV) design, two quadrupole magnets 
($PATSY$ and $FELICIA$ see Table~\ref{tab:magnets}) were used. 

GEANT simulation shows that for 11~GeV era power of two and even all three existing M\o{}ller 
quadrupole magnets is not enough. In order to cover the new beam energy range of 0.8~--~11~GeV we 
proposed to move the first quadrupole 40~cm downstream and to install the fourth
quadrupole with its center at 70~cm from the M\o{}ller target. 

\begin{figure}
   \begin{center}
        \includegraphics[width=0.6\linewidth]{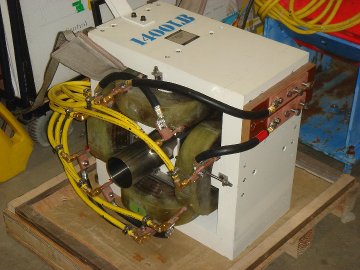}
   \end{center}
    \caption{ New quadrupole magnet for M\o{}ller polarimeter.}
\label{fig:new_quad}
\end{figure}

The new quadrupole magnet was designed by Robin Wines. The magnet is shown on Fig.~\ref{fig:new_quad}. 
 The new quadrupole has been field mapped by Ken Bagget ~\cite{bagget} before 
installation on the Hall A beam line and the results for the new magnet are presented in 
Table~\ref{tab:magnets}. A new bench was designed and manufactured to install the new quadrupole 
and to shift the first magnet ($PATSY$). A distance between the 
M\o{}ller target and the new quadrupole magnet center is 0.684~m. A distance between the M\o{}ller 
target and the first M\o{}ller quadrupole magnet is 1.334~m. The second and the third M\o{}ller 
quadrupole magnets position is unchanged.

Available $Danfysik$ power supply from Accelerator Division will be used to power the 
new M\o{}ller quadrupole magnet to save money. It is already installed and working, but the EPICS controls
have not been done. It is a work in progress.

\begin{table}[ht]
   \caption{Parameters of the M\o{}ller quadrupole magnets.}
\begin{center}
\begin{tabular}{|l|c|c|c|c|} \hline
 M\o{}ller notation            & Q0        &  Q1        &  Q2          &  Q3            \\  \hline
 MCC notation                  & $MQO1H01$ & $MQM1H02$  &  $MQO1H03$   & $MQO1H03A$     \\  \hline
 Name                          & $new$     & $PATSY$    & $TESSA$      & $FELICIA$      \\  \hline
 Bore, cm                      & 10.16     & 10.16      & 10.16        & 10.16          \\
 Effective length, cm          & 36.58     & 44.76      & 35.66        & 35.66          \\
 Maximum current, A            & 300       & 300        & 280          & 280            \\
 Pole tip field at 300~A, kGs  & 6.39      & 5.94       & 6.03         & 6.14           \\  \hline
\end{tabular}
\end{center}
\label{tab:magnets}
\end{table}

\paragraph{Dipole bending angle} 
\label{sec:b_angle}

The old M\o{}ller electrons bending angle in the dipole is 10$^{\circ}$. A dipole 
current of about 700~A and a field of about 19.2~kGs is needed to keep this bending angle at 
11~GeV. The maximal magnetic field measured in this dipole in Los Alamos was 17.5~kGs. 
The present dipole power supply provides the maximal current of 550~A. This current is not enough 
to provide for the beam bending angle in dipole of 10$^{\circ}$ at the beam energy 11~GeV.  
This limitation, along with the problem of shielding the beam area at high fields, described 
below, is mitigated by reducing the bending angle from 10$^{\circ}$ to 7.0$^{\circ}$. The smaller bending 
angle allows to keep the existing M\o{}ller dipole and its power supply. The reduction of the bending 
angle requires a new detector position, as it will be described below.

\paragraph{Dipole shielding insertion design}
\label{sec:shield} 

The dipole shielding insertion attenuates the strong dipole magnetic field in the region where 
the main electron beam passes through the dipole. It was designed for the dipole magnetic 
field up to 10~kGs. This field is enough to bend the M\o{}ller electrons to the M\o{}ller detector 
at a beam energy of 6~GeV. For a higher beam energy and a stronger magnetic field the shielding insertion 
becomes saturated leading to a strong residual field and a large deflection of the electron beam..

\begin{figure}[htb]
   \begin{center}
      \includegraphics[width=0.7\linewidth]{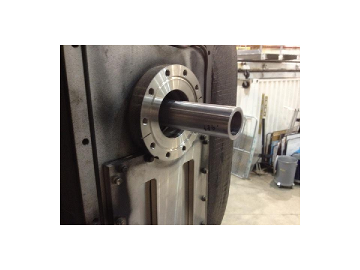}
   \end{center}
\caption{ M\o{}ller dipole assembly with additional shielding pipe in the shielding insertion. }
\label{fig:pipe_dipole}
\end{figure}

The diameter of the bore in the shielding insertion is 4.0~cm.  The diameter of the electron beam line 
before and after the M\o{}ller polarimeter is 2.54~cm. A coaxial magnetically isolated pipe, made of magnetic steel AISI-1010, was placed inside the bore (see Fig.~\ref{fig:pipe_dipole}) 
to increase the attenuation of the shielding insertion. The inner pipe diameter 
is 2.5~cm and the outer diameter is 3.4~cm. The shielding pipe consists of eight assembled together sections to reduce the cost. The shielding pipe is 
centered in the shielding insertion bore with  seven isolating rings made of a non-magnetic aluminum 
6061-T6. The total shielding pipe length is 202.4~cm. It is about 15~cm longer than the shielding insertion length in order to reduce the influence of the fringe field outside of the shielding 
insertion. 

The new design allows to attenuate to an acceptable level the dipole magnetic field up to 14.8~kGs. 
A field of 14.0~kGs (and power supply current 513~A) corresponds to the beam energy of 11~GeV and 
the dipole bending angle 7.0$^{\circ}$. This field can be provided with the existing power supply. 

\begin{figure}[htb]
   \begin{center}
      \includegraphics[width=0.6\linewidth]{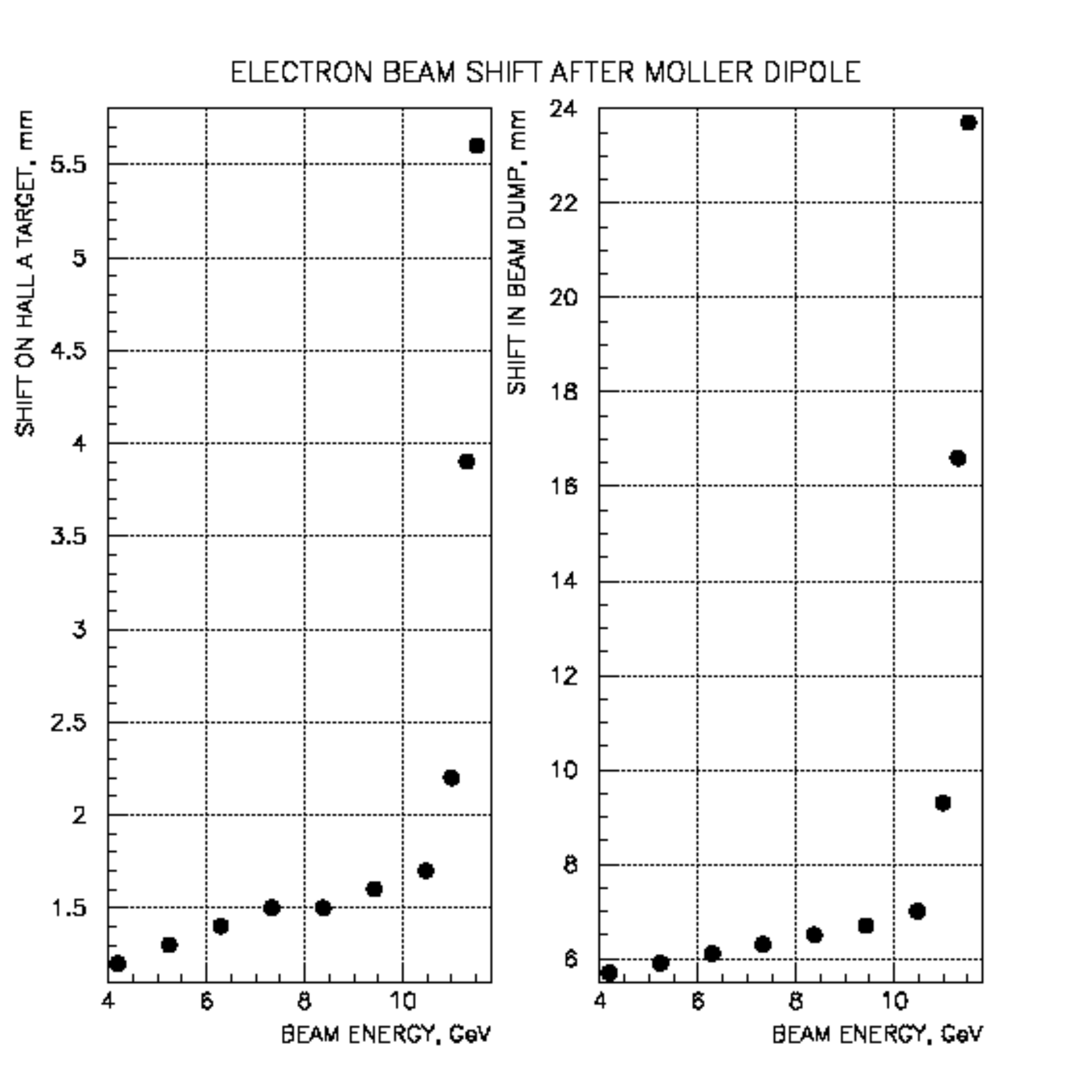}
   \end{center}
\caption{ TOSCA result for the M\o{}ller dipole with the 10~cm extended shielding pipe. 
The electron beam shift on the Hall A target (left picture) and in the Hall A beam dump 
(right picture). }
\label{fig:shift1}
\end{figure}

The TOSCA simulated fields in the dipole gap, in the shielding pipe and the expected electron beam 
shift on the Hall A target and in the beam dump are shown in Fig.~\ref{fig:shift1}. A new vertical corrector is installed downstream of the M\o{}ller dipole 
(see Sec.~\ref{sec:girder}) to compensate the beam shift at high beam energies.

\paragraph{Detector position and shielding } 
\label{sec:detect} 

\begin{figure}[htb]
   \begin{center}
      \includegraphics[width=0.6\linewidth]{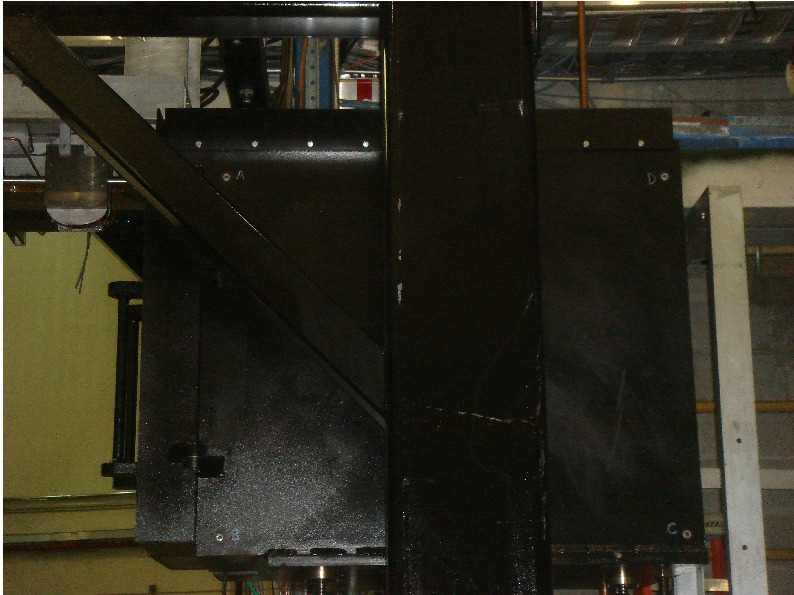}
   \end{center}
\caption{ A new M\o{}ller detector shielding box on the Hall A beam line. }
\label{fig:shield_box}
\end{figure}

Because of the smaller bending angle of the M\o{}ller electrons the detector has to lifted by 10cm.
The beam line downstream of the dipole also has to be modified.
Originally, it was planned to re-use the old detector shielding box with some modifications. 
It occurred that the design of the old box was in conflict with the design of a new beam line girder
downstream of the M{\o}ller detector.
A new shielding box was designed, manufactured and 
installed at the new position on the Hall A beam line (see Fig.~\ref{fig:shield_box}).

Before the upgrade the beam pipe diameter after the M\o{}ller dipole was 6.35~cm (2.5~inches). 
The beam pipe diameter over the detector shielding box was 10.16~cm (4~inches), and after that 
(girder area) 2.54~cm (1~inch). After the upgrade one 6.35~cm (2.5~inches) pipe is used between 
the M\o{}ller detector and the beam line girder.

Lead bricks on the top of the shielding box and and along the beam line downstream of the M\o{}ller 
dipole have been reassembled in accordance with the new beam line design.

\paragraph{New girder design downstream of the M\o{}ller dipole }
\label{sec:girder}

Precise knowledge of the beam position and angle on the M\o{}ller target is important for the
optimal beam tuning and for understanding of the systematic errors of the beam polarization measurements. 
The old beam line provided only three BPMs for the position/angle measurements:

\begin{itemize}
\item BPM $IPM1H01$ - in 1~m upstream of the M\o{}ller target;
\item BPM $IPM1H04A$ - upstream of the Hall A target (in 17~m downstream of the M\o{}ller target);
\item BPM $IPM1H04B$ - in the Hall A beam dump.
\end{itemize} 

There were three (at least two) M\o{}ller quadrupole magnets, M\o{}ller dipole, two quadrupole magnets downstream of the M\o{}ller detector and a few beam position correctors between BPM $IPM1H01$ and BPM $IPM1H04A$. Because of that precise information about the beam position 
and especially beam angle on the M\o{}ller target and good beam tuning was not available. 

\begin{figure}[htb]
   \begin{center}
      \includegraphics[width=0.6\linewidth]{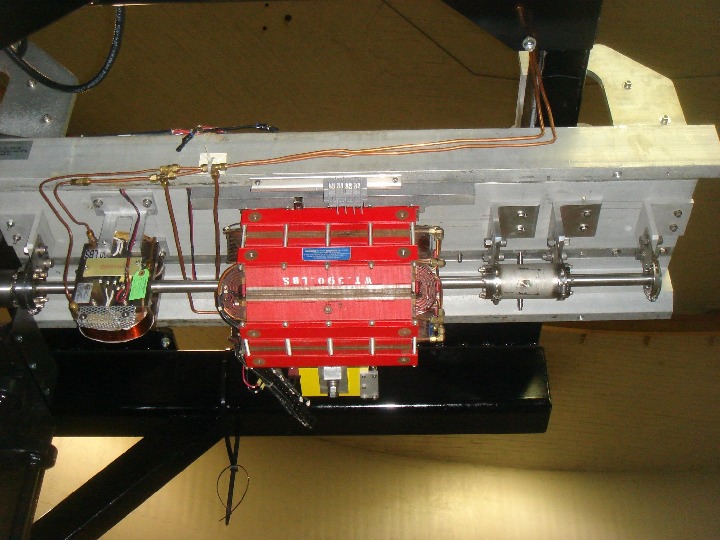}
   \end{center}
\caption{ A new girder downstream of the M\o{}ller detector shielding box on the Hall A beam line. 
From left to right: new vertical corrector $MBD1H04$, focusing quadrupole magnet $MQAH04$ and new 
beam position monitor $IPM1H04$. }
\label{fig:girder}
\end{figure}

In the new beam line design a new BPM (see Fig.~\ref{fig:girder}) is installed on the girder 
downstream of the M\o{}ller 
detector. The new BPM is located $\sim{}$7~m downstream of the M\o{}ller target. Centering of the beam with the M\o{}ller quadrupole magnets and dipole should provide correct beam tuning for 
the beam polarization measurement and precise information about the beam position and angle on the 
M\o{}ller target. 

At high energies the shielding insertion in the M\o{}ller dipole is saturated and the residual field deflects the beam 
down (see Fig.~\ref{fig:shift1}). A new vertical corrector 
(see Fig.~\ref{fig:girder}) was installed on the girder to compensate for this effect.

\subsubsection{M\o{}ller polarized electron targets}
\label{sec:targets}

Magnetized ferromagnetic materials are used to provide polarized electrons in the target.
The average electron polarization in such targets is about 7-8\%. It is not theoretically
calculable with an accuracy sufficient for polarimetry, and has to be somehow measured. The
uncertainty of this value is typically the dominant systematic error of the the M{\o}ller polarimetry. 
Two different techniques to magnetize ferromagnetic targets are used.. The first one  - the ``low field'' technique - 
uses a thin ferromagnetic foil tilted at a small angle to the beam and magnetized in the foil's plane by
a relatively weak magnetic field ($\sim{}$20~mT) directed along the beam.  
The second one - the ``high field'' technique - uses a thin ferromagnetic foil positioned
perpendicular to the beam and polarized perpendicular to its plane by a very strong field ($\sim{}$3~T).
Description and comparison of both types of the polarized 
electron targets can be found in ~\cite{cimento}.
The Hall A M\o{}ller polarimeter is a unique polarimeter which uses both this techniques.
This allows a better understanding of the systematic error associated with target polarization.

\paragraph{``Low field'' polarized electron target status}
\label{sec:old_target}

A detailed description of the ``low field'' target is done in ~\cite{report_2010}. The target 
was used with the Hall A M\o{}ller polarimeter in 2005~-~2009. The target consists of six foils,  
of Supermendure and iron with different thickness from 6.8~$\mu$m to 29.4~$\mu$m, fixed at an angle 
of 20.5$^\circ$ to the beam in the $YZ$ plane, magnetized by a $B_Z\sim{}0.035$~T field. 

\begin{figure}
   \begin{center}
        \includegraphics[width=0.6\linewidth]{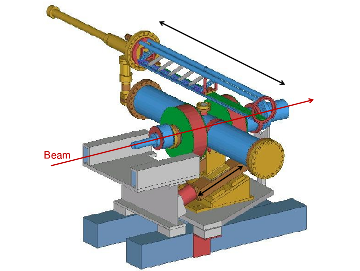}
   \end{center}
\caption{ The ``low field'' target holder design. The electron beam direction and directions of the
         target motion in two projections are shown.}
\label{fig:target}
\end{figure}

The target holder design is shown on Fig.~\ref{fig:target}. The holder can move the targets across 
the beam in two projections:
transversely - along $X$, and longitudinally - along the longer sides of the foils
(a line in the $YZ$-plane, at 20.5$^\circ$ to $Z$). The goal is to study the observed effects 
of non-uniformity of the target magnetic flux, measured by a small pickup coil  at different 
locations along the foil. Systematic error budget for the M\o{}ller polarimeter with the 
``low field'' polarized electron target is presented in Tab.~\ref{tab:comparison}

In the beginning of 2011 after PREX and DVCS experiments the ``low field'' target was restored back to 
the M\o{}ller polarimeter for the beam polarization measurements for g2p experiment. There were a few reasons to choose the ``low field'' target for g2p experiment:
\begin{itemize}
\item g2p experiment does not require high precision of the beam polarization measurement;
\item g2p experiment was running with very low beam current 0.1~$m$A. A maximal efficient thickness of the ``high field'' target is $\sim{}$10~$\mu$m. A maximal efficient thickness of the ``low field'' target is $\sim{}$90~$\mu$m. Thus, using of ``low field'' target allows to reduce essentially time required for 
the beam polarization measurement with the same statistical error;
\item operation of the ``low field'' target is cheaper because it does not require 
expensive cryogenics;
\item operation of the ``high field'' target at present requires a daily accesses 
to the Hall to feed the target superconducting magnet.
\end{itemize}

The ``low field'' target was successfully used during the running of the g2p experiment.
The ``low field'' target is installed on the 
Hall A beam line now and it will be used for the M\o{}ller polarimeter commissioning after the
11~GeV upgrade.

\paragraph{``High field'' polarized electron target status}
\label{sec:new_target}

Experiment PREX required a polarimeter accuracy of $\sim 1\%$. As it is shown 
in Tab.~\ref{tab:comparison}, the M\o{}ller polarimeter with the ``low field'' target can not meet the requirement. 
Instead, a new ``high field'' polarized electron target for the Hall A M\o{}ller was built. 
The ``high field'' technique {\cite{Steiner}} uses a strong 
magnetic field - larger than the magnetic field inside of the ferromagnetic domains.
The field should orient the magnetization in the domains along the field direction and drive the magnetization into saturation.
In the polarimeter, the magnetic field is parallel to the beam direction. 
The foil is perpendicular to the field, in order to minimize the effects of the
magnetization in the foil plane, and is magnetized perpendicular to its plane. 
The value of the magnetization (and of the average electron polarization) at saturation
depends only on the material properties, and for pure iron can be derived from
the existing world data {\cite{Scott:1969:pr184}}. 

\begin{table}[htb]
  \caption{Systematic errors for the Hall A M{\o}ller polarimeter with the ``low field'' and the ``high field'' polarized electron targets.}
   \begin{center}
     \begin{tabular}{|l|c|c|}
    \hline
    Variable  &  ``Low field''  &  ``High field'' \\
    \hline
    Target polarization  &  $1.5\%$  &  $0.35\%$ \\
    \hline
    Analyzing power  &  $0.3\%$  &  $0.3\%$ \\
    \hline
    Levchuk-effect  &  $0.2\%$  &  $0.3\%$ \\
    \hline
    Background  &  $0.3\%$  &  $0.3\%$ \\
    \hline
    Dead time  &  $0.3\%$  &  $0.3\%$ \\
    \hline
    High beam current  &  $0.2\%$  &  $0.2\%$ \\
    \hline
    Others  &  $0.5\%$  &  $0.5\%$ \\
    \hline
    Total  &  $1.7\%$  &  $0.9\%$ \\
    \hline
    \end{tabular}
   \end{center}
  \label{tab:comparison}
\end{table}

\begin{figure}[htb]
   \begin{center}
      \includegraphics[width=0.8\linewidth]{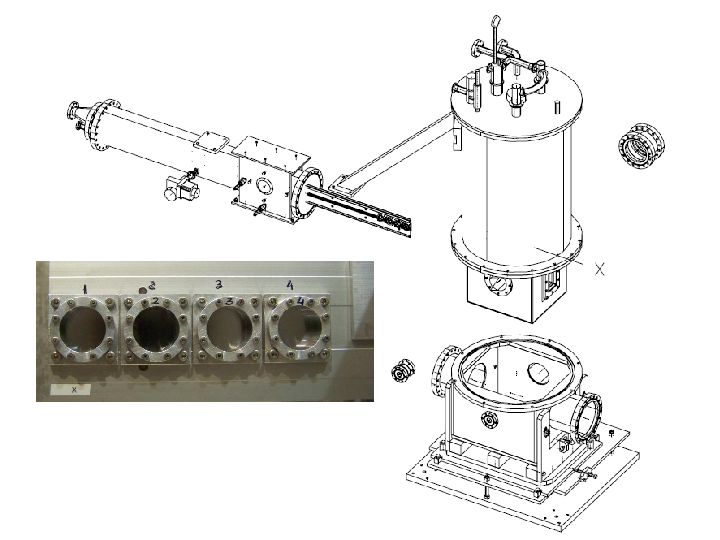}
   \end{center}
\caption{ Design of the ``high field'' polarized electron target. The target holder with four pure 
iron foils is shown on photo. }
\label{fig:brute_force_layout}
\end{figure}

Design of the ``high field'' polarized electron target is shown on Fig.~\ref{fig:brute_force_layout}. 
The target consists of:

\begin{itemize}
\item a superconducting magnet for a maximal magnetic field of 4~T. The magnet needs 
liquid He$^4$ at low pressure;
\item a target holder with a set of four iron foils with the purity of 99.85$\%$ and 99.99$\%$. The foils 
thicknesses are 1,4,4 and 10~$\mu$m to study possible sources of systematic errors (see 
Fig.~\ref{fig:brute_force_layout});
\item a mechanism of target foils orientation along the magnetized field;
\item a mechanism for targets motion into the beam;
\item a mechanism of the magnetic field orientation along the beam.
\end{itemize}

The ``high field'' target was used in 2010 for the beam polarization measurements during the PREX and DVCS experiments running. As it is seen from Tab.~\ref{tab:comparison} using of the 
``high field'' target allows to increase the accuracy of the beam polarization measurements by a factor of two. 
It should be noted that a successful operation of the ``high field'' requires considerable efforts:

\begin{itemize}
\item improvements of the target foils and magnetized field alignment;
\item gaining the target operation experience;
\item a systematic error study;
\item building a supply line for liquid He$^4$.
\end{itemize}

\subsubsection{M\o{}ller polarimeter DAQs}
\label{sec:daq}

The Hall A M\o{}ller polarimeter has two DAQs:

\begin{itemize}
\item old DAQ based on combination of CAMAC and VME modules;
\item new DAQ based on FADC.
\end{itemize}

The old DAQ is fully operational with both polarized electron targets, well understood but 
slow, occupies a few crates and uses a few hundred cables to connect modules etc. New DAQ based on 
FADC is fast, generates two two types of triggers, compact, but not fully operational yet. Running of two different DAQs in parallel and comparison of the results gives a unique opportunity to study possible sources of systematic errors.

\paragraph{Old M\o{}ller DAQ upgrade status}
\label{sec:old_daq}

Present DAQ for the Moller polarimeter detector has been designed in the mid-90th. It uses a lot of 
 slow modules not available in stock anymore. The main goals of the electronics upgrade for the 
Moller polarimeter are: 

\begin{itemize}
\item to increase bandwidth (up to 200~MHz) of the detector system;
\item to reduce readout time from ADC and TDC modules;
\item to replace the old PLU module LeCroy-2365 that is not available in stock anymore.
\end{itemize}

The list of modules to be replaced:

\begin{itemize}
\item to increase bandwidth:
\begin{itemize}
\item PLU module LeCroy-2365, bandwidth $<75$~MHz, CAMAC replaced with PLU module based on 
CAEN V1495 board (bandwidth 200~MHz, VME);
\item Discriminator   Ortec-TD8000, input rate $<150$~MHz, CAMAC replaced with P/S 706 
(300~MHz, NIM), modified for remote threshold setup with DAC type of VMIC4140;
\end{itemize}
\item to reduce readout time:
\begin{itemize}
\item ADC LeCroy 2249A, 12 channels, CAMAC replaced with QDC CAEN V792 (32 channels, VME);
\item TDC LeCroy 2229, CAMAC replaced with TDC V1190B (64 channels, 0.1~ns, VME).
\end{itemize}
\end{itemize}

\begin{figure}[htb]
   \begin{center}
      \includegraphics[width=0.6\linewidth]{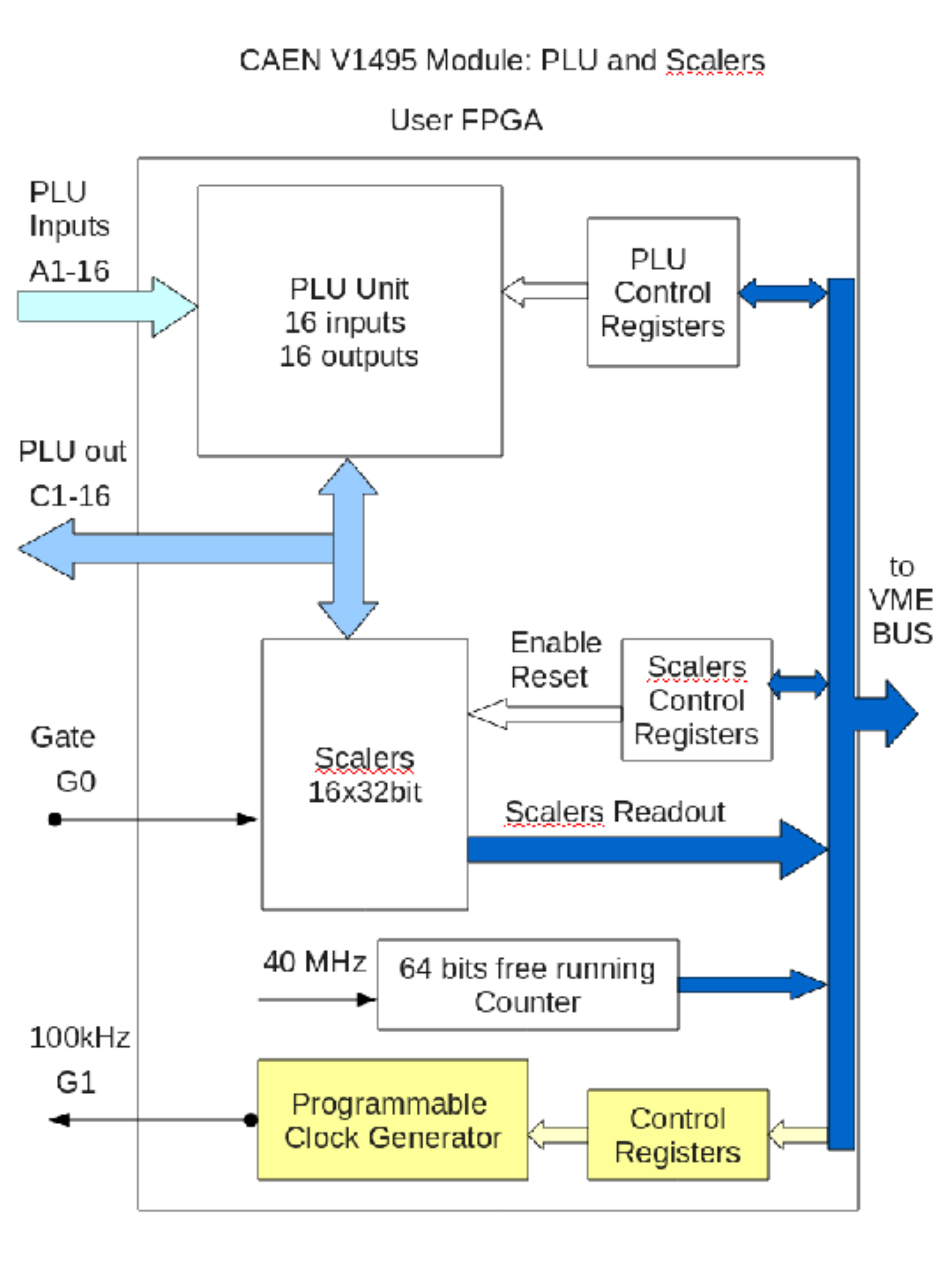}
   \end{center}
\caption{ PLU diagram for CAEN V1495 module. }
\label{fig:plu_scheme}
\end{figure} 

Diagram for new PLU unit based on CAEN V1495 module is shown on Fig.~\ref{fig:plu_scheme}. 
The module CAEN V1495 has the following parameters:

\begin{itemize}
\item Input bandwidth $\sim{}$200~MHz;
\item 2 input ports x32 bits;
\item 1 output port x32 bits;
\item 2 input/output front LEMO connectors;
\item The FPGA ``User'' can be reprogrammed by the user using custom logic functions.
\end{itemize}
 
Firmware for the PLU module is under development and will consists of the following units:

\begin{itemize}
\item Programmable Logical Unit (PLU): 16 inputs, 16 outputs;
\item Scalers unit: 16 channels, 32 bit, gate input, connected to PLU outputs; 
\item Free running 64 bit timer with base frequency 40~MHz.
\end{itemize}

All the modules required for the upgrade have been procured. The work is in progress. We plan to use the old DAQ after the upgrade at least until the new DAQ based on flash-ADC will be fully 
operational with both the low and high field targets (see details below in 
Sec.~\ref{sec:fadc}). Also, running of two different DAQs in parallel provides a unique opportunity 
to study systematic errors.

\paragraph{Status of FADC DAQ for the Moller detector}
\label{sec:fadc}

\begin{figure}[htb]
   \begin{center}
      \includegraphics[width=0.6\linewidth]{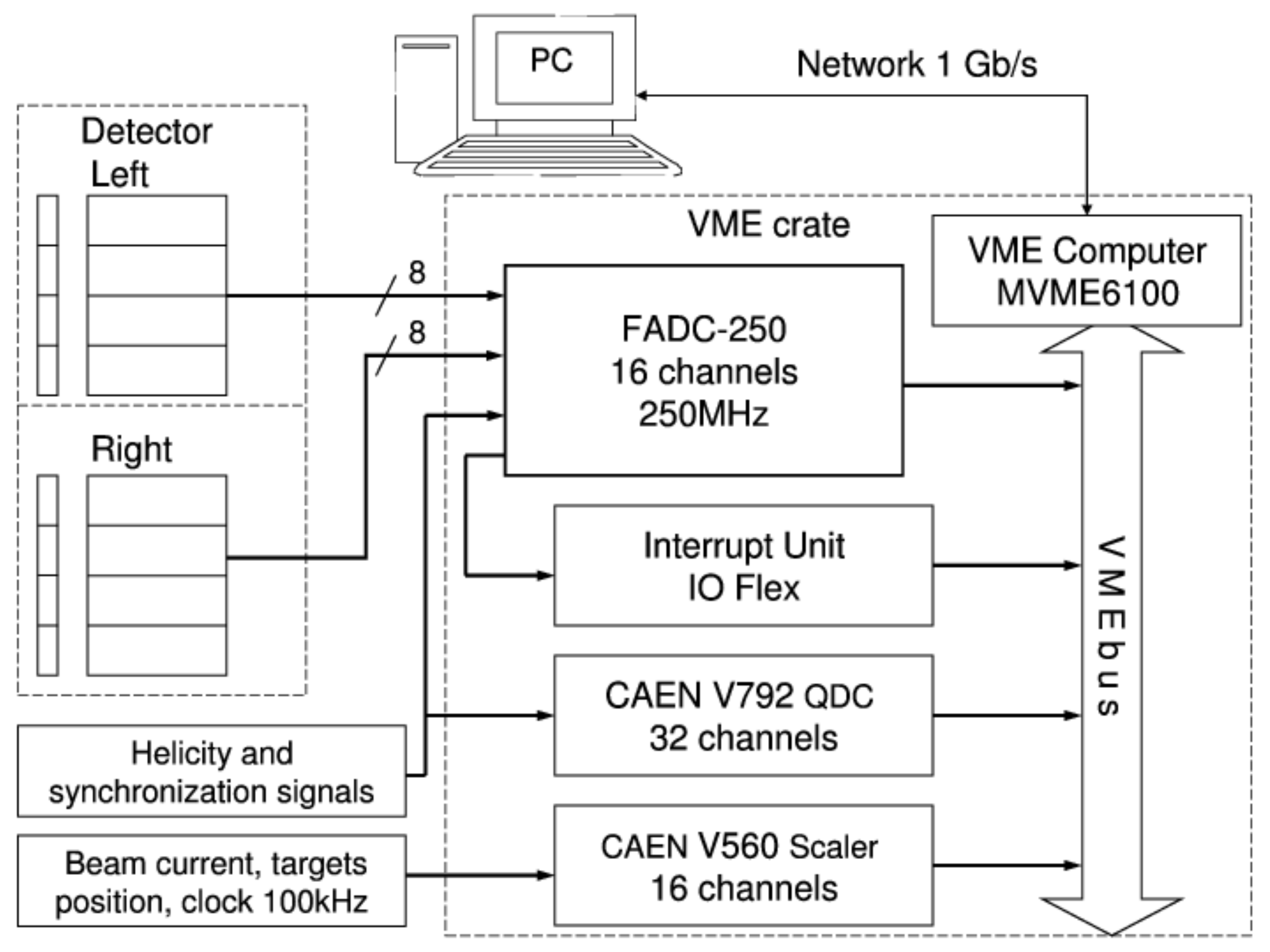}
   \end{center}
\caption{ Scheme of a new M\o{}ller DAQ based on FADC. }
\label{fig:new_daq}
\end{figure}

A new DAQ based on the JLab-built FADC was created in 2009 for PREX experiment to be operated with the new high field polarized electron target (see ~\cite{an_report}, ~\cite{review}). 
The schematics of the new DAQ is shown on Fig.~\ref{fig:new_daq}. 

There are some 
 differences between the old and the new DAQ due to differences between the low field and the 
high field targets operation. For the low field target, the target polarization is a function of 
the particular foil, the foil coordinate and the magnetic field of the magnetized Helmholtz coils. The direction of the magnetic field is flipped every run to reduce the systematic error.
For each run the old DAQ with analyzer is doing the following: 

\begin{itemize}
\item ramps up the current in the Helmholtz coils;
\item reads out the value of the current;
\item starts the data taking when the field is established;
\item reads out of the foil number and the coordinate of the foil on the beam line;
\item reads out of the beam position;
\item turns off the current in the Helmholtz coils when the required number of events has been acquired;
\item calculates the foil polarization for the particular place of the foil and for the particular magnetizing field and the field direction;
\item uses the calculated foil polarization for the beam polarization calculation.
\end{itemize}

There are two versions of analyzers to run the old DAQ with the low field and the high field targets.

As it was mentioned above, FADC was built to run with the high field target. For this 
configuration the target is fully saturated and the target polarization is a constant for any foil, foil coordinate and magnetic field. Magnetizing field in the superconducting magnet is turned on 
in the beginning of the beam polarization measurement and turned off in the end of the measurements. The FADC DAQ does not perform some functions needed for the low field running
and, therefore, for the low field running the old DAQ is mandatory while the new DAQ is optional.

The DAQ based on the FADC generates two types of triggers:

\begin{enumerate}
\item helicity flipping triggers (integral mode / scalers);
\item data triggers (single events).
\end{enumerate}

\begin{figure}[htb]
   \begin{center}
      \includegraphics[width=0.8\linewidth]{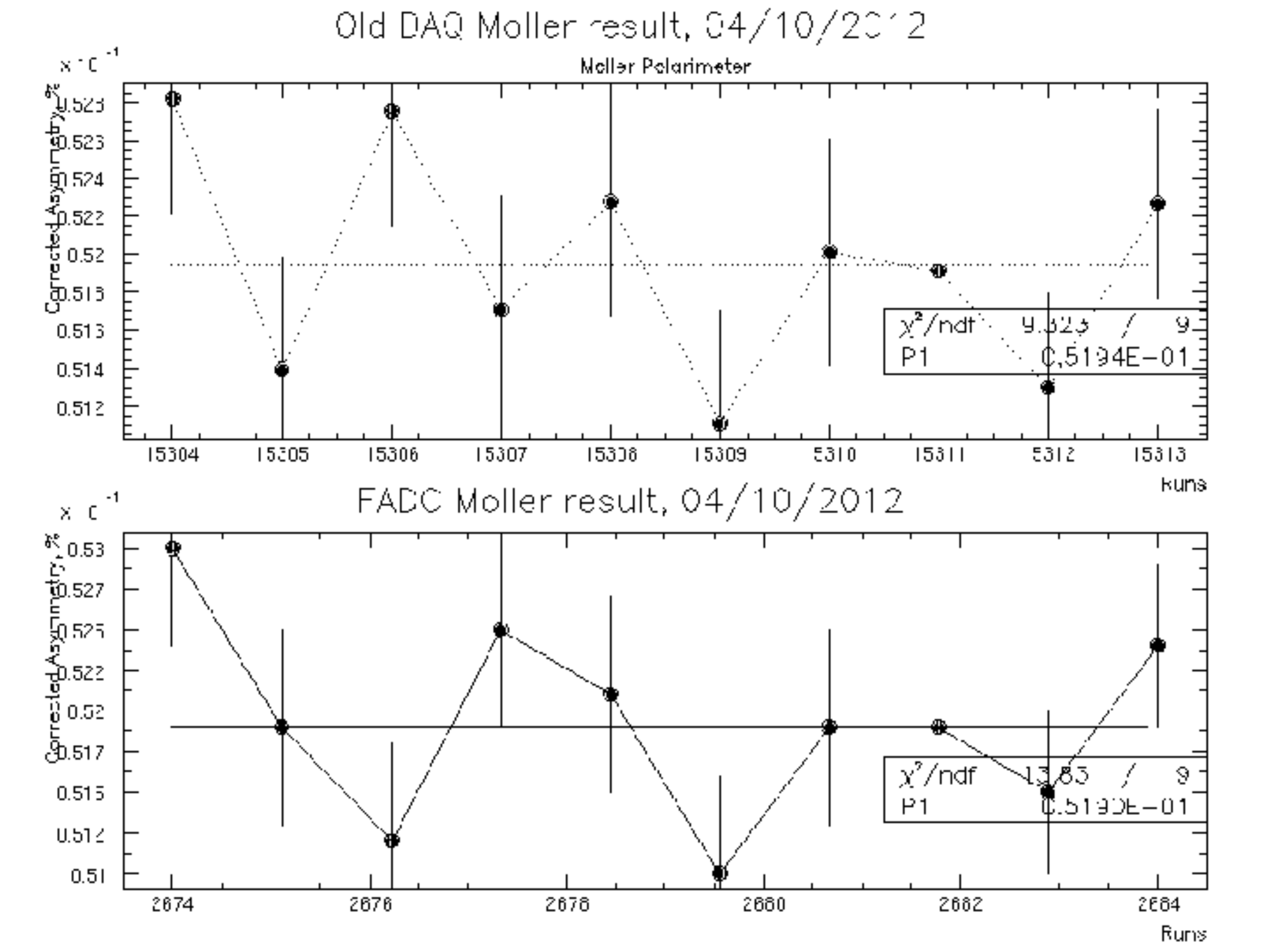}
   \end{center}
\caption{ Comparison of results of the beam polarization measurements with the old and the new 
M\o{}ller DAQs in the scaler mode. }
\label{fig:compar}
\end{figure}

There is a good agreement between the old and the new DAQs in scalers mode (see Fig.~\ref{fig:compar}.)

Running of the FADC in the data trigger mode is important to study the systematic errors. The triggers data should help to:

\begin{itemize}
\item improve the GEANT model of the polarimeter;
\item increase the accuracy of the evaluation of the average analyzing power;
\item study the Levchuk-effect.
\end{itemize}

At the moment, the work on the event data analysis is in progress.

\subsubsection{Summary}
\label{sec:summary}

The beamline part of the M\o{}ller polarimeter 11~GeV upgrade is completed. The polarimeter can be operated in the beam energy range of 0.8~-~11.0~GeV. The M\o{}ller polarimeter
 is ready for commissioning with the beam. The remaining work includes modifications and checkout of the DAQ system, the 
``high field'' target, a cryogenics line to feed the ``high field'' magnet, and 
the documentation for the M\o{}ller polarimeter operations after upgrade.

\clearpage
\subsection{Compton Polarimeter}

\begin{center}
\bf The Compton Polarimeter Upgrade
\end{center}

\begin{center}
contributed by Sirish~Nanda.
\end{center}
\subsubsection{Overview}
\begin{figure}[htp]
\begin{center}
\includegraphics[width=4in]{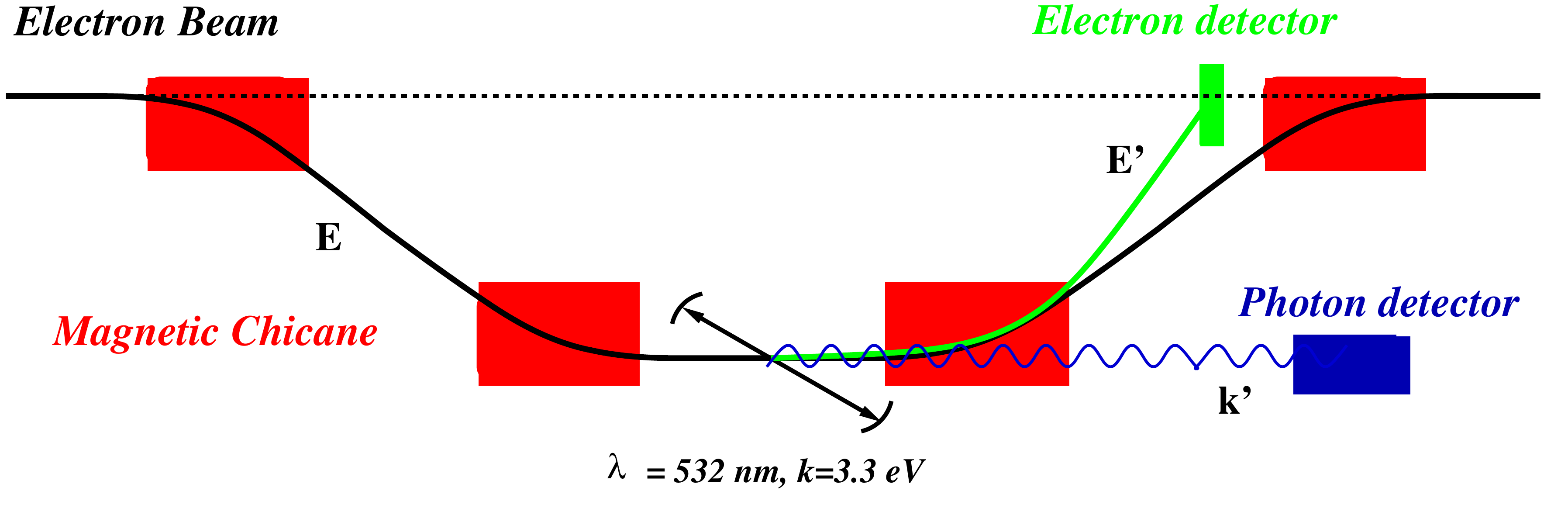}
\caption{\it Schematic layout of the Hall A Compton polarimeter.}
\label{fig:cpt_sch_setup}
\end{center}
\end{figure}

The Hall A Compton
Polarimeter provides  electron beam polarization measurements in a continuous and non-intrusive manner
using  Compton scattering of polarized electrons from polarized photons. A schematic layout of the
Compton polarimeter is shown in Fig.\ref{fig:cpt_sch_setup}. The electron beam is 
transported through a vertical magnetic chicane consisting of four dipole magnets. A
high-finesse Fabry-Perot (FP) cavity located at the lower straight section of the chicane with the cavity axis at  an angle of 24~mrad with respect to  the electron beam,  serves as the photon target.  The electron beam interacts with the photons trapped in the FP cavity at the Compton Interaction Point (CIP) located at the center of the cavity.
The Compton back-scattered photons are detected in an electromagnetic
calorimeter. The recoil electrons, dispersed  from the 
primary beam by the third dipole of the chicane are detected in a silicon micro-strip
detector. The electron beam polarization is deduced from the counting rate asymmetries of 
the detected particles. The electron and the photon arms provide redundant measurement of the electron beam polarization.

In the recent years the Compton polarimeter has undergone a major upgrade\cite{cpt_upgrade} to green optics, in order to improve accuracy of polarimetry  for  high precision parity violating experiments at lower energies such as PReX\cite{cpt_prex}. The conceptual design of the green  upgrade  utilizes much of the 
the existing infrastructure of the present Compton polarimeter. The original Saclay built 1064 nm FP cavity has been replaced with  a high power 532~nm system. In addition, the 
electron detector, photon calorimeter, and data acquisition system  have been upgraded 
to achieve beam polarimetry accuracy of 1\% at ~1~GeV beam energy. 
The new systems have been  operating  successfully in Hall A beam line with about 3 kW of cavity power for the past two years. Electron beam polarimetry was carried out  successfully during the PReX experiment with the upgraded polarimeter. Preliminary results indicate ~1.5\% accuracy in  the electron beam polarization has been achieved. Recently, the cavity power has been boosted to over 10kW with new low loss mirrors. The higher power cavity was successfully commissioned with beam during the g2p experiment in  2012.

Additionally, as part of the CEBAF 12~GeV upgrade, the Hall A Compton polarimeter is being  upgraded\cite{cpt_12gev_upgrade}  to accommodate 11~GeV beam envisioned for Hall A. The construction of the 12 GeV upgrade is proceeding well with  expected completion in 2013 and commissioning with beam in early 2014. At higher energies in the 12~GeV era, Compton polarimetry with 1064~nm infrared light has sufficient analyzing power to be  an attractive option that provides  higher photon density with less complications. Preliminary investigation into a high power infrared cavity has been started at the Compton polarimetry laboratory. Similarly, development of a high speed data acquisition system to keep pace with the higher luminosities of the upgraded polarimeter has been started as well. Discussions on designs for new electron and photon detectors optimized for the 12 GeV era, are in preliminary stages. 
\subsubsection{Fabry-Perot Cavity}
\begin{figure}[hpt]
\begin{center}
\includegraphics[width=3.3in,angle=-90]{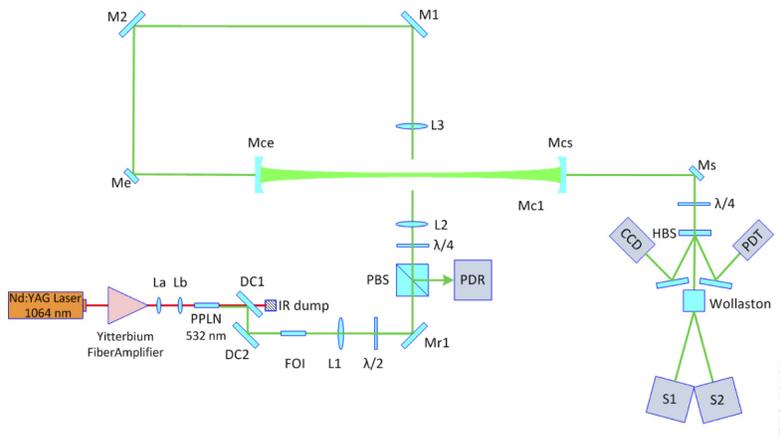}
\caption{Optical setup of the green Compton polarimeter.}
\label{fig:cpt_optics}
\end{center}
\end{figure}

The newly installed optics  replaces the original infrared cavity with a  high gain
532~nm green cavity capable of delivering 3~kW of intra-cavity power. Recent advances in the manufacturing of high reflectivity and low loss dielectric mirrors as well as availability of narrow line width green lasers facilitates the feasibility of our challenging design goal.   
High gain cavities at 532~nm have been successfully constructed by 
the PVLAS\cite{cpt_bregant} group with geometry and gain comparable to our proposed design. A schematic layout of the optical setup for the upgrade is shown in Fig.~\ref{fig:cpt_optics}.  

Our  solution for the green laser system begins with a narrow line  CW fiber coupled  Nd:YAG seed laser operating at 1064~nm (Innolight Mephisto S \cite{cpt_mephisto}).  The beam from the seed laser is then amplified by a Ytterbium doped fiber amplifier (IPG Photonics\cite{cpt_ipg}) which can produce up to 10~W of CW beam while maintaining the  line-width and the tunability of the seed laser. The amplified infrared beam is then shaped with lenses L$_a$ and L$_b$ to pump a Periodically Poled Lithium Niobate (PPLN) crystal supplied by HC Photonics \cite{cpt_hcp}.  The PPLN crystal is placed  in a temperature controlled housing equipped with a thermo-electric heat pump to maintain the temperature of the PPLN crystal at about 60$^o$ with better than .05$^o$ regulation. This temperature corresponds to the quasi-phase matching condition for the PPLN necessary to generate  the second harmonic of the pump beam at 532~nm. Typically about 2~W of 532~nm beam is generated with about 5~W of 1064~nm pump beam.

The green beam from the PPLN laser is then separated from the infrared pump with a pair of dichroic mirrors (DC1 and DC2), and transported through polarization conditioning optics and mode-matching lenses (L$_1$ - L$_3$) to produce circularly polarized light with the same Gaussian beam profile as the TEM$_{00}$ mode of the FP cavity. The beam is then injected to the 850~mm long cavity using conventional beam steering optics   to properly couple the beam to the cavity. The cavity, constructed out of  Invar, has dielectric mirrors mounted on adjustable gimbaled mounts with special ports for the transport of the electron beam. The structure, held in ultra-high vacuum,  is part of the electron beam line in Hall A. 

Part of the laser beam reflected from the cavity is steered by a polarizing beam splitter to a photo-diode receiver PDR. The PDR signal is used to lock the cavity on resonance using the well known Pound-Drever-Hall locking scheme. The part of the laser beam transmitted through the cavity is converted back to linearly polarized light and   analyzed in  a Wollaston polarimeter. The intensities of the analyzed horizontal and vertical polarization components of the beam  are measured in integrating spheres S$_1$ and S$_2$. In addition, a small part of the transmitted  beam, separated  with a holographic beam splitter, is used for beam monitoring instruments.

The  green laser systems and FP cavity have been in development in the Compton Lab for the past few years with participation from many graduate students from collaborating institutions. The development work was successful in late 2009 with stable lock acquisition with dielectric mirrors supplied by Advanced Thin Films\cite{cpt_atf}  (ATF) and homemade locking electronics. The system was then  installed and commissioned  in the Hall A beam line in early 2010 in preparation for the PReX experiment. During the commissioning,  calibration of the laser beam power and polarization transfer functions were carried out in order to accurately determine the power and polarization of the light trapped inside the cavity. 

\begin{figure}[htp]
\begin{center}
\includegraphics[width=4.2in]{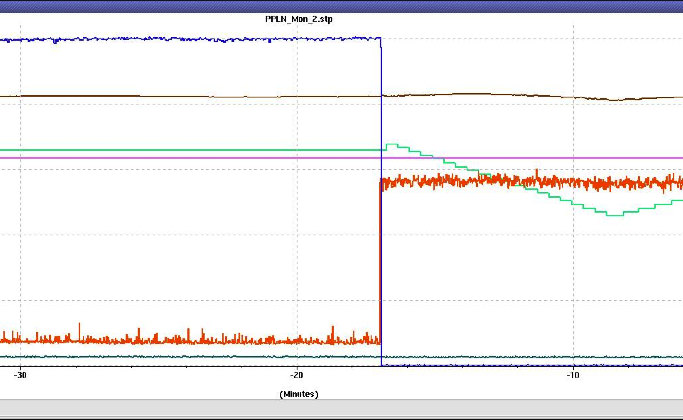}
\caption{Stable lock acquisition with 10~kW intra-cavity power in the green cavity. The  The blue line shows the power in the cavity making a transition from 10~kW locked state to zero in unlocked state.  The orange line show the corresponding power reflected by the cavity. }
\label{fig:cpt_green_10kW}
\end{center}
\end{figure}

As reported in last year's annual report, this cavity power saw a major boost in power in 2011 in preparation for the g2p experiment with 1$\mu$A beam current. In early 2012, taking advantage of schedule delays of the experiment,  the green cavity was dismantled again and  the cavity mirrors were changed to new set of mirrors supplied by ATF.  During this down time, the PPLN setup was realigned to restore its conversion efficiency. The power and polarization transfer functions were measured again. With the entire system retuned, lock was acquired in the  cavity with the new mirrors at 10~kW, far exceeding our expectation. Shown in Fig.~\ref{fig:cpt_green_10kW} are strip-charts of various cavity parameters as a function of time during this lock acquisition. The  blue line shows the power in the cavity whereas the orange line shows the power reflected by the cavity, both making a sharp transitions upon lock acquisition. For the 10~kW performance, the infrared  laser was set at 5 mW to seed the fiber amplifier which produced 4~W of 1064~nm light to pump the the PPLN subsystem resulting in 1~W of 532~nm green light. With 20\% transport losses, only 0.8~W of the green light was injected into the cavity. The cavity gain was about 1.2$\times$10${^4}$ resulting  in intra-cavity power of 10~kW. 

Following the successful running of the FP cavity during the g2p experiment, at the beginning of the long shutdown of CEBAF in June 2012, the cavity performance was further improved before preparing the optics table for the 12~GeV upgrade. Optics data for the final setup were recorded.
Preliminary results have indicated that more power is possible before instabilities set in. These data are under analysis.

Test setup in the Compton Lab is being established in order to continue   further  development work on laser systems and FP cavities. In particular, the optical setup in the Compton lab is being modified to handle both 1064~nm infrared and 532~nm  green beams with minimal setup changes. In the 12GeV era of CEBAF, the infrared system becomes competitive since the cavity power in the infrared is generally significantly higher than that in the green, while the  analyzing power remains adequate. The old Saclay cavity has been set up in the Compton Lab to provide a platform cavity development.  
The cavity ends have been modified with new mirror mounts to accommodate the smaller ATF mirrors while providing adjustment  of their angular alignment. 
With the new mechanism 
the mirrors can be manually aligned with respect to the cavity axis prior to establishing vacuum in the cavity. Preliminary results indicate the alignment concept works well. Successful establishment of fundamental TEM$_{00}$ resonant mode under vacuum was achieved in 2012. Locking exercises  are yet to be performed.

\subsubsection{ Detectors }
 
\begin{figure}[htp]
\begin{center}
\includegraphics[width=4.0in]{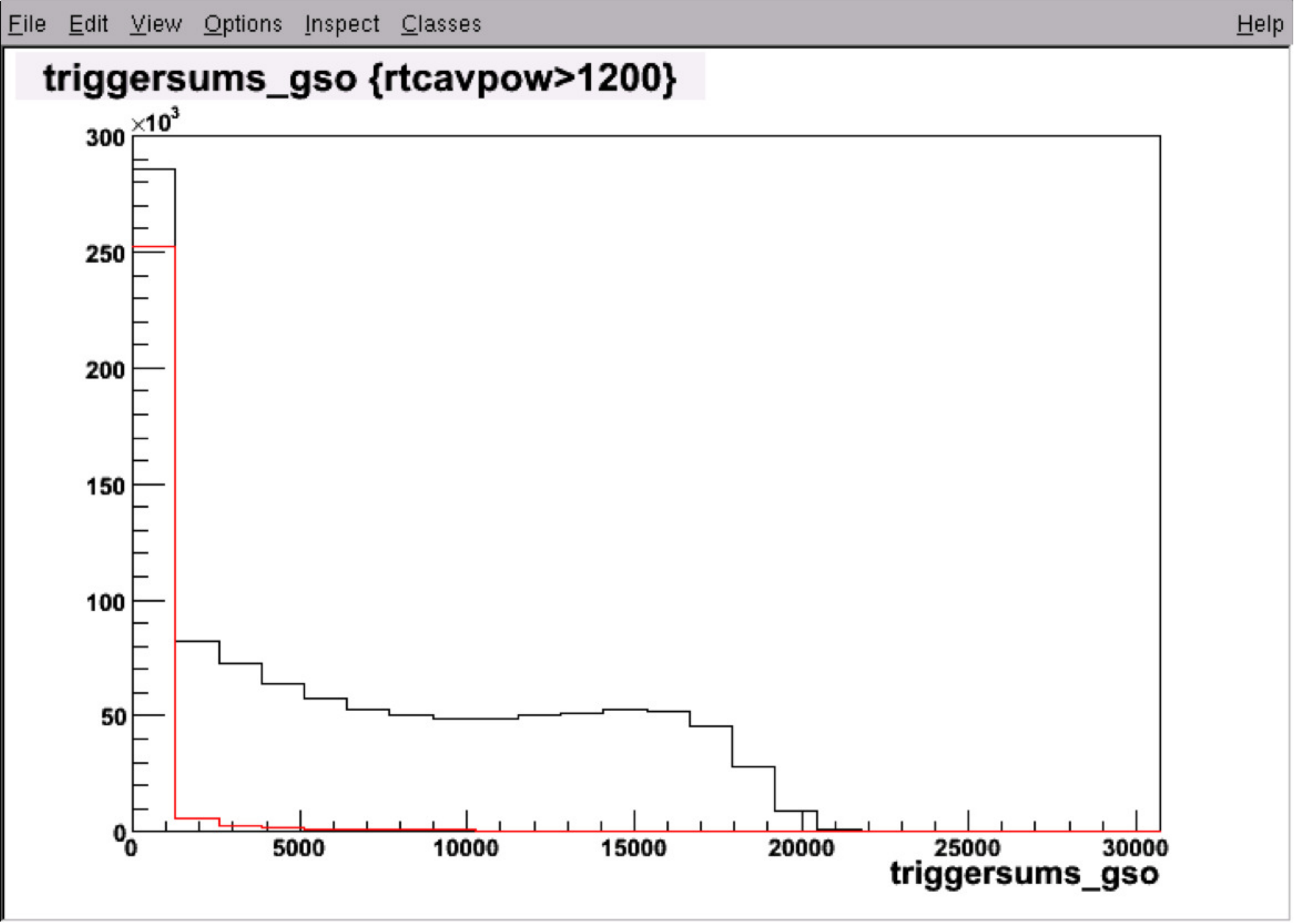}
\caption{ The first Compton scattering spectrum obtained with the GSO calorimeter  with the 10~kW cavity. The black solid line, obtained with laser on, is the counting rate as a function of photon energy showing a sharp Compton edge corresponding to back scattered photons. The solid red line is with laser off showing the background rate. }
\label{fig:cpt-gso-10kW}
\end{center}
\end{figure}

The  photon detector, a Carnegie Mellon University (CMU) responsibility, consists of a single GSO crystal, 60~mm in diameter and 150~mm in length, fabricated  by Hitachi Chemicals Ltd. This detector was removed from Hall A setup for tests in Hall C in 2011. It was successfully reinstalled in Hall A in early 2012 with help from M. Friend and A. Camsonne for the g2p test run. As illustrated in  Fig.~\ref{fig:cpt-gso-10kW}  during the g2p experiment, the GSO calorimeter obtained a high quality Compton scattering spectrum with the 10~kW cavity. Better than 100 signal-to-background ratio was obtained  with minimal effort in beam tuning. Observed counting rates were in accordance with a 10~kW of stored photon power confirming the power in the cavity previously calculated with optical measurements.

For higher photon energies in the 12~GeV era, a single crystal PbWO$_4$ calorimeter is under study. With high density of 8.3 g/cc, PbWO$_4$ offering  0.9~cm radiation length  and 2.2~cm Moliere radius, could be  an ideal solution for a compact calorimeter.  GEANT simulation are being carried out by Franklin {\it et al.} at CMU to study applicability of this material for the 12GeV Compton polarimeter and determine optimum geometry. Discussions are being held with the Shanghai Institute of Ceramics\cite{cpt_siccas} for the feasibility of  fabricating large diameter PbWO$_4$ cylindrical single crystal. 

The present data acquisition system has the capability to count photons and electrons up to 100~kHz rate at 30~Hz electron beam spin-flip rate. With the higher luminosity of the new cavity and higher spin-flip rates offered by the CEBAF polarized electron source, higher counting rate capabilities are required. This  long standing necessity to upgrade the counting data acquisition system to 1~MHz counting rate at 1~kHz spin flip rate has been under taken by Michaels {\it et al.}. Recent bench tests of a prototype system with a pulser signal show that more than 1~MHz rate will be achievable.

The electron detector supplied  by
Laboratoire de Physique Corpusculaire IN2P3 Universite Blaise Pascal, Clermont-Ferrand has 4 planes of 192 silicon micro-strip of 0.5~mm thickness with 240  $\mu m$ pitch. 
The expected resolution is about 100 $\mu m$.  A high precision vertical motion of 120 mm for the detector has been incorporated to the design so as to facilitate covering the recoil electrons 
corresponding to the Compton edge over a broad range of energies. 
The  electron detector along with its associated   mechanical  structures and electronics were installed 
Although  Compton scattering spectra and asymmetry were successfully obtained with 3~GeV electron beam and the old 1064~nm FP cavity. However, the detection efficiency of the micro-strips was found to be unacceptable,  at about 10-20\%,  due to poor signal-to-noise ratio. The detector was removed from the beam-line and sent back to Clermont-Ferrand  to study the signal-to-noise characteristics of the micro-strips with a cosmic tray test setup.  A 1~mm thick Si micro-strip detector was procured  from Canberra systems to study its signal compared to the 0.5~mm micro-strips in the cosmic tests. 

\begin{figure}[htp]
\begin{center}
\includegraphics[width=4.2in]{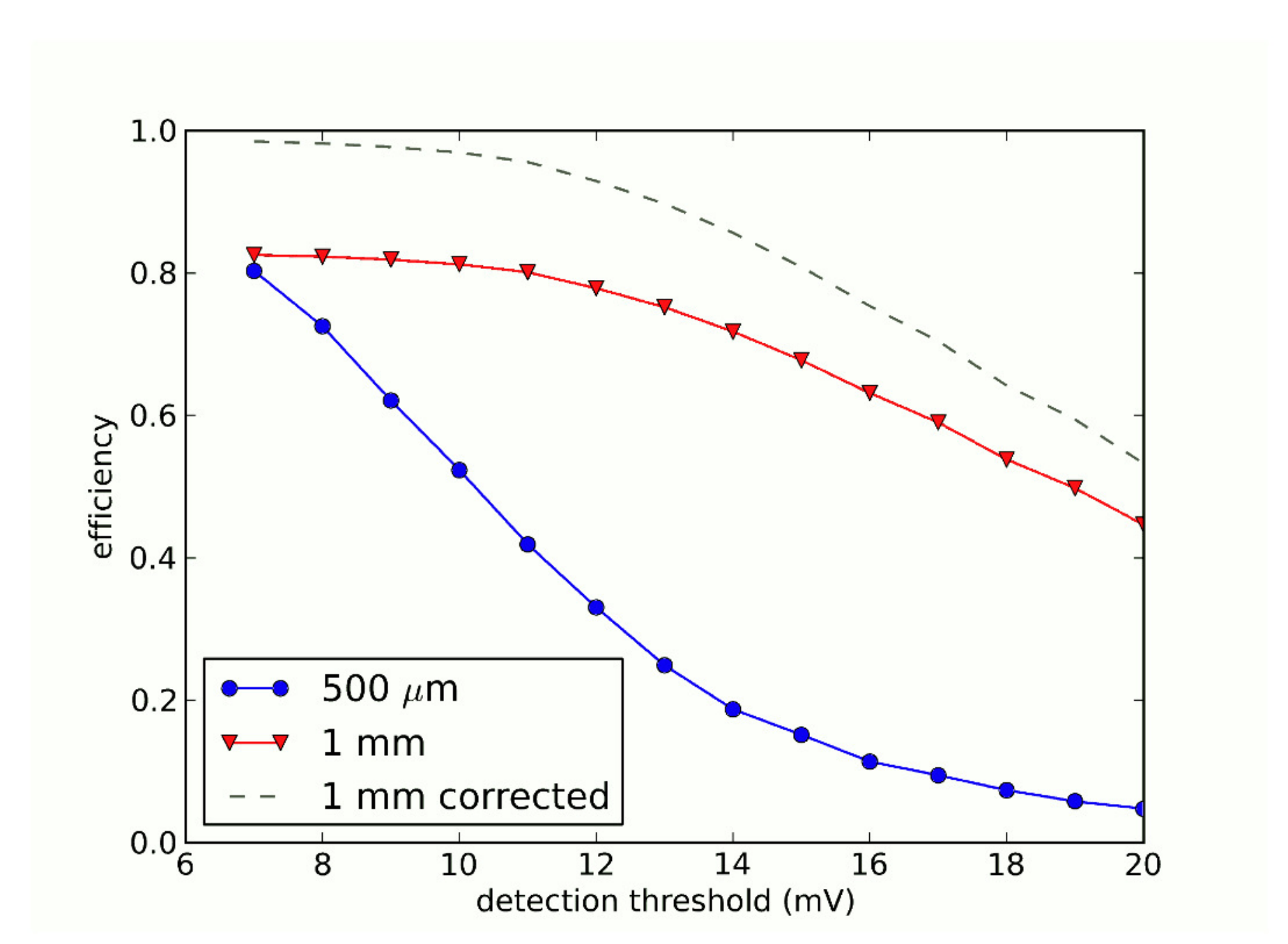}
\caption{ Threshold scan of detection efficiency for the 0.5~mm thick (blue) and 1 mm thick (red)Si  micro-strip detectors with cosmic rays. The solid red curve is for the 1 mm Si strips where as the solid blue is for 0.5~mm thick Si. The dashed curve is for the actual efficiency of 1~mm Si corrected for mis-positioning in the experiment }
\label{fig:cpt-edet-cosmic-ths}
\end{center}
\end{figure}

The Clermont-Ferrand team Joly {\it et al.}\cite{cpt_edet_cosmic_test} concluded their cosmic studies in early 2012.  As expected the thicker Si yielded twice the signal of the thinner ones. In principle, either detector should perform with good efficiency  if  noise levels, electronics or  electron beam related,  in Hall A environment is less than 7~mV. With higher noise levels where one has to set the detection thresholds higher, as illustrated in Fig.~\ref{fig:cpt-edet-cosmic-ths}, the thin strips lose efficiency rapidly with increasing thresholds while the 1~mm thick detector  retain higher efficiency till about 14 mV. The electron detector was re-installed in Hall A in early 2012 for the g2p test run. However, we ran out of beam time before conclusive beam tests could be made. During the long shutdown of CEBAF, cosmic tests are planned in Hall A. Furthermore, discussions are under way with University of Idaho for possible beam tests with a low energy electron machine at Idaho.  

Ideally, a redesign of the front end electronics is necessary to yield acceptable signal-to-noise ratios.  With declining support  from Clermont-Ferrand University for the electron detector, a new front end is unlikely. Discussions are under way with University of Manitoba and Mississippi State University to address the electron detector issues in the 12 GeV era. In the interim, the thick silicon micro-strips remain as the viable option for the electron detector subject to successful beam tests.

\subsubsection{The 12 GeV Upgrade}
\begin{figure}[htp]
\begin{center}
\includegraphics[width=5.2in,angle=0]{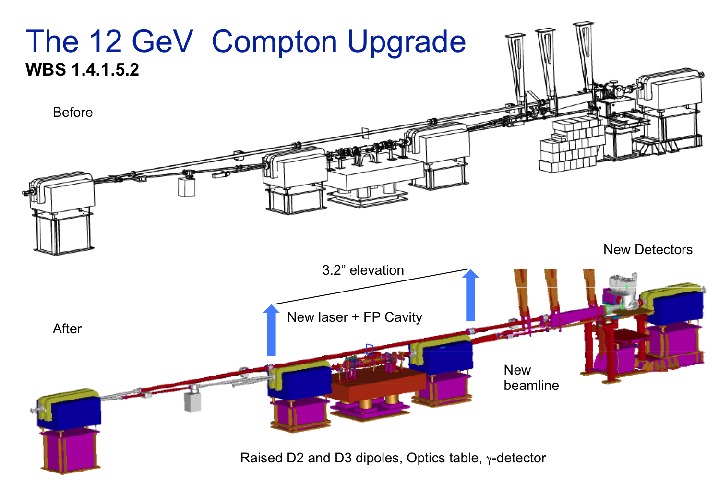}
\caption{ Model of the 12 GeV Compton polarimeter in Hall A Beam line before and after the upgrade.  Shown in blue are the dipole magnets of the chicane.  The two middle dipoles, the optics table, and the photon detector  are being raised by 85~mm for the 12GeV Upgrade }
\label{fig:cpt-12GeV}
\end{center}
\end{figure}

The Hall A Compton Polarimeter is being upgraded to  11~GeV  as part of the CEBAF 12~GeV upgrade project. Before the upgrade, the dipole magnets, the main transport elements of the electron beam line chicane, were configured to produce a 300~mm vertical displacement at 1.5~T for 8~GeV electron beam. In order to minimize cost, the conceptual design of the 12 GeV upgrade simply reuses the same magnets with a reduced bend angle to produce a 218~mm chicane displacement.  In the final design, this displacement has been further reduced to 215~mm. 
Shown in Fig.~\ref{fig:cpt-12GeV} is a computer model of the Compton polarimeter  beam line before and after the upgrade. As shown in the figure, the 2nd and the 3rd dipole magnets, the optics table, and the photon detectors will be raised up by 85~mm. The scope of the upgrade (WBS~1.4.1.5.2) consists of reconfiguration the electron beam chicane, changes to the optical setup, electron and photon detectors to be compatible with 11~GeV configuration. Additionally, the section of the beam pipes between the 3rd and the 4th dipoles is being enlarged substantially to accommodate the larger acceptance of the scattered electrons with a green FP cavity.

\begin{figure}[htp]
\begin{center}
\includegraphics[width=5.in]{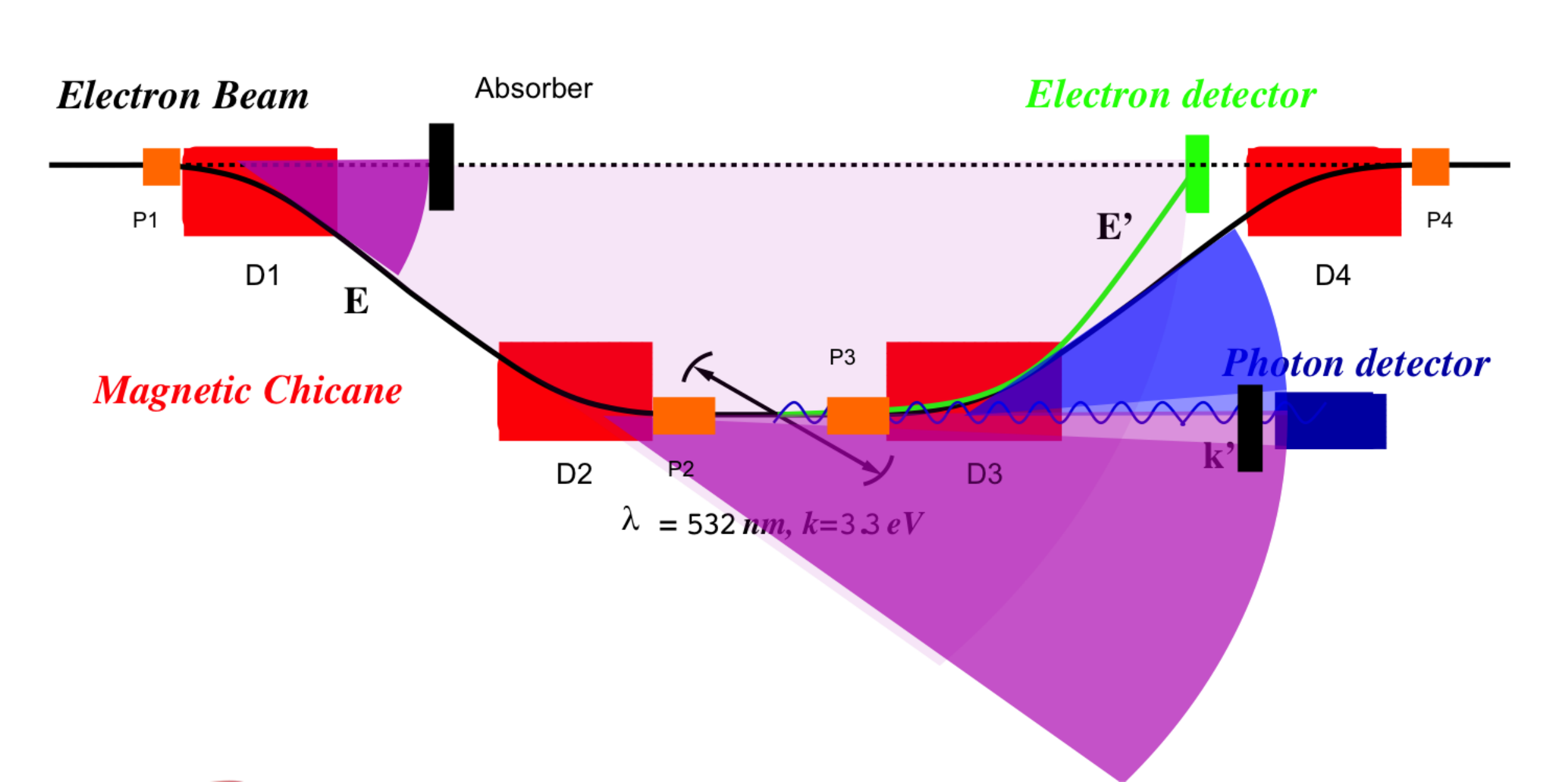}
\caption{ Illustration of suppression of synchrotron radiation background with fringe field modifying field plates P1-P4 attached to dipole magnets D1-D4. A combination of reduced magnetic field seen by the photon detector and absorbing material attenuates synchrotron radiation flux to negligible  levels~\cite{ref:beneschtalk}. }
\label{fig:cpt-sr-sch}
\end{center}
\end{figure}

\begin{figure}[htp]
\begin{center}
\includegraphics[width=4.3in,angle=0]{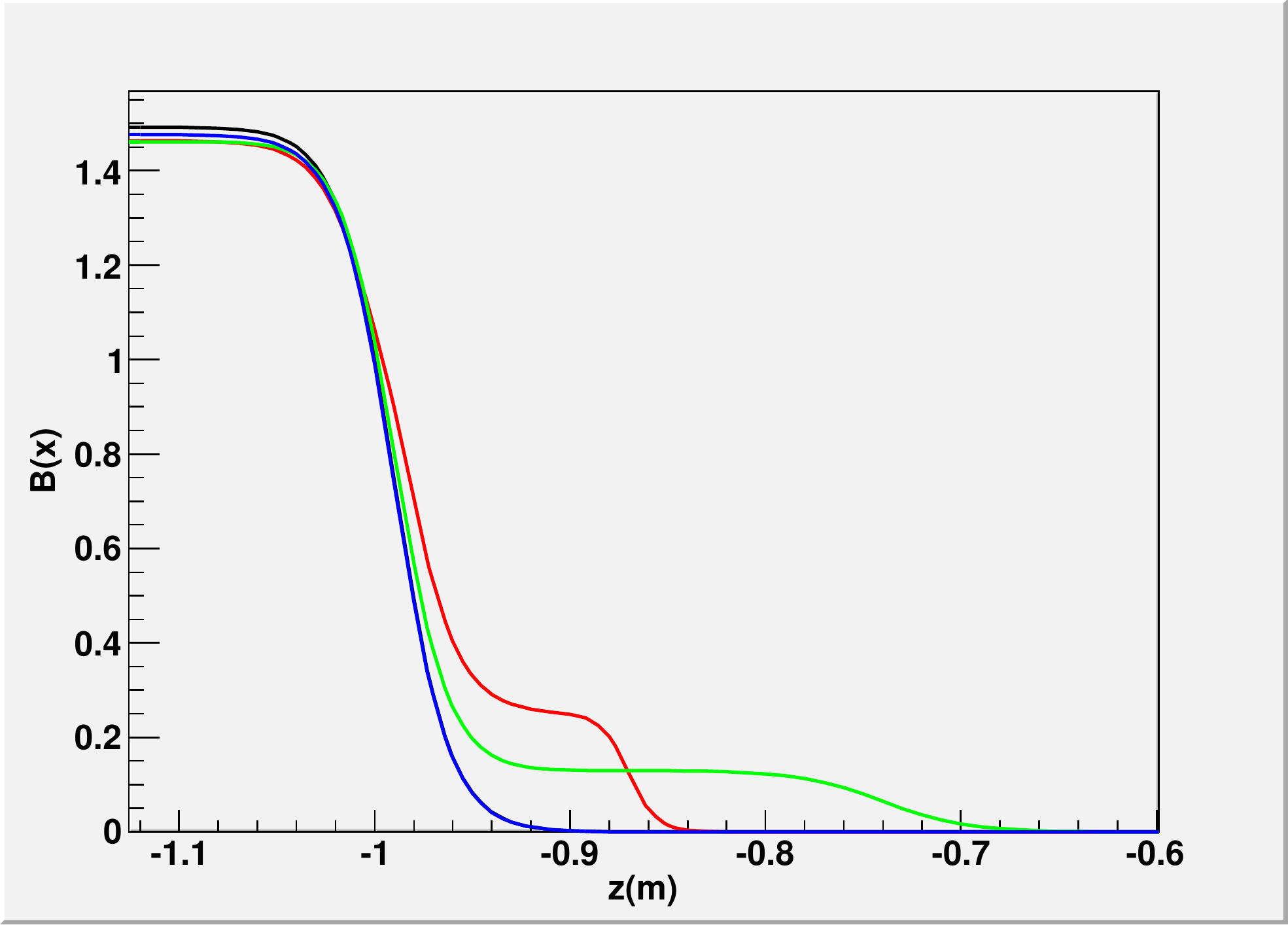}
\caption{  The effect on the fringe field of the dipole magnets due to the addition of the field plates. Shown are magnetic field strengths in T seen by the electron beam at 11~GeV for  the basic dipole in blue, addition of short plates P1/P4 in red, and addition of long plate P3/P4 in green.}
\label{fig:cpt-fp-ff}
\end{center}
\end{figure}

With higher energy, synchrotron radiation in the Compton chicane increases dramatically both in flux and hardness. Simulation by Quinn {\it et al.} show that at 11 GeV with 100 $\mu$A beam, energy deposited from synchrotron radiation will be about 10-20\% of that from Compton scattered electrons from a 10kW green cavity. This poses a major  dilution of asymmetry for the integrating photon detector. However, TOSCA simulation by Benesch shows that addition of  passive iron plates in the fringe field region of the dipole magnets will reduce the magnetic field seen by the photon detector by an order of magnitude, thus reducing synchrotron radiation background  to negligible level. Shown in Fig.~\ref{fig:cpt-sr-sch} is a schematic representation of the synchrotron radiation background and its suppression scheme. Dipole magnet D1 poses a potential source of synchrotron radiation for the electron detector via  the straight through beam line. The radiation  will be softened with the addition field plate P1 and reduced in flux with an absorber. Dipole magnets D2 and D3 are being modified with fringe field plate P2 and P3 which reduces the hardness of the radiation by more than three orders of magnitude. An absorber in front of the photon detector easily attenuates the softer photons. Shown in  Fig.~\ref{fig:cpt-fp-ff} are the effects on the fringe field of the dipole magnets due to the addition of the field plates. The curves correspond to the   magnetic field strengths in T seen by the electron beam at 11~GeV for the basic dipole in blue, addition of short plates P1/P4 in red, and addition of long plate P3/P4 in green. The addition of the field plates modify the overall integral field of the dipole magnets at negligible levels. Nonetheless, we plan to map the fields for one of the dipoles, D3, to confirm the design.

\begin{figure}[htp]
\begin{center}
\includegraphics[width=4.1in,angle=0]{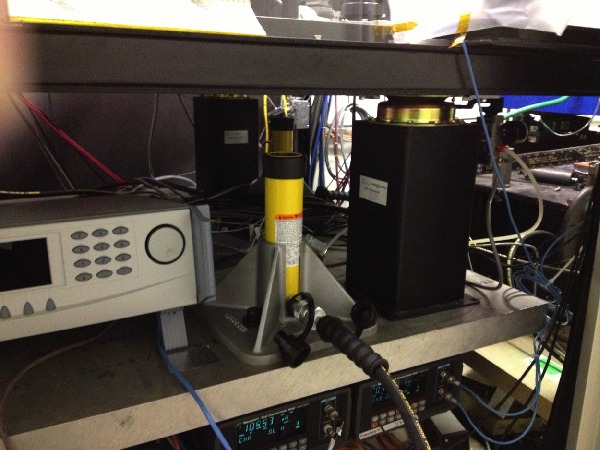}
\caption{ The optics table has been raised to the 12~GeV upgrade configuration. New vibration isolators with the correct height replace old leaky isolator legs.}
\label{fig:cpt-raised-optics}
\end{center}
\end{figure}

\begin{figure}[htp]
\begin{center}
\includegraphics[width=5.2in,angle=0]{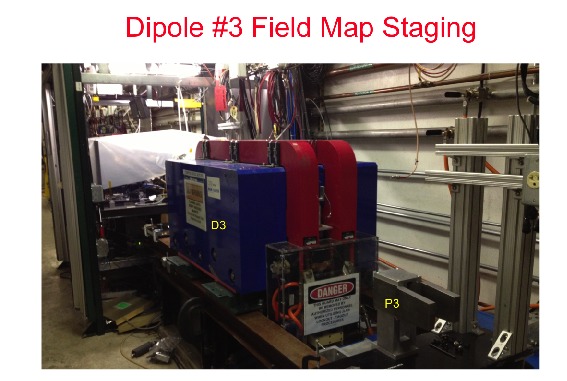}
\caption{ The 3rd dipole magnet  in the Compton Polarimeter chicane being staged for field mapping. Shown in the foreground are the field plates P3 for synchrotron radiation suppression.}
\label{fig:cpt-d3-stg}
\end{center}
\end{figure}

Engineering design of the  12 GeV upgrade has been completed. Fabrication of all components is complete and all  parts are in house. The old beam line has been dismantled.  As shown in Fig.~\ref{fig:cpt-raised-optics}, the optics table has been boxed up and raised to its 12GeV configuration with new isolator legs. The 2nd dipole magnet has been raised to its final location. The third dipole magnet has been rolled out of its place and is being staged for field measurements as shown in Fig.~\ref{fig:cpt-d3-stg}. Three sets of both integral and differential field measurements are planned. The basic dipole will be mapped first followed by the addition of P1 and P3 respectively. The integral field measurements are being carried out by Bagget {\it et al.} of the JLab magnetic measurement group using a stretched wire technique. Differential measurements with a hall probe 3D mapper will be carried out by Jones {\it et al.}, University of Virginia, following the integral measurements.

The 12~GeV upgrade is on track to be completed in 2013. Commissioning with the upgraded CEBAF first beam delivery to Hall A are planned for early 2014.

\subsubsection{Conclusion}
As we wind down the 6~GeV operations of the Hall A Compton polarimeter and welcome the 12 GeV era, it is worthy of note that the green laser  upgrades of the Compton polarimeter for 6~GeV operations have been immensely successful. The upgraded polarimeter was put into operation successfully for the recent PReX, DVCS, and g2p experiments. The green  FP cavity far exceeds design goal by achieving upwards of 10~kW intra-cavity power. The green cavity along with the GSO photon calorimeter with integrating data acquisition system  has provided the first set of high precision polarimetry results. Recent development in high speed counting data acquisition will further enhance the accuracy of the Compton polarimeter. Lackluster performance  of the electron detector will get a boost with  new collaborators from University of Manitoba and Mississippi. The 12~GeV upgrade project construction is proceeding well with initial operation expected in early 2014.

\clearpage
\newpage
\subsection[3He Target]{Polarized $^3$He Target}

\begin{center}
\bf Polarized $^3$He Target: Status, Progress and Future Plan
\end{center}

\begin{center}
contributed by J. P. Chen for the polarized $^3$He group.\\
\end{center}
\par
The polarized $^3$He target $^{\cite{web}}$ 
was successfully used for fourteen 6 GeV 
experiments GDH $^{\cite{e94010}}$, GMn$^{\cite{e95001}}$, A1n$^{\cite{e99117}}$, g2n$^{\cite{e97103}}$, Spin-duality$^{\cite{e01012}}$, 
Small-Angle-GDH$^{\cite{e97110}}$, GEn
$^{\cite{e02013}}$, Transversity)$^{\cite{transversity}}$,
$A_y$-DIS$^{\cite{AyDIS}}$,
d2n$^{\cite{d2n}}$, $A_y$-QE$^{\cite{Ay}}$, $(e,e'd)$$^{\cite{target:e05102}}$ and $A_y$-$(e,e'n)$$^{\cite{e08005}}$. 

The polarized $^3$He target initially used optically pumped
Rubidium vapor to polarize $^3$He nuclei via spin exchange.
Typical in-beam (10-15 $\mu$A) polarization steadily increased from $~30\%$ $^{\cite{slifer}}$
to over $40\%$ $^{\cite{target:Zheng}}$ when target cells were extensively tested and selected. 
A new hybrid technique for spin-exchange which 
uses a K-Rb mixture$^{\cite{hybrid}}$ increased the in-beam 
polarization to over
$50\%$ (close to $60\%$ without beam), due to the much higher K-$^3$He spin 
exchange efficiency~\cite{ref:baranga}. The new hybrid cells also achieved significantly shorter spin-up times ($<5$ hours compared to $<10$ hours for pure Rb cell)~\cite{ref:jaideep}. Further 
improvement in polarization was achieved 
recently for the transversity series of experiments by using the newly 
available high-power narrow-width diode lasers (Comet) instead
of the broad-width diode lasers (Coherent) that have been used in the 
previous experiments.
The target polarization improved significantly to
$55\%$ with up to 15 $\mu$A beam and 20-minute spin-flip 
and over $60\%$ in the pumping chamber. Without beam the target polarization reached over $70\%$.

The earlier experiments used two sets of Helmholtz coils, which provided a 
 holding field of 25-30 Gauss for any 
direction in the scattering (horizontal) plane. The transversity experiment
also required vertical polarization. A third set of coils provide the field in 
this direction. These three sets of coils allow polarization in any direction 
in 3-d space. Target cells were up to
40-cm long with a density of about 10 amg (10 atm at $0^\circ$). 
Beam currents on target ranged from 10 to 15 $\mu$A to keep the beam 
depolarization effect
small and the cell survival time reasonably long ($> 3$ weeks). The luminosity
reached was about $10^{36}$ nuclei/s/cm$^2$.

Fast target spin reversals are needed for the Transversity experiment (every 
20 minutes). The fast spin reversal was achieved with Adiabatic 
Fast Passage (AFP) technique. The polarizing laser spin direction reversal was
accomplished with rotating 1/4-wave plates. 
The polarization loss due to fast spin reversal is less than $10\%$ relative 
depending on the AFP loss and the spin up time. 
The improvement of spin up time with the hybrid cell
has significantly reduced the polarization loss due to the fast spin reversal.
With the BigBite magnet nearby (1.5 m) and
a large shielding plate, the field gradients are at the level of 20-30 mg/cm,
which is about a factor of 2 larger than that without the 
BigBite magnet.
These field gradients lead to about $0.5-0.7\%$ AFP loss. Correction coils 
could reduce the field gradients. However, it was found that when the field 
gradients reduced to less than $10-15$ mg/cm, masing effects$^{\cite{romalis}}$ 
started, which caused a
significant drop in the target polarization (from $70\%$ to $<60\%$). 
We decided to leave the field gradients high by tuning off the correction 
coils to avoid the masing effect.

Two kinds of polarimetry, NMR and EPR (Electron-Paramagnetic-Resonance),
were used to measure the polarization of the target. While the EPR measurement 
provides absolute polarimetry, the NMR measurement on $^3$He is only a relative 
measurement and needs to be calibrated, often with NMR on water. Water NMR provides an absolute calibration since the proton polarization at room temperature is known. The water signal is very small and it is often a challenge to control systematic uncertainties below a few percent level. The uncertainty achieved
with Rb only cell (before GEn) was $3\%$ for both EPR and NMR with water, 
while for hybrid cell (GEn and transversity series) was $5\%$ relative. The main reasons for the larger uncertainty with the hybrid cell were due to 1) the higher operating temperature ($230^\circ$C) while the existing measurements of the EPR calibration constant ($\kappa_0$) were only performed at below $200^\circ$C; 2) the large uncertainty in the diffusion due to the longer transfer tube. With the exception of the GEn experiment, 
all experiments have both EPR and NMR with water calibration and the two 
methods agree well within errors. The GEn experiment used a magnet box 
instead of the standard Helmholtz coils
to provide the main holding field. 
Online polarimetry during GEn was accomplished using the NMR technique of adiabatic fast passage,
as had been done previously.  The calibration of the NMR, however, was done solely using EPR.
This was because the study of NMR signals from water were unsuccessful, due largely to 
hysteresis effects associated with the iron-core magnet.  The lack of a water calibration, however, did 
not contribute significantly to the final errors of the experiment.

\begin{figure}
  \centerline{\includegraphics[width=\textwidth]{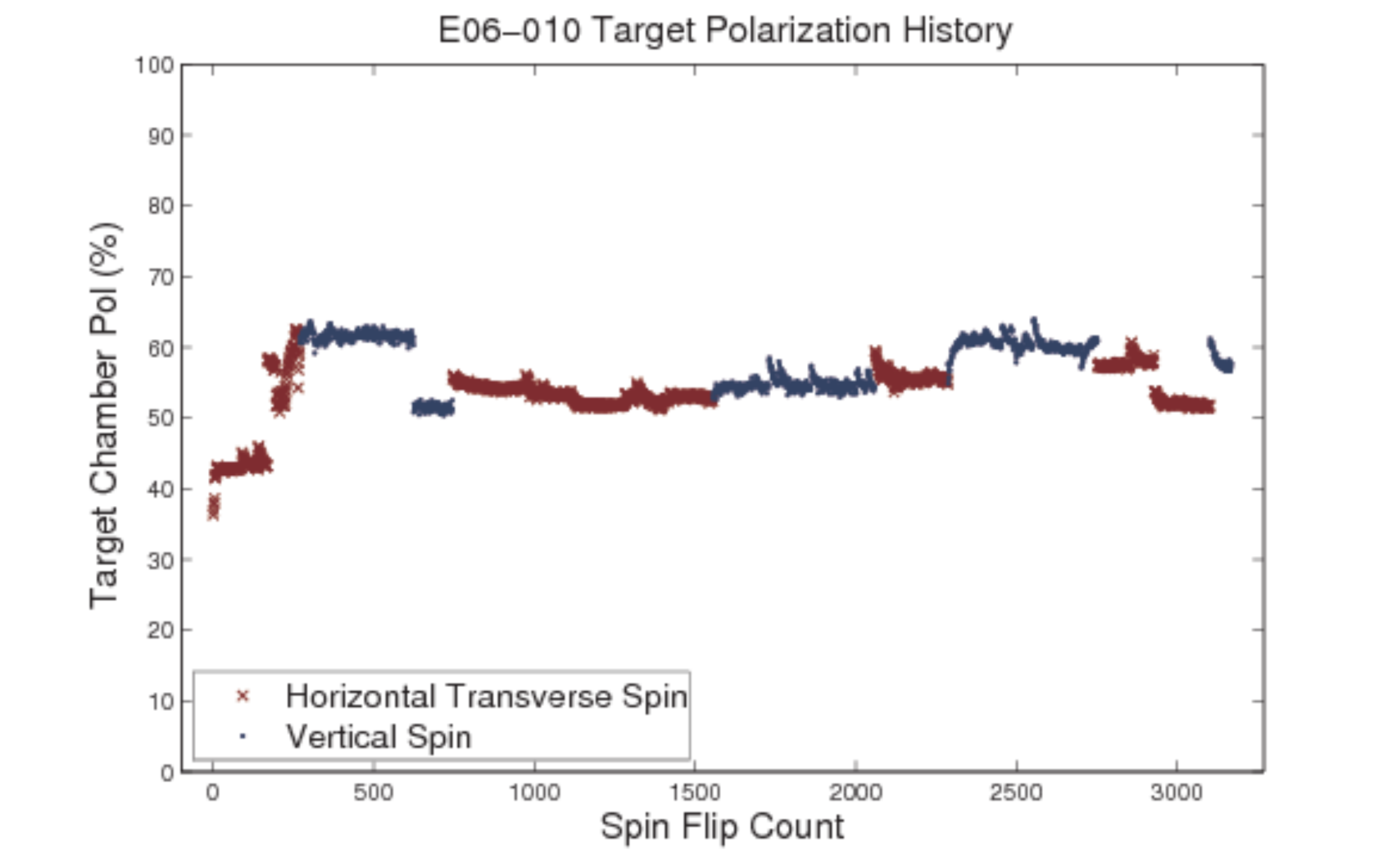}}
  \caption[Polarization]{The polarized $^3$He target performance
with $10-15 \mu$A and spin-flip during the transversity experiment.}
   \label{fig:polarization} 
\end{figure}

The latest analyses of the polarimetries (Fig. \ref{fig:polarization}) were done by Yi Zhang$^{\cite{YiZ}}$
(with cross check by Jin Huang and help from Yi Qiang)
for the transversity experiment, by Yawei Zhang for $A_y$ and by Matthew Posik for the $d_2^n$ experiment. The polarization results were 
cross-checked with a measurement of the e-$^3$He elastic asymmetry,  showing 
an agreement at a level of about $5\%$.

After the last set of polarized $^3$He target experiments in the 6 GeV era (the transversity series), the target was moved back to the target lab in the EEL building 
and was set up to continue tests and R$\&$D. EPR measurements with D1 line was tested and performed successfully for the first time. Fast spin reversal with 
field rotation was tested and it was proved in principle that spin reversal
speed could be increased significantly by using field rotation.
Several tests were performed aiming to 
help polarimetry analysis. Careful study was done to understand the diffusion effect (the polarization changes from the pumping chamber to the target chamber due to temperature gradients), which was one of the dominate uncertainties
in the polarimetry. Systematic studies were also performed to study the temperature dependence of polarization decay (life) time and spin up time. Masing effects were
also studied. These studies provide some basic information for our 
understanding and control of systematic uncertainties.

Further R\&D efforts were focussed on meeting the need the requirements 
for the planned 12 GeV experiments. Seven polarized $^3$He experiments are approved with high scientific rating (three A and four A-). Two experiments using 
the SoLID spectrometer$^{\cite{SoLID}}$  
will be a few years away and only require the already achieved performance.
The following discussion describes the requirements of remaining five experiments. 
The first group of experiments (A1n-A$^{\cite{A1nA}}$, d2n-C$^{\cite{d2nC}}$, SIDIS-SBB$^{\cite{SIDIS-SBB}}$) ask for improvements 
of about a factor of 3-4 in figure of merit over the best achieved performance 
and the second group (A1n-C$^{\cite{A1nC}}$ and GEN-II$^{\cite{GENII}}$) demand
improvement of about a factor of 6-8 in figure of merit.
With the limited resources (engineering/design manpower and funding) available
to meet the challenge of the 12 GeV experiments, a plan has been developed 
to have a two-stage approach for the target upgrade. In the first stage, we 
aim to have a 40-cm 10 amg target capable of handling 30 $\mu$A of beam current
with an in-beam target polarization of $60\%$. Polarization measurements aim to reach a precision of $3\%$. 
This target will meet at least the 
acceptable requirements of the first group. To reach this goal, we plan to make
full use of the existing polarized $^3$He setup from the last set of 6 GeV 
running and make the necessary modifications as outlined in the following 
steps:
\begin{itemize}
\item	Using cells with convection flow
\item	Using a single pumping chamber with 3.5-inches diameter sphere
\item	Shielding the pumping chamber from radiation damage
\item	Using pulsed NMR, calibrated with EPR and water NMR
\item	Measure EPR calibration constant $\kappa_0$ to higher temperature 
range covering the hybrid cell operation temperature (user responsibility)
\item	Metal end-windows desirable (optional for $30 \mu$A, must for higher current, another user R\&D project).
\end{itemize}

The first running of a polarized 3He experiment is expected to be in 2016. Our goal is to have the first-stage (moderately-upgraded) target system ready by 
early 2016.  The goal can be reached with a moderate cost and a moderate 
engineering/designing/supporting manpower. 

The second stage is to meet the needs of A1n-C and GENII-A 
experiments. It will be a 60-cm 10 amg target capable of handling 
60 $\mu$A beam with a polarization reaching $65\%$. 
It requires at least to double the pumping volume by using a 
double-pumping-chamber design. Metal end-window or metal cell
will be needed. To increase the target cell length from 40 cm 
to 60 cm, the holding field coils need to be re-arranged. The small set of 
coils needs be replaced by the large set of coils (used for vertical pumping during transversity).

The polarized $^3$He program at JLab has been very successful 
in the 6 GeV era, mainly due to a collaborative effort among the
users and JLab. R\&D efforts at the user polarized $^3$He 
laboratories have contributed critically to the continuous improvements over
the years. This successful model is continuing into the 
12-GeV program. R\&D by the user groups have already made significant progress. The UVa (Gordon Cates) group has developed and tested
the convection cell concept and demonstrated that it works~\cite{ref:dolph}. They also 
developed a pulsed NMR system and tested it at UVa. R\&D on metal 
end-window is also underway by the UVa group. One important item for polarized $^3$He is the narrow width laser. Comet laser production has been discontinued. Investigation is on-going to find a suitable replacement. William and Mary (Todd Averett) group has performed first tests of a new laser from a new vendor QPC.

\begin{figure}
  \centerline{\includegraphics[width=\textwidth]{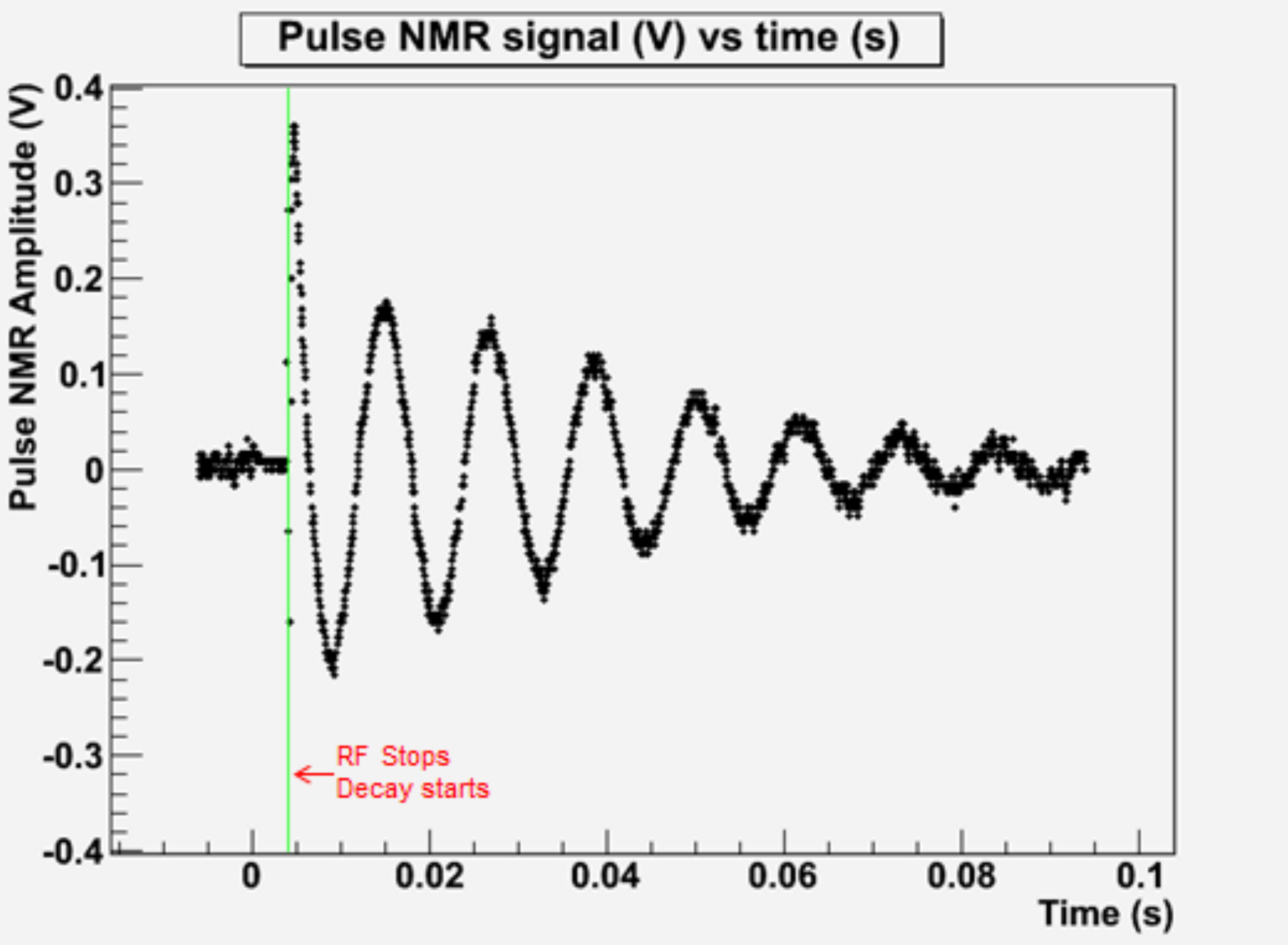}}
  \caption[Pulsed NMR signal]{Pulsed NMR signal from JLab target lab test.}
   \label{fig:pulsedNMR} 
\end{figure}

\begin{figure}
  \centerline{\includegraphics[width=\textwidth]{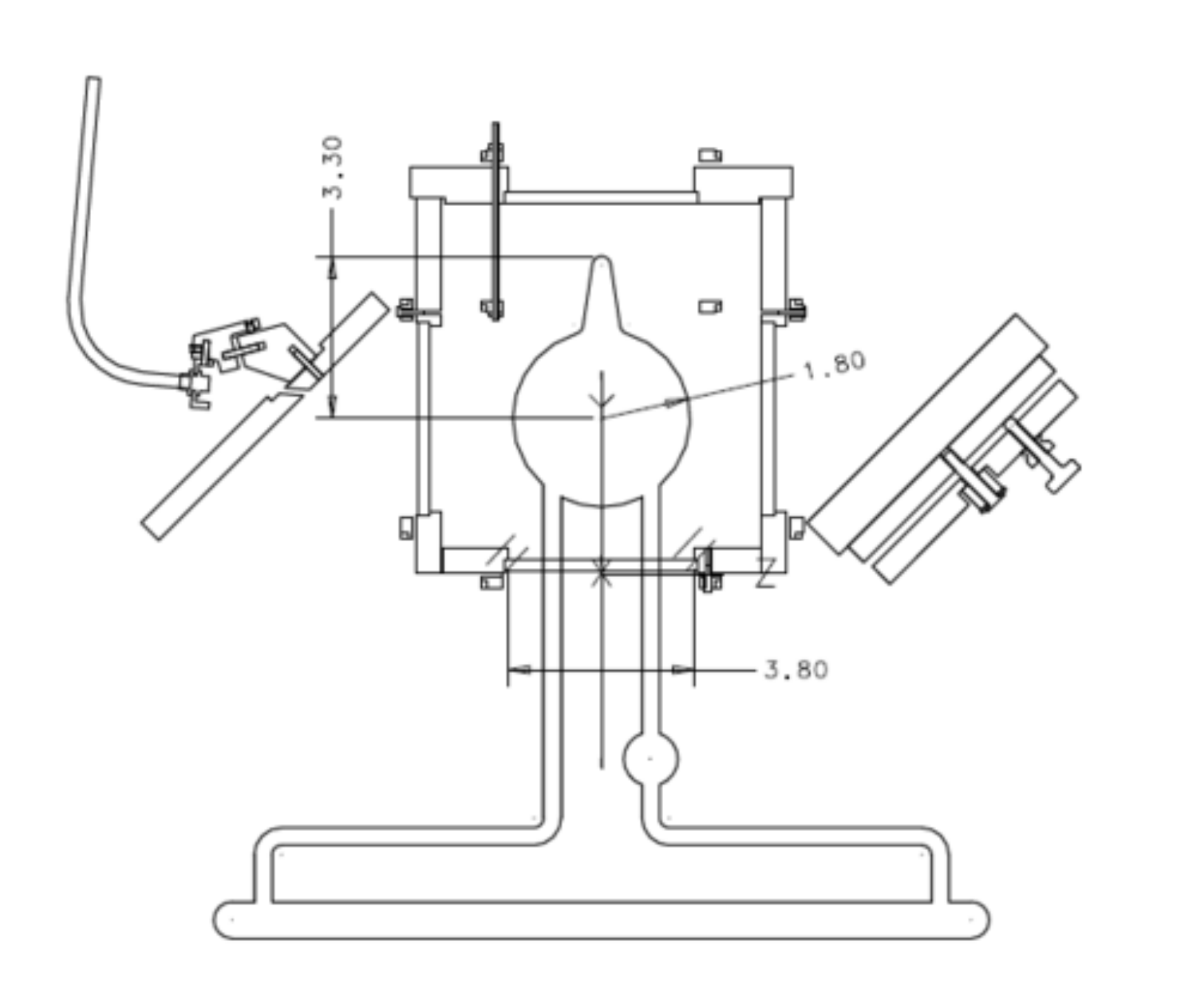}}
  \caption[Convection cell test setup ]{Convection cell test setup at JLab target lab.}
   \label{fig:setup} 
\end{figure}

R\&D activities at the JLab target lab has been on-going for
the upgrade. 
A graduate student (Jie Liu) and a postdoc 
(Zhiwen Zhao), both from Xiaochao Zheng's group at UVa, have been actively working in the target lab. 
In addition to the tests necessary to understand systematics 
for target polarization measurements, a pulsed NMR system was developed at the JLab target lab. It has been tested 
to work well (Fig. \ref{fig:pulsedNMR}). Systematic study to understand the pulsed NMR system
is underway. A new laser from QPC is undergoing tests. 
One convection cell with single 3.5 inch diameter
pumping chamber has been manufactured. After initial 
tests at UVa, it is now being set up for full tests at JLab (Fig. \ref{fig:setup}).

\clearpage
\newpage
\subsection{Data Analysis}
\label{sec:software}

\begin{center}
\bf Data Analysis
\end{center}

\begin{center}
contributed by J.-O. Hansen.
\end{center}

\subsubsection{Podd (ROOT/C{\tt ++} Analyzer) Status}
\label{sec:podd}

For several years now, the main Hall A data analysis package, ``Podd''
\cite{ref:podd_homepage},
has been in a stable production state. All experiments that ran during
2011/2012 have used Podd for analysis. Over the past year, updates to
Podd have consisted mostly of small bug fixes and usability improvements.
However, there have been a few noteworthy upgrades and changes:
\begin{itemize} \setlength{\parskip}{0ex}
  \item Helicity decoder classes for the Qweak helicity scheme were
    contributed by Julie Roche.
  \item Additional VME front-end module decoders were included
    by Larry Selvy.
  \item A system for automatic run-time string replacement in the various
    text input files was implemented, encapsulated in the
    {\tt THaTextvars} class.
  \item Previously existing string size limits in the {\tt EPICS} classes
    have been completely lifted.
  \item Support current ROOT and Linux versions: ROOT versions up to
    5.34 and gcc compilers up to version 4.7 (as used in Fedora 17, for
    example) have been tested.
  \item Experimental support for Mac OS X has been added.
  \item Version management of the code has been moved from the
    obsolete CVS system to {\tt git}.
\end{itemize}

The {\tt THaTextvars} system mentioned above allows users to write 
generic text input files (database, cut and output definitions)
that are independent of actual spectrometer and module names used in
the replay setup. These files may now contain macro names that are
replaced at run time with values set in the replay script.
Moreover, replacement macros may contain entire
lists of module names, not just single ones. As a result, it is no
longer necessary to create multiple text files, or duplicate sections
within a text file, if these parts only differ by module name, 
which is frequently the case.
This system is optional and fully backwards compatible 
with existing input files.

Additionally, a much-improved version of the HRS VDC analysis code was
developed for the APEX test run in 2010. This code
finally addresses the issue of ambiguities due to multiple hits in the VDCs,
which occur at high singles rates, largely due to accidental coincidences
during the relatively wide (ca.~400~ns) drift time window. The new
algorithm is able to suppress background from accidental coincidences
by at least a factor of 5 by exploiting the timing information from
the group of wires in a hit cluster. A non-linear, 3-parameter fit to
the drift times is performed to extract the approximate time offset of
the cluster hit with respect to the main spectrometer trigger.
This new VDC class is not yet part of the official Podd release, but
is available in the {\tt git} repository.  It will be included in
a future releases, once the code has been cleaned up and documented.

The current version of Podd is 1.5.24. Downloads and documentation is
available, as usual, from the main Hall A web page.

\subsubsection{External Software Review}

In June 2012, an external review of the status of the laboratory's 
online analysis software and computing facilities was held at 
Jefferson Lab. The review focused on the suitability of
the different halls' software and analysis plans for the 12~GeV era.
With respect to Hall A, the committee found us generally well prepared,
and considered our plans sound. Naturally, there were several recommendations,
which are summarized here:
\begin{enumerate}
  \item In general, all halls were encouraged to continue and improve
    good software development practices, such as automatic code
    builds, use of standard code evaluation tools, and development of standard
    software validation procedures.
    \label{software-review-item1}
  \item The main suggestions for our Hall A analysis package Podd were
    \label{software-review-item2}
    \begin{itemize}
      \item Try to implement automatic event-level parallelization
        in keeping with the prevalent industry trend toward increasingly 
        large-scale multi-core processing.
      \item Collaborate with Hall~C to develop a common core 
        analysis package.
    \end{itemize}
  \item Finally, it was recommended to carry out a more detailed
    evaluation of performance and limitations of 
    the track reconstruction algorithm for SuperBigBite.
    \label{software-review-item3}
\end{enumerate}

Many of the suggestions under item \ref{software-review-item1} are
already being employed in Hall A, although we can do better in some
areas, such as developing a more comprehensive test suite.
The recommendation under item \ref{software-review-item2} are well
taken and have actually been on our ``like to do'' list for years, along with
other items, such as overall performance improvement and more modularity.
They will be addressed with the upcoming release of Podd 2.0, described
below. Item \ref{software-review-item3} is project-specific.

\subsubsection{Collaboration with Hall~C}
Already early in 2012, Hall~C decided to rewrite their analysis software 
for 12~GeV experiments in an objected-oriented way, using C{\tt ++} and ROOT,
and to base this project on our Podd package. This decision gained
further momentum by the outcome of the software review, discussed above.
For several months now, we have been consulting with
Hall~C in implementing the code necessary to support their hall's 
equipment within the framework of the Podd analyzer. This collaboration
has proven fruitful, as expected. Several limitations of the
current Podd design were exposed, and Hall~C is similarly
discovering inefficiencies in their existing analysis methods and
algorithms. Effectively, the porting of the Hall~C analyzer to the Podd
framework has sparked a thorough code review and general rethinking
of analysis methods for both halls. The end result will clearly be
improved software quality.

There is now a task list of
architectural improvements to Podd that will permit the use of a
core library with hall-specific libraries as add-ons. Development
of a number of smaller new features is in progress, which will be included
in Release 1.6 of Podd, expected in  early 2013.

\subsubsection{12~GeV Software}

In preparation for 12~GeV data taking, we plan to develop a major new
version of Podd, Release 2.0, for the second half of 2013. This is intended
to be the core software to be used for the first 12~GeV experiments in 
Halls A and C. Podd 2.0 will probably include most and hopefully all of
the following:

\begin{itemize}
  \item Automatic event-level parallelization of the core analyzer.
  \item Significantly improved speed of writing ROOT output files.
  \item Decoders for the pipelined JLab 12~GeV electronics.
  \item SQL database backend.
  \item VDC multicluster analysis, mentioned in section \ref{sec:podd}
    above.
\end{itemize}

At this time, it appears as if most of the 12~GeV experiments planned
for Hall A anticipate to use Podd as the basis of their offline analysis.
(The one notable exception is the M{\o}ller parity-violation experiment.)
Consequently, the considerable effort we will spend over the next year 
or two on updating our analysis software will be a sound
investment in the future.

\subsubsection{Computing Infrastructure}
Many improvements to the lab's and the hall's computing infrastructure
occurred in 2012. A few highlights, which may be of particular
benefit to Hall A users, follow:

\begin{itemize}
  \item A new {\tt /volatile} disk area was commissioned by the Computer
    Center. This is a very large and very fast storage area with
    an automatic cleanup policy \cite{ref:volatile_disk}. Its behavior
    is a mix of that of the old {\tt /scratch} area, with no storage
    guarantees, and that of {\tt /work}, with a guaranteed safe space.
    Like {\tt /work}, {\tt /volatile} is not backed up. This new space
    is ideal for large output files 
    that need to survive for a few weeks to a few months. It is readable
    and writable from the compute farm.
    
    Currently, 170~TB of space are available for all halls together (some 
    dedicated to certain projects who paid for the disk hardware).
    Hall A currently has a reservation of 10~TB (guaranteed space)
    and a 30~TB quota. The quota is soft, {\it i.e.} can be briefly
    exceeded if unused space is available. More space will be requested
    once our demand rises.

  \item Several old machines in the Hall A counting house were
    (or will soon be) replaced in favor of modern hardware.
    The function of the two DAQ machines, {\tt adaql1} and {\tt adaql2}, 
    is being migrated to {\tt adaq1} and {\tt adaq2}, each with over 3~TB of 
    very fast raw data staging disks and 6-core processors.
    The local file server, {\tt adaql10}, has been replaced with a newer
    machine, {\tt chafs}, that currently offers 20~TB of work space
    for online analysis.
    Any new server functions should be put onto this machine.

  \item Hall A now has a dedicated {\tt MySQL} 
    database server, {\tt halladb.jlab.org},
    which is accessible from anywhere within the lab. (We may also
    request offsite access, if needed.) This server currently
    contains a copy of the online databases that used to stored on the
    now-decommissioned counting house computer {\tt adaql10}.
    A new backup server has been set up in the counting house as well;
    it will be synchronized with the central server.

\end{itemize}

Due to increased cyber security demands, the counting house computers
as well as the hall networks will stop being accessible with direct {\tt ssh}
from the general computing networks ({\it i.e.}, user desktops, 
CEBAF Center, etc.) by late December 2012. Access from these ``remote''
networks will require a two-factor authentication token and login via a 
gateway, as is already the case for the accelerator networks. This
change only affects remote access; users at terminals in the counting
house or in the hall will see no changes.

\clearpage
\newpage
\subsection{HRS Detectors}
\label{sec:HRS_detector}
\begin{center}
\bf Detector package of the High Resolution Spectrometers
\end{center}

\begin{center}
contributed by B.~Wojtsekhowski, Hall A, Jefferson Lab. \\
\end{center}

\subsubsection*{Introduction}
The  detector package of the High Resolution Spectrometers is
described in the Hall A instrumentation paper~\cite{hallA}. 
Here we have listed upgrades which are underway in 2012-2013. 
After 15 years of operation, some of the detectors are losing performance due to aging.
Some improvements in the detectors have become possible due to progress in electronics.
A few years ago we constructed and implemented the S2m scintillator
hodoscopes in each HRS, which have superior time resolution enabling
reliable PID in several experiments from 2004 to 2010.  
A program of HRS detector ``hardening'' was approved in 2010.  
It includes an upgrade of the VDC front-end electronics, replacement of
the aged S1 hodoscope, and refurnishing of the A1 aerogel counter (n=1.015).  
The initial Hall A FY12 plan anticipated significant advances in the hardening program.  
However, the final plan, due to a funding redistribution in FY12,
provided only minimum resources for the HRS hardening.

In 2013 the VDC electronics upgrade will be implemented by re-using the A/D
cards donated by W\&M after completion of the Qweak experiment.
Two new S1 hodoscopes are 40-90\% completed, but during 2013 they will be on
hold due to a lack of funding for manpower and PMT procurement.
The A1 refurnishing has been postponed indefinitely because none of the currently 
approved experiments plans to use HRS(s) for hadron detection.

\subsubsection{VDC electronics}
\label{sec:VDC}

The previously used front-end VDC electronics is a LeCroy 2735DC card
based on a 20-year-old design.  
Due to the open design of the VDC frames (all of the them are non-conductive),
the output signals of the A/D cards induce large feedback and often cause
so-called oscillation.  A high threshold level in the A/D cards required
for suppression of such oscillation leads to a high value of HV on the chamber.
For the Ar-Ethane(62/38) gas mixture, the operational HV is of 4000~V.
The threshold level defines the required gas multiplication factor and
average anode current at the given rate of the tracks.  
At the current level of 100~$\mu$A the VDC operation become unstable. 
For the GEn (E02-013) experiment, the A/D cards based on the MAD chip were
constructed for use in the BigBite spectrometer wire chambers~\cite{ADcard}.
Two key advantages of those cards are a low amplitude of the output
signal (LVDS) and reduced noise of the front elements.   
The resulting reduction of the threshold is about 5 for the input charge.

In addition to the upgrade of the VDC front-end cards, we are going to
implement the sparsification window by using 1877S FastBus TDCs (procured
by Carnegie-Mellon University from CLEO for a symbolic \$50 per module).
These modules were tested in 2012 and have already been installed in the HRS DAQ crates.
The sparsification window of 400~ns will allow significant reduction of the event size
at a large rate of the particles and lower dead time of DAQ.

\subsubsection{Front FPP chambers}
\label{sec:FPP}

The HRS Focal Plane Polarimeter, FPP is installed in the HRS-left spectrometer.  
The FPP includes four large drift chambers of straw tube design.
Two chambers constitute the front group and two others the rear group.
The adjustable thickness carbon analyzer is installed between the groups.
Many highly rated experiments have been performed using the HRS FPP system. 
Sometimes operation of FPP was difficult, mainly due to gas leakage from
the chambers, which was very high.  Even with a huge gas flow of 50 l/h,
the chambers were unstable due to frequent HV trips.

Analysis of the problem performed in 2010 showed that the main leak is due
to misalignment in the attachment of the gas lines into the rubber plug of the straw tubes.
The proper modification was developed and all parts have been procured.
Additional improvements have been made in the HV distribution panel (a single
board) and the gas distribution configuration (a parallel input and an
open output).  
In 2013 these upgrades of the front FPP chambers will be implemented.
The first use of the upgraded chambers will be in the GMp experiment (E12-07-108)~\cite{GMp}.

\subsubsection{Shower Calorimeter Trigger}
\label{sec:Shower}
The HRS-right shower calorimeter was constructed together with the
custom summing electronics for trigger purposes by the University of Maryland.
However, those sums were never used in the experiments, and the electronics were later removed.
In 2011 we restored the summing electronics and have a plan to use both 
the pre-shower and the total energy signals in the trigger.
These signals allow us an additional measurement of the trigger
efficiency and efficient pion rejection on the trigger level.
Similar electronics will be assembled on the HRS-left.

\subsubsection{Status of the S1/S2 hodoscopes}
\label{sec:S1}
The HRS original trigger counters were made of thin plastic scintillator (5~mm BC408)
to allow detection of low energy hadrons.
Each S1 and S2 hodoscope has six counters with about a 5~mm overlap between them.
A large aspect ratio (of 60:1) of the scintillator cross section requires a long
twisted light guide resulting in reduced light collection efficiency. 
In their virgin state, these counters had in the cosmic ray signal about
120 photo-electrons per PMT, and the time resolution per counter was about
0.30~ns ($\sigma$), which currently degrades to about 0.6-0.7~ns.

The S2 hodoscope, which in fact doesn't need to be thin, was replaced in
2003 by a 16-paddle hodoscope, S2m(odified) made of 50~mm fast plastic
scintillator EJ-230~\cite{Eljen} (see for details~\cite{S2m}) 
with a time resolution of 0.10~ns ($\sigma$)~\cite{S2m_res}.
Last year we finished construction of a new 16-paddle hodoscope, 
S1m(odified) made of 10~mm plastic scintillator (EJ-200) and XP2262B. 
The optimized shape of the light guide and doubled thickness resulted 
in a time resolution of 0.20~ns.
The mounting frame for this hodoscope has also been designed and constructed. 
For the second, S1f(ast) hodoscope, we selected a fast scintillator
(EJ-230), a novel design of the light guide, and the high performance PMT R9779~\cite{Ham}.
This allowed us to get about 300 photo-electrons per PMT and reach a time
resolution of 0.08~ns with only a 5~mm thickness of the scintillator.  
The components of these counters have been produced.
However, completion of the S1f hodoscope requires procurement of the
32~PMTs and manpower for assembling the counters and mounting frame 
which will not be available in 2013.

\subsubsection{Status of the Gas Cherenkov counters}
\label{sec:GC}

The performance of the HRS Gas Cherenkov counters, GC, has degraded by a factor
of 2 since commissioning.  During the last 6-GeV experiment, the average number of
photo-electrons was only about 5.  The plan of GC refurnishing includes
re-coating of the mirrors and implementation of the 5'' Hamamatsu PMTs.
The latter is not possible in 2013, and we will select the best available 5'' PMTs
from those used in the two existing aerogel counters.

\subsubsection{Active sieve slit counter}
\label{sec:SciFi}
A long-standing problem of the high energy magnetic spectrometers is the sieve slit
calibration for the positively charged particle configuration.
For the negatively charged particle configuration, the sufficiently thick
sieve slit allows selection of the holes when the electron tracks are identified in the focal plane.
In contrast, for the positively charged particle configuration, the flux of positrons is too  
small to compete with the secondary positrons produced in the sieve slit body.

We have developed an active sieve slit device based on a scintillator fiber hodoscope.
It includes 32 fibers in the X-direction and 32 fibers in the Y-direction,
all connected to the 64-channel maPMT.
The detector is placed in a vacuum between the target and the first magnet
of the spectrometer.   The position of the detector is remotely
controlled. 
The first use of this device will be in the APEX experiment (E12-10-009) for
calibration of the positron arm magnetic optics.

\clearpage
\newpage
\subsection[SBS]{Super Bigbite}

\begin{center}
\bf The SBS Program in Hall A
\end{center}

\begin{center}
contributed by John J. LeRose.
\end{center}

\subsubsection{Introduction}

The Super BigBite Spectrometer (SBS) while in principle consisting of a very simple, single dipole based magnetic spectrometer is actually a collection of many devices that will be used in conjunction with SBS to achieve the desired Physics goals. What follows describes the plans, as presented to and approved\footnote{DOE approved the program management plan. However, the program is to be funded from JLab Physics Division funding with no additional funding explicitly intended for SBS.} by DOE, to construct the basic spectrometer and a limited set of detectors needed to get started on the envisioned research plan.

\subsubsection{Program Description}

Using the Super Bigbite Spectrometer Program Management Plan Jefferson Lab and its collaborative partners will produce key components of the research equipment required to conduct a series of elastic nucleon electromagnetic form factor measurements.

Each of the measurements will use the electron beam from the upgraded 12~GeV\, CEBAF accelerator. The three experiments include:

\begin{itemize}
\item{A measurement of  to $Q^2=10~\mathrm{GeV}^2$ using the double polarization beam-target technique (E12-09-016)}

\item{A measurement of to $Q^2=12~\mathrm{GeV}^2$ using the double polarization beam-recoil-polarimeter technique (E12-07-109)}

\item{A measurement of  to $Q^2=13.5~\mathrm{GeV}^2$ by determining the cross-section ratio for the two reactions $d(e,e'n)$ and $d(e,e'p)$ (E12-07-109).}
\end{itemize}

These experiments along with precise data from the  experiment (E12-07-108) will determine all four elastic electromagnetic nucleon form factors, as well as making possible a flavor separation.  Because all the elastic form factors drop off so quickly at high values of $Q^2$, the three experiments all depend critically on both high luminosity as well as relatively large acceptance.

The three measurements each require a somewhat different experimental set-up. They are designed so that they can be accomplished using largely common components that can be rearranged into the required configurations.  Collectively, this set of components is called the ``Super Bigbite Spectrometer''. The research program will also take advantage of the equipment associated with the existing BigBite spectrometer that resides in Jefferson Lab’s experimental Hall A, and some hardware under construction for other approved experiments.

The SBS focuses on large-acceptance detection, makes use of an existing magnet, and will utilize a detector package with Gas Electron Multiplier (GEM) tracking detectors. GEM technology has advanced significantly in recent years. A collaboration with INFN-Rome has already begun development of the front-end GEM's and data-acquisition electronics such that the final design is essentially in-hand.

\subsubsection{Program Approach}

The SBS Program will consist of three separate, but interrelated Projects.

	The first Project, {\bf SBS Basic}, involves the acquisition of an existing magnet (48D48 magnet from Brookhaven National Lab) and the associated work of preparing it for use during the SBS research program. The effort includes modifications to the magnet, including machining a slot in the yoke for beam passage, field clamps, and a solenoid to reduce the transverse magnetic field on the beam line, the design and development of the infrastructure needed to run the magnet, and the construction of the platform on which it will stand.

	The second Project, {\bf Neutron Form Factor}, involves the construction of twenty-nine GEM detector modules with associated front-end and DAQ modules to meet the requirements of the approved neutron form factor measurements.

	The third and final Project, {\bf Proton Form Factor}, involves the construction of thirty-five GEM detector modules with associated front-end and DAQ modules, a sophisticated trigger system, and the addition of pole shims for increased magnetic field integral to meet the requirements of the approved proton form factor measurements.

The three Projects that comprise the SBS Program include construction and commissioning of the individual hardware systems. Installation in the Hall, as with essentially all other experiments, will follow the as yet undetermined beam schedule and will use Hall A Operations funding.

\paragraph{The SBS Basic Project}

The SBS spectrometer itself is a large dipole made from a modified 48D48 magnet that will be obtained from Brookhaven National Laboratory for the cost of disassembling and moving it. The modifications included in this project are:

\begin{itemize}
\item{Machining a slot in the return yoke to allow passage of the beam at small scattering angles}

\item{New coils to accommodate the new geometry}

\item{Field clamp to reduce the field at the target}

\item{Providing a small solenoid to minimize the magnetic field in the beam slot around the field clamp}
\end{itemize}

The magnet will require support structures to hold it and the detectors. The support structure will be versatile enough to accommodate the various angles anticipated by the future experimental program. The magnet will also require a power supply with all necessary infrastructures to power the magnet (water, AC power, busses, safety interlocks).

For this project, spread over fiscal years 2013-2015, \$1,693k, including 28\% contingency, has been budgeted from JLab Physics Division capital funds.

At this time (12/3/2012) design work has begun on the magnet modifications (including yoke modifications and new coils), and the support structure to hold the magnet. It is expected that the steel and coil modification drawings will be completed by February 1, 2013. This will allow us, by the present schedule, two months to negotiate a contract with a vendor to make the needed modifications to the steel and have the steel shipped directly to the vendor from Brookhaven.

\paragraph{The Neutron Form Factor Project}

In this project the following will be constructed: 

\begin{itemize}
\item{Gas Electron Multiplier (GEM) tracking detectors (UVa), 29 modules}

\item{Front-end and data-acquisition electronics (UVa) to accompany these modules}

\item{Electronics Hut, Detector Frames, and Materials needed to construct the Lead Tube Shielding, and the Lead Tube Platform for the beamline (JLab)}
\end{itemize}

At the present time (12/3/2012), although not formally a part of the Neutron Form Factor Project, an intensive effort is underway at UVa, under the guidance of Nilanga Liyanage, to develop the techniques and technology needed to successfully manufacture these detectors. It is expected that those techniques and technologies will be finalized by April 2013 at which time ordering of the needed parts for production will begin.

For this project, spread over fiscal years 2013-2015, \$1,572k, including 30\% contingency, has been budgeted from JLab Physics Division capital funds.

\paragraph{The Proton Form Factor Project}

In this project the following will be constructed:
\begin{itemize}

\item{Gas Electron Multiplier (GEM) tracking detectors (UVa) (35 modules)}

\item{Front-end and Data Acquisition Electronics to accompany these GEM modules (UVa)}

\item{A Trigger (RU)}

\item{Pole shims (JLab)}

\item{Exit field clamp (JLab)}
\end{itemize}

For this project, spread over fiscal years 2014-2017, \$1,582k, including 27\% contingency, has been budgeted from JLab Physics Division capital funds.

\subsubsection{The Future}

At the end of these three projects Hall A will have a unique and versatile, large acceptance ($\Delta \Omega \approx 70~\mathrm{mstr}$ at $15^\circ$, $-12.5\% < \delta < 12.5\%$), high luminosity (up to $10^{39}~\mathrm{s}^{-1}\mathrm{cm}^{-2}$) capability suitable for many future experiments.

\clearpage
\newpage

\subsection[ECal]{Electromagnetic Calorimeter}
\label{sec:HCAL}

\begin{center}
\bf Electromagnetic Calorimeter
\end{center}

\begin{center} 
contributed by C.F. Perdrisat, C. Ayerbe Gayoso, College of William and Mary,

M.K. Jones, Jefferson Lab,

V. Punjabi, NSU, E. Brash, CNU.
\end{center}


\subsubsection{Detector}
\label{sec:Detector}
\begin{figure}[hbt]
\begin{center}
\includegraphics[width=60mm,angle=90]{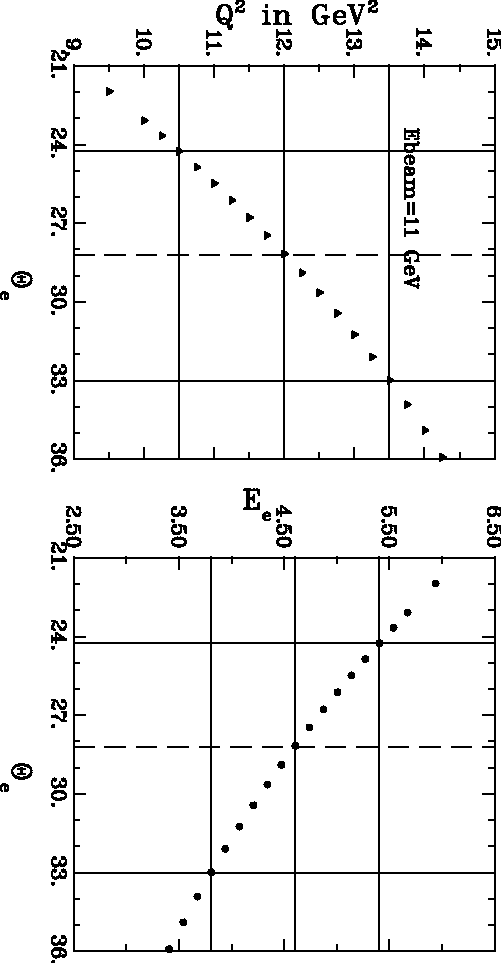}
\caption{\small{Left figure: Plot of Q$^2$ versus the scattered electron's angle, $\theta_e$. Right figure: The scattered electron's energy,E$_e$, versus angle. 
 In both plots, the vertical dashed line highlights the value of $\theta_e$ at Q$^2$ = 12~GeV$^2$. }}
\label{fig:kinematics}
\end{center}
\end{figure}

An  electromagnetic calorimeter, EMC,  is required for the GEp(V) experiment, a measurement 
of the $G_{Ep}/G_{Mp}$
 form factor ratio up to Q$^2$=15~GeV$^2$ (original proposal, 2007) by the recoil polarization method
 $\vec e p\rightarrow e' \vec p$. The subsequent rating of the proposal resulted in a 
decrease of the maximum
 Q$^2$ to 12~GeV$^2$. In GEp(V) both electron and proton need to be detected, and 
a trigger defined with high
 threshold on the electron (and proton) energy. Needing a threshold of 80-90\% of the electron
energy to reduce the DAQ rate presents a challenge as illustrated 
in Fig. \ref{fig:kinematics}: the electron energy range for Q$^2$=12~GeV$^2$ is 3.5 to 6~GeV; the gain of the PMTs will have to be 
adjusted according to the distribution of
energy over the front face of the calorimeter, and according 
to the radiation dose received by the individual
 bars, which is not uniform. To reduce the low-energy background on the EMC, a 20-cm aluminum block is placed
in front of the lead glass. This has the unfortunate side effect of degrading the energy resolution of the
calorimeter, since the shower start in the aluminum block. This leads to energy resolution which is only 1.5$\sigma$ of
the desired electron energy threshold and loss of about 14\% of the triggers.
Off-line analysis will require accurate determination
 of the angle of the electron, to sharpen background rejection using the two-body angle correlation
 in elastic {\bf ep}. The BigCal spatial resolution has been 
typically 0.5 cm (1$\sigma$).
A new calorimeter will be equipped with
 a coordinate detector, originally designed as two plane of GEMs, with readout of the vertical
 coordinate only; the resolution in the vertex position defined by the proton arm (and the 40 cm long target) is not
 sufficient to justify reading the horizontal coordinate.   

The beam intensity and energy, target length and detector angles required for Q$^2$=12~GeV$^2$ are such
 that the rate of gain loss  for the lead glass of GEp(III) calorimeter will be \~13 times larger than in GEp(III). This
has lead to the estimate that recovery of the leadglass transparency after 7 hours of beam requires
one hour a UV irradiation curing. After the curing, a check of the PM gain is required; together, irradiation 
and PM re-tuning represents a minimum of 2 hour loss of data taking every 8 hours, or 25\%.

In view of this significant loss of beam time or efficiency, we have been considering other options for
 a new electron calorimeter. Of particular interest is the sampling calorimeter design, in which the
 electron shower is sampled in a stack of scintillator and lead (or iron) plates, with the light
 collected in WLS plastic fibers running through the whole length off the stack; often called shashlik. Many such
 calorimeters exist, but we have concentrated our attention to the HERA-B setup, which contains
 a very large stack of shashlik elements that  are not currently being used. 
This calorimeter was built at ITEP in Moscow. 
It consists of modules 11.2x11.2 cm$^2$ in cross section, and a thickness of 20 radiation lengths (X$_o$).
A very good design for the new calorimeter would consist of either 300 or 432 such modules; the calorimeter, 
NewCal, would be located at an appropriate distance from the target to insure that all electrons elastically
 scattered with a
 proton accepted in the SBS proton arm, are contained in the e-calorimeter. To insure that the analog signals
 from all elements containing a shower are summed independently of where the electron hits, systematic
 overlap between neighbor modules, both horizontally and vertically, is required. Various schemes to
 achieve this requirement have been studied.  

\vspace{0.5in}
\begin{figure}[hbt]
\begin{minipage}[b]{0.45\linewidth}
\begin{center}
\hspace{-0.25in}\
\vspace{0.05in}
\includegraphics[width=75mm]{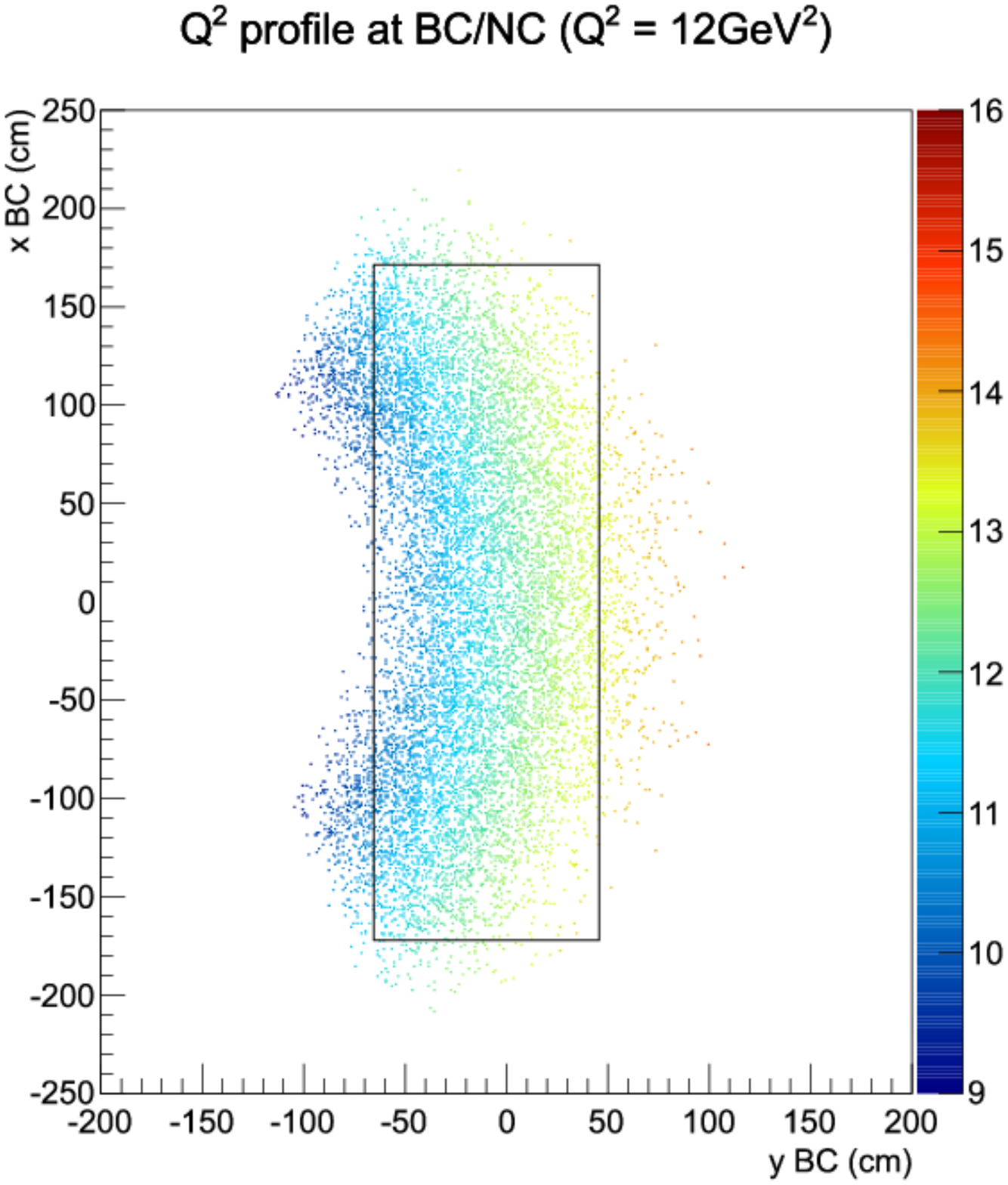}
\caption{\small{Q$^2$ profile at the NewCal for Q$^2$ = 12~GeV$^2$}.}
\label{fig:gepgd}
\end{center}
\end{minipage}\hfill
\begin{minipage}[b]{0.45\linewidth}
\begin{center}
\hspace{-0.25in}
\vspace{0.05in}
\includegraphics[width=75mm]{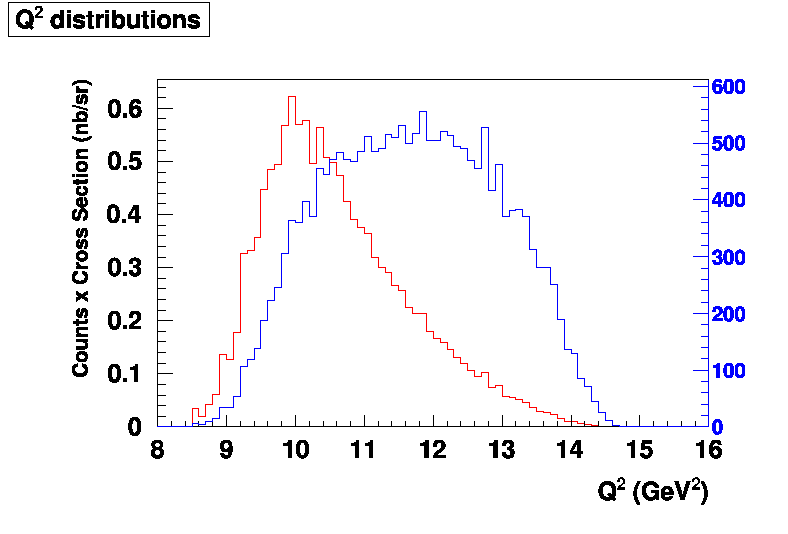}
\caption{\small{The unweighted (blue) and cross-section weighted (red) yields for Q$^2$ = 12~GeV$^2$ }}
\label{fig:gmpgd}
\end{center}
\end{minipage}
\end{figure}






\clearpage
\newpage
\subsection[HCAL]{Hadron Calorimeter}
\label{sec:inst:HCAL}

\begin{center}
\bf Hadron Calorimeter, HCAL-J
\end{center}

\begin{center} 
contributed by G.~B.~Franklin and V. Mamyan, Carnegie-Mellon University,\\ for the CMU, Catania, JINR, JLab collaboration.
\end{center}

The DOE review report of October, 2011, stated that, "while not formally part of the SBS program, the HCAL calorimeter and its good
performance are essential for the envisioned suite of experiments."  They found the design concept,
based on the existing COMPASS HCAL1 detector appropriate and noted that "the HCAL design appears to have
adequate segmentation to provide sufficient angular constraints for the neutron program."  The report recommended  the development of relevant HCAL specifications and quality control procedures.  In response, specifications for the HCAL-J detector have been developed and numerous performance tests of prototype modules and materials have been performed as the first steps of developing realistic quality control specifications.   

\subsubsection{Detector Specifications}
\label{sec:inst:Detector}

The design specifications of HCAL-J are determined by the requirements of the GEp, GMn, and GEn experiments.  
They are summarized as follows:

{\bf Size:}
The detector should have an active area of 160 cm x 330 cm to match the acceptance of the SBS.
If allowed by budget considerations, a detector larger than this minimum size will be constructed 
to maintain resolution near the acceptance boundaries and to provide the improved performance 
associated with a larger time-of-flight distance.

{\bf Energy Resolution:}
The module design will be optimized to allow the use of a high threshold trigger to reject background 
events while maintaining high trigger efficiency for the real events.   The goal will be to achieve 
efficiency greater than 95\% with a trigger threshold set at 25\% of the average signal.

{\bf Time Resolution:}
Simulations show timing requirements, combined with an RMS time resolution of 1.0~ns from the HCAL
detector, would result in a trigger efficiency of 80\% for the GEn experiment.   This is acceptable
as the minimum performance criteria, but the HCAL-J module geometry, scintillator materials, and 
waveshifters will be optimized to obtain a time resolution better than 1.0~ns RMS with an overall 
goal of achieving a time resolution closer to 0.5~ns RMS.

{\bf Angular Resolution:}{
The detector will be designed to provide a spatial resolution of 8 cm RMS and thus the desired 
angular resolution of 5 mrad can be achieved by placing the detector 17 m from the target.

{\bf Gain monitoring:}
The detector will have an LED or laser-based pulser system that will allow in-situ monitoring of 
the PMT gains and facilitate data-acquisition testing.

\subsubsection{Module Design}
\label{sec:Module}
The HCAL-J detector design will be based on a modular construction similar to the existing HCAL1 detector used in the COMPASS experiment.  Each module will consist of alternating layers of iron and scintillator.   In use, the hadronic showers are formed in the iron and the scintillators sample the energy of the showers.  The resulting scintillation light is director to PMTs using wavelength-shifter/light guides that run the length of each module.

Each module will have a 15~cm$ \times $15~cm front face  and will be approximately 1 meter in length.  With this design, the 160 cm x 330 cm active area design criteria can be achieved by stacking 288 modules into an array 12 wide by 24 high.  The individual modules will be constructed to allow this stacking by using the iron plates to transfer the load of the upper modules through the lower modules.

In order to make calorimeter suitable for the JLab energy range, 2-10 GeV, 
the thickness of scintillator plates will be  increased from 0.5 cm to 1.0 cm 
while keeping the total length and number of plates the same. This increases the light yield but 
reduces the thickness of iron plates (from 2.0 cm to 1.5 cm) and therefore hadron detection efficiency. 
The energy resolution of a device with this design was simulated realistically by taking account light propagation in the wavelength shifter (WLS)
and the PMT efficiency. For a hadron momentum of 2.7 GeV/c,  energy resolution was found to be 42.3\% and it gradually improves with 
increasing hadron momentum.
The Geant4 simulation shows that 1.5 cm thick iron plates are thick enough to provide more 
than 95\% hadron detection efficiency in the 2-10 GeV energy range when the hadron calorimeter 
threshold is 1/4 of the average  deposited energy.  

The position of the incident hadron is determined  using an energy-weighted sum of the positions of the 
modules containing the corresponding hadronic shower.
The resulting position resolution predicted from simulation is 5.5 - 3.0 cm for hadrons in the momentum range 2-10 GeV/c.  The angular resolution of 5 mrad  $\sigma$ can be achieved by placing the calorimeter about 12 m from the target.  The predicted coordinate resolution at 40 GeV hadron momentum agrees with the resolution achieved by the COMPASS experiment (1.5 cm RMS). 

\subsubsection{Time Resolution Optimization}
Time resolution is critical for rejecting background during form factor experiments. 
The goal resolution is 0.5~ns sigma and is most important for background separation at higher 
hadron momentum due to short time of flight.

To achieve the required time resolution, we will deviate from the COMPASS design in the choice of scintillator and wavelength shifter (WLS) materials.
The choice of plastic scintillators is constrained by several factors, most important of which is 
the need to match emission spectrum of the scintillator with the absorption spectrum of the fast wavelength shifter .
Fast wavelength shifters ( for example EJ-299-27 and BC-484 ) tend to have absorption spectra
near the end of the UV and start of the violet range (350-400 nm). Since wavelength shifters are sensitive to light 
in the 350-400 nm range absorption, absorption of light in scintillator becomes problematic due to increased light 
absorption by the base material (PVT) of scintillators.  Another problem is absorption and re-emission of light by scintillator fluor (PPO).
The variation in the efficiency of the re-absorption and re-emission process as a function
of spatial position of a shower track
increases non-linearity of the calorimeter signal and can cause coordinate and energy resolution degradation. 
).  
\begin{figure}[ht]
\centerline{
\includegraphics[scale=0.5]{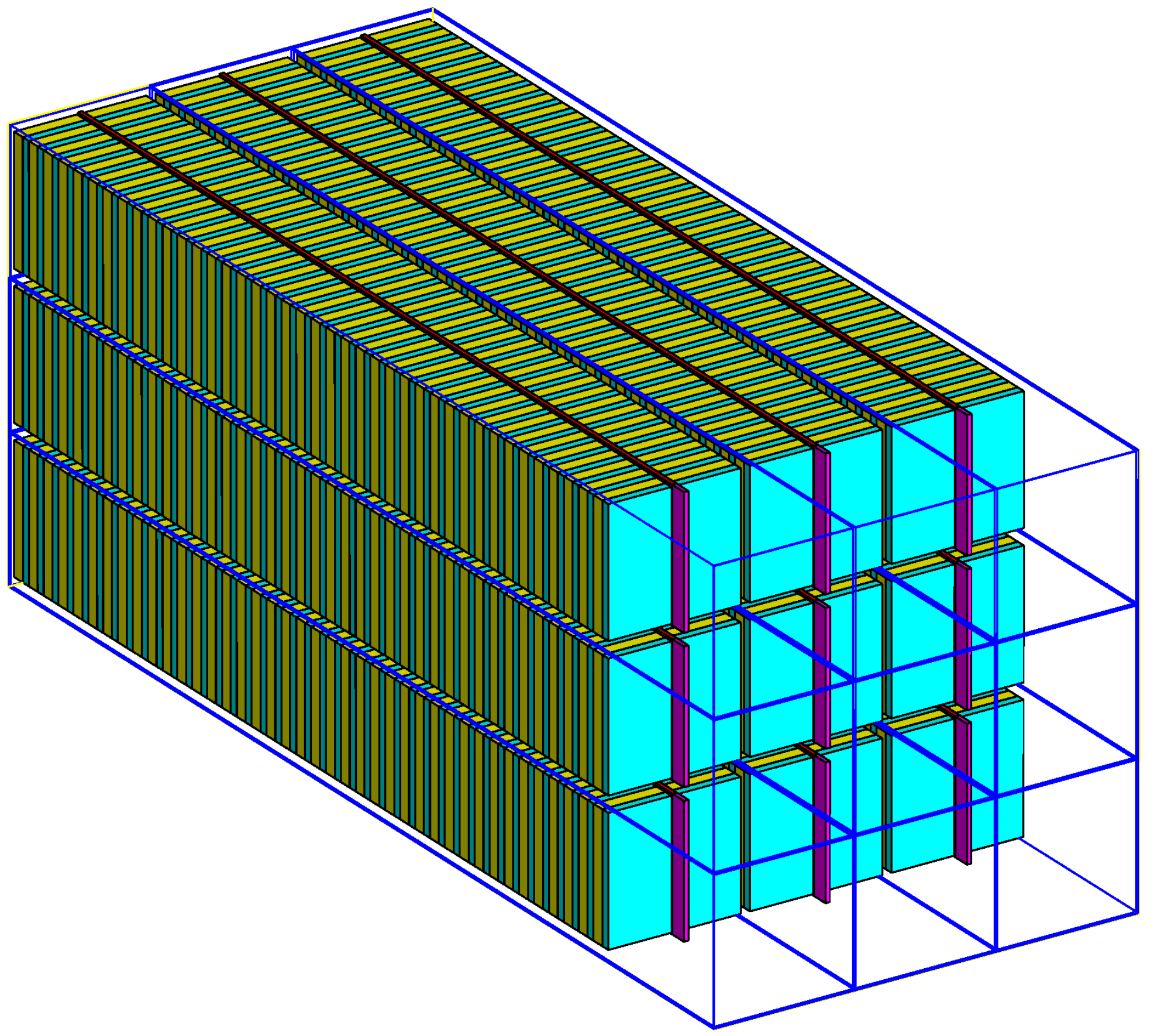}
}
\caption{ Geant4 representation of 9 calorimeter modules showing the WLS (in red) positioned vertically in the middle of the each module. }
\label{fig:MODULE}
\end{figure}

One of shortcomings of EJ-232 scintillator is the light absorption length (13.5 $\pm$ 0.5 cm) 
in the spectra range useful for exciting the EJ-299-27 WLS. To mitigate this problem 
the WLS will be positioned in the middle of the calorimeter module, as shown in Fig.~\ref{fig:MODULE}, instead of along one side.
The advantage of this design is that light travels half the length in the scintillator 
and is absorbed less. As a result. the light yield is increased by almost a factor of two. Another advantage is 
the resulting symmetric distribution of reconstructed particle position (a source of systematic error in angle reconstruction) 
and reduction of non-linearity of calorimeter response. 

In simulation, the time of flight is found by analyzing the signal from the module with the most energy deposited.  A flash ADC with sampling step of 2~ns (500 MHz) is simulated and the time is found when 
the signal reaches at 1/4 of its amplitude.  With EJ-232 scintillator and EJ-299-27 WLS simulation predicts 0.65~ns RMS time resolution. 
Muon tests performed at JLab showed that simulation is able to predict muon time resolution 
with 5\% precision. Based on this the predicted hadron time resolution is quite realistic.

\subsubsection{Light Guides}
Light guides are required to transport the light from the WLS to the PMT.   Production of 288 or more adiabatic
light guides based on a traditional twisted-strip design would take considerable time and resources.
However, we are exploring the alternative light guide design shown in Fig.~\ref{fig:LGUIDE}.  This design is well-suited
to injection molding techniques and, 
after extensive searchers, we have found companies that are capable of manufacturing acrylic light guide of this design with appropriate surface quality.   The ability to produce the light guides with injection molding will greatly reduce the cost of production and gives additional design options, such as a lip to strengthen the scintillator/WLS glue joint, without significant increase
in cost.

A Geant4 based simulation of the light guide has been developed to optimize light collection efficiency. 
The efficiency is estimated with and without aluminum foil wrapping. Without aluminum foil
light collection efficiency is 65\%, with aluminum foil wrapping (assuming 90\% reflection) 
light collection efficiency is 75\%.

\begin{figure}[ht]
\centerline{
\includegraphics[scale=0.7]{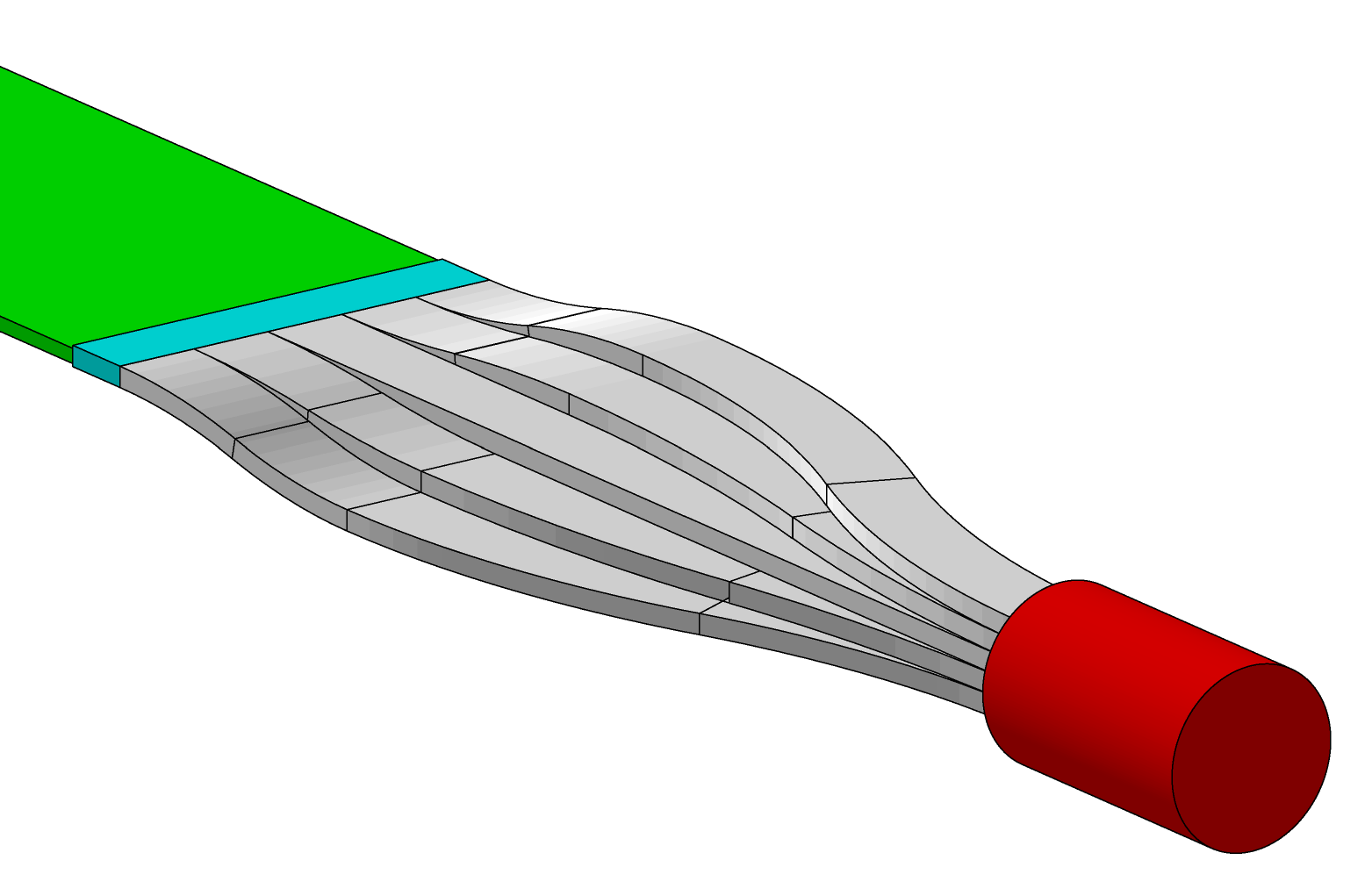}
}
\caption{ Light guide design from Geant4. }
\label{fig:LGUIDE}
\end{figure}

\subsubsection{Phototubes}
\label{sec:PMT}

Simulation shows that the light produced and propagated in ELJEN-299-27 WLS has wavelength in the 410-500 nm range. 
Over 95\% of photons have wavelength in 410-460 nm range. PMTs that have bi-alkali based cathodes are well suited 
to this photon spectrum range (they have max sensitivity around 420 nm). 
Since the XP2262-type PMTs from BigHAND detector have bi-alkali photo-cathodes, they can be coupled with ELJEN-299-27 
and BC-484.

\subsubsection{DAQ and Trigger}
\label{sec:DAQ}
The energy from a single hadronic shower will be typically spread over 9 modules or more.  To provide an effective trigger signal, the signals from groups of modules, corresponding to each possible hadronic shower location, must be summed prior to discrimination.   This can be done using a complex system of linear fan-ins/fan-outs.  However, we are also considering the use of low-resolution FADCS and executing the necessary summing and trigger logic using FPGAs.

\subsubsection{Results of Component Tests}
\label{sec:Tests}

Oleg Gavrishchuk performed cosmic tests at JLab using one of the COMPASS modules provided by JINR, Dubna.
The tests had the following goals: measure trigger time resolution of muons, 
estimate number of photo-electrons produced by minimum-ionizing particles (muons), and to calibrate and test our Geant4 simulation. 
The Geant4 comparison was done in the following steps:

\begin{itemize}

\item { The number of photo-electrons found from cosmic muon passing across the module was inputted into the simulation (80 photo-electrons). }
\item { Simulation was performed with muon going opposite to the light propagation direction 
        and muons going in the direction of light propagation. } 
\item { The simulated timing results were compared with cosmic test results. }

\end{itemize}
It was found that simulation and test results agree with good precision. 
With muons going in the direction of light propagation muon time resolution was 
0.6~ns while, when going opposite to light propagation direction, the trigger time resolution was 0.77~ns. 
In both cases agreement of simulation with data was within 5\%. 

Cosmic tests with EJ-232 scintillator coupled with EJ-299-27 WLS were performed at CMU.
The purpose of these tests were to study light attenuation length in the WLS, attenuation length  
in EJ-232 scintillator and to estimate the number of photo-electrons produced by coupling EJ-232 scintillator and EJ-299-27 WLS. 
Test results show that the attenuation length of light in the WLS is $\approx$ 1.5 m. 
The somewhat short attenuation length is not indicative of the optical quality of the WLS surface, but is a result of 
self absorption in the wavelength shifter and can not be avoided. 
The tests done in the lab were also simulated in Geant4 to estimate the accuracy of the simulation. 
Optical properties of the EJ-232 scintillator and EJ-299-27 wavelength shifter provided 
by ELJEN technology were input into the simulation. The simulation predicts 14 cm absorption length for EJ-232 scintillator, which 
is in good agreement with test results of 13.0 $\pm$ 0.5 cm. 
The mean number of photo-electrons produced by muons passing through the 0.635 cm thick EJ-232 scintillator 
is found to be 10 (light-guide efficiency is not taken account). 
The test setup is simulated in Geant4 and the scintillation yield per MeV of deposited energy is adjusted to 
match this result. 

\subsubsection{Plan and Summary}
\label{sec:Summary}

The overall geometric design of the calorimeter is complete; the only remaining  issue is determining the optimal scintillator/WLS combination.  The goal is to achieve good time resolution by using scintillators and wavelength shifters with
relatively fast decay-times while, at the same, time, minimizing the reduction of light and spatial uniformity that will arise from
short attenuation lengths.

The light guide simulation shows that we have a design that will give suitable light collection efficiency. 
Our plan is to produce these light guides using injection molding and our immediate goal is to obtain quotes from at least three companies for this project.  If possible, we will obtain sample products from companies to provide a means of matching the surface properties obtained from injection molding to our Geant4 optical photon simulations.

Due to high cost of commercial scintillators ($\sim$\$1M) new options for the scintillators are 
necessary. It appears that both Fermilab and JINR are able to produce scintillators with properties necessary for HCAL-J. 
We have written an MOU with Fermilab to produce scintillator samples with several concentration (0, 0.5, 1.0, 2.0, 3.0 percent)  of PPO fluor.  These samples will be tested at CMU to determine their absorption lengths 
and photon statistics.  We expect additional samples to be provided by JINR in the near future.   The results of these tests will be used in simulation to evaluate HCAL-J timing  performance.

The final details of the HCAL-J design will be fixed when the tests of the prototype scintillator materials have been completed.


\clearpage
\newpage

\section{Summaries of Experimental Activities}
\subsection[E99-114 - RCS]{E99-114 - Real Compton Scattering}

\label{sec:e99114}

\begin{center}
\bf Cross section measurement of $\pi^0$-photoproduction in RCS experiment
\end{center}

\begin{center}
J.~Annand, D.~Hamilton, B.~Wojtsekhowski\\
and the E99-114 Collaboration.\\
contributed by Johan Sj\"{o}gren.
\end{center}

\subsubsection{Overview of Experiment}\label{sec:e99114overview}
The Real Compton Scattering experiment, E99-114, was designed to measure the Compton cross section, the
polarization transfer in that reaction and the $\pi^0$-photoproduction cross section, all for wide angle
scattering/photoproduction on the proton at multiple kinematic settings. The experiment ran in January and
February 2002.

Liquid hydrogen was used as the target and photons were created through bremsstrahlung in a 6\% Cu radiator. The
left high resolution spectrometer detected the scattered proton, while the $e/\gamma$ was detected in an
electromagnetic calorimeter. For the $\pi^0$ case one of the decay products($2\gamma$) was detected, primarily the
forward boosted photon. The calorimeter\cite{calo} consisted of 704 blocks of lead glass and it was assembled
specifically for the experiment. A deflection magnet was installed between the target and the calorimeter to
slightly offset electrons allowing for a kinematical distinction between electrons and photons.

The experiment offers a test of constituent counting rules and pQCD predictions of the scaling of the cross
sections. It is also possible to infer information about connected form factors. Likewise the polarization
transfer is a good test of the predictions of pQCD and GPD techniques.  Results for the Compton channel were
published in 2005\cite{david} and 2007\cite{areg}.
 
\subsubsection{Analysis update}\label{sec:e99114analysis}
The analysis is now focused on the $\pi^0$-channel, the remaining channel of interest. The $\pi^0$-channel was
partially analyzed for background subtraction during the study of the Compton channel. This preliminary analysis
is outlined in Ref. \cite{aregthesis}.

Since the experiment was performed while the ROOT-based Hall A analyzer code was still being developed the old
analysis software is a mix of the FORTRAN based ESPACE and the early C++ analyzer. This has now been changed and
the software has been updated/ported into the contemporary standard.

The calibration of the calorimeter is very important for accurate extraction of the $\pi^0$ yield. It was noted in
the beginning of this year that the final calorimeter calibrations were in part missing, presumably lost in the
abyss of time, and the existing parts were not sufficiently well made. Hence, the calibration is being remade
aiming to improve timing, position- and energy-resolution. Figure \ref{fig:calocomp} shows the improvements made
for the energy resolution for a selected kinematic point. This kinematic point was chosen as an example due to the
fact that the old calibration displayed both a constant offset and an unnecessarily large spread. 

The goal of the calibration is to maximize visibility of the characteristic correlation between decay angle and
energy for the photons originating from $\pi^0$ decay. By knowing the angle it is easy to calculate the expected
energy of the photon. The difference between this calculated energy and the measured energy in the calorimeter
should for $\pi^0$ events be a Gaussian distribution, see Fig. \ref{fig:depion}. This proves a solid identification
of pion events.

\begin{figure}[hbt]
\center{\includegraphics[width=0.7\textwidth]{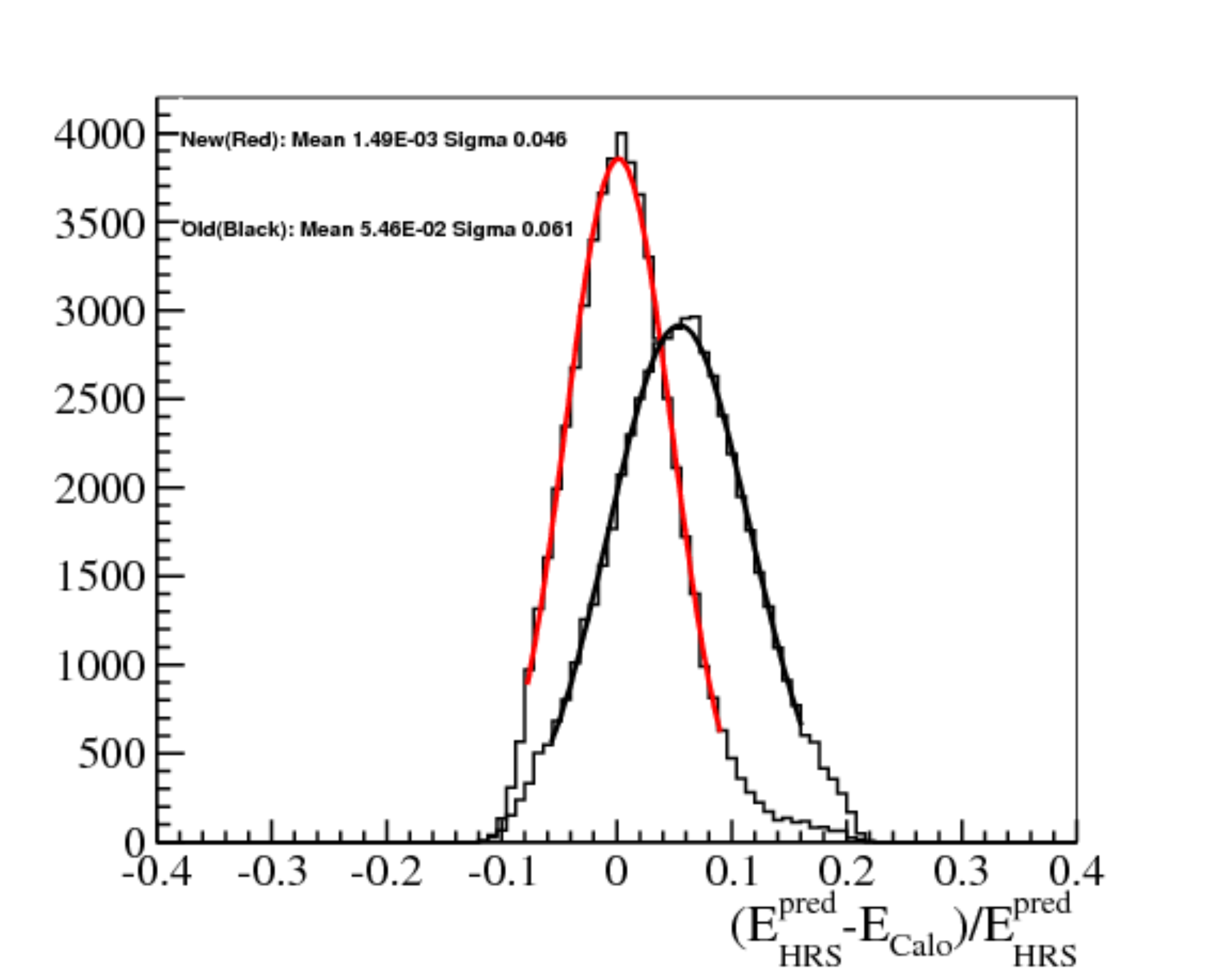}}
\caption[E99114: Energy calibration comparison]{ A comparison between the old energy reconstruction calibration
  and the remade one. $E_{HRS}^{pred}$ is the energy calculated from HRS variables assuming $ep$ scattering and
  $E_{Calo}$ is the energy measured by the calorimeter. }
\label{fig:calocomp}
\end{figure}

\begin{figure}[hbt]
  \center{\includegraphics[width=0.7\textwidth]{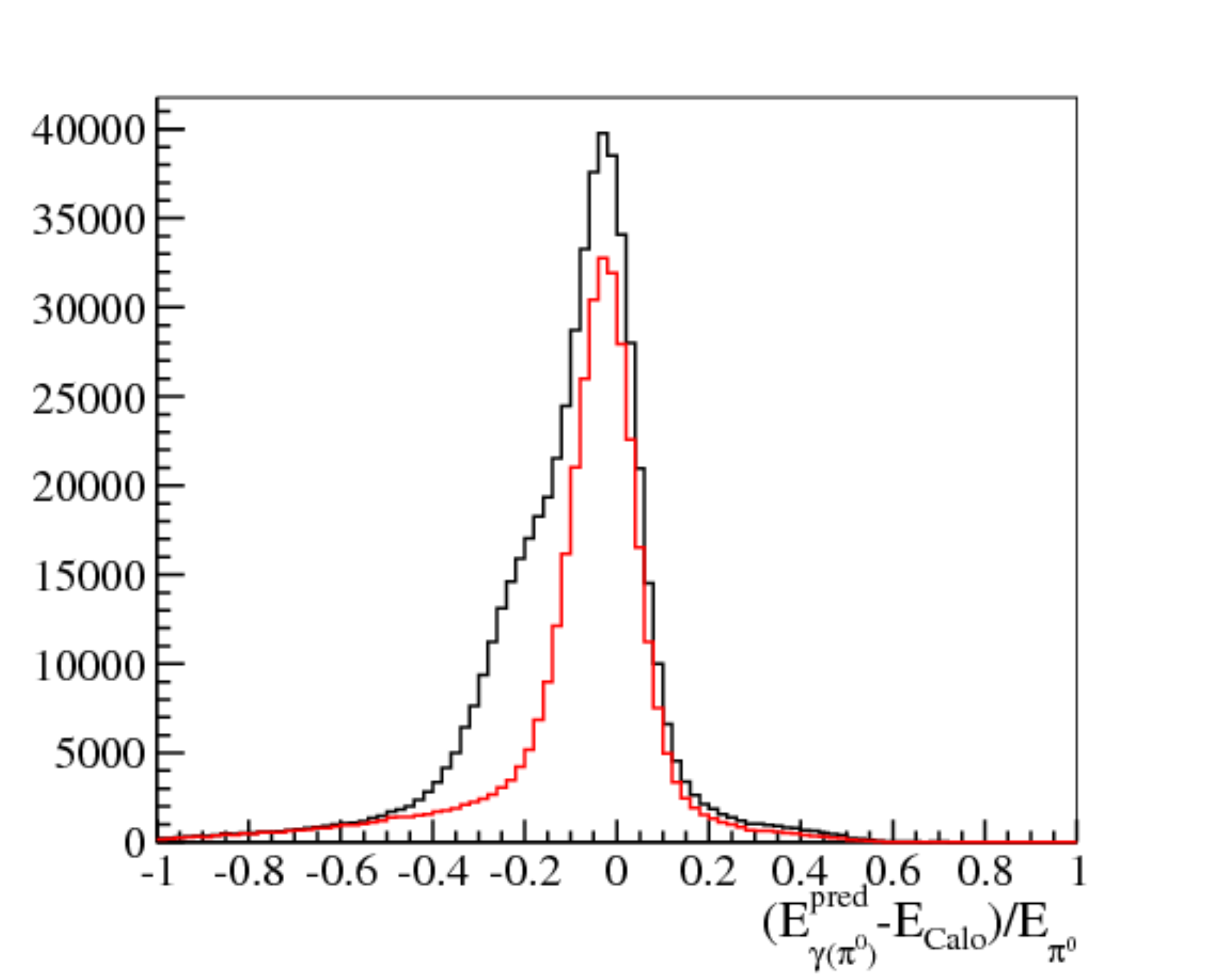}}
  \caption[E99114: dE distribution]{ The energy difference, between the energy measured by the calorimeter $E_{Calo}$ and
    the expected photon energy $E_{\gamma(\pi^0)}^{pred}$ calculated from the decay angle, as a fraction of the
    pion energy $E_{\pi^0}$. Red represents data with cuts to remove elastic electron events, black is without
    those cuts.}
  \label{fig:depion}
\end{figure}


\clearpage \newpage

\subsection[E04-007 - pi0]{E04-007 - $\pi^0$}

\begin{center}
\bf Precision Measurements of Electroproduction of $\pi^0$ near Threshold:\\
A Test of Chiral QCD Dynamics
\end{center}

\begin{center}
J.R.M.~Annand, D.W.~Higinbotham, R.~Lindgren, B.~Moffit, B.~Norum, V.~Nelyubin, spokespersons, \\
M. Shabestari and K. Chirapatimol, students\\
and\\
The Hall A Collaboration.\\
contributed by Richard Lindgren and Cole Smith.\\
\end{center}
 
 \subsubsection{Introduction}
The experiment is designed to measure the electroproduction reaction $p(e,e'p)\pi^0$ of neutral pions off the proton at the lowest possible invariant mass W. Results from previous electroproduction measurements at Mainz with four-momentum transfers of $Q^2$~=~0.10~GeV/c$^2$~\cite{Distler_98} and $Q^2$~=~0.05~GeV/c$^2$~\cite{Merkel_02} were in disagreement with the $Q^2$ dependence predicted by Heavy Baryon Chiral perturbation theory (HBChPT) and also inconsistent with the predictions of the MAID model~\cite{e04007:MAID}.  If the Mainz discrepancies remain unresolved, they will constitute a serious threat to the viability of Chiral Dynamics as a useful theory of low energy pion production.  Our experiment has measured absolute cross sections as precisely as possible from threshold to $\Delta W$~=~30~MeV above threshold at four-momentum transfers in the range from $Q^2$~=~0.050~GeV/c$^2$  to $Q^2$~=~0.150~GeV/c$^2$ in small steps of $\Delta Q^2$~=~0.01~GeV/c$^2$. This will cover and extend the Mainz kinematic range allowing a more sensitive test of chiral corrections to Low Energy Theorems for the S and P wave pion multipoles. In addition, the beam asymmetry was measured, which can be used to test predictions for the imaginary components of the of S wave $E_{0+}$ and  $L_{0+}$ pion multipoles, which are sensitive to unitary corrections above the $n \pi^+$ threshold.  Mainz recently repeated the electroproduction measurements and now report~\cite{Merkel_09} that the new results are more consistent with HBChPT predictions, but are in disagreement with their own previous measurements~\cite{Merkel_02}. In view of the importance of knowing whether or not HBChPT is valid in this domain, it is imperative that an independent set of measurements be reported.
  
\subsubsection{Experimental Results}
The experiment was performed using the Hall A Left High Resolution Spectrometer (LHRS) to detect the electron and the large acceptance BigBite spectrometer instrumented with  MWDC followed by $E-\Delta E$ scintillation paddles to detect and identify the proton.  A model independent partial wave analysis was made by fitting the measured cross section to a  Legendre polynomial series assuming s and p-wave dominance. The total cross section $\sigma_{0}$ was extracted from the fit and is plotted as orange squares in Figure 1 as a function of $Q^2$= from 0.047~GeV$^2$ to 0.105~GeV$^2$ for $\Delta W$~=~from 0.5 MeV to 11.5 MeV, which extends the kinematic range previously explored in threshold electroproduction.  The size of the orange square data point is typical of the statistical error and excludes the systematic error at this time. The red curve is a HBChPT calculation using parameters taken from fitting the old Mainz $Q^2$=0.1~GeV/c$^2$ data \cite{Distler_98}. The green curve and and the blue curves are calculations from the phenomenological models DMT and MAID07. The latest Mainz results  below 4 MeV~\cite{Merkel_09} shown as green points in Figure 1 are in agreement with our data. The HBChPT calculation~\cite{Wang_12}  fits the trend of the data below $Q^2$=  0.10~GeV$^2$ from threshold  up to  about $\Delta W$~=~11.5 MeV.  Above  $Q^2$=  0.10~GeV$^2$ the HBChPT calculation increases linearly with $Q^2$ substantially departing from the data. 

We are continuing to study improvements in the LHRS simulation to better understand our W resolution at threshold and the effects of target straggling and bin migration due to resolution smearing. We are also working to improve the VDC calibration and energy loss corrections in LHRS to achieve the best performance at threshold.  In order to get a handle on our systematic errors associated with electron detection with the LHRS such as angular acceptance, momentum calibration, pointing, charge integration, detector efficiency, and absolute beam energy and energy drifts, single arm measurements were made on carbon and tantalum foils before and after every change in kinematics. The carbon foil was enriched to 99.95$\%$ $^{12}$C and was 83.88 mg/cm$^2$ in thickness. The angular distribution of elastic cross section measurements on carbon with moderately tight cuts on the vertical acceptance $\Delta\theta_{tg}$ is shown in Figure 2. The data are compared with DWBA phase-shift calculations~\cite{Heisenberg_12} using an 18 term Fourier-Bessel charge density determined from fitting previous NIKHEF data~\cite{Offerman_91}  over a similar $Q^2$ range but at lower electron beam energies. The radiative corrected green data points were taken at an average beam energy of 1.192~GeV at angles of 12.5, 14.5, 16.5 and 20.5~degrees. The yellow points are the results of the DWBA calculation using a charge density fit to the NIKHEF data. A closer look at the small differences between our data and the calculation show that the differences average around 5$\%$ excluding the minimum where energy dependent dispersion effects may not be accounted for. This excellent agreement suggest that any systematic errors associated  with the central region of the acceptance, dead time, charge integration, detector efficiencies, etc. are of the order of 5\% or less on average.

\begin{figure}[htp]
\begin{center}
\includegraphics[scale=0.4]{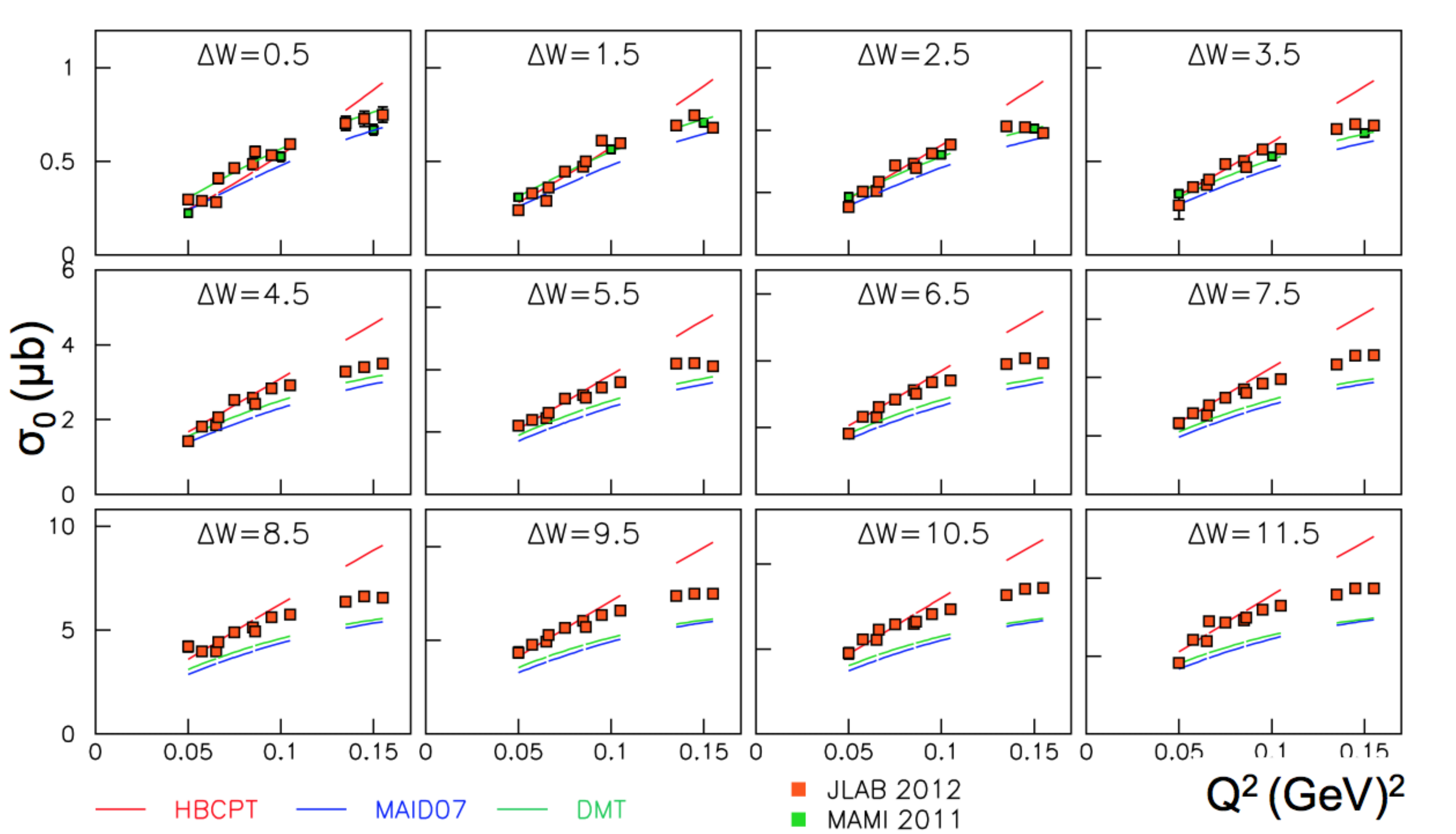}
\label{fig:asymmetry}
\caption{The total  $p\pi^0$ C.M. cross section, $\sigma_{0}$($\mu$b), is plotted versus
 $Q^2$(GeV$^2$). Each plot corresponds to an invariant mass bin $\Delta W$ measured in MeV relative to threshold. The orange points correspond to the JLAB 2012 data and the green points to the MAMI data points~\cite{Merkel_09}. The red, green, and blue curves correspond to calculations of expected results from different models.}
\end{center}
\end{figure}
 
\begin{figure}[htp]
\begin{center}
\includegraphics[scale=0.8]{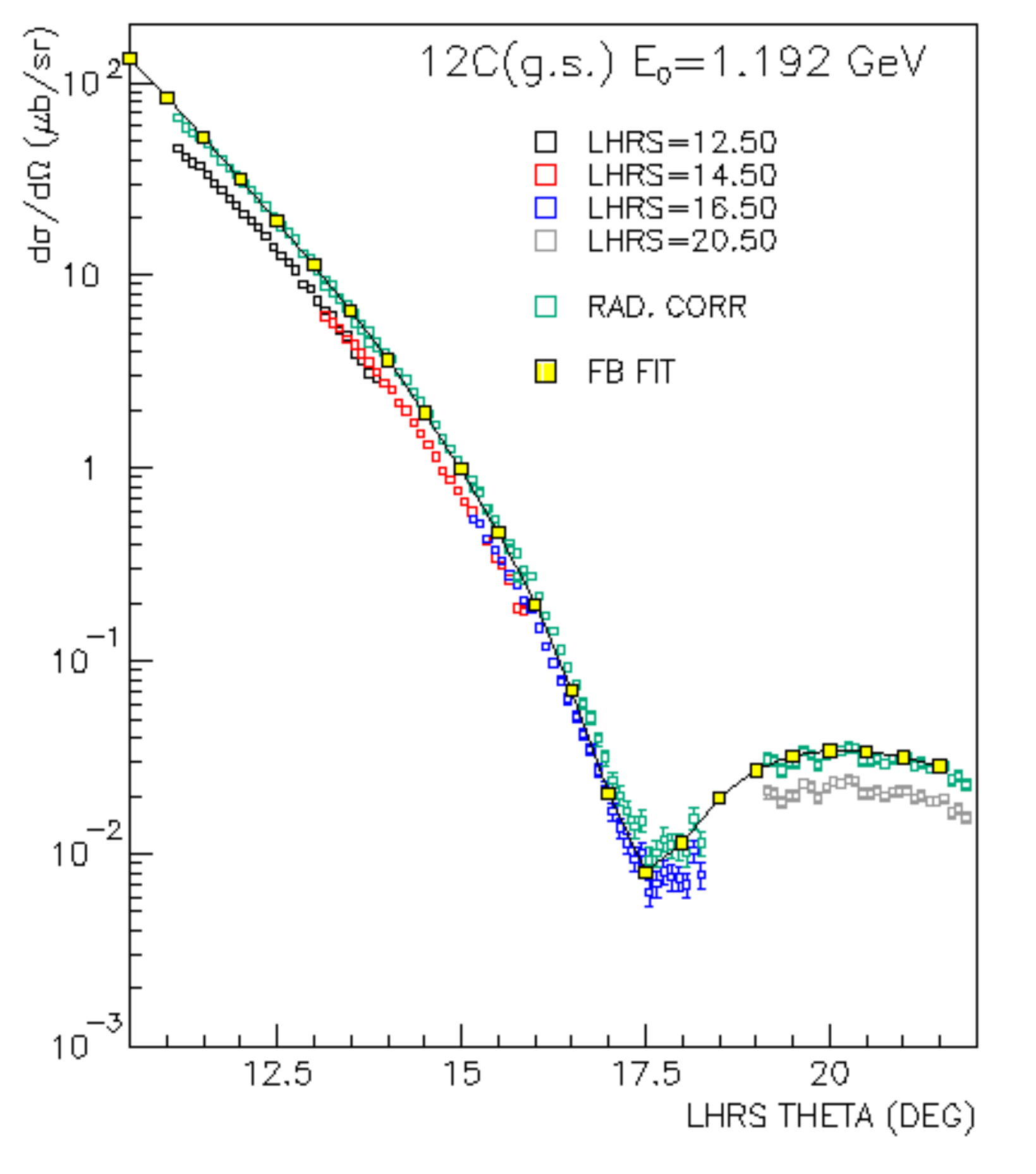}
\label{fig:cross section}
\caption{The elastic differential scattering cross section measurements and model calculations are plotted versus the LHRS Hall A spectrometer electron scattering angle. The black, red, blue, and gray points are absolute cross section measurements (uncorrected for radiation effects) after cuts on the vertical LHRS acceptance angle $\theta_{tg}$. The different colors represent the LHRS central angles of 12.5, 14.5, 16.5 and 20.5 degrees. The green points represent the radiation corrected measurements. The black curve and the yellow data points represent the DWBA calculation described in the text.}
\end{center}
\end{figure}

\clearpage \newpage

\subsection{E05-015 - $A_y$ in ${}^3\mathrm{He}$(e,e')}

\begin{center}
\bf Measurement of the Target Single-Spin Asymmetry in Quasi-Elastic $^{3}$He$^{\uparrow}(e,e')$
\end{center}

\begin{center}
T.~Averett, J.-P.~Chen and X.~Jiang, spokespersons, \\
and \\
the Hall A Collaboration.\\
contributed by Y.-W.~Zhang.
\end{center}

\subsubsection{Motivation}

In the past several years, nuclei and nucleon structures are often studied by measuring form factors using the Born approximation, which assumes one photon exchange neglecting the exchange of multiple photons. However, as new precision data on cross sections and polarization observables becomes available, the importance of two-photon exchange contributions cannot be ignored. For elastic scattering, $A_{y}$ is expected to be zero in the one-photon exchange approximation due to time-reversal invariance, but it can receive a non-zero contribution from the interference between the single-photon exchange amplitude and the imaginary part of the two-photon exchange amplitude. Although the importance of observing $A_{y}$ has been realized for many years, a non-vanishing $A_{y}$ has never been clearly established.

\subsubsection{The experiment}

JLab experiment E05-015 measured the neutron single spin asymmetry (SSA), $A_{y}^{n}$, formed by scattering unpolarized electrons from a polarized effective neutron target. The target polarization vector was normal to the electron scattering plane. The experiment used a longitudinally polarized electron beam at energies of 1.2, 2.4 and 3.6 GeV with an average current 12 $\mu$A. However, the polarized electron beam was summed over the two helicity states to achieve an unpolarized beam. The residual beam charge asymmetry was less than 100 ppm for a one-hour run.

The polarized target used in this experiment was a 40 cm long glass cell filled with 10.92 atm of $^{3}$He and a small amount (0.13 atm) of N$_{2}$ gas to reduce the depolarization effects. The target was polarized through spin-exchange optical pumping (SEOP)~\cite{Babcock} of a Rb-K mixture. In order to reduce systematic uncertainty, the target polarization vector was automatically flipped every 20 minutes using adiabatic fast passage (AFP). The polarization was monitored during each spin-flip using nuclear magnetic resonance (NMR) measurements. Electron paramagnetic resonance (EPR) measurements were also used periodically throughout the experiment in order to calibrate the NMR signal. The target polarization during the experiment is shown in Fig.~\ref{fig:target_pol}, an average in-beam target polarization of (51.4$\pm$0.4$\pm$2.9)$\%$ was achieved in this experiment.
\begin{figure}[hbt]
\begin{center}
\includegraphics[width=90mm]{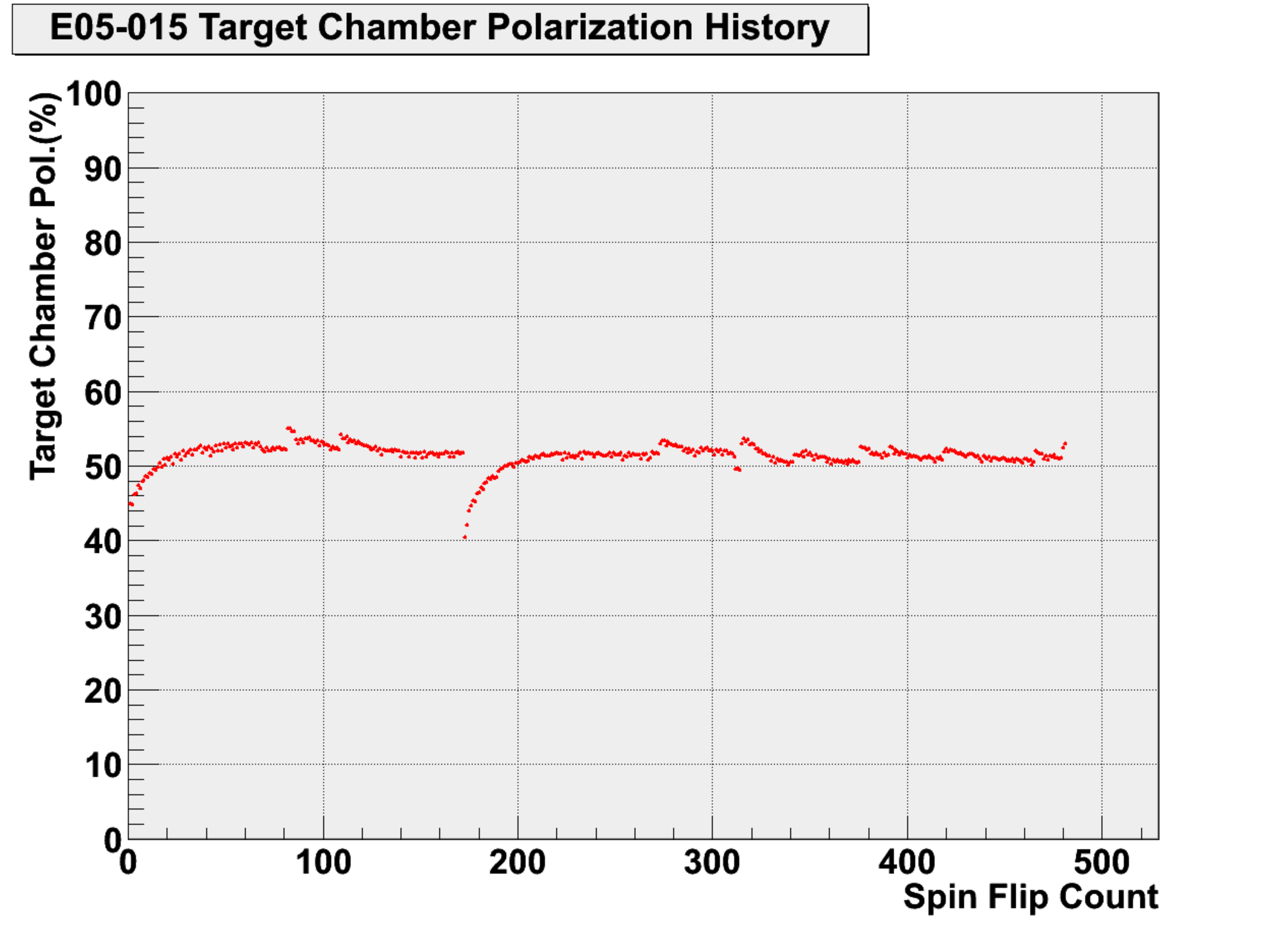}
\caption{E05-015: $^{3}$He target performance during the experiment.}
\label{fig:target_pol}
\end{center}
\end{figure}

The beam passed through the glass cell and was rastered with a 3 mm $\times$ 3 mm pattern to reduce the possibility of local beam-induced density changes in the target. Scattered electrons from the target were observed using the two Hall A high resolution spectrometers (HRS)~\cite{Alcorn}, left HRS at 17$^{\circ}$ beam-left and right HRS at 17$^{\circ}$ beam-right. Both spectrometers were configured to detect electrons in single-arm mode using nearly identical detector packages consisting of two dual-plane vertical drift chambers for tracking, two planes of segmented plastic scintillator for trigger formation, and a CO$_{2}$ gas Cherenkov detector and Pb-glass total-absorption shower counter for pion rejection. The left HRS and right HRS were synchronized during the experiment in order to increase the statistics and as a method to cross-check results. The kinematics settings during this experiment are provided in Table~\ref{table:kine}.

\begin{table}[ht]
\caption{Kinematics setting for the A$_{y}$ measurements. Listed are the central four-momentum transfer, $\langle$Q$^{2}\rangle$; beam energy, E$_{0}$; HRS central angle, $\theta$; the HRS central momentum, P$_{0}$.}
\begin{center}
\begin{tabular}{c|c|c|c}
$\langle$Q$^{2}\rangle$(GeV$^{2})$ & E$_{0}$(GeV) & $\theta$($^{\circ}$)  & P$_{0}$(GeV/$c$)\\\hline
0.13 & 1.245 & 17.0 & 1.176\\
0.46 & 2.425 & 17.0 & 2.181\\
0.96 & 3.605 & 17.0 & 3.086\\
\end{tabular}
\label{table:kine}
\end{center}
\end{table}

\subsubsection{Analysis progress}

The raw experimental asymmetries were extracted as a function of energy transfer, $\omega$. For each data bin, they were formed as
\begin{equation}
A_\mathrm{raw} = \frac{Y^{\uparrow} - Y^{\downarrow}}{Y^{\uparrow} - Y^{\downarrow}}
\label{eqn:A_raw}
\end{equation}
where the electron yield, $Y^{\uparrow(\downarrow)}$, is the number of electrons, $N^{\uparrow(\downarrow)}$, in the target spin-up (spin-down) state that pass all the PID cuts, normalized by accumulated charge, $Q^{\uparrow(\downarrow)}$, and DAQ live-time, $LT^{\uparrow(\downarrow)}$:
\begin{equation}
Y^{\uparrow(\downarrow)} = \frac{N^{\uparrow(\downarrow)}}{Q^{\uparrow(\downarrow)}\cdot LT^{\uparrow(\downarrow)}}
\label{eqn:Y}
\end{equation}

The raw asymmetry was then corrected for nitrogen dilution and target polarization. The nitrogen dilution factor is defined as
\begin{equation}
f_\mathrm{N_{2}}\equiv\frac{\rho_\mathrm{N_{2}}\sigma_\mathrm{N_{2}}}{\rho_\mathrm{^{3}He}\sigma_\mathrm{^{3}He}+\rho_\mathrm{N_{2}}\sigma_\mathrm{N_{2}}}
\label{eqn:fn}
\end{equation}
where $\rho$ is the density and $\sigma$ is the unpolarized cross-section. The nitrogen density was measured when filling the target cell and the cross-section was determined experimentally by electrons elastically scattered from a reference cell filled with a known quantity of N$_{2}$. The denominator was taken from the production data.

\subsubsection{Preliminary results}

The physics asymmetry A$_{y}^\mathrm{^{3}He}$ was finally obtained after corrections for radiative effects. Results for A$_{y}^\mathrm{^{3}He}$ are shown in Fig.~\ref{fig:result_he3_q2} for all three kinematic settings of the experiment and A$_{y}^\mathrm{^{3}He}$ as a function of $\omega$ for all three kinematic settings of the experiment are shown in Fig.~\ref{fig:result_he3_bin}. The error bars on the data points are statistical only with the total experimental systematic error indicated as an error band. The systematic uncertainty in A$_{y}^\mathrm{^{3}He}$ includes contributions from inconsistencies of LHRS and RHRS results, live-time asymmetry, target polarization, target misalignment, target density fluctuation, nitrogen dilution and radiative corrections.
\begin{figure}[hbt]
\begin{center}
\includegraphics[width=90mm]{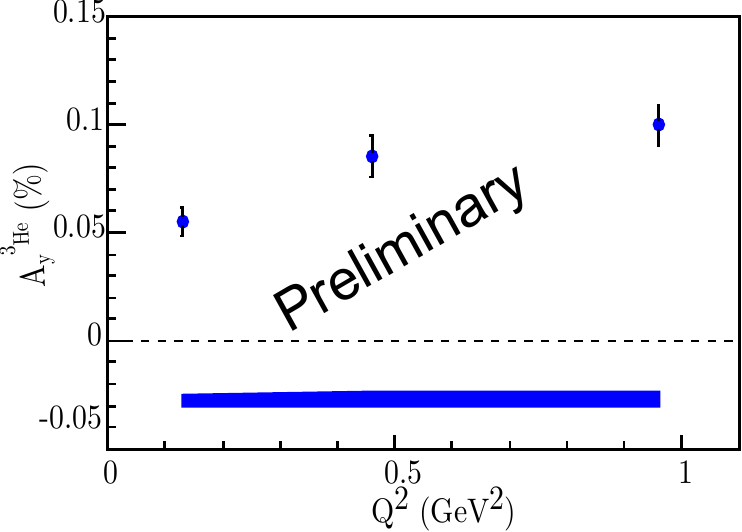}
\caption{\label{fig:result_he3_q2} The vertical asymmetry of polarized $^{3}$He at $\langle$Q$^{2}\rangle$=0.13, 0.46 and 0.96 GeV$^{2}$.}
\end{center}
\end{figure}
\begin{figure}[hbt]
\begin{center}
\includegraphics[width=90mm]{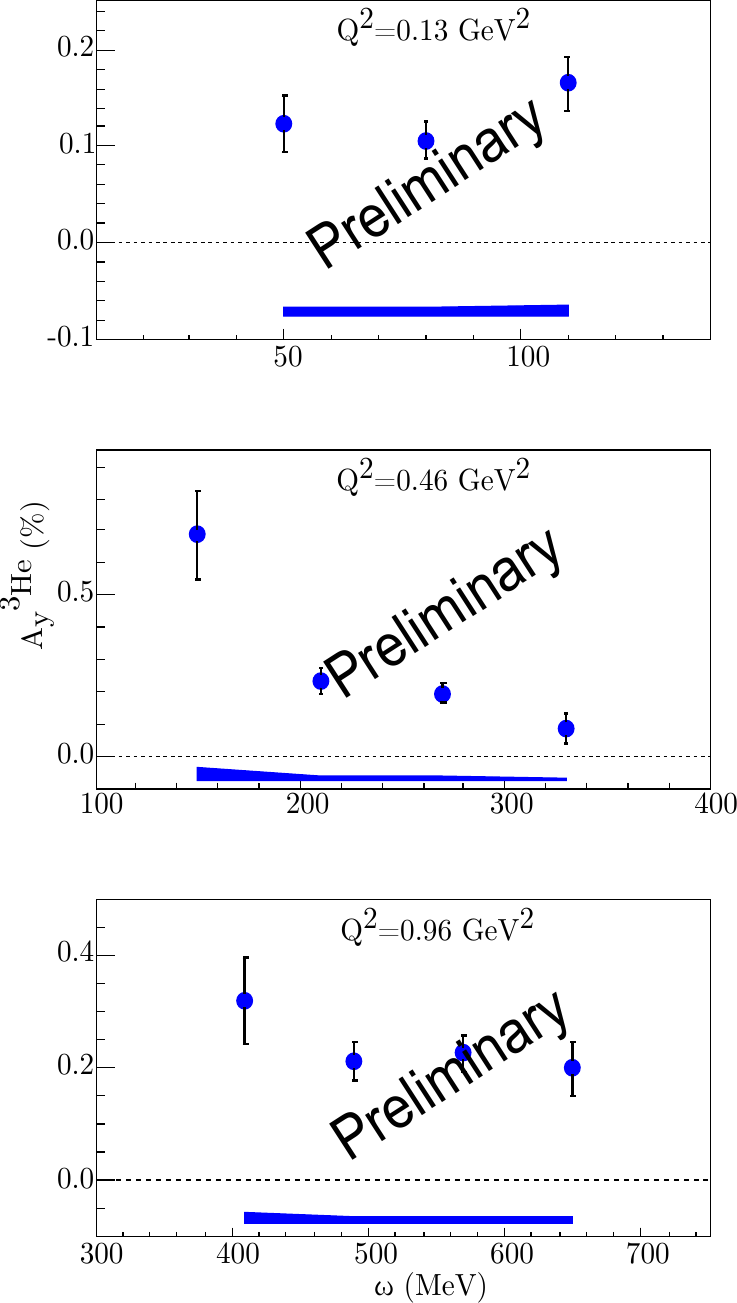}
\caption{\label{fig:result_he3_bin} The vertical asymmetry of polarized $^3$He as a function of $\omega$ at $\langle$Q$^{2}\rangle$=0.13, 0.46 and 0.96 GeV$^{2}$.}
\end{center}
\end{figure}

\subsubsection{Remaining tasks}

 Two main tasks remain before publication. First the data must also be corrected for proton dilution if we are to show $A_{y}^{n}$ rather than $A_{y}^\mathrm{^{3}He}$. The neutron asymmetry, $A_{y}^{n}$, can be extracted from the $^{3}$He asymmetry using the effective polarization approximation, given by
\begin{equation}
A_{y}^{n}=\frac{1}{(1-f_{p})P_{n}}(A_{y}^\mathrm{^{3}He}-f_{p}P_{p}A_{y}^{p})
\label{eqn:pdilution}
\end{equation}
where the proton dilution factor $f_{p}=2\sigma_{p}/\sigma_\mathrm{^{3}He}$ . The effective neutron and proton polarizations in $^{3}$He are given by P$_{n}$=0.86$_{-0.02}^{+0.036}$ and P$_{p}$=-0.028$_{-0.004}^{+0.009}$~\cite{Zheng}, respectively. If we know $f_{p}$ and $A_{y}^{p}$, we can extract $A_{y}^{n}$.
Second, the systematic uncertainties budget must be finalized. Generating reliable estimates for most systematics will be straightforward for this experiment.



%
%

\clearpage \newpage

\subsection{E05-102 - $A_x$, $A_z$ in ${}^{3}\mathrm{He}$(e,e'd)}

\begin{center}
\bf Measurement of $A_x$ and $A_z$ asymmetries in the quasi-elastic ${}^{3}\overrightarrow{\mathrm{He}}$(e,e'd) reaction
\end{center} 

\begin{center}
S.~Gilad, D.W.~Higinbotham, W.~Korsch, B.E.~Norum, S.~\v{S}irca spokespersons, \\
contributed by Miha Mihovilovic and Simon \v{S}irca.
\end{center}

The E05-102 experiment~\cite{e05102} is devoted to a detailed study of the
$^3\vec{\mathrm{He}}(\vec{\mathrm{e}} \mathrm{e}' \mathrm{d})$,
$^3\vec{\mathrm{He}}(\vec{\mathrm{e}},\mathrm{e}'\mathrm{p})\mathrm{d}$, 
$^3\vec{\mathrm{He}}(\vec{\mathrm{e}},\mathrm{e}'\mathrm{p})\mathrm{pn}$
processes at low $Q^2$.  The experiment in which double-polarization
(beam-target) transverse and longitudinal asymmetries in these
exclusive channels have been measured, has been performed in Summer 2009.
Its main purpose is to approach the ground-state
structure of the $^3\mathrm{He}$ nucleus by studying
the missing-momentum ($p_\mathrm{miss}$) dependence
of the asymmetries.  These dependences are rather strong;
in particular, it is seen that various components
of the $^3\mathrm{He}$ ground-state wave-function have
quite distinct signatures that become apparent only
with a sufficient lever arm in $p_\mathrm{miss}$.

\paragraph*{Theoretical input for acceptance averaging}

The bulk of the detector calibrations has been performed, including
the optics calibration of the BigBite spectrometer \cite{BBNIM}.
The data analysis is now at the stage when the theoretical results
are being averaged over the experimental acceptance,
and the largest progress has been made in the
$^3\vec{\mathrm{He}}(\vec{\mathrm{e}},\mathrm{e}'\mathrm{d})$ channel.
We have received the results from the Bochum/Krakow group
that span our whole accepted range of $p_\mathrm{miss}$ 
(discussed below), as well as preliminary results from
the Hannover/Lisbon group which are currently being extended
to a larger range of $p_\mathrm{miss}$.  Since these calculations
are extremely CPU-intensive, we rely on a $35$-point grid
in the $(E',\theta_\mathrm{e})$ plane to cover the complete
acceptance.  The grid corresponding to the data taken
at $E=2425.5\,\mathrm{MeV}$ and $\theta_\mathrm{e} = 12.5^\circ$
(central) is shown in Fig.~\ref{pEptheta}.

\begin{figure}[htbp]
\begin{center}
\includegraphics[width=9cm]{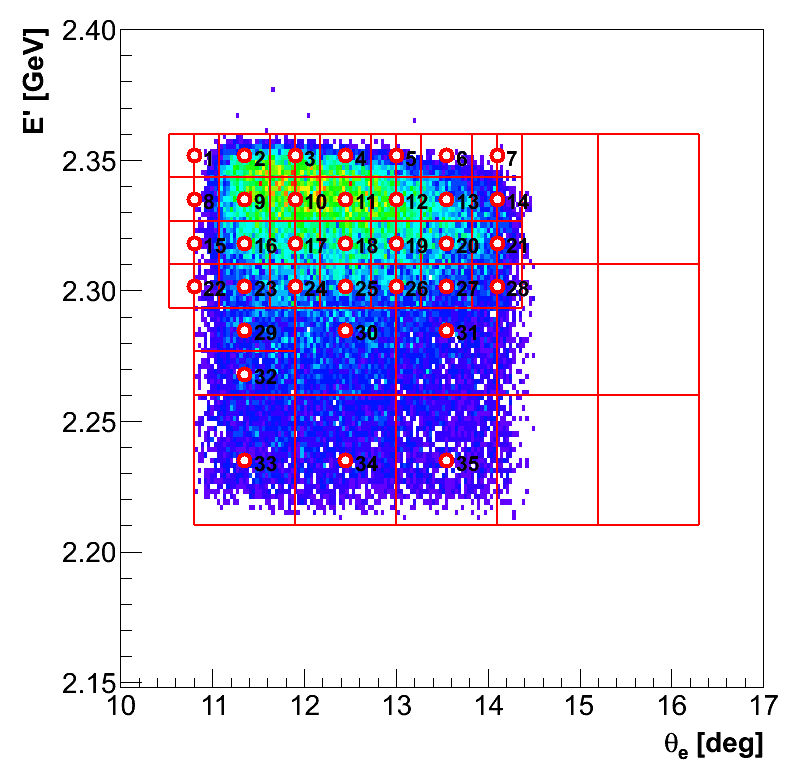}
\end{center}
\vspace*{-5mm}
\caption{Acceptance in $E'$ versus $\theta_\mathrm{e}$ for
$E=2425.5\,\mathrm{MeV}$ ($Q^2 \approx 0.25$), with the mesh
of $35$ kinematics points used in the averaging procedure.}
\label{pEptheta}
\end{figure}

Presently we are binning the theory in 25 $p_\mathrm{miss}$ bins
ranging from $6\,\mathrm{MeV}/c$ to $294\,\mathrm{MeV}/c$ to ensure
smooth averaging.  The theoretical asymmetries are available
only for the centers of each electron kinematics bin;
due to the kinematic restrictions posed by energy and momentum
conservation, it becomes physically impossible (in some bins)
to find a corresponding calculated asymmetry for all $p_\mathrm{miss}$.
On the other hand, real data span the whole bin and in general contain
all missing momenta, even such that are forbidden at the exact center
of the bin.  To prevent such acceptance mismatches that could skew
the resulting asymmetries, we consider only those data that lie
within the missing-momentum range accessible by the theory.

By using the codes provided by the theorists we have generated --- 
in each of the $35$ regions shown above --- the asymmetries
for each $p_\mathrm{miss}$ bin (or the corresponding
$\theta_{dq}$, which is equivalent) as functions of $\phi_{dq}$.
We use $25$ bins in $\phi_{dq}$ to cover the entire range
between $0^\circ$ and $360^\circ$.  This means that for each
electron kinematics the asymmetries at $625$ different
$(p_{\mathrm{miss}},\phi_{dq})$ points were determined.

\paragraph*{Interpolation in $(p_{\mathrm{miss}},\phi_{dq})$}

The asymmetries corresponding to intermediate values of
$p_\mathrm{miss}$ and $\phi_{dq}$, where the asymmetries
were not calculated, were then obtained by two-dimensional
interpolation between these $625$ points.  The same procedure
was used for each of the $35$ electron kinematics.  An example
of the interpolated 2D asymmetry for electron kinematics \#12
is shown in Fig.~\ref{figure2}.

\begin{figure}[!ht]
\begin{center}
\includegraphics[width=0.50\textwidth]{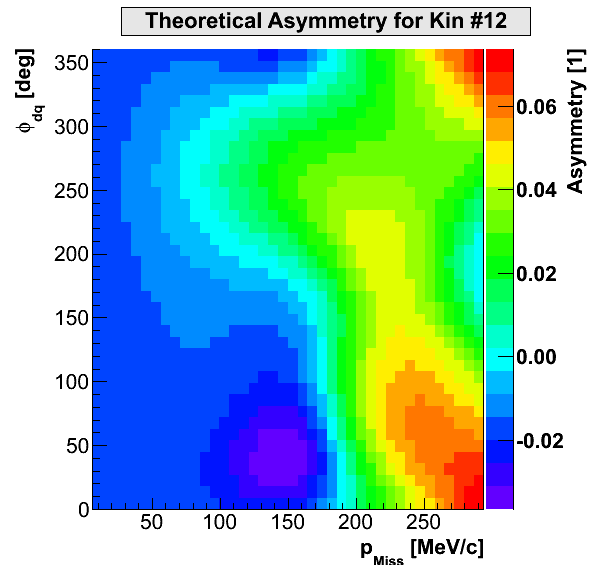}
\vspace*{-3mm}
\caption{The theoretical asymmetry $A(\theta^{*}=73^\circ, \phi^{*}=0^\circ)$
as a function of $\phi_{dq}$ and $p_{\mathrm{miss}}$ obtained
by interpolation between the values computed at $625$
different $(p_{\mathrm{miss}},\phi_{dq})$ points (kinematics \#12).}
\label{figure2}
\vspace*{-5mm}
\end{center}
\end{figure}

\paragraph*{Averaging the theoretical asymmetries}

The averaging of the theoretical asymmetries over the kinematic
acceptance of the electron spectrometer was then performed
on an event-by-event basis.   Each event was assigned to
one of the $35$ possible boxes in the 
$(E',\theta_\mathrm{e})$ plane shown in Fig.~\ref{pEptheta}.
Once the electron kinematics of an event was fixed,
the measured values of $p_{\mathrm{miss}}$ and $\phi_{dq}$
were used to pick the corresponding value of the interpolated
asymmetry.  The obtained value was then stored in a histogram 
(see Fig.~\ref{figure3}~(left)). 

\begin{figure}[!ht]
\begin{center}
\includegraphics[width=1\textwidth]{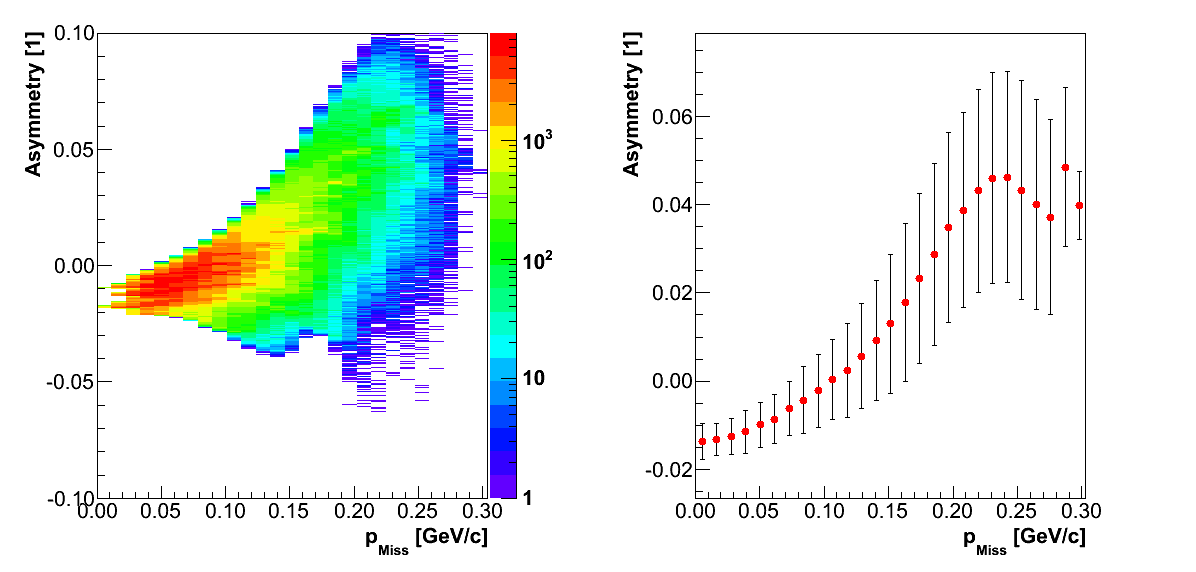}
\vspace*{-3mm}
\caption{ [Left] The distribution of events in terms of
the predicted asymmetry $A(\theta^{*}=73^\circ, \phi^{*}=0^\circ)$
and $p_{\mathrm{miss}}$. [Right] Acceptance-averaged theoretical
asymmetry as a function of $p_{\mathrm{miss}}$.
The mean value of the asymmetry for each $p_{\mathrm{miss}}$
has been obtained by calculating the weighted average
of the asymmetries in the 2D histogram in the corresponding
$p_{\mathrm{miss}}$ bin.}
\label{figure3}
\vspace*{-5mm}
\end{center}
\end{figure}

The final acceptance-averaged theoretical asymmetries for each
$p_{\mathrm{miss}}$ bin and their variation were then obtained
by evaluating a weighted average of the contents of the 2D histogram
(a sort of weighted projection on the $p_{\mathrm{miss}}$ axis).
An example is shown in Fig.~\ref{figure3}~(right).
The error bars on the asymmetries are not a measure
of the theoretical uncertainty.  Rather, they reflect
the scatter of data in any given $p_{\mathrm{miss}}$ bin:
the population of the asymmetry is more condensed
at low $p_{\mathrm{miss}}$ (red blob in the 2D histogram)
while it spreads out widely at high $p_{\mathrm{miss}}$
(green and blue regions).

\paragraph*{Work in progress}

Apart from obtaining the results from further theory groups,
we are presently striving to devise a smoother initial
$(E',\theta_\mathrm{e})$ configuration so that a multi-dimensional
interpolation (not only in the hadron momenta and angles
but also in the electron variables) would be possible,
thus allowing us to exploit the full statistics.

Although the E05-102 experiment was dedicated to a precision
measurement of the asymmetries, we have also been working
on the extraction of the absolute cross-sections.
This requires a rather intricate simultaneous use of MCEEP
and the existing theory results on the discrete grid,
which is work in progress.
We are also trying to incorporate both proton knock-out channels, 
$^3\vec{\mathrm{He}}(\vec{\mathrm{e}},\mathrm{e}'\mathrm{p})\mathrm{d}$ and
$^3\vec{\mathrm{He}}(\vec{\mathrm{e}},\mathrm{e}'\mathrm{p})\mathrm{pn}$,
since the beam energy was too high for a clean separation
of the two-body and three-body breakup channels.

\clearpage \newpage

\subsection{E06-002 - PREX}
\label{sec:e06002}

\begin{center}
\bf PREx: Precision Measurements of the Neutron Radius of Lead
\end{center}

\begin{center}
K.S.~Kumar, R.W.~Michaels, P.A.~Souder, G.M.~Urcioli, spokespersons, \\
and \\
the PREX Collaboration.\\
contributed by Seamus Riordan.
\end{center}

The PREX experiment completed in Spring 2010 with the purpose of measuring the RMS neutron radius of lead through the use of parity-violating electron scattering.  The results for this were published in early 2012~\cite{e06002:apvprl} and showed the existence of a neutron skin with a 95\% confidence level for the first time.  Plans to improve the precision on this measurement to $0.06~\mathrm{fm}$  have been approved for after the 12~GeV upgrade~\cite{PREX-II}.

The main parity-violating neutron radius results were presented in the previous annual report.  In the time since then, additional analysis has been performed on separate, ancillary measurements which are non-parity violating in nature and required to constrain the systematic error involving possible transverse asymmetry components.  This component, $A^m_n$, is described by
\begin{equation}
A^m_n = A_n \vec{P}_e \cdot \hat{k}
\label{e06002:eq:at}
\end{equation}
where $A_n$ is the maximal value of the asymmetry, $\vec{P}_e$ is the electron polarization vector, and $\hat{k}$ is the vector normal to the electron scattering plane.

Such components are introduced if the beam is not entirely longitudinally polarized.  Because this asymmetry is non-parity violating, it is in principle different from the desired measured quantity and could also become fairly large.  To minimize the effects from this, running the two high resolution spectrometers in Hall A in a symmetric configuration cancels out the azimuthal modulation given in Eq.~\ref{e06002:eq:at}.  However, even with this, it is necessary to perform direct measurements to provide a constraint on the size of the systematic effect.

The value of the transverse asymmetry is interesting by itself.  Due to time reversal symmetry, in the single photon exchange approximation it is identically zero, meaning that this becomes a probe of multi-photon exchanges.  Understanding such effects is crucial particularly in the study of nucleon form factors and are notoriously difficult to calculate due to the presence of the off-shell propagator in the multiphoton exchange diagrams.  Additionally, in doing such calculations, effects such as coulomb distortions must also be taken into account for nuclei of high $Z$.

Results from the PREX experiment and previous HAPPEX measurements were recently published for a variety of nuclei~\cite{e06002:atprl}.  It was observed that for three light nuclei, $\mathrm{H}$, ${}^4\mathrm{He}$, and ${}^{12}\mathrm{C}$ that the results were in good agreement with predictions using a dispersion relation approach~\cite{e06002:misha}.  However, for similar measurements on ${}^{208}\mathrm{Pb}$ there is a severe discrepancy, and the asymmetry was observed to be compatible with zero, Fig.~\ref{e06002:fig:atres}.  At present there is no definitive explanation for such an effect, though the calculations of~\cite{e06002:misha} do not include Coulomb distortions which grow rapidly with $Z$ and are critical in understanding the elastic parity-violating asymmetry on ${}^{208}\mathrm{Pb}$.

\begin{figure}
\begin{center}
\includegraphics[width=0.65\textwidth]{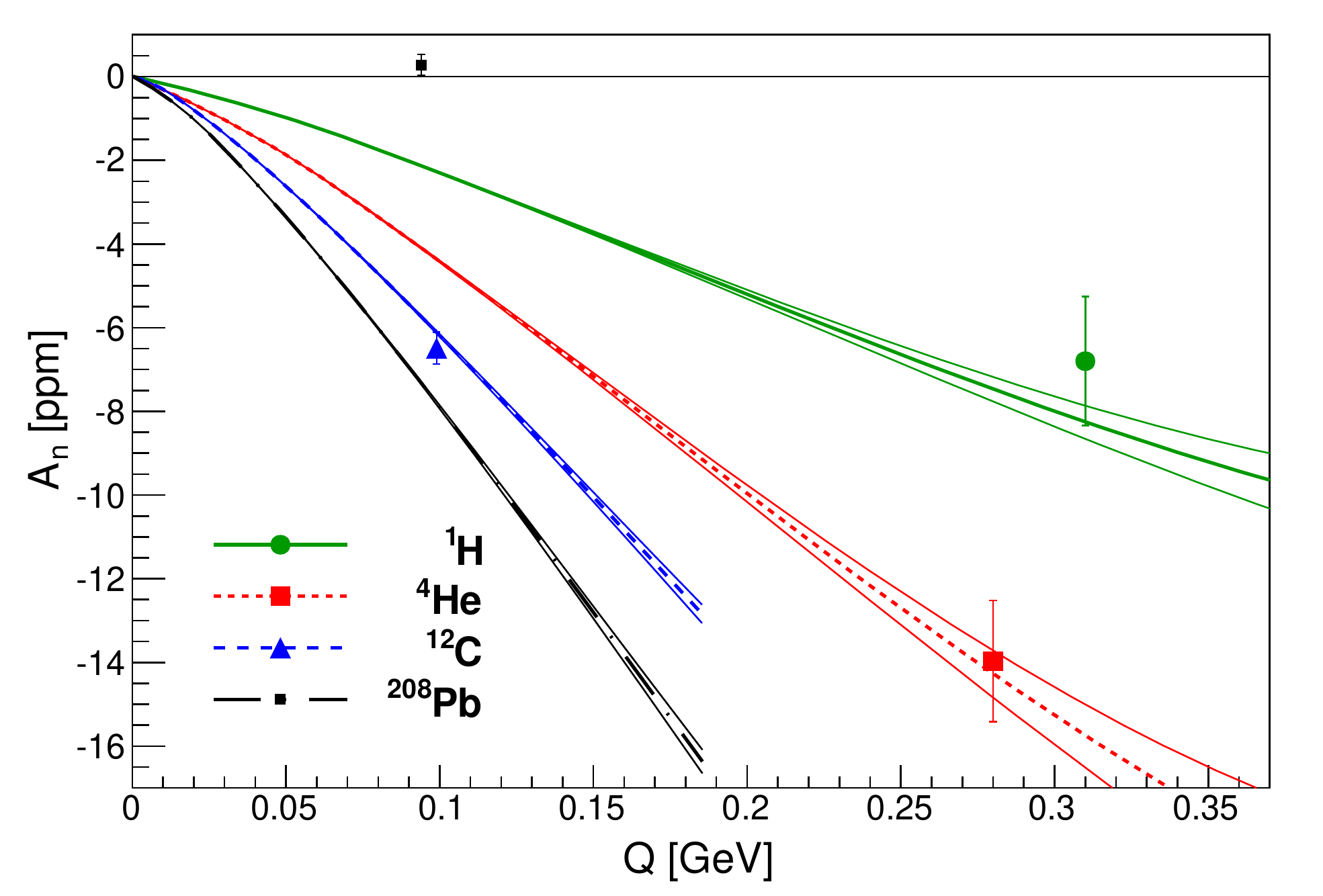}
\caption{Transverse asymmetry results from both the PREX and HAPPEX experiments on four different nuclei.}
\label{e06002:fig:atres}
\end{center}
\end{figure}

\clearpage \newpage

\setlength{\textwidth}{6.5in}
\setlength{\oddsidemargin}{0in}
\setlength{\evensidemargin}{0in}
\setlength{\textheight}{9in}
\setlength{\topmargin}{0in}
\setlength{\headheight}{0in}
\setlength{\headsep}{0in}
\newcommand{\Fig}[1]{(Fig.~\ref{fig:#1})}
\newcommand{\Figure}[1]{Figure~\ref{fig:#1}}
\newcommand{\SubFig}[1]{\subref{fig:#1}}
\newcommand{\Sect}[1]{(Sect.~\ref{sect:#1})}
\newcommand{\Section}[1]{Section~\ref{sect:#1}}
\newcommand{\Equation}[1]{Equation~\ref{eq:#1}}
\newcommand{\Table}[1]{Table~\ref{tab:#1}}
\newcommand{\HeliumThree}{\ensuremath{^3}He}
\newcommand{\NTwo}{\ensuremath{\textrm{N}_{2}}}
\newcommand{\Es}{\ensuremath{E_s}}
\newcommand{\Ep}{\ensuremath{E_p}}
\newcommand{\dtwon}{\ensuremath{\text{d}_{\text{2}}^{\text{n}}}}
\newcommand{\gOne}{\ensuremath{g_{1}}}
\newcommand{\gTwo}{\ensuremath{g_{2}}}
\newcommand{\AOne}{\ensuremath{A_{1}}}
\newcommand{\AOneN}{\ensuremath{A_{1}^{n}}}
\newcommand{\AOneHeThree}{\ensuremath{A_{1}^{^{3}\textrm{He}}}}
\newcommand{\SigRad}{\ensuremath{\sigma_{\textrm{rad}}}}

\subsection{E06-014 - $d_2^n$} \label{sect:e06014}
\begin{center}
   \bf A Precision Measurement of $d_2^n$: Probing the Lorentz Color Force 
\end{center}

\begin{center}
   S. Choi, X. Jiang, Z.-E. Meziani, B. Sawatzky, spokespersons, \\
   and                                                           \\
   the $d_2^n$ and Hall A Collaborations.                        \\
   Contributed by D. Flay, D.~Parno and M.~Posik.
\end{center}
\subsubsection{Physics Motivation} \label{sect:phys_mot}
\paragraph{$d_2^n$: Quark-Gluon Correlations in the Nucleon}

To date, extensive work has been done investigating the spin structure function $g_1$ within the 
context of the Feynman parton model and pQCD.  However, far less is known about the $g_2$ structure 
function.  It is known to contain quark-gluon correlations.  It follows from a spin-flip Compton 
amplitude and may be written as

\begin{equation} \label{eq:g_2}
g_2 \left( x, Q^2 \right) = g_2^{WW}\left( x, Q^2 \right) + \bar{g}_2\left( x, Q^2 \right), 
\end{equation}

\noindent where $g_2^{WW}$ is the Wandzura-Wilczek term, which may be expressed entirely in terms 
of $g_1$~\cite{ww}

\begin{equation} \label{eq:g_2ww}
   g_2^{WW}\left( x, Q^2 \right) = - g_1 \left( x, Q^2 \right) 
                                   + \int_x^1 \frac{g_1 \left( y, Q^2 \right)}{y} dy. 
\end{equation}

\noindent The second term is given as

\begin{equation} \label{eq:g_2bar}
   \bar{g}_2\left( x, Q^2 \right) = - \int_x^1 \frac{1}{y}\frac{\partial}{\partial y}  
                                    \left[ \frac{m_q}{M} h_T \left( y, Q^2 \right) 
                                    + \xi \left( y, Q^2 \right) \right] dy,  
\end{equation} 

\noindent where $h_T$ is the transverse polarization density, and $\xi$ is a term arising from quark-gluon 
correlations.  Here, $h_T$ is suppressed by the ratio of the quark mass $m_q$ to the 
target mass $M$.  Therefore, a measurement of $\bar{g}_2$ provides access to quark-gluon interactions inside 
the nucleon~\cite{rlj}. \*

Additionally, a measurement of both $g_1$ and $g_2$ allows for the determination of the quantity $d_2^n$, 
which is formed as the second moment of a linear combination of $g_1$ and $g_2$

\begin{equation} 
d_2^n \left( Q^2 \right) = \int_0^1 x^2\left[ 2g_1^n \left( x,Q^2 \right) + 3g_2^n \left( x, Q^2 \right)\right]dx                         
= 3 \int_0^1 x^2 \bar{g}_2^n \left( x,Q^2 \right) dx. \label{eq:d2n} 
\end{equation}

\noindent $d_2^n$ also appears as a matrix element of a twist-3 operator in the operator product expansion~\cite{fil_ji}

\begin{equation} \label{eq:d2pps}
   \langle P,S \mid \bar{\psi}_q \left( 0 \right) gG^{+y}\left( 0 \right) {\gamma}^{+} \psi_q \left( 0 \right) \mid P,S \rangle = 2MP^{+}P^{+}S^{x}d_2^n,
\end{equation}

\noindent where $G^{+y} = \frac{1}{\sqrt{2}} \left( B^{x} - E^{y} \right)$.  We see from Equations~\ref{eq:g_2bar}--\ref{eq:d2pps} 
that $d_2^n$ is a twist-3 matrix element that measures quark-gluon interactions.
 
Recent work has shown~\cite{mb_1,mb_2} that at high $Q^2$, $d_2^n$ is seen as a color Lorentz force averaged over 
the volume of the nucleon. This is given by the expression of the transverse (color) force on the active quark 
immediately following its interaction with a virtual photon

\begin{equation}
   F^{y} \left( 0 \right) \equiv - \frac{\sqrt{2}}{2P^{+}} 
                                 \langle P,S \mid \bar{\psi}_q \left( 0 \right) gG^{+y}\left( 0 \right) {\gamma}^{+} \psi_q \left( 0 \right) \mid P,S \rangle 
                          = - \frac{1}{2}M^2 d_2^n. \label{eq:tlcf}
\end{equation}

\noindent This theoretical interpretation reveals how $g_2$ and subsequently $d_2^n$ will allow us to examine 
the color interactions of the constituents inside the nucleon.

While bag and soliton model calculations of $d_2$ for the neutron yield numerical values consistent with those of lattice QCD, 
current experimental data differs by roughly two standard deviations (see the highest $Q^2$ data in \Figure{d2n_current}).  
One of the goals of our experiment is to improve the experimental error on the value of $d_2^n$ by a factor of four.  
It subsequently provides a benchmark test of lattice QCD calculations, shown in \Figure{d2n_current}.  

\begin{figure}[hbt]
 \begin{centering}
   \includegraphics[scale=0.5]{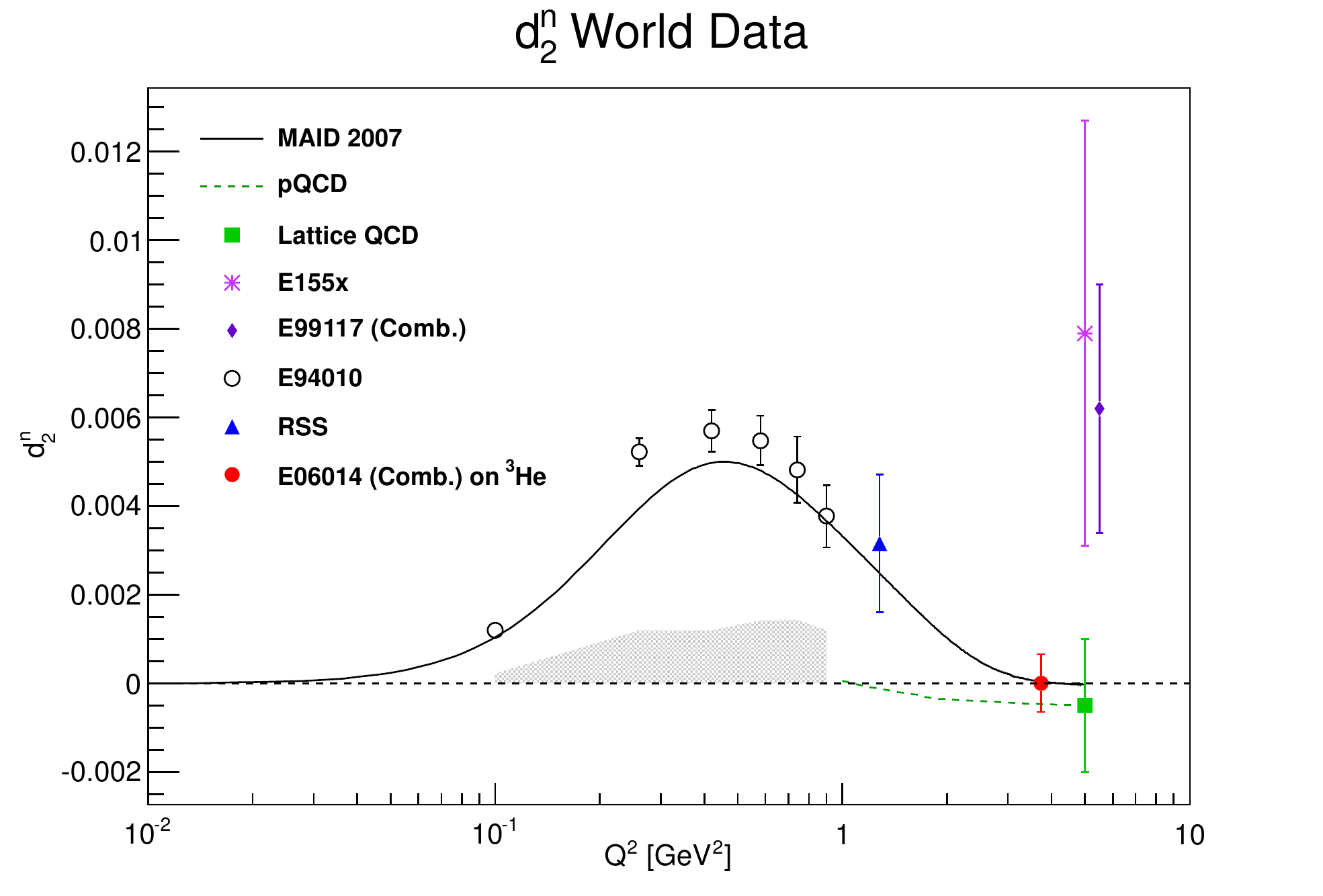} 
   \caption{\dtwon{} as a function of $Q^2$.  All the data shown with the exception of the SLAC E155x data are dominated by resonance
            contributions. E06-014 data will observe mostly the deep inelastic scattering (DIS) contribution. The projected error 
            on from E06-014~\cite{PAC29} is shown, along with the lattice QCD result~\cite{LQCD}. The dashed green curve shows the 
            pQCD evolution from the lattice point~\cite{pQCDEvol} based on the calculations of~\cite{ShuryakAndVain,JiAndChou}. Data from JLab 
            experiments E94-010~\cite{E94010} and RSS~\cite{RSS} are included in the plot. For comparison to the resonance contribution, 
            a MAID model~\cite{MAID} is plotted. Also plotted is the total $d_2$ from SLAC experiment E155x~\cite{E155x}.}
   \label{fig:d2n_current}
 \end{centering}
\end{figure}

\paragraph{$A_1$: The Virtual Photon-Nucleon Asymmetry}

Another quantity of interest is the virtual photon-nucleon longitudinal spin asymmetry $A_1$.  It provides insight into the 
quark structure of the nucleon and can be defined as

   \begin{equation}
      A_1\left( x, Q^2 \right) \equiv \frac{\sigma_{1/2} - \sigma_{3/2}}{\sigma_{1/2} + \sigma_{3/2}}, 
   \end{equation}
 
\noindent where the subscript 1/2 (3/2) gives the projection of the total spin of the virtual photon-nucleon system
along the virtual photon direction corresponding to the nucleon's spin anti-parallel (parallel) to the virtual photon. 
Constituent quark models (CQM) and pQCD models predict $A_1$ to be large and positive at large $x$.  
\Figure{A1n_current} shows the current world data compared to these models.  It is seen that the CQM (yellow band~\cite{CQM}) describes
the trend of the data reasonably well.  The pQCD parameterization with hadron helicity conservation (dark blue curve~\cite{pQCDHHC})---assuming 
quark orbital angular momentum to be zero---does not describe the data well.  However, the pQCD model allowing for quark 
orbital angular momentum to be non-zero (green curve~\cite{pQCDNoHHC}) describes the data well, pointing perhaps to the 
importance of quark orbital angular momentum in the spin structure of the nucleon.  

  \begin{figure}
      \centering
      \begin{subfigure}{0.5\textwidth}
         \centering
         \includegraphics[width=1.0\textwidth]{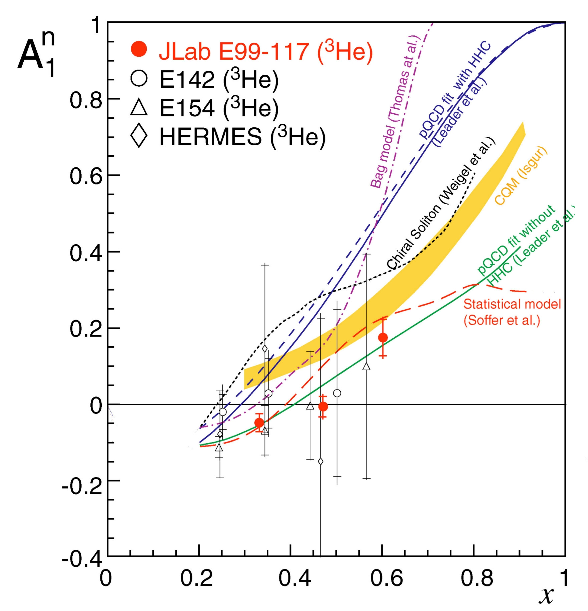}
         \caption{$A_1^n$}
         \label{fig:A1n_current}
      \end{subfigure}\begin{subfigure}{0.5\textwidth}
         \centering
         \includegraphics[width=1.0\textwidth]{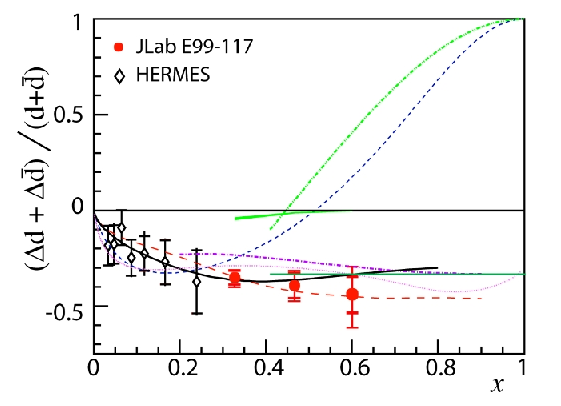}
         \caption{$\Delta d/d$}
         \label{fig:delta_d_current}
      \end{subfigure}

      \caption{Current data for $A_1^n$ and $\Delta d/d$. 
               (\subref{fig:A1n_current}): The current world data for the neutron $A_1$ from SLAC E143~\cite{E143} and E154~\cite{E154}  
               and HERMES~\cite{HERMES}, along with JLab E99-117~\cite{E99117}.  Also shown are CQM models and various 
               pQCD models; 
               (\subref{fig:delta_d_current}): the corresponding models and data 
               from HERMES and JLab for $\Delta d/d$.}
      \label{fig:A1_current}
   \end{figure}

Combining $A_1^n$ data measured on a polarized effective neutron target with $A_1^p$ data measured on a
polarized proton target allows access to $\Delta u/u$ and $\Delta d/d$. Recent results from Hall A~\cite{E99117} and 
from CLAS~\cite{CLAS} showed a significant deviation of $\Delta d/d$ from the pQCD predictions, which have that ratio 
approaching 1 in the limit of $x \rightarrow 1$ \Fig{delta_d_current}. As part of the 12 GeV program, two approved experiments 
(one in Hall A~\cite{E1206122} and one in Hall C~\cite{E1210101}) will extend the accuracy and $x$ range of this measurement, but a 
measurement of $A_1^n$ at the kinematics of this experiment (E06-014) will provide valuable support (or refutation) 
of prior JLab results, while producing additional input for theoretical models in advance of the coming experiments at 12 GeV.

\subsubsection{The Experiment} \label{sect:exp}

The experiment ran in Hall A of Jefferson Lab from February to March of 2009, with two beam energies of 
$E = 4.73$ and $5.89$ $\textrm{GeV}$, covering the resonance and deep inelastic valence quark regions, characterized 
by $0.2 \leq x \leq 0.7$ and $2 \textrm{ GeV}^2 \leq Q^2 \leq 6$ $\textrm{GeV}^2$. The coverage in the $x$ and $Q^2$ plane is 
shown in \Figure{coverage}. \*

 \begin{figure}[hbt]
  \begin{centering}
    \includegraphics[width=0.6\textwidth]{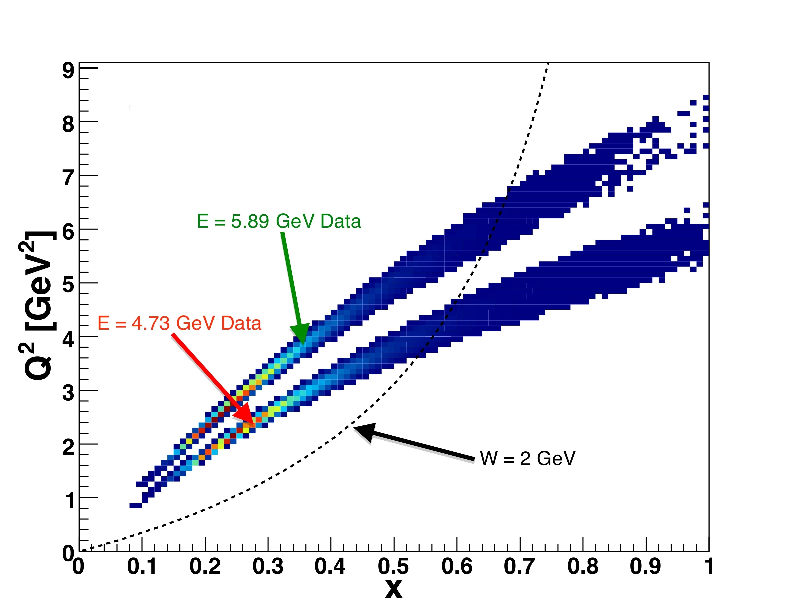}
    \caption{The E06-014 kinematic coverage in $Q^2$ and $x$. The lower band is the 4.73 GeV data set and the upper
             band is the 5.89 GeV data set. The black dashed line shows W = 2 GeV. The data to the left and right of this line 
             corresponds to DIS and resonance data, respectively.
    }
    \label{fig:coverage}
  \end{centering}
 \end{figure}

In order to calculate $d_2^n$, we scattered a longitudinally polarized electron beam off of a $^3$He target, in two polarization
configurations -- longitudinal and transverse.  $^3$He serves as an effective polarized neutron target since roughly $86\%$ of the 
polarization is carried by the neutron.  This is due to the two protons in the nucleus being primarily bound in a spin singlet 
state~\cite{jlf,fb_awt_ira}. \*

We measured the unpolarized total cross section $\sigma_0$ and the asymmetries $A_{\parallel}$ and 
$A_{\perp}$.  The cross section was measured by the Left High-Resolution Spectrometer (LHRS), while the asymmetries were measured 
by the BigBite Spectrometer.  The LHRS and BigBite were oriented at scattering angles of $\theta = 45^{\circ}$ to the left 
and right of the beamline, respectively.  \*

Expressing the structure functions entirely in terms of these experimental quantities, we have the expression for $d_2^n$

\begin{equation}
   d_2^n = \int_0^1 \frac{MQ^2}{4{\alpha}^2} \frac{x^2 y^2}{\left( 1 - y\right) \left( 2 - y\right)}{\sigma}_0
           \left[ \left( 3 \frac{1 + \left( 1 - y \right)\cos \theta}{\left( 1 - y \right)\sin \theta} + 
                  \frac{4}{y} \tan \left( \theta/2 \right)\right)A_{\perp} + \left( \frac{4}{y} - 3 \right)A_{\parallel} \right] dx,
\end{equation}

\noindent where $x = Q^{2}/2M\nu$, $\nu = E - E'$ is the energy transfer to the target, $E'$ is the scattered electron energy, and $y = \nu/E$ is the 
fractional energy transfer to the target.  The asymmetries are given by

\begin{eqnarray*}
A_{\parallel} = \frac{ N^{\downarrow \Uparrow} - N^{\uparrow \Uparrow}}{ N^{\downarrow \Uparrow} + N^{\uparrow \Uparrow} } 
\quad \textrm{and} \quad
A_{\perp} = \frac{ N^{\downarrow \Rightarrow} - N^{\uparrow \Rightarrow}}{ N^{\downarrow \Rightarrow} + N^{\uparrow \Rightarrow} }, 
\end{eqnarray*}

\noindent where $N$ is the number of electron counts measured for a given configuration of beam helicity (single arrows) and target 
spin direction (double-arrows). 

While $d_2^n$ was the main focus of the experiment, the measurement of the asymmetries allowed for the extraction of $A_1^n$, according to

        \begin{equation}
           A_1^n = \frac{1}{D\left( 1 + \eta\xi \right)}A_{\parallel}^n - \frac{\eta}{d\left( 1 + \eta\xi \right)}A_{\perp}^n,  
        \end{equation} 

\noindent where $D$, $\eta$, $\xi$ and $d$ are kinematic factors~\cite{Anselmino}.


\subsubsection{Beam Polarization}

E06-014 used a polarized electron beam at energies of 4.73 and 5.89 GeV. The polarization of the electron 
beam was measured independently through Compton and M\o ller scattering. During the running of E06-014, 
there were several M\o ller measurements performed while Compton measurements were taken continuously 
throughout the experiment. Figure~\ref{fig:beam_pol} shows the beam polarization as a function of BigBite 
run number for the M\o ller and Compton results. The beam polarization data was split into four run sets 
and the average polarization for each run period was then computed by taking into account both the Compton 
and M\o ller data. The final beam polarizations can be seen in Table \ref{tab:beam_pol}~\cite{parno_thesis}.

\begin{table}[h!]
\centering
\begin{tabular}{|c|c|c|c|c|}
\hline
 \textrm{Run Set} & \textrm{Beam Energy (GeV)} & \textrm{$P_e$ from Compton} & \textrm{$P_e$ from M\o ller} & \textrm{Combined $P_e$} \\ \hline
\hline
1 & 5.90 & $0.726 \pm 0.018$ & $0.745 \pm 0.015$ & $ 0.737 \pm 0.012$ \\ \hline
2 & 4.74 & $0.210 \pm 0.011$ & -                 & $0.210 \pm 0.011$ \\ \hline
3 & 5.90 & $0.787 \pm 0.020$ & $0.797 \pm 0.016$ & $ 0.793 \pm 0.012$ \\ \hline
4 & 4.74 & $0.623 \pm 0.016$ & $0.628 \pm 0.012$ & $ 0.626 \pm 0.010$ \\ \hline
\end{tabular}
\caption[E06-014:Final beam polarization]{Final beam polarization for E06-014, corrected for beam fluctuations. 
         For run set 2 there was no M\o ller measurement. ~\cite{parno_thesis}}
\label{tab:beam_pol}
\end{table}

\begin{figure}[bt]
\center{\epsfig{figure=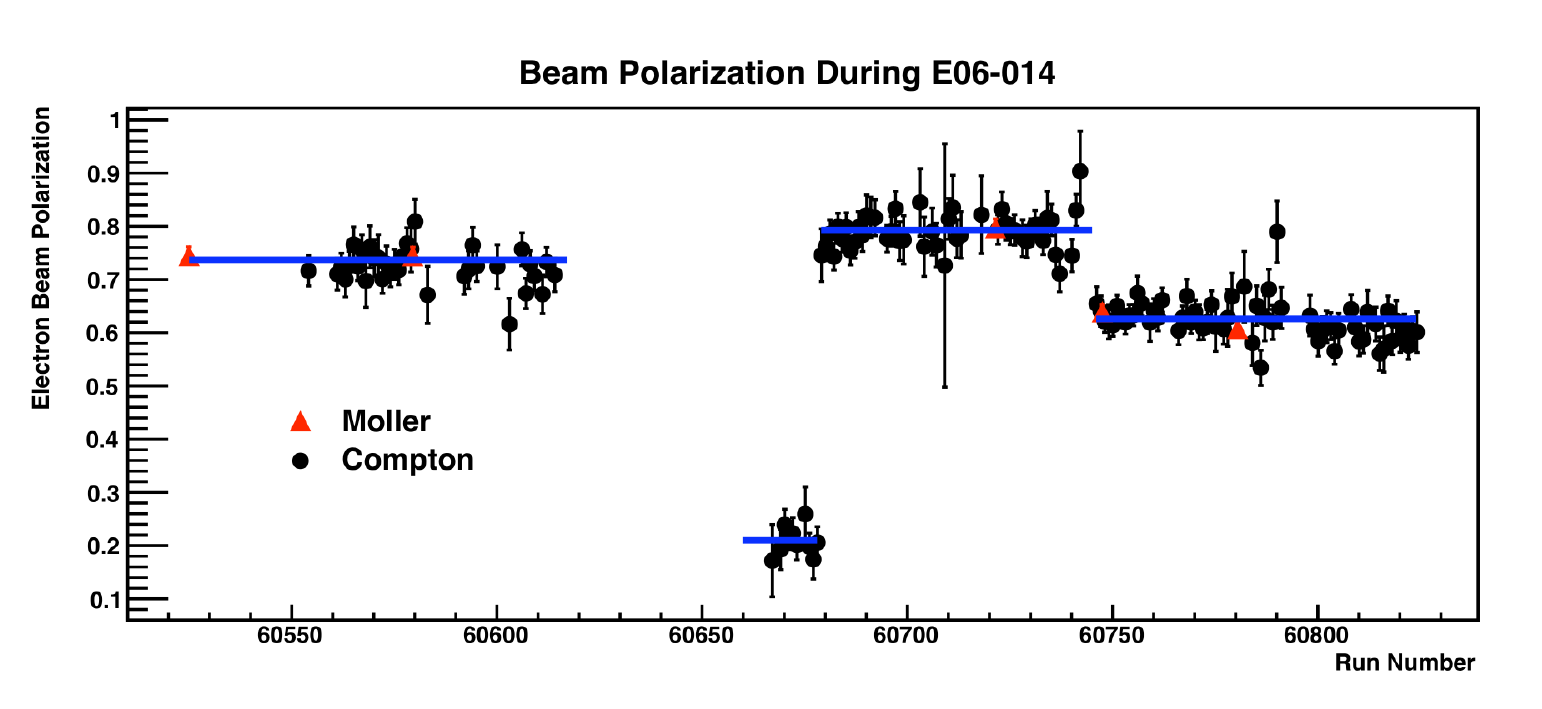,width=0.7\textwidth}}
\caption[E06-014:Beam polarization]{ Final electron beam polarization from M\o ller and Compton measurements for E06-014. 
         Note there was no M\o ller measurement for the second run set~\cite{parno_thesis}.}
\label{fig:beam_pol}
\end{figure}

\subsubsection{\HeliumThree{} Target Density} 

A complete understanding of the target density is essential, since the calculation of the target 
polarization from the EPR and NMR measurements depends on the $^3$He density. The number density of 
$^3$He was measured in both the pumping and the target chambers. This measurement was achieved by exploiting 
the fact that collisions with $^3$He atoms broaden the D1 and D2 absorption lines of rubidium~\cite{Kom}. 
The $^3$He number density at room temperature, $n_0$, can be obtained by measuring the width of the D1 
and D2 absorption lines and subtracting a 1\% N$_2$ contribution.

The full analysis to determine the \HeliumThree{} density may be found in~\cite{2011Report}.

\subsubsection{Polarized \HeliumThree{} Target} \label{sect:target}

Knowledge of the target polarization is crucial when performing a double-spin asymmetry experiment. 
E06-014 used the standard Hall A polarized $^3$He target with two holding field directions: longitudinal 
and transverse in plane, with respect to the electron beam direction.  The target polarization 
was extracted through electron paramagnetic resonance (EPR). The longitudinal polarization was cross 
checked using nuclear magnetic resonance (NMR) measurements. EPR measurements were taken every several 
days during the experiment, while NMR measurements were taken every few hours. 

\paragraph{EPR Calibration}

The frequency shift of potassium level transitions in the presence of $^3$He was measured using EPR.
This frequency shift $\Delta\nu_{EPR}$ can be related to the target polarization, $P_{^3\mathrm{He}}$

\begin{equation}
 \Delta\nu_{EPR} = \frac{4\mu_0}{3}\frac{d\nu_{EPR}}{dB}\kappa_0\mu_{^3\mathrm{He}}n_{pc}P_{^3\mathrm{He}},
 \label{eq:epr}
\end{equation}

\noindent where $\mu_0$ is the vacuum permeability; $\mu_{^3\mathrm{He}}$ is the magnetic moment; 
$\frac{d\nu_{EPR}}{dB}$ is the derivative of the EPR frequency with respect to the magnetic field; 
$\kappa_0$ is the enhancement factor, and $n_{pc}$ is the pumping chamber number density. EPR measurements 
give the absolute $^3$He polarization in the pumping chamber. However, it is the $^3$He polarization in the 
target cell that needs to be extracted. A polarization gradient model is used in order to determine the polarization 
between the two chambers. The change in polarization in the two chambers is given by

 \begin{equation}
     \frac{dP_{T}}{dt} = d_{P}\left(P_{T} - P_{P}\right)+\gamma_{SE}\left(P_{Rb} - P_{P} \right) - \Gamma_{P}P_{P},
     \label{eq:pt}
  \end{equation}

  \begin{equation}
     \frac{dP_{P}}{dt} = d_{T}\left(P_{P} - P_{T} \right)+ \Gamma_{T}P_{T},
     \label{eq:pp}
 \end{equation}

\noindent where $P_{T,P,Rb}$ is the polarization of the target chamber, pumping chamber $^3$He and rubidium atoms. 
$\Gamma_{T}$ is the depolarization rate of the $^3$He; $\gamma_{SE}$ is the spin exchange rate between $^3$He and 
rubidium atoms, and $d_{P,T}$ are diffusion constants that depend on the target cell geometries and $^3$He density. 
Taking the equilibrium solution, we obtain an expression that relates the polarizations between the two chambers

 \begin{equation}
   P_{T} = \frac{1}{1+\frac{\Gamma_{T}}{d_{T}}}P_{P}. 
   \label{eq:eq_diff}
 \end{equation} 

The calculated diffusion constant, $d_T$, is shown in \Table{dt} for both target spin directions.
The depolarization rate is a sum of various depolarization rates caused by different sources as shown in~\Equation{depol}. 
$\Gamma^{He} + \Gamma^{wall}$ are determined by measuring the target cell polarization live time. The depolarization 
rate due to the beam, $\Gamma^{beam}$, was found by using a model~\cite{beamdepol}. $\Gamma^{\nabla B}$ was calculated 
by measuring the gradient magnetic holding fields which polarize the target, and were found to be negligible. 
$\Gamma^{AFP}$ was also found to be negligible. \Table{depol} shows the results of the depolarization rates.

 \begin{equation}
   \Gamma_T = \Gamma^{He} + \Gamma^{wall} + \Gamma^{beam} + \Gamma^{AFP} + \Gamma^{\nabla B}
   \label{eq:depol}
 \end{equation} 

  \begin{table}
     \centering
     \caption{$d_T$ diffusion constant for both target spin directions.}
     \label{tab:dt}
     \begin{tabular}{|c|c|c|c|c|}
     \hline
     Parameter & Target Spin & Value & Units & Uncertainty [\%]\\
     \hline
     $d_T$ &  Long. & 0.892 & hour$^{-1}$ & 15.04 \\
     \hline
     $d_T$ & Trans. & 0.889 & hour$^{-1}$ & 15.06\\
     \hline
     \end{tabular}
  \end{table}

  \begin{table}
     \centering
     \caption{List of parameters used to calculate $\Gamma_{T}$}
     \label{tab:depol}
     \begin{tabular}{|c|c|c|c|c|}
     \hline
     Parameter & Value & Units & Uncertainty [\%]\\
     \hline
     $\Gamma^{He}+\Gamma^{wall}$ & 0.0714  & hour$^{-1}$ & 35    \\
     \hline
     $\Gamma^{beam}$             & 0.0794  & hour$^{-1}$ & 10.45 \\
     \hline
     $\Gamma^{AFP}$              & neg.    & hour$^{-1}$ & neg.  \\
     \hline
     $\Gamma^{\nabla B}$         & neg.    & hour$^{-1}$ & neg.  \\ 
     \hline
     \hline
     $\Gamma_{T}$                & 0.1508  & hour$^{-1}$ & 36.53\\ 
     \hline
     \end{tabular}
  \end{table}
  
During EPR measurements, a NMR measurement was done simultaneously, allowing us to calibrate NMR measurements during 
production with the EPR measurements by taking the ratio of the target polarization measured by EPR, $P_{T}$, and the 
measured NMR amplitude, $h$. A conversion factor $c'$ can then be formed that allows NMR measurements to be converted into an 
absolute $^3$He polarization.

After applying the $c'$ factor to all NMR measurements, a linear interpolation was done as a function of run time. 
This allowed the extraction of a target polarization on a run-by-run basis. The pumping and target chamber polarizations were 
extracted via EPR measurements, shown in \Figure{target_polarization}.
 
\begin{figure}[hbt]
\centering
\includegraphics[width=0.8\textwidth]{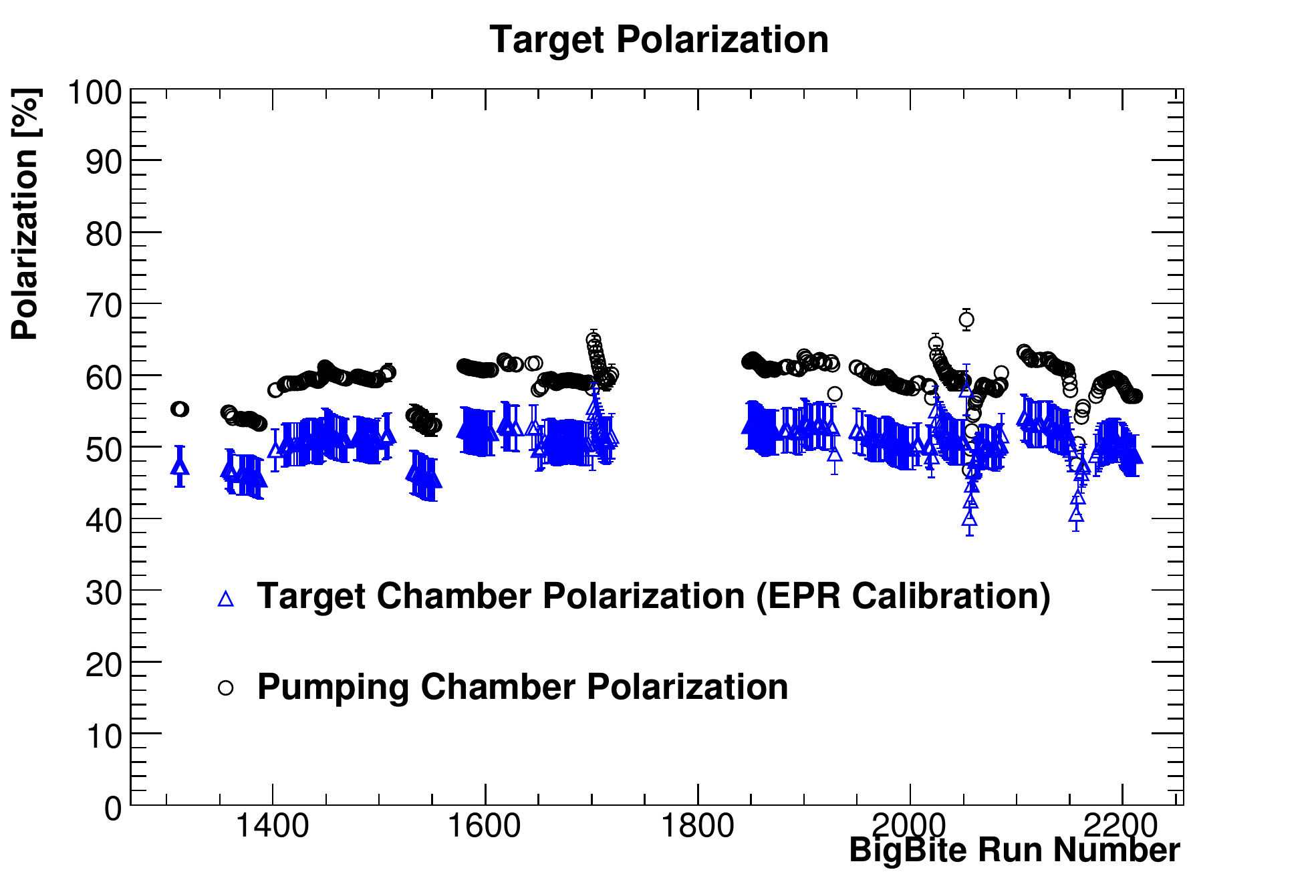}
\caption{$^3$He polarization in the pumping and target chambers. Some $^3$He polarization is lost while traveling between the two chambers.}
\label{fig:target_polarization}
\end{figure}

\paragraph{Water Calibration}

In addition to calibrating the NMR using EPR measurements, NMR measurements on a water sample can also 
be used to calibrate $^3$He NMR signals. The polarization of the protons in the water, when placed in a 
known magnetic field, can be solved exactly. The water polarization was measured by performing NMR 
measurements on a target cell filled with water. The water target cell was similar in geometry to the $^3$He 
filled cells. The water NMR signal was detected in two sets of pick-up coils that extended the length of the 
target (40\,cm) on both sides. A NMR cross-calibration factor needs to be applied when using the water calibration, 
because the NMR measurement for the water cell and $^3$He target cell took place in two different locations and 
the signals were measured in two different pick-up coil sets. The cross-calibration factor can be calculated by 
selecting a $^3$He target spin direction and then taking the ratio of the $^3$He NMR signal measured in the pick-up 
coils at the water cell position and the $^3$He NMR signal measured at the $^3$He target cell position during a 
production run. This could in principle be done for all target spin directions, longitudinal and transverse. 
Unfortunately, there was no transverse NMR measurements with the $^3$He in the water cell position; as a result, 
there is a large systematic uncertainty on the transverse target polarization. With this in mind, the water calibration 
for the longitudinal direction is used to cross-check the target polarization extracted from the longitudinal EPR calibration.      

Due to the fact that the polarization of water is small ($\approx 7 \times 10^{-9}$), a water polarization model was used in 
order to fit the water NMR signal and accurately extract the NMR signal height. The time evolution of the water 
polarization can be described by the Bloch equations given as

  \begin{eqnarray}
  \label{eq:blochx}
      \frac{dP_x(t)}{dt} &=& -\frac{1}{T_2}P_x(t) +\gamma\left(H(t)-H_0\right)P_y(t) +\frac{1}{T_2}\chi H_1,\\
  \label{eq:blochy}
      \frac{dP_y(t)}{dt} &=& -\gamma\left(H(t)-H_0\right)P_x(t) - \frac{1}{T_2}P_y(t) + \gamma H_1 P_z(t),\\
  \label{eq:blochz}
      \frac{dP_z(t)}{dt} &=& -\gamma H_1 P_y(t) - \frac{1}{T_1}P_z(t) +\frac{1}{T_1}\chi H(t),
  \end{eqnarray}

\noindent where $P$ is the water polarization in a particular direction; $t$ is the time; $T_1$ and $T_2$ are the 
longitudinal and transverse spin relaxation times; $H_0$ is the resonance field; $H_1$ is the transverse field component; 
$H(t) = H_0 + \alpha t$ is the field component along the z-axis; $\alpha = 1.2$ G/s is the field sweep speed; $\gamma$ 
is the gyro-magnetic ratio of the proton; $\chi = \frac{\mu_{p,H_2O}}{k_BT}$, with $\mu_{p,H_2O}$ being the magnetic moment 
of a proton in water; $k_B$ is the Boltzmann constant and $T$ is the target chamber temperature.    

Using the Bloch equations, an effective polarization, $P_{\mathit{eff}} = \sqrt{P_x^2 + P_y^2 + P_z^2}$, can be calculated and leads 
to the integral equation shown in~\Equation{peff}.  This equation was solved numerically using Mathematica.  However, 
an analytic function is needed to fit the water NMR signal, so approximations to $P_{\mathit{eff}}$ were made. \Figure{water_nmr} 
shows the water NMR fit results for 6,189 NMR sweeps.

  \begin{equation}
  \label{eq:peff}
  \centering
     P_{\mathit{eff}}(t) = e^{-(t-t_i)/T_1}\left[ P_{eq}(t_i)+\frac{1}{T_1} \int_{t_i}^t e^{(u-t_i)/T_1}P_{eq}(u) du\right]
  \end{equation}
 
    \begin{figure}[hbt]
     \begin{centering}
         \includegraphics[scale=0.5]{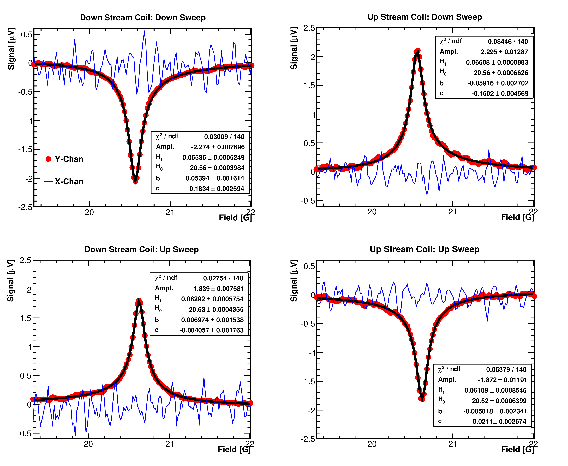}
         \caption{Presented are the sweep up and sweep down signals for the downstream and upstream coils. 
                  The Y lock-in channel is shown as red markers with water fit shown as a black line. 
                  The X lock-in channel is shown as a blue line.}\label{fig:water_nmr}
     \end{centering}
    \end{figure}
  
While the geometries of the water and $^3$He cells are similar, they are not identical. To correct for this discrepancy, 
the ratio of the magnetic flux through the $^3$He and water cells was calculated.  With this information, a water calibration 
constant can be formed, shown in~\Equation{water_const}.

  \begin{equation}
   \label{eq:water_const}
    c_{w} = \left(\frac{P_w}{S_{w}}\right)\left(\frac{G_w}{G_{He}}\right)\left(\frac{\mu_p}{\mu_{He}}\right)
            \left(\frac{n_p\Phi_w}{n_{He}^{pc}\phi_{He}^{pc} + n_{He}^{tc}\phi_{He}^{tc}}\right)\left( \frac{S^{He}_{pick-up}}{S^{He}_{prod.}}  \right),
  \end{equation} 

\noindent where $w(p)$ means water target (proton), $He$ means $^3$He target, $P$ is the polarization, $S$ is the NMR signal height 
and $\mu$ is the magnetic moment. $S_{pick-up}$ is the NMR signal with the $^3$He target measured at the pick-up coil location 
where the water NMR was done. $S_{prod.}$ is the NMR signal measured with the $^3$He in the production position. Applying this 
constant to the interpolated NMR measurements, a run-by-run $^3$He target polarization can be extracted. By comparing the 
longitudinal target polarizations extracted from the EPR and water calibrations, both methods were found to give consistent results. 

\subsubsection{The Left High-Resolution Spectrometer}
\paragraph{Unpolarized Total Cross Sections} \label{sect:uxs}

The Left High-Resolution Spectrometer (LHRS) was used to measure the unpolarized total cross section. 
The analysis for the extraction of the experimental cross section, \SigRad, for the E = 4.73 GeV 
and 5.89 GeV data sets is shown in~\cite{2011Report}. 

\paragraph{Radiative Corrections} \label{sect:rad_cor}

Electrons lose energy due to interactions with material.  This includes the material
before and after the target, and the target material itself.  These interactions will alter
the electron's {\it true} incident energy and also its {\it true} scattered energy.
This ultimately results in a different cross section than the true value.  These effects
are characterized by ionization (or Landau straggling) and bremsstrahlung.  There are also
higher-order processes at the interaction vertex that must be considered.  Collectively,
the correction of these effects is called {\it radiative corrections}.

A first correction that must be done {\it before} carrying out the radiative corrections
is to subtract the elastic radiative tail, since it affects all states of higher invariant mass 
$W$~\cite{MT}.  For these kinematics, the elastic tail is negligible and was not subtracted from the data.

The \HeliumThree{} quasi-elastic tail, however, has a larger contribution and needs to be subtracted.
The tail was built up from calculating the {\it elastic} tail of the proton and neutron using
ROSETAIL~\cite{Rosetail} and adding them together as $2p + n$, to account for two protons
and one neutron in \HeliumThree.  The systematic effect of the subtraction on the resulting
cross section, \SigRad, was $\leq$ 0.5\%.

In considering the effects mentioned above, the {\it measured} cross section is realized in
terms of a triple-integral

\begin{equation} \label{eq:triple-int}
   \SigRad\left(\Es,\Ep \right) = \int_0^T \frac{dt}{T} \int_{\Es^{min}}^{\Es} d\Es' \int_{\Ep}^{\Ep^{max}}d\Ep' 
                         I\left( \Es,\Es',t \right) \sigma_r\left(\Es',\Ep'\right) I\left( \Ep,\Ep',T-t \right),
\end{equation}

\noindent where \SigRad{} is the measured (radiated) cross section, $\sigma_r$ is the
{\it internally}-radiated cross section. \Es{} is the incident electron energy, \Ep{} is the
scattered electron energy.  $I\left(E_0,E,t\right)$ is the probability of finding an
electron with incident energy $E_0$ that has undergone bremsstrahlung with final energy $E$
at a depth $t$ inside a material~\cite{MT,Stein}.

In order to {\it unfold} the Born cross section, an iterative procedure is carried
out in RADCOR~\cite{RadCor}.  It amounts to an ``energy-peaking'' approximation, resulting 
in the calculation of

\begin{equation} \label{eq:unfold} 
   \sigma_b^i = \frac{1}{\textrm{C}}\left[ \sigma_{\textrm{rad}} 
              - \int \left( \ldots \right)\sigma_b^{i-1}dE_{s}' 
              - \int \left( \ldots \right)\sigma_b^{i-1}dE_{p}'\right],  
\end{equation}

\noindent where C and the two integrals are defined in Equation IV.2 in~\cite{MT}.
$\sigma_b^i$ is the Born cross section obtained for the $i^{\textrm{th}}$ iteration of
the code, \SigRad{} is the radiated cross section to be corrected.  $\sigma_b^i$ is then
re-inserted into equation for the next iteration. It was found that the calculation
converges within the first 3--4 iterations.  \Figure{BornXS} shows the preliminary Born 
cross sections.

In E06-014, we took data for only two \Es{} values of 4.73 GeV and 5.89 GeV.  However,
we need enough data to properly calculate the integrals above.  Therefore, we used a
suitable cross section model~\cite{Bosted} to fill in the rest of the phase space for
each data set.

   \begin{figure}
      \centering
	\begin{subfigure}{0.5\textwidth}
         \centering
         \includegraphics[width=1.0\textwidth]{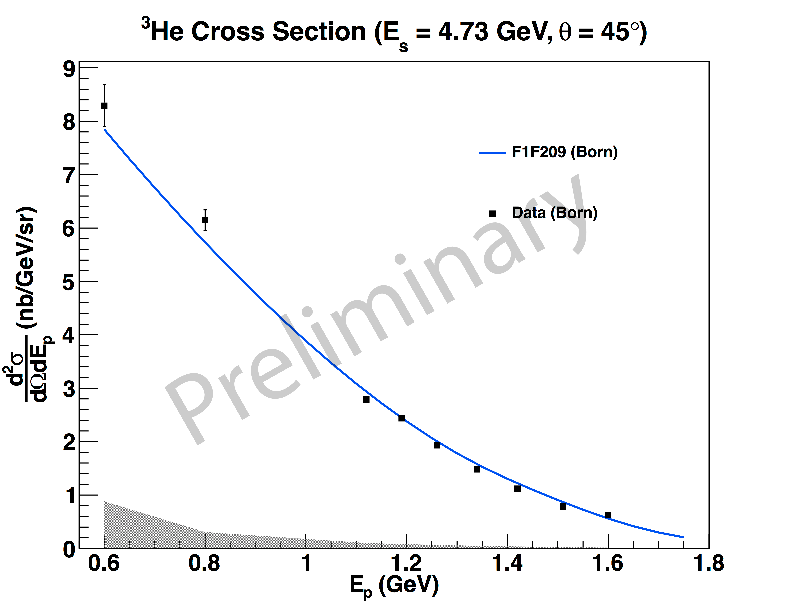}
	 \caption{E~=~4.73~GeV}
         \label{fig:born_xs_4}
       \end{subfigure}
	\begin{subfigure}{0.5\textwidth}
         \centering
         \includegraphics[width=1.0\textwidth]{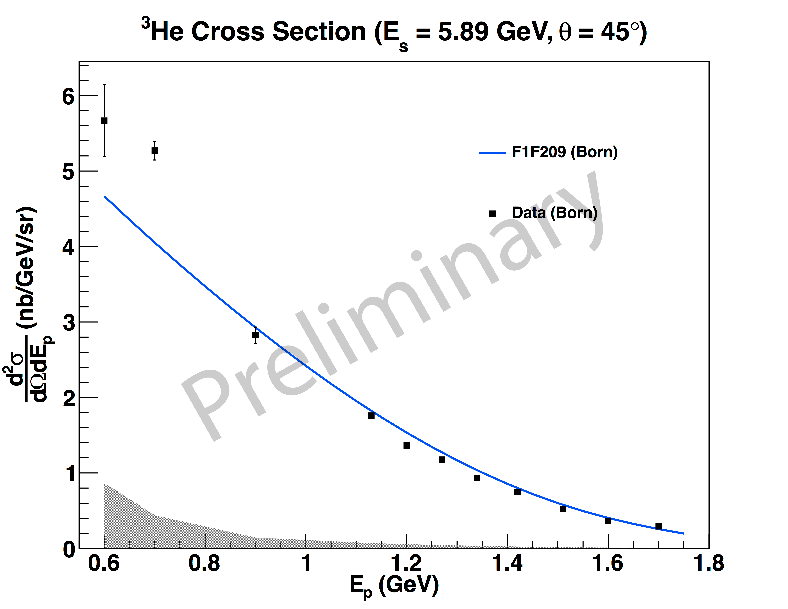}
	 \caption{E~=~5.89~GeV}
         \label{fig:born_xs_5}
      \end{subfigure}
      \caption{Preliminary results for the unpolarized Born cross sections, compared to P. Bosted and V. Mamyan's
               F1F209 model~\cite{Bosted} for E $=$ 4.73 GeV~(\subref{fig:born_xs_4}) and 5.89 GeV~(\subref{fig:born_xs_5}).  
               The error bars show the statistical errors, while the bands show the in quadrature sum of systematic errors from
               experimental cuts, radiative corrections and other uncertainties \Sect{XSErr}, and errors due to  
               the subtraction of background signals from nitrogen scattering and positron production in scattering from \HeliumThree. 
               The systematic errors and radiative corrections are still tentative.}  
      \label{fig:BornXS}
   \end{figure}

\paragraph{Systematic Errors} \label{sect:XSErr}

\Table{SysErr} shows the systematic errors determined from the data as compared
to the projected errors in the E06-014 proposal~\cite{PAC29}.  One large contribution
comes from the cuts on the target variables; the cut on the horizontal scattering angle,
$\phi$, contributes at the $\sim$ 2\% level.  This is not surprising since the Mott
cross section is most sensitive to this quantity.  Another large contribution
comes from the radiative corrections.  The source of this is due to the dependence
on the cross section model used and  how accurately we know the material thicknesses
in the electron's path before and after scattering.  These two radiative correction
errors combine for an error of $<$ 4\%.

The largest contribution to the systematic errors comes from the uncertainty in
subtracting the background signals in scattering from nitrogen and pair production in
scattering from \HeliumThree.  The presence of nitrogen in the cell is to prevent 
depolarization effects on \HeliumThree~\cite{Kom}.  The largest error from the background 
signals comes from the positrons, contributing at a level of $\approx$ 0.8 nb/GeV/sr at 
the lowest bin in \Ep. These errors are still tentative.  Adding these contributions in 
quadrature with the values listed in \Table{SysErr} gives the error bands shown in \Figure{BornXS}.

\begin{table}[hbt]
  \centering
  \small{\begin{tabular}{|c|c|c|}
            \hline
            {\bf Type} & {\bf Proposal (\%)}  & {\bf Experiment (\%)}   \\[0.5ex]
            \hline
            PID Efficiency                  & $\approx$ 1 & 1             \\ \hline
            Background Rejection Efficiency & $\approx$ 1 & 1             \\ \hline
            Beam Charge                     & $<$ 1       & $\approx$ 0.3 \\ \hline
            Acceptance Cut                  & 2--3        & 2.7           \\ \hline
            Target Density                  & 2--3        & 2.2           \\ \hline
            Dead Time                       & $<$ 1       & $<$ 1         \\ \hline
            Radiative Corrections           & $\leq$ 10   & $<4$          \\ \hline
         \end{tabular}
   }
   \caption{The systematic errors on the Born cross section compared to the estimates
            from the proposal.  The largest contributions here come from the radiative 
            corrections and the target cuts.  However, all values are within the limits 
            specified in the proposal.}
   \label{tab:SysErr}
\end{table}

\subsubsection{The BigBite Spectrometer}
\paragraph{The Double-Spin Asymmetries}

The BigBite spectrometer was used to measure the parallel and perpendicular double-spin asymmetries 
between longitudinally polarized electrons and a longitudinally or transversely polarized \HeliumThree{} 
target.  These asymmetries were then corrected for imperfect beam and target polarizations.  Corrections
were also made for dilution effects due to the presence of \NTwo{} in the target~\cite{Kom}.  The full 
details of this analysis may be found in~\cite{2011Report}.   

\paragraph{Positron Contamination Correction}

In addition to \NTwo{} contamination, pair-produced electrons from $\pi^{0}$ and $\gamma$ decay 
can also contaminate the asymmetry.  This is treated as a {\it dilution} to the electron asymmetry, 
and is corrected for by applying a dilution factor $D$: 

\begin{eqnarray}
   A_{\textrm{c}}^{e^{-}} &=& \frac{1}{D} A_{\textrm{m}}^{e^{-}}    \\
   D                      &=& 1 - \frac{ N_{e^{+}} }{ N_{e^{-}} } 
\end{eqnarray}  

\noindent where $A_{\textrm{c}}^{e^{-}}$ is the corrected electron asymmetry;  $A_{\textrm{m}}^{e^{-}}$
is the electron asymmetry with target and beam dilution corrections, but no positron dilution corrections;
$N_{e^{+}}/N_{e^{-}}$ is the positron to electron ratio, measured using BigBite.  With BigBite in 
negative polarity, electrons bend up into the detector, whereas positrons bend downwards.

After applying this correction to the parallel and perpendicular asymmetries, we obtain the values
shown in \Figure{phys_asym_4} for E = 4.73 GeV and \Figure{phys_asym_5} for E = 5.89 GeV.
Radiative corrections have not been applied. Work is ongoing to determine the pair-produced electron
asymmetry and its effect on the desired electron asymmetry.  

   \begin{figure}
      \centering
      \begin{subfigure}{0.5\textwidth}
         \centering
         \includegraphics[width=1.0\textwidth]{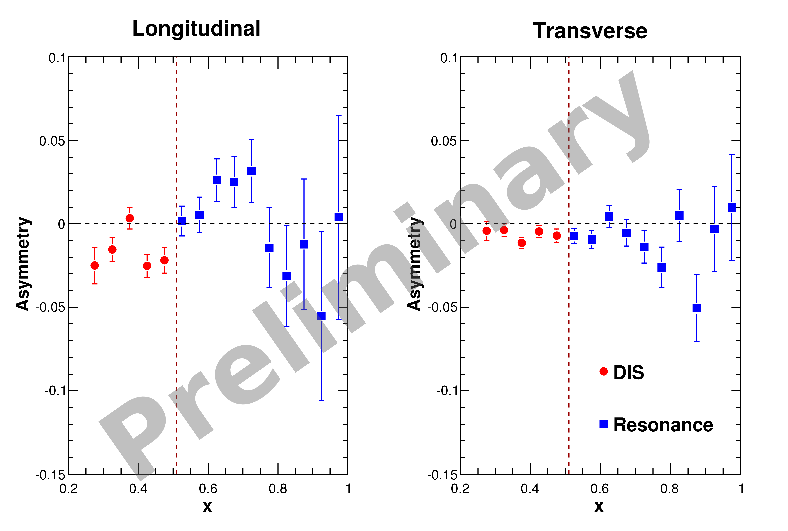}
	 \caption{E = 4.73 GeV}
         \label{fig:phys_asym_4}
      \end{subfigure}
      \begin{subfigure}{0.5\textwidth}
         \centering
         \includegraphics[width=1.0\textwidth]{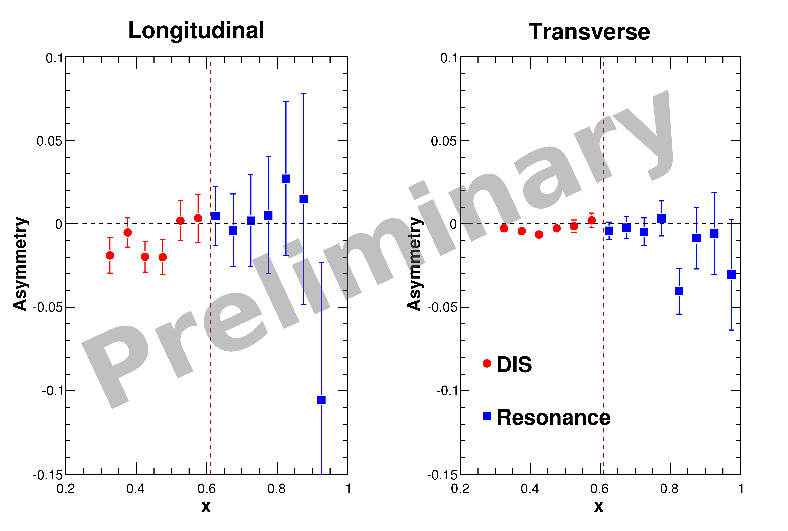}
	 \caption{E = 5.89 GeV}
         \label{fig:phys_asym_5}
      \end{subfigure}
      \caption{Physics asymmetries with positron corrections.  The magenta line shows the DIS threshold, 
               below which is the DIS region. The error bars represent the statistical errors. 
               No radiative corrections.  The positron corrections have not been finalized.    
               (\subref{fig:phys_asym_4}): E = 4.73 GeV data; 
               (\subref{fig:phys_asym_5}): E = 5.89 GeV data.} 
      \label{fig:PhysAsym}
   \end{figure}

\subsubsection{Preliminary Physics Results}

In this section, we present our preliminary physics results for asymmetry \AOneHeThree{} and the spin
structure functions \gOne{} and \gTwo{} on \HeliumThree.  These results are preliminary because 
work is being done on the radiative corrections to the asymmetries along with a Geant4 simulation to 
further investigate the difference between the bend-up and bend-down acceptances in the BigBite spectrometer.
Work concerning the corrections to the asymmetries due to asymmetry of pair-produced electrons is also ongoing. 
 
The extraction of $d_2^{^{\textrm{3}}\textrm{He}}$ and \dtwon{} along with the neutron asymmetry \AOneN{}, 
and the spin structure functions $g_{1,2}$ are also underway; however, the extraction is model-dependent. 
Previous experiments~\cite{E99117} have used Bissey et al.'s complete model in the DIS regime~\cite{Bissey}.  
However, E06-014's data spans both the DIS and resonance regions. A consistent treatment of both DIS and 
resonance data requires careful consideration of structure-function smearing~\cite{Kulagin}.  We are 
working with W. Melnitchouk to extract neutron quantities across our entire kinematic range.  

\paragraph{The Virtual Photon-Nucleon Asymmetry} \label{sect:A1_analysis}

\Figure{A1He3_4} and \Figure{A1He3_5} shows the preliminary result for \AOneHeThree{} at 
E = 4.73 and 5.89 GeV, respectively.  Also shown is world data from SLAC E142~\cite{E142} and 
JLab E01-012~\cite{E01012} and E99-117~\cite{E99117}.  The red (blue) data points indicate the 
DIS (resonance) data for this experiment. No radiative corrections have been applied to these data.  
The data from this experiment are consistent with the world data across a wide range in $x$, despite 
the larger error bars in the resonance region.    

   \begin{figure}
      \centering
      \begin{subfigure}{0.5\textwidth}
         \centering
         \includegraphics[width=1.0\textwidth]{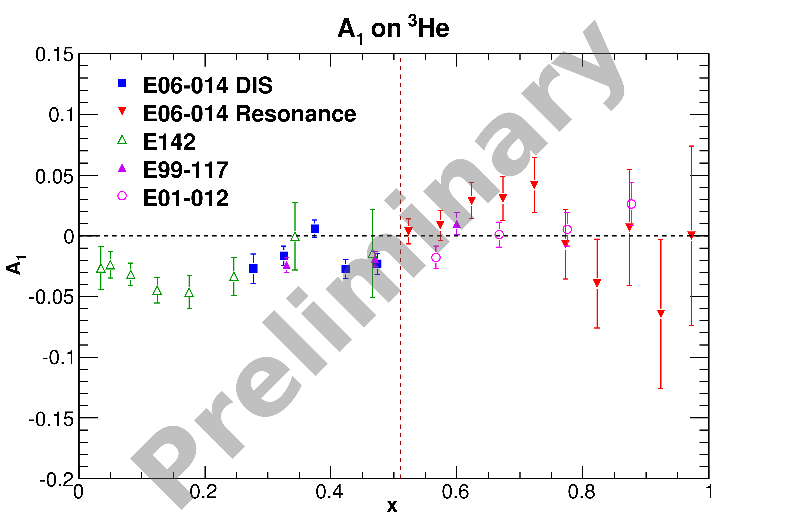}
	 \caption{E = 4.73 GeV}
         \label{fig:A1He3_4}
      \end{subfigure}
      \begin{subfigure}{0.5\textwidth}
         \centering
         \includegraphics[width=1.0\textwidth]{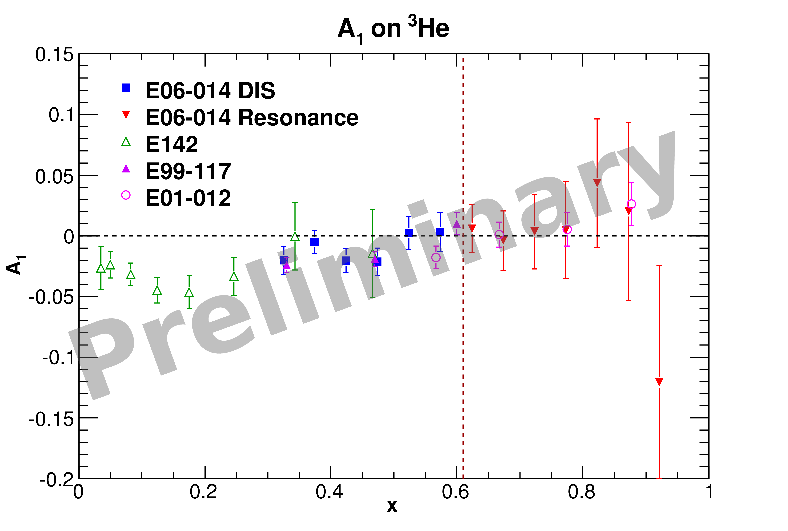}
	 \caption{E = 5.89 GeV}
	 \label{fig:A1He3_5}
      \end{subfigure}
      \caption{\AOneHeThree{} compared to the world data from SLAC E142~\cite{E142} and JLab 
               E01-012~\cite{E01012} and E99-117~\cite{E99117}. The error bars on our data are statistical only.
               The radiative corrections to our data have not been applied, and the positron corrections are still 
               under investigation.  
               (\subref{fig:A1He3_4}): E = 4.73 GeV data; 
               (\subref{fig:A1He3_5}): E = 5.89 GeV data.}
      \label{fig:A1He3}
   \end{figure}

\paragraph{The Spin Structure Functions} \label{sect:spin_struc_func}

En route to extracting $d_2^n$, the spin structure functions \gOne{} and \gTwo{} can be obtained 
according to: 

\begin{eqnarray}
  g_1 &=& \frac{MQ^2}{4{\alpha}^2}\frac{2y}{\left( 1 - y\right) \left( 2 - y\right)}\sigma_0 
          \left[ A_{\parallel} + \tan \left(\theta/2 \right)A_{\perp}\right] \\
  g_2 &=& \frac{MQ^2}{4{\alpha}^2}\frac{y^2}{\left( 1 - y\right) \left( 2 - y\right)}\sigma_0 
          \left[ - A_{\parallel} + \frac{1 + \left( 1 - y \right)\cos \theta}
          {\left( 1 - y \right)\sin \theta}A_{\perp}\right], 
\end{eqnarray}

\noindent where $M$ is the nucleon mass; $\alpha$ is the electromagnetic fine structure constant; 
$y = \nu/E$, the fractional energy transfer to the target; $\theta$ is the electron scattering angle; 
$\sigma_0$ is the unpolarized total cross section; $A_{\parallel}$ ($A_{\perp}$) is the parallel (perpendicular) 
double-spin electron asymmetry. 

The preliminary results for \gOne$^{^{\textrm{3}}\text{He}}$ and \gTwo$^{^{\textrm{3}}\text{He}}$ are shown
in \Figure{pol_sf}, which compares the data to various models~\cite{DSSV,Gamberg,Soffer,Stratmann} and the world data.  Radiative 
corrections have been applied {\it only} to the unpolarized total cross sections for the data from this experiment.

\begin{figure}
      \centering
      \begin{subfigure}{0.5\textwidth}
         \centering
         \includegraphics[width=1.0\textwidth]{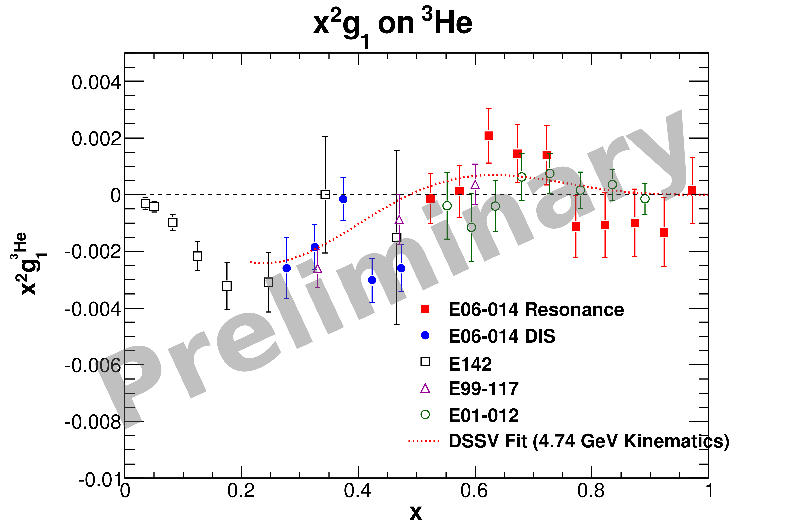}
         \label{fig:g1He3_4}
         \caption{\gOne$^{^{\textrm{3}}\textrm{He}}$ at E = 4.73 GeV}
      \end{subfigure}\begin{subfigure}{0.5\textwidth}
         \centering
         \includegraphics[width=1.0\textwidth]{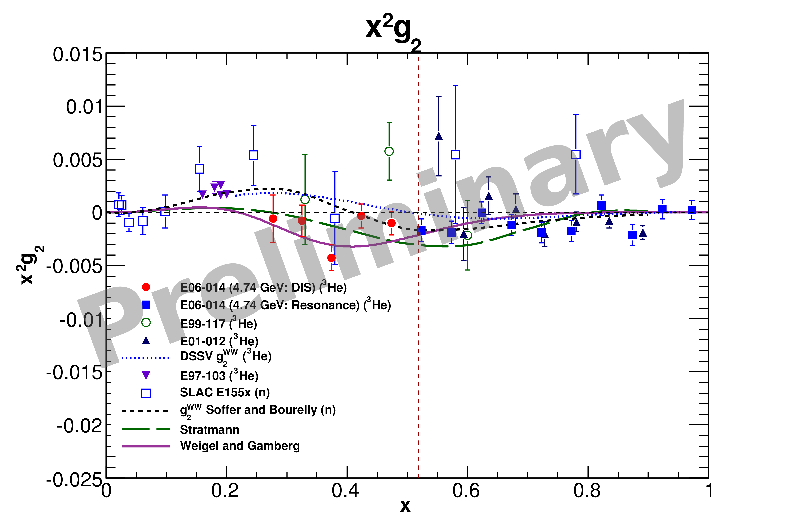}
         \label{fig:g2He3_4}
         \caption{\gTwo$^{^{\textrm{3}}\textrm{He}}$ at E = 4.73 GeV}
      \end{subfigure}

      \begin{subfigure}{0.5\textwidth}
         \centering
         \includegraphics[width=1.0\textwidth]{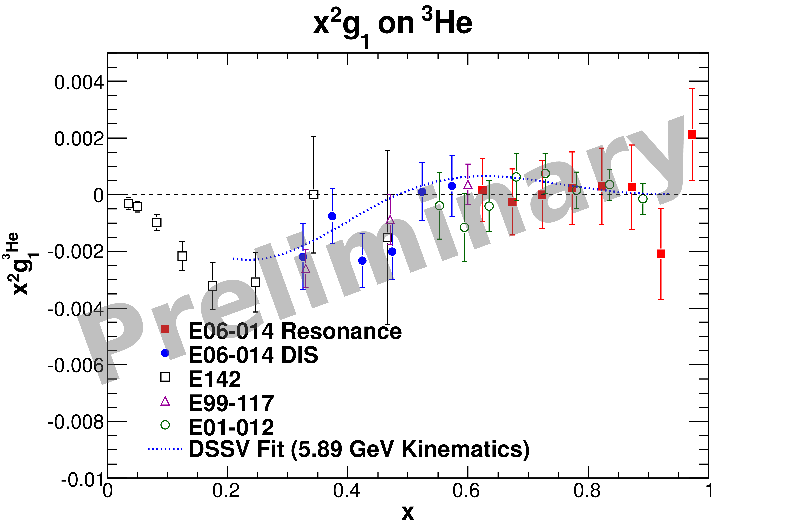}
         \caption{\gOne$^{^{\textrm{3}}\textrm{He}}$ at E = 5.98 GeV}
         \label{fig:g1He3_5}
      \end{subfigure}\begin{subfigure}{0.5\textwidth}
         \centering
         \includegraphics[width=1.0\textwidth]{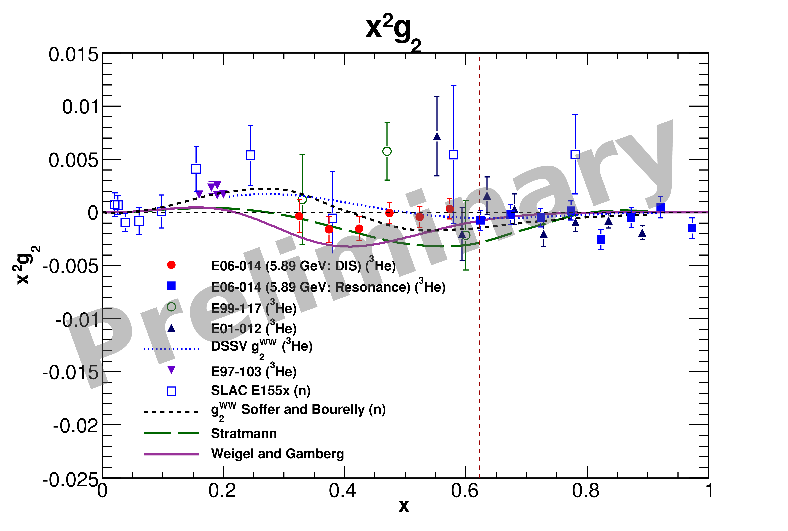}
         \caption{\gTwo$^{^{\textrm{3}}\textrm{He}}$ at E = 5.98 GeV}
         \label{fig:g2He3_5}
      \end{subfigure}

      \caption{Preliminary results for the spin structure functions \gOne{} and \gTwo{} on a \HeliumThree{} target
               for E = 4.73 and 5.89 GeV compared to the world data~\cite{E155x,E99117,E142,E01012,E97103}
               and the DSSV model~\cite{DSSV} and models from Weigel and Gamberg~\cite{Gamberg}, 
               Bourelly and Soffer~\cite{Soffer}, and Stratmann~\cite{Stratmann}.  The error bars on
               our data are statistical only.  The radiative corrections to our data have only been applied to 
               the cross sections and not the asymmetries.  The positron corrections to the asymmetries are still under investigation.
               (\subref{fig:g1He3_4}) and~(\subref{fig:g2He3_4}): \gOne$^{^{\textrm{3}}\textrm{He}}$ and \gTwo$^{^{\textrm{3}}\textrm{He}}$ 
                                                      for a beam energy of E = 4.73 GeV.
               (\subref{fig:g1He3_5}) and~(\subref{fig:g2He3_5}): \gOne$^{^{\textrm{3}}\textrm{He}}$ and \gTwo$^{^{\textrm{3}}\textrm{He}}$ 
                                                      for a beam energy of E = 5.89 GeV.}
      \label{fig:pol_sf}
   \end{figure}


\clearpage \newpage

\subsection[E07-006 - SRC]{E07-006 - Short Range Correlations}
\label{sec:e07006}

\begin{center}
\bf Triple Coincidence $^4$He(e,e'pn) Measurements
\end{center}

\begin{center}
S. Gilad, D. Higinbotham, V. Sulkosky, and J. Watson spokespersons, \\
and \\
the Hall A Collaboration.\\
contributed by I. Korover and N. Muangma.\\
\end{center}

Experiment E07-006~\cite{e07006} was conducted to continue the study of nucleon-nucleon (NN) short-range correlations (SRC).  During the first high luminosity triple coincidence experiment at JLab, E01-015~\cite{Subedi:2008}, we measured the $\left(\mathrm{e,e'pp}\right)$ and $\left(\mathrm{e,e'pn}\right)$ reactions on $^{12}$C over the $\left(\mathrm{e,e'p}\right)$ missing momentum range from 275 to 550~MeV/$c$.  These measurements were sensitive to the short-range NN tensor force.  The recently completed triple coincidence experiment measured these reactions on $^4$He over the missing momentum range from 400 to 875~MeV/$c$ in order to study the short-range repulsive part of the NN interaction and investigate the transition from a tensor-force-dominated region.  The kinematic conditions $\left(Q^{2} = 2~\mathrm{GeV}^{2} \hspace{0.1cm} \mathrm{and} \hspace{0.1cm} x_B > 1 \right)$ allow us to extract the abundance of $pn$- and $pp$-correlated pairs with minimal interference from final state interactions, meson exchange currents and resonance production.

During the experimental run period, the detection of the neutrons in the 
triple coincidence $^4$He(e,e'pn) measurement was done by using the Hall A 
Neutron Detector (HAND), and the protons in the $^4$He(e,e'pp) measurement 
were detected in the BigBite hadron spectrometer~\cite{Mihovilovic:2012hi}.  
The analysis so far has focused on calibration of the two high resolution 
spectrometers (HRSs), the neutron detector and the BigBite spectrometer.  
However, these calibrations are nearing completion, and preliminary results 
will soon be available.

\subsubsection{$^4$He(e,e'p) Coincidence Time}
\label{sec:HRS_CT}

The experiment requires good timing information from the HRSs. During the 
production measurement, the left HRS detected the scattered electron and 
provides the reference time. The relative time between the left HRS and right HRS 
is used to identify the recoil
neutrons or protons in the third arm (HAND / BigBite), which are in 
coincidence with (e,e'p) events.  These are the recoil particles from the 
breakup of pn-SRC/pp-SRC pairs respectively.  The coincidence time 
distribution between the HRSs for one of the three kinematics is shown in 
Fig.~\ref{fig:hrs_ctof} with a resolution of 0.6~ns.  The same resolution is 
observed in the other two kinematical setups.

\begin{figure}[htb]
\center{\includegraphics[width=9.0cm]{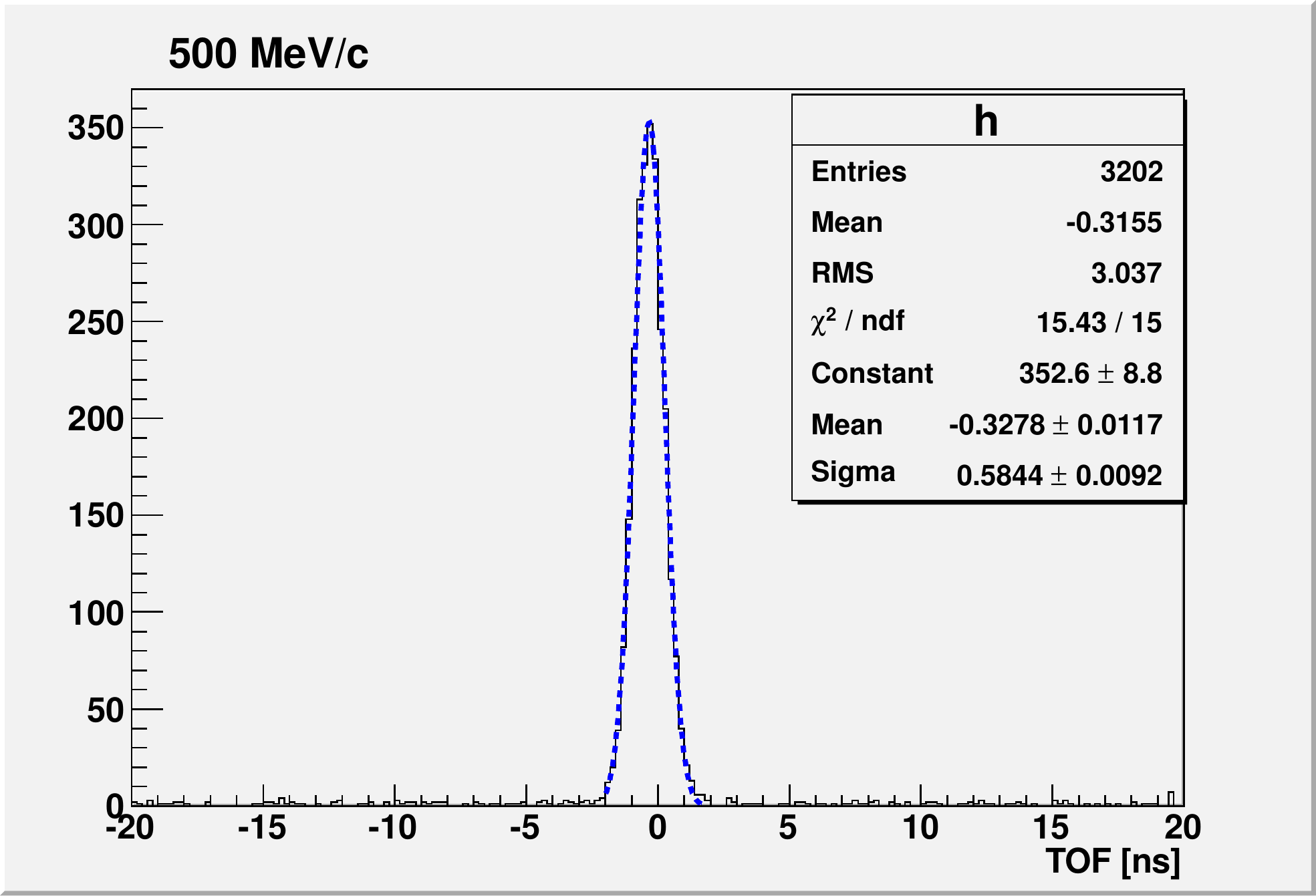}}
\caption{Coincidence time between the HRSs at 500 MeV/c}
\label{fig:hrs_ctof}
\end{figure}

The first steps in the analysis of the (e,e'pn) events were
\begin{enumerate}
  \item The time calibration of HAND.
  \item The determination of the absolute neutron detection efficiency.
\end{enumerate}

\subsubsection{Time calibration of HAND}
\label{sec:handcalib}

HAND consists of 112 (1-m long, 10-cm thick) scintillator bars arranged 
in 6 layers and an additional front layer of 64 thin (2~cm) counters that 
were used as a veto. In the first part of the timing calibration, 
time alignment between all bars was performed.  We used the H(e,e'p) reaction
to calibrate HAND in which electrons were detected by the left HRS and 
the protons in the neutron detector, located at 15~m from the target. 
Figure~\ref{fig:HAND_tof} shows the difference between the measured TOF and 
the calculated one based on the known proton momentum. As can be seen, the 
achieved resolution is $\sigma \approx$ 0.5~ns. 

\begin{figure}[htb]
\center{\includegraphics[width=9.0cm]{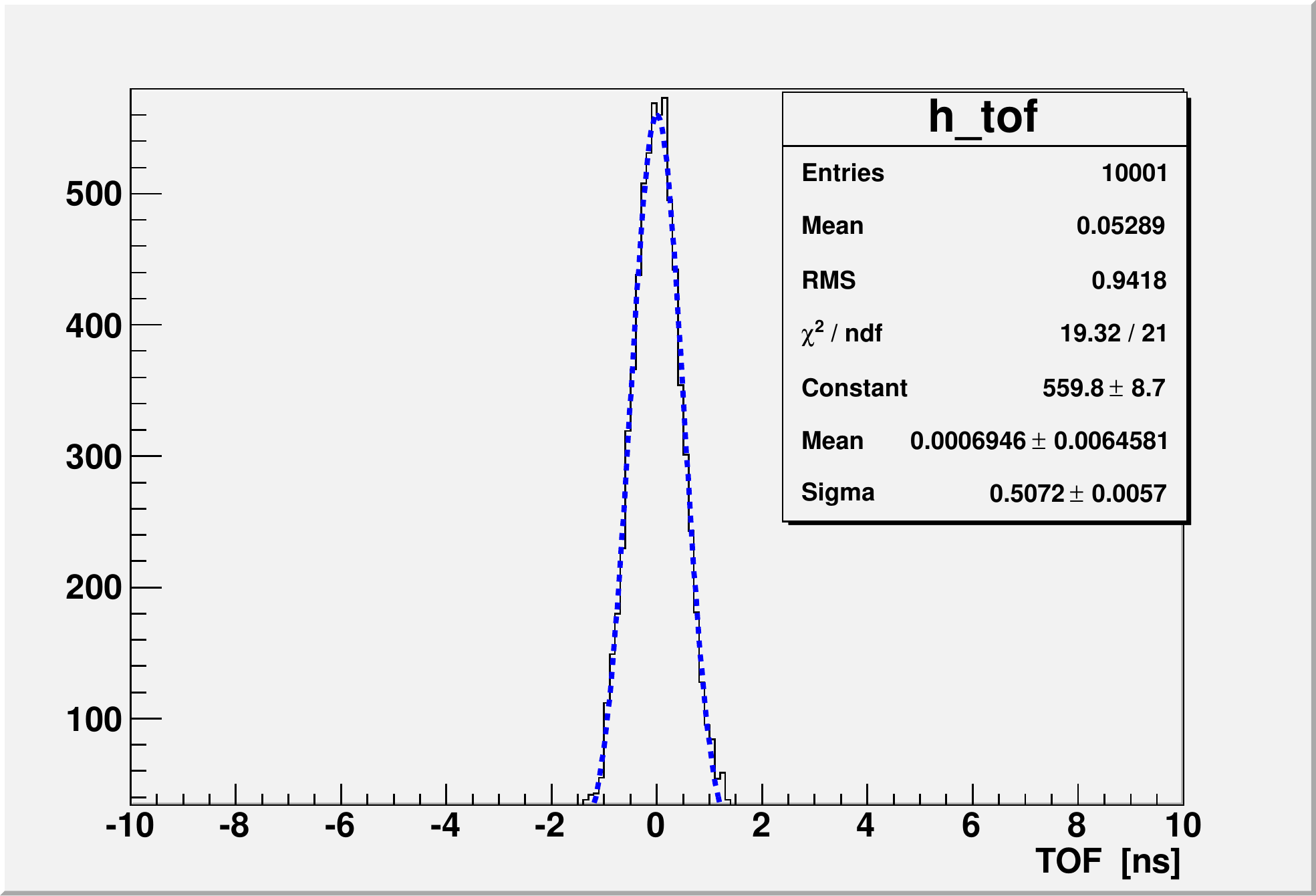}}
\caption{TOF for protons from the $^4$He(e,e'p) reaction as measured by the 30 detectors in the first layer of HAND.}
\label{fig:HAND_tof}
\end{figure}

The calibration procedure adjusts the relative timing between the different bars. The absolute timing  and the TOF resolution was determined using the liquid deuterium target measuring the d(e,e'pn) exclusive reaction. The electron and proton were measured by the HRSs and the known momentum of the neutron by HAND. 
The difference between the measured TOF and that calculated based on the 
neutron known momentum is shown in Fig.~\ref{fig:HAND_LD2_tof}.  The 
resolution is $\sigma \approx$ 1.7~ns.

\begin{figure}[htb]
\center{\includegraphics[width=9.0cm]{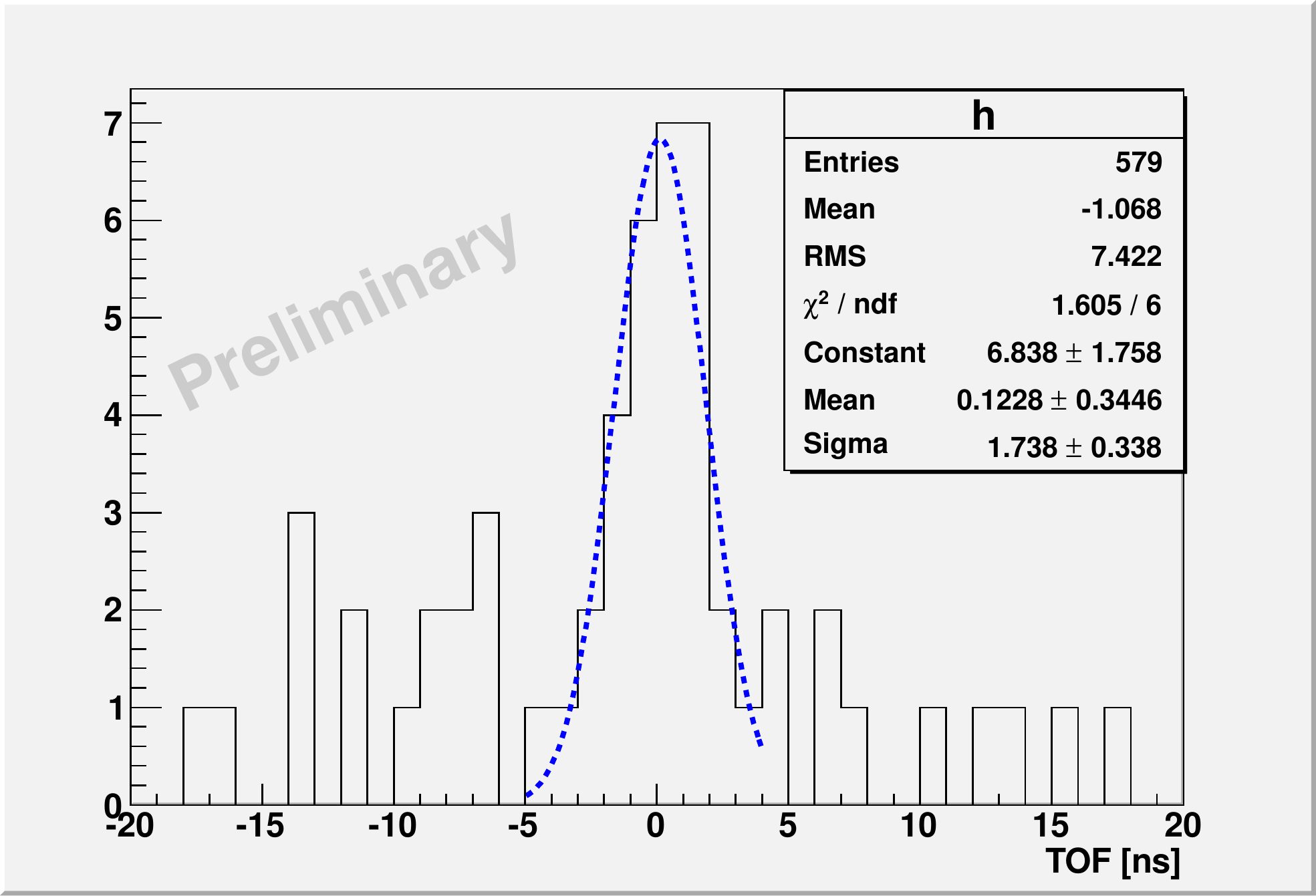}}
\caption{The difference between the Neutron TOF measured and calculated, see text for details.}
\label{fig:HAND_LD2_tof}
\end{figure}

\subsubsection{Neutron detection efficiency}
\label{sec:hand_deteff}

The absolute neutron detection efficiency was determined using the exclusive 
d(e,e'pn) reaction at two different kinematical settings corresponding to 
neutrons with mean momenta of 250 MeV/$c$ and 450 MeV/$c$. The measured efficiency 
compared to a simulation~\cite{ref1} is shown in Fig~\ref{fig:HAND_deteff}.

\begin{figure}[htb]
\center{\includegraphics[width=11.0cm]{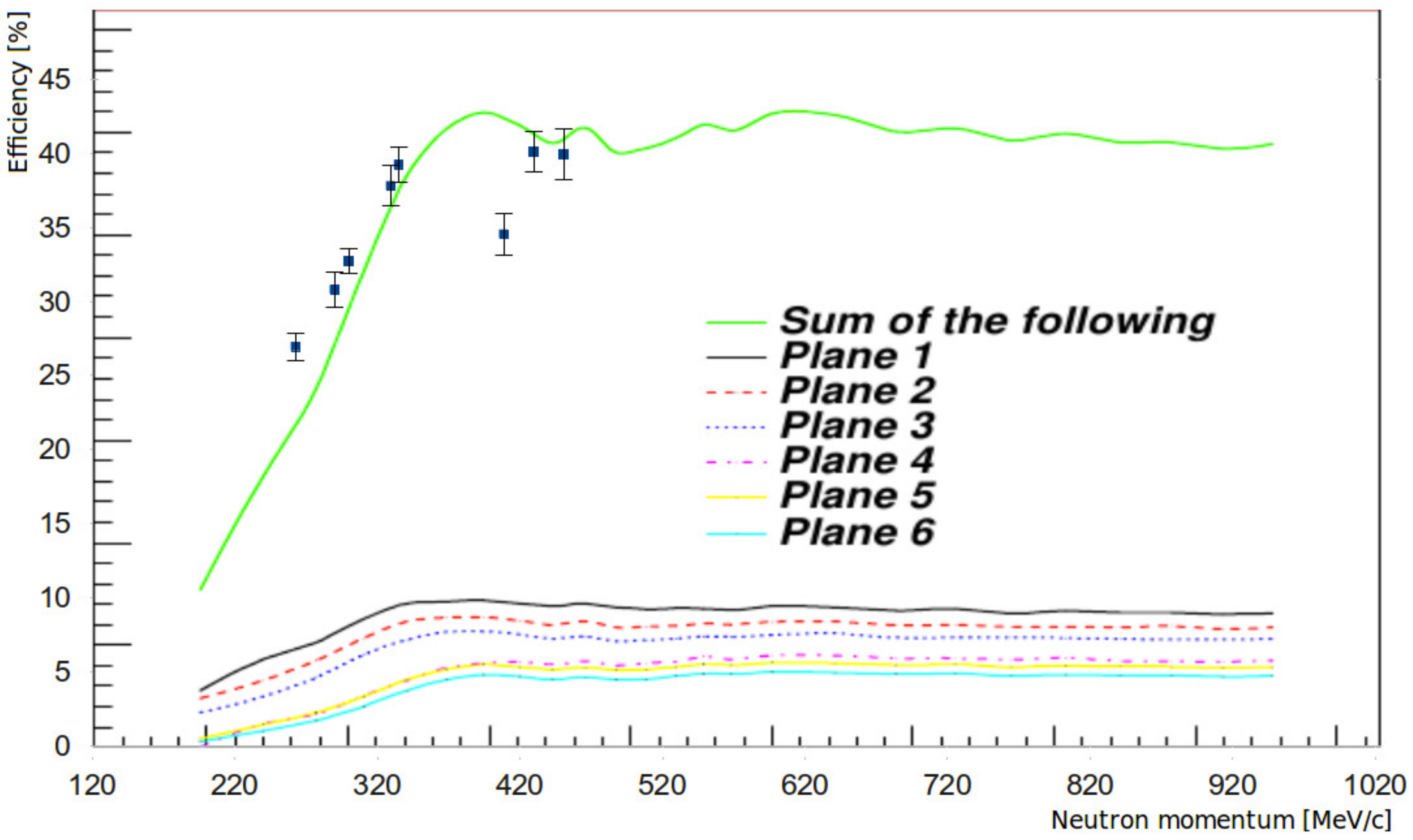}}
\caption{Neutron detection efficiency as a function of the neutron momentum, measured and simulated.}
\label{fig:HAND_deteff}
\end{figure}

Following the calibration, the TOF spectrum for the  
750~MeV/$c$ missing momentum is shown in Fig~\ref{fig:HAND_750_tof}. 
\begin{figure}[htb]
\center{\includegraphics[width=9.0cm]{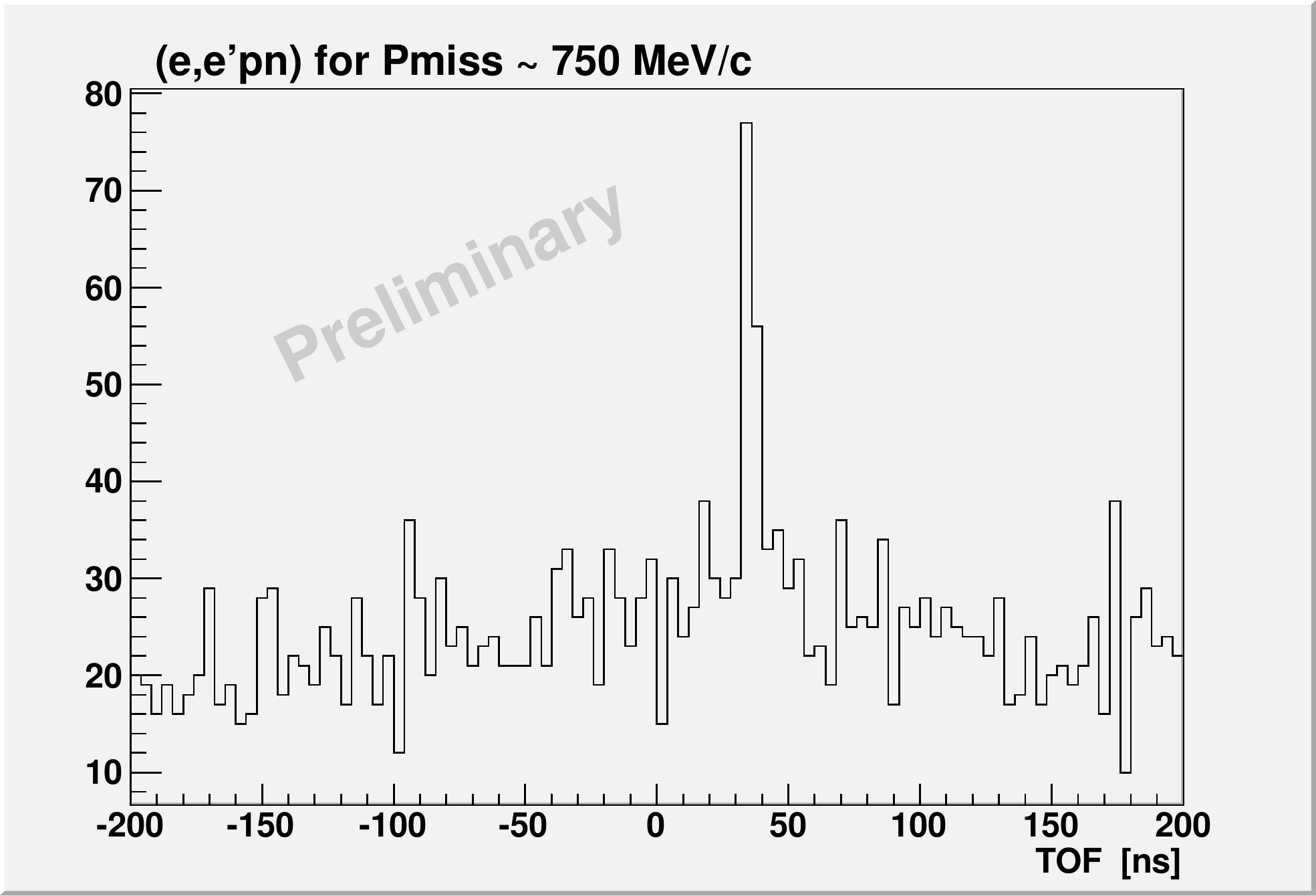}}
\caption{Neutron TOF distribution for the (e,e'p) missing momentum of 
750~MeV/$c$.}
\label{fig:HAND_750_tof}
\end{figure}
A clear coincidence peak on top of random coincidence background can be 
identified at all of the measured missing momenta.
The background level can be estimated by fitting to a constant 
background in the off-time region and also by using an event mixing procedure. 
The event mixing is done by taking the (e,e'p) information from one event and 
the neutron information from another event.  The accepted events pass all the 
physical cuts. The angular correlation between the initial proton and the 
recoil neutron is shown in Fig~\ref{fig:HAND_angle}.  A clear back-to-back 
angular correlation of the short range np - pair can be seen above the 
background.

\begin{figure}[htb]
\begin{centering}
\begin{minipage}{.48\textwidth}
\includegraphics[width=\textwidth]{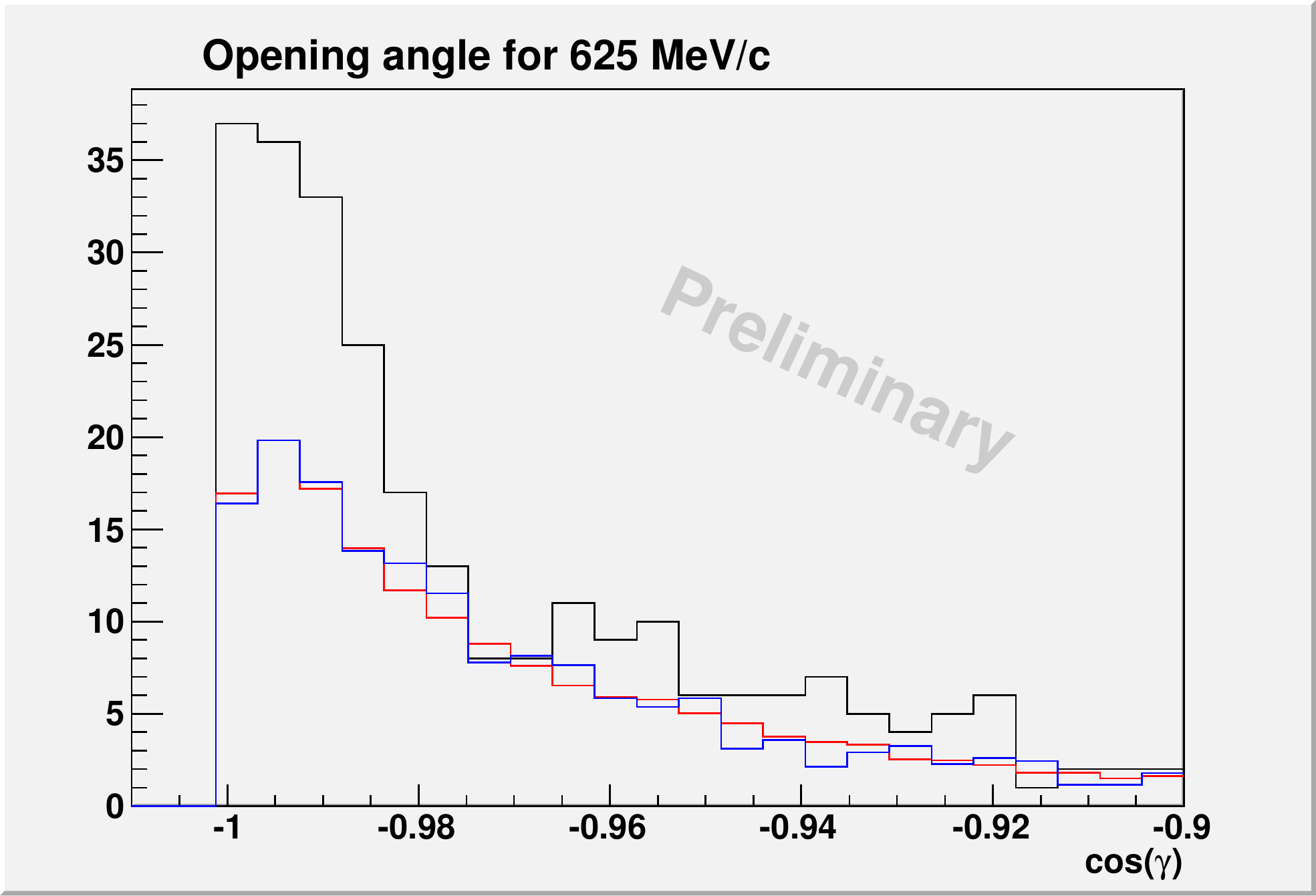}
\end{minipage}
\begin{minipage}{.48\textwidth}
\includegraphics[width=\textwidth]{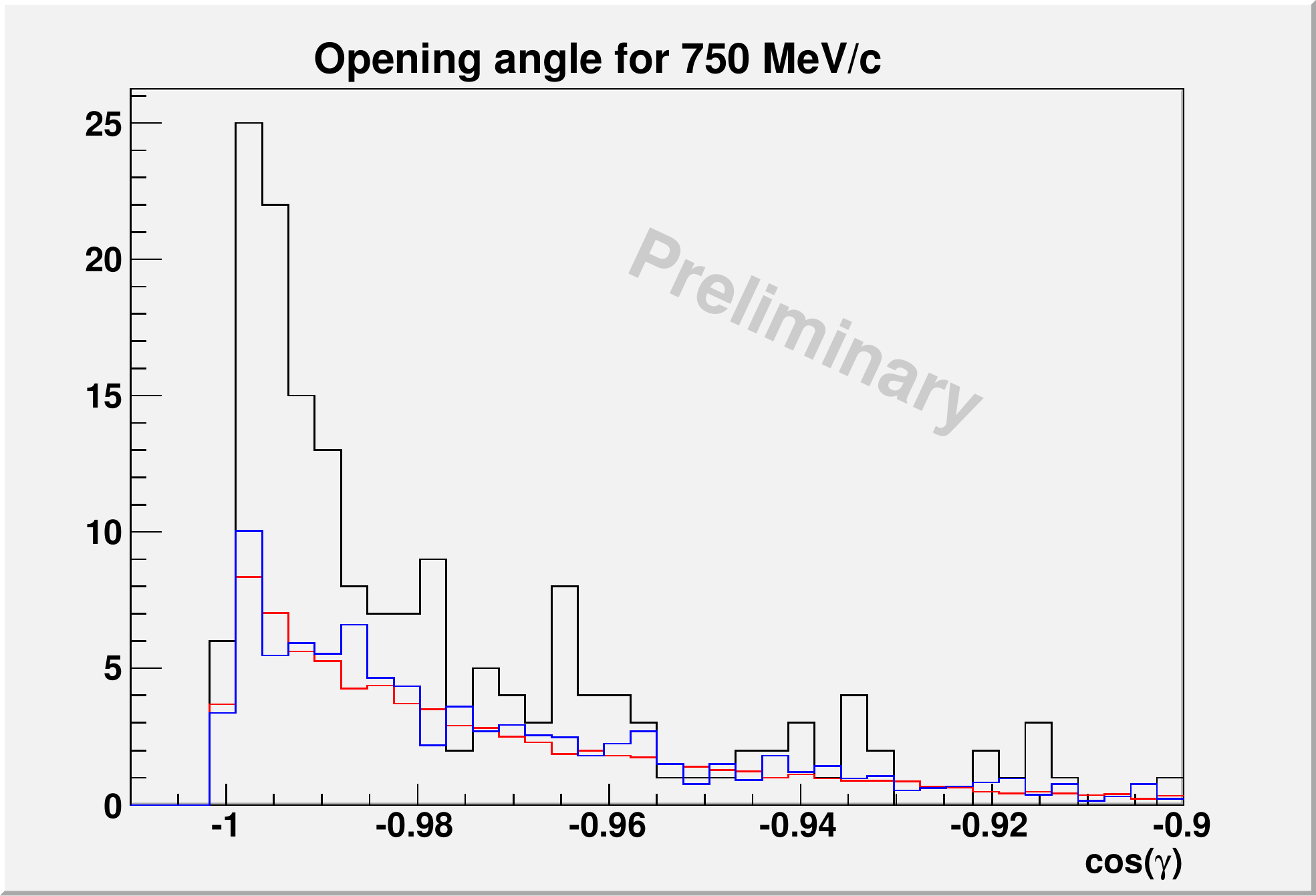}
\end{minipage}
\caption{Cosine of the opening angle between the proton and the neutron. Black: Signal + Background, Blue: Background estimated by event mixing, 
Red: Background from the off time region.}
\label{fig:HAND_angle}
\end{centering}
\end{figure}

\subsubsection{BigBite Spectrometer Calibrations}
\label{sec:BB_calibs}

The BigBite Spectrometer is used to detect the recoil proton partners.  The BigBite 
Spectrometer consists of a single dipole, two Multi-Wire Drift Chambers (MWDCs), 
and two scintillator planes, named the dE and E planes.   The Particle 
Identification (PID) process requires the clear separation of protons from 
deuterons and Minimum Ionizing Particles (MIPs).  The momentum is calibrated 
against $\left|\vec{q}\right|$ from the left HRS using the elastic hydrogen data 
for the momentum range of 300--500~MeV/$c$.  After a first iteration of 
calibrations, the resolution ($\sigma$) is 7.7~MeV/$c$, providing a better than 
2\% momentum resolution.  The 
coincidence time between the left HRS (electron arm) and BigBite is used to 
eliminate  backgrounds including random coincidence events as shown in 
Fig.~\ref{fig:LHRS-BB_CT} with an achieved resolution $<$ 1.4~ns.

\begin{figure}[htb]
\center{\includegraphics[width=9.0cm]{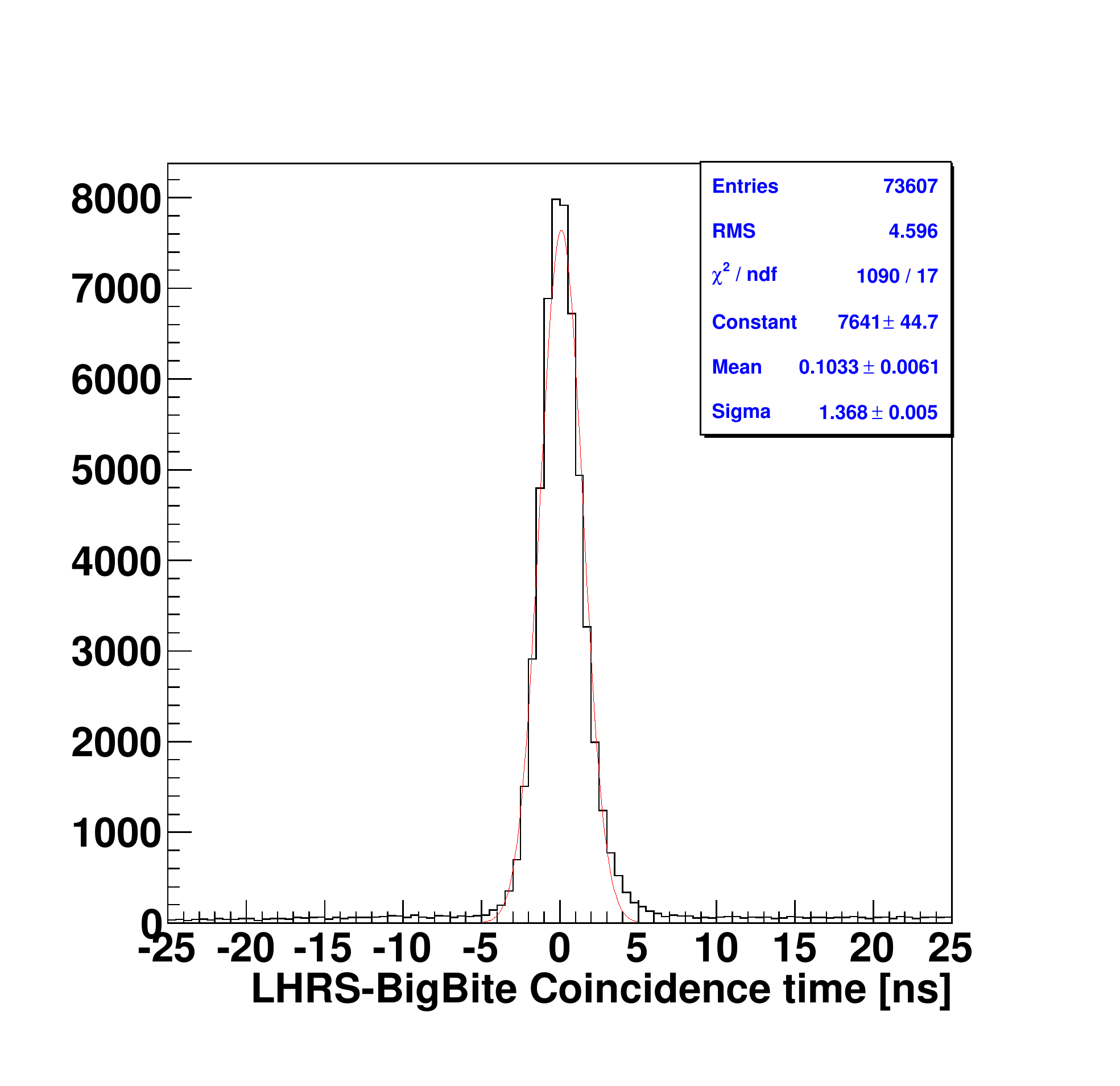}}
\caption{The coincidence time between the left HRS and BigBite for the 
 elastic d(e,e'p) data.  The red curve shows a fit to the data.}
\label{fig:LHRS-BB_CT}
\end{figure}

Different methods can be utilized via some combination 
of the energy deposited in the dE/E planes, the reconstructed momentum, and the 
coincidence time between the left HRS and BigBite.  For the first method, the 
energy deposited in the dE plane vs. the energy deposited in E plane is shown in 
Fig.~\ref{fig:BB_dEvsE} for an elastic deuterium run.
\begin{figure}[hbt]
\begin{centering}
\begin{tabular}{cc} 
\includegraphics[width=7.0cm]{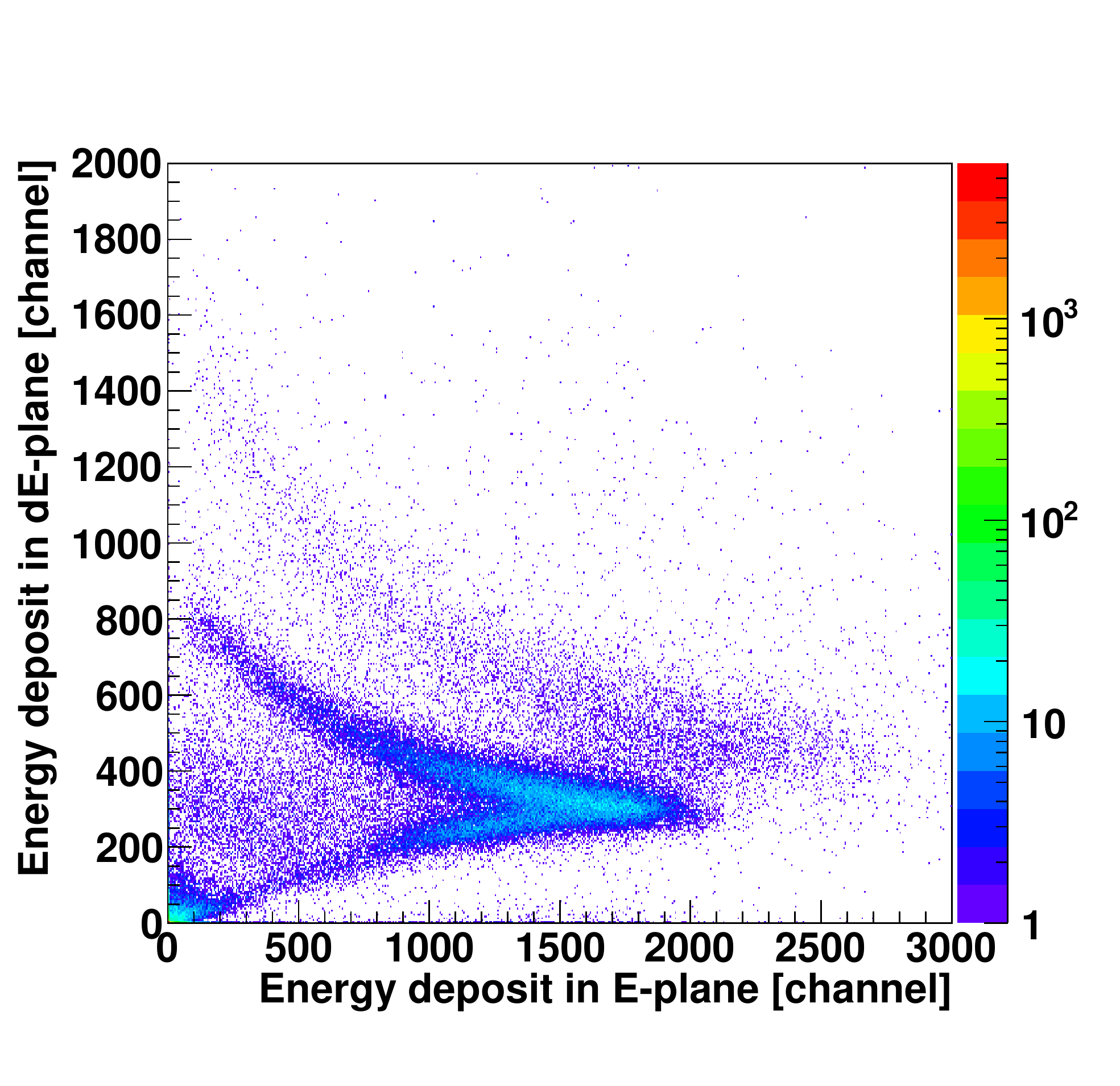}
&\includegraphics[width=7.0cm]{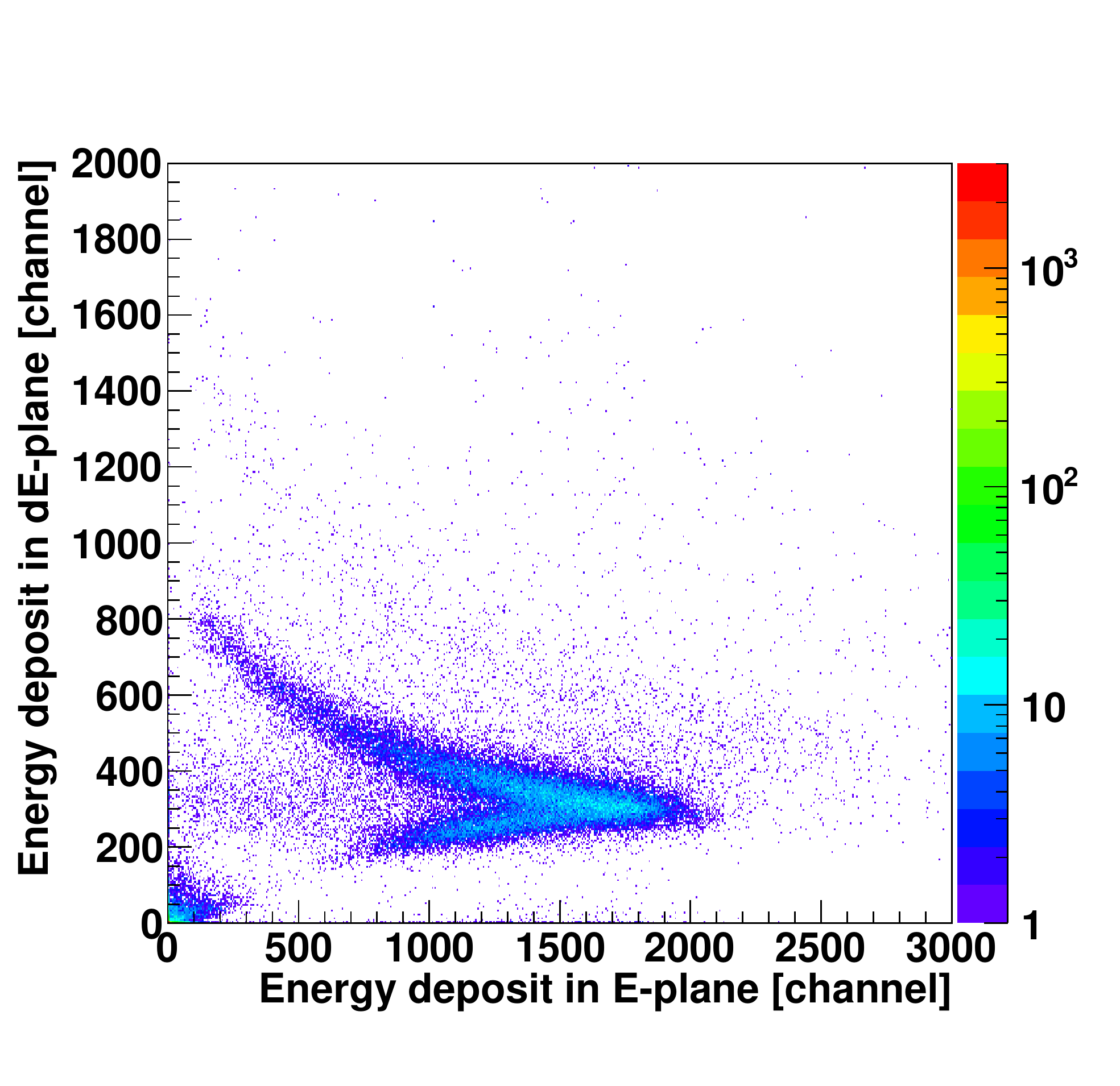}
\end{tabular}
\caption{Energy deposited in the dE plane versus the energy deposited in the 
E plane.  The left panel is without a timing cut on the left HRS and BigBite
coincidence time, whereas the right panel is with a coincidence time cut on 
Fig.~\ref{fig:LHRS-BB_CT} to remove deuterons.}
\label{fig:BB_dEvsE}
\end{centering}
\end{figure}
\noindent This method separates well the protons and deuterons from the MIPs but 
cannot separate the proton from the deuterons at high energy.  By adding a 
coincidence time cut on the data, most of the deuterons can be removed as shown
in the right-side panel of Fig.~\ref{fig:BB_dEvsE}.

In the second method, the momentum vs. the energy deposited in each scintillator 
plane can also be used to separate different particle species as illustrated in
Fig.~\ref{fig:BBmom_vs_dEnE}.
\begin{figure}[hbt]
\begin{centering}
\begin{tabular}{cc} 
\includegraphics[width=7.0cm]{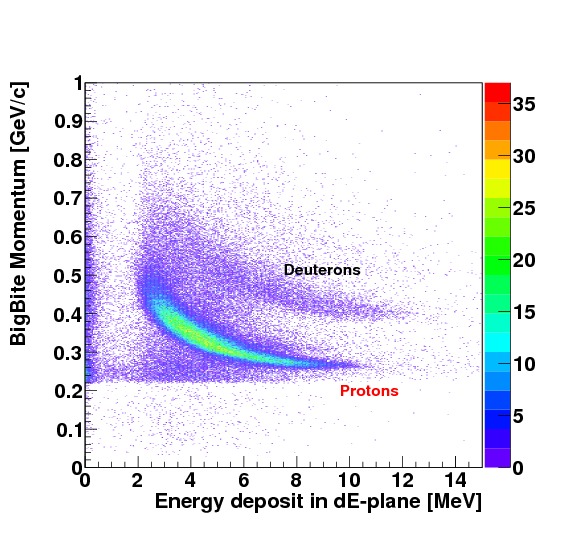}
&\includegraphics[width=7.0cm]{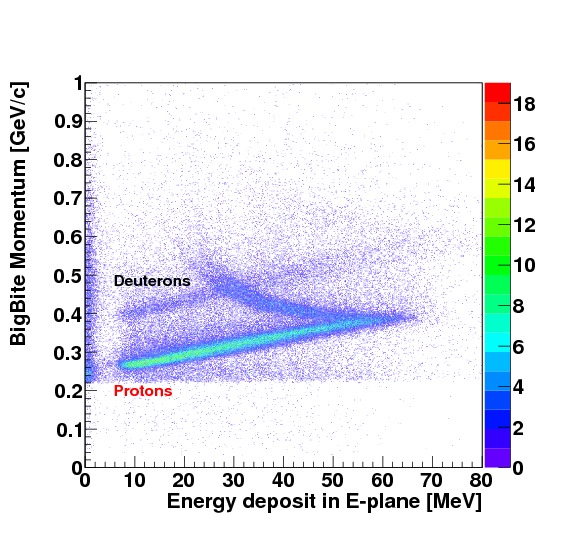}
\end{tabular}
\caption{The left panel shows the reconstructed BigBite momentum from the MWDC vs.
the energy deposited in the dE plane.  The right panel shows the reconstructed 
BigBite momentum vs. the energy deposited in the E plane.}
\label{fig:BBmom_vs_dEnE}
\end{centering}
\end{figure}
\noindent This method shows the distinction of the proton strip from the deuteron strip, and there is a cleaner separation for the dE plane.

\subsubsection{Summary}
\label{sec:SRC_summary}

At this point, the analysis is focused on extraction of the cross section 
ratios of $^4$He(e,e'pn) to $^4$He(e,e'pp). 
This ratio was found in the 300 -- 500 MeV/c missing momenta range to be about 
18:1~\cite{Subedi:2008}. As the missing momentum increases, the dominance of the 
tensor force is expected to be reduced and the short range 
repulsive force is expected to be more important, making the 
(e,e'pn)/(e,e'pp) ratio smaller.  The measured ratio of the $^4$He(e,e'pn) to 
the $^4$He(e,e'pp) can hopefully teach us about the interplay between the tensor 
and the repulsive part of the short distance nucleon-nucleon force.

\clearpage \newpage

\subsection{E07-007,  E08-025 - DVCS}
\label{sec:e07007}

\begin{center}
\bf Deeply Virtual Compton Scattering
\end{center}

\begin{center}
P. Bertin, C. Mu\~noz Camacho, A. Camsonne, C. Hyde, M. Mazouz, J. Roche, spokespersons, \\
and \\
the Hall A Collaboration.\\
contributed by C. Hyde.
\end{center}

\subsubsection{Introduction}\label{sec:DVCSintro}

Generalized Parton Distributions (GPDs) describe the transverse spatial distribution of 
quarks and gluons inside a hadron, as a function of the longitudinal momentum fraction
of the the parton.  Deeply virtual Compton scattering
($ep \rightarrow ep\gamma$)
is the simplest reaction (at least from the point of view of QCD theory) in which to study
GPDs. To lowest order in QED, the exclusive electroproduction reaction is described by
the coherent superposition of the virtual Compton scattering (VCS) and Bethe-Heitler
amplitudes. The deeply virtual
limit is defined as high $Q^2$, high $W^2$,  but small net invariant momentum transfer $-t$ to the
target proton. In this generalization of the Bjorken
limit of Deep Inelastic Scattering (DIS), the virtual Compton amplitude simplifies to a
power series in $1/Q^2$, with the leading 'Twist-2' term determined by the
GPDs.

Our Hall A DVCS program is focussed on the measurement of  precision cross
sections as a function of $Q^2$, at fixed $x_{\rm Bj}$ and  $t$
\cite{Munoz Camacho:2006hx}.   
This allows us to isolate 
the Twist-2 GPD terms from the higher order terms.
In Oct--Dec 2010, we completed data taking for two DVCS experiments in Hall A:
proton DVCS E07-007, and D$(e, e'\gamma)pn$ E08-025. In
these experiments, we measured the $(e, e'\gamma)$ and $(e, e'\pi^0)$ reactions as a function of $t$
at a fixed $x_{\rm Bj}= 0.36$,
at three values of $Q^2 $ and at two incident beam energies for each $Q^2$. The main goal
of these experiments is to isolate the $ |DVCS |^2$ and $\Re{\rm e}[DVCS^\dagger BH]$ terms in the cross
section. This is possible, because the kinematic pre-factors of these two terms vary with
the incident energy ($k$), with the ratio
$\left|DVCS\right|^2 : \Re{\rm e}\left[ DVCS^\dagger BH\right] \sim k .$
This is a generalized Rosenbluth separation. A secondary goal of the experiment is to
perform a Rosenbluth separation of $\sigma_L$ and $\sigma_T$ in  the exclusive $N(e, e'\pi^0)N$ channels.
This will help to clarify the role of the recently discovered transversity GPDs \cite{Goloskokov:2011rd}.

\subsubsection{Analysis}\label{sec:DVCSanalysis}
In these two experiments, we detected the scattered electron in the HRS-L, and the emitted $\gamma$-ray
in a dedicated calorimeter of 208 PbF$_2$ crystals.  
The $\gamma$-ray shower in the 208 element PbF$_2$ calorimeter is digitized by a custom 1 GHz analog sampling circuit
based on the ANTARES ARS chip
\cite{Feinstein:2003vi}.
Sample single pulse and double pulse fits are displayed in Fig.~\ref{fig:DVCS-ARS}.
The $(e,e'\gamma)$ coincidence timing, after all corrections, is displayed in Fig.~\ref{fig:DVCS-t}.
The PbF$_2$ energy response is calibrated with dedicated elastic H$(e,e_{\rm calo} p_{\rm HRS})$ runs.
The resulting H$(e,e'\gamma)X$ missing mass-squared spectrum is displayed in Fig.~\ref{fig:DVCS-MX2}.
Note that accidental coincidences and the background of H$(e,e'\gamma)\gamma X'$ events 
(where the two photons come from a neutral pion have not
yet been subtracted.  A simulation of the latter is in progress, based on the measured sub-sample of 
H$(e,e\pi^0)X'$ events in which the $\pi^0\rightarrow\gamma\gamma$ decay is fully reconstructed in the
calorimeter.

The mass distribution of $\gamma\gamma$ events is displayed in the left-hand plot of Fig.~\ref{fig:DVCS-pi0}.
Events in the $\pi^0$ peak are used to fine-tune the calorimeter calibration.  The result can be seen
in the missing mass squared distribution of H$(e,e'\pi^0)$X events of the right hand plot
of Fig.~\ref{fig:DVCS-pi0}.  

All calibrations for  E07-007 and E08-025 are now complete.

\begin{figure}[hbp]
\center{\includegraphics[width=0.475\textwidth]{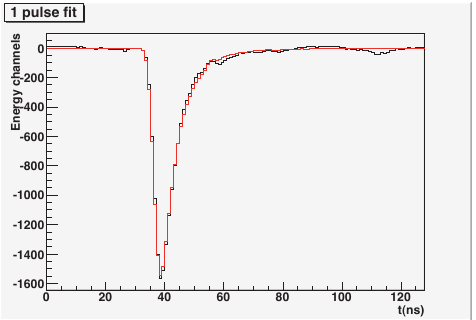}\hfill
\includegraphics[width=0.475\textwidth]{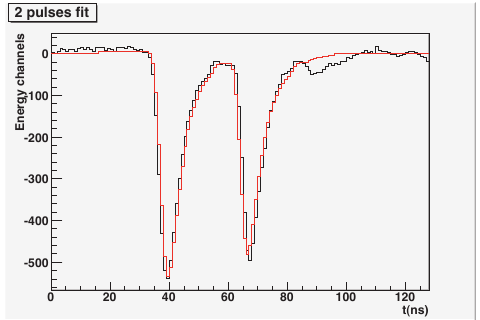}}
\caption{
Sample 1 GHz waveforms for $\gamma$ shower in one crystal of PbF$_2$ calorimeter.
{\bf Left:} Single pulse fit; {\bf Right:} Double pulse fit.  Roughly $8\%$ of all events have at least one crystal
with a double pulse signal.}
\label{fig:DVCS-ARS}
\end{figure}

\begin{figure}[hbp]
\begin{minipage}{0.50\textwidth}
\includegraphics[width=\textwidth]{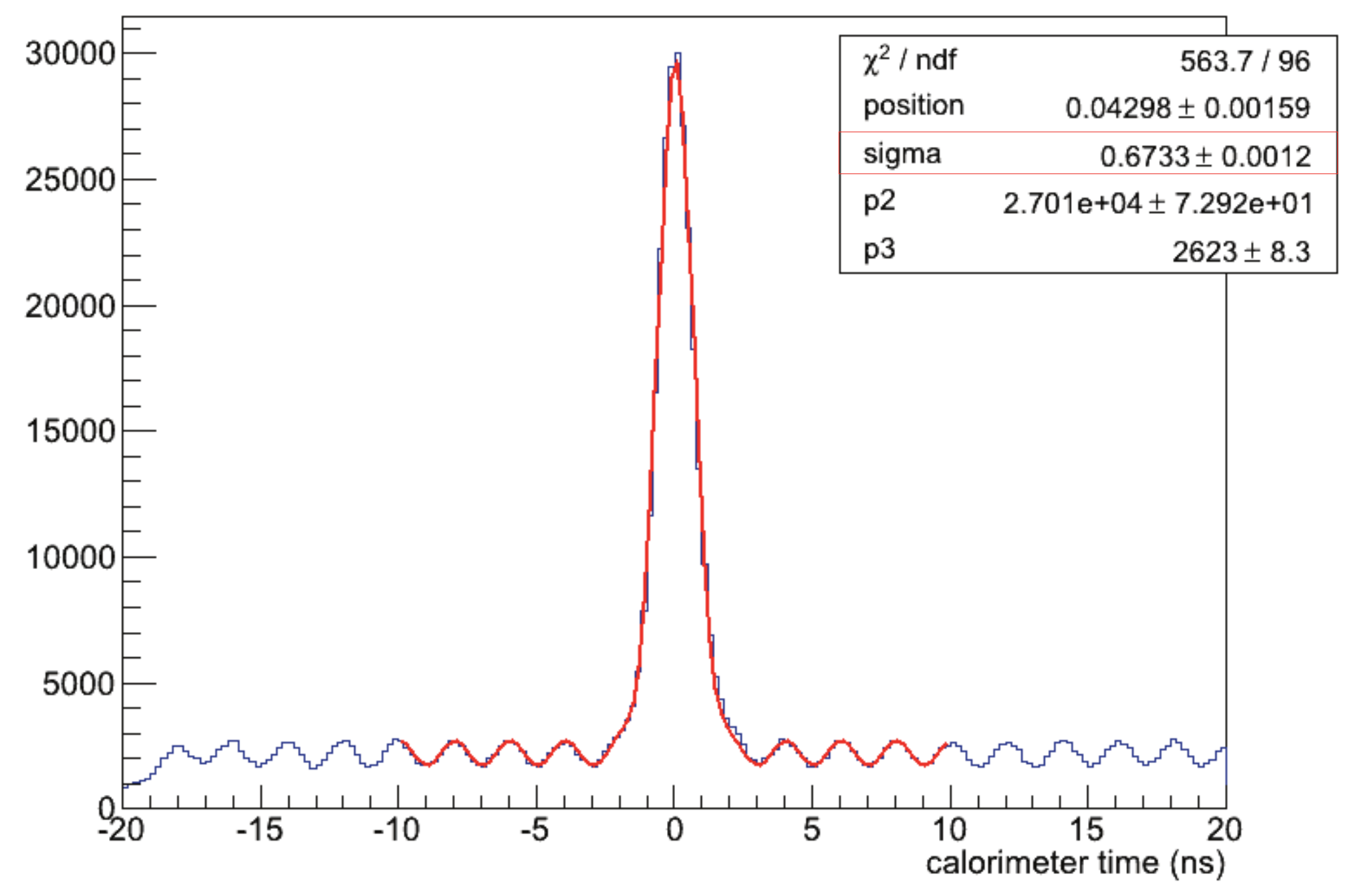}
\end{minipage}\hfill\begin{minipage}{0.475\textwidth}
\caption{
Coincidence time spectrum for $(e,e'\gamma)$ events 
The $\gamma$ signal is a weighted sum of all PbF$_2$ crystals in the shower.
The electron timing is corrected for time-walk in the HRS scintillator paddles, path length in the spectrometer, and other effects.
The one-$\sigma$ timing resolution is 0.7 ns, the 2 ns beam structure is clearly visible.  Within a $\pm 3$ ns coincidence window,
the accidental to true coincidence ratio is $3\%$.  All calorimeter signals of  more than 0.1 GeV are included in this spectrum.
}
\label{fig:DVCS-t}
\end{minipage}
\end{figure}

\begin{figure}[hbp]
\begin{minipage}{0.50\textwidth}
\includegraphics[width=\textwidth]{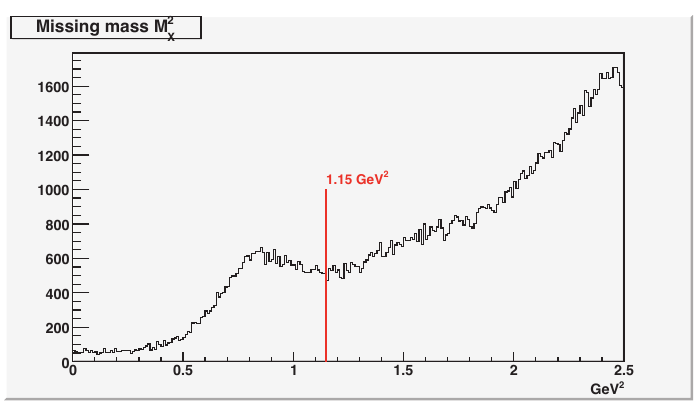}
\end{minipage}\hfill\begin{minipage}{0.475\textwidth}
\caption{
Missing mass squared $M_X^2$ distribution of H$(e,e'\gamma)X$ events.
The exclusive peak is at $M_X^2=M_p^2 = 0.88$ GeV$^2$.  The hadronic H$(e,e'\gamma)N\pi$ continuum
starts at $M_X^2 = (M_p+m_\pi^0)^2 = 1.15$ GeV$^2$.  Accidental coincidences
and the statistical sample of H$(e,e'\gamma)\gamma X'$ events are not yet subtracted.
}
\label{fig:DVCS-MX2}
\end{minipage}
\end{figure}

\begin{figure}[hbp]
\center{\includegraphics[width=0.475\textwidth]{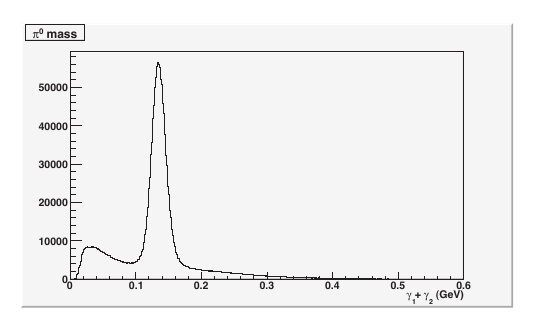}\hfill
\includegraphics[width=0.475\textwidth]{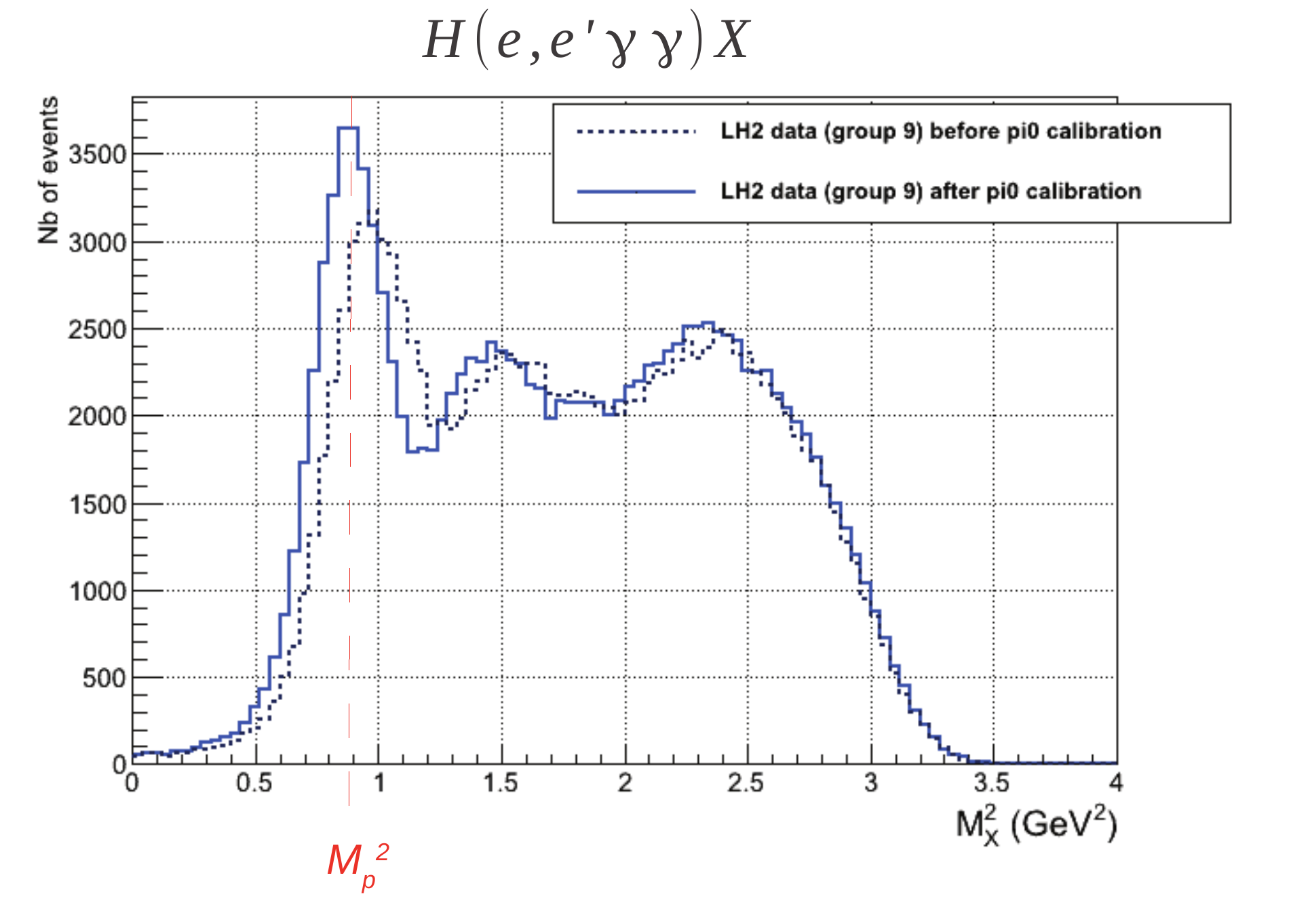}}
\caption{
{\bf Left:}  ${\gamma\gamma}$ mass spectrum of H$(e,e'\gamma\gamma)X$ events.
The 1-$\sigma$ width of the $\pi^0$ peak is 11 MeV.
{\bf Right:} Missing mass squared $M_X^2$ of H$(e,e'\pi^0)X$ events before(dotted) and after (solid) refining
the calorimeter calibrations based on the $\pi^0$ mass reconstruction.}
\label{fig:DVCS-pi0}
\end{figure}


%
%

\clearpage \newpage

\subsection[E05-102, E08-005 - $A_T$, $A_L$, $A_y^0$ in $^3$He($e,e'n$)]{E05-102, E08-005 - $A_T$, $A_L$, $A_y^0$ in $^3$He($e,e'n$)}
\label{sec:quasi-elastic}

\begin{center}
\bf Measurements of the Double-Spin Asymmetries $A_T$ and $A_L$ and the Target Single-Spin Asymmetry $A_y^0$ in the Quasi-Elastic $^3$He$^{\uparrow}$($e,e'n$) Reaction
\end{center}

\begin{center}
S. Gilad, D.W. Higinbotham, W. Korsch, B.E. Norum, S. \v{S}irca, E05-102 spokespersons, \\
T. Averett, D. Higinbotham, V. Sulkosky, E08-005 spokespersons, \\
and \\
the Hall A Collaboration.\\
contributed by E. Long (Kent State University).
\end{center}

\subsubsection{Progress of $^3$He($e,e'n$) Asymmetries}\label{sec:details}
Progress has been made on the $^3\vec{\mathrm{He}}$($\vec{e},e'n$) double-spin asymmetries for experiment E05-102, where the target was polarized in the transverse direction and longitudinal directions. The $A_T$ and $A_L$ observables are correlated to neutron electromagnetic form factors. 

Data from the RHRS were used to isolate the quasi-elastic peak and apply basic kinematic cuts. It was also used to identify scattered electrons. Neutrons were identified using the Hall A Neutron Detector. The raw asymmetries for $A_T$ and $A_L$ are presented against the energy transfer, $\nu$, in Figures \ref{at-asyms} and \ref{al-asyms}, respectively. Each asymmetry was measured at $Q^2$ values of 0.51 (GeV/$c)^2$ and 0.95 (GeV/$c)^2$.

\begin{figure}[hbt]
\center{\includegraphics[width=8.75cm]{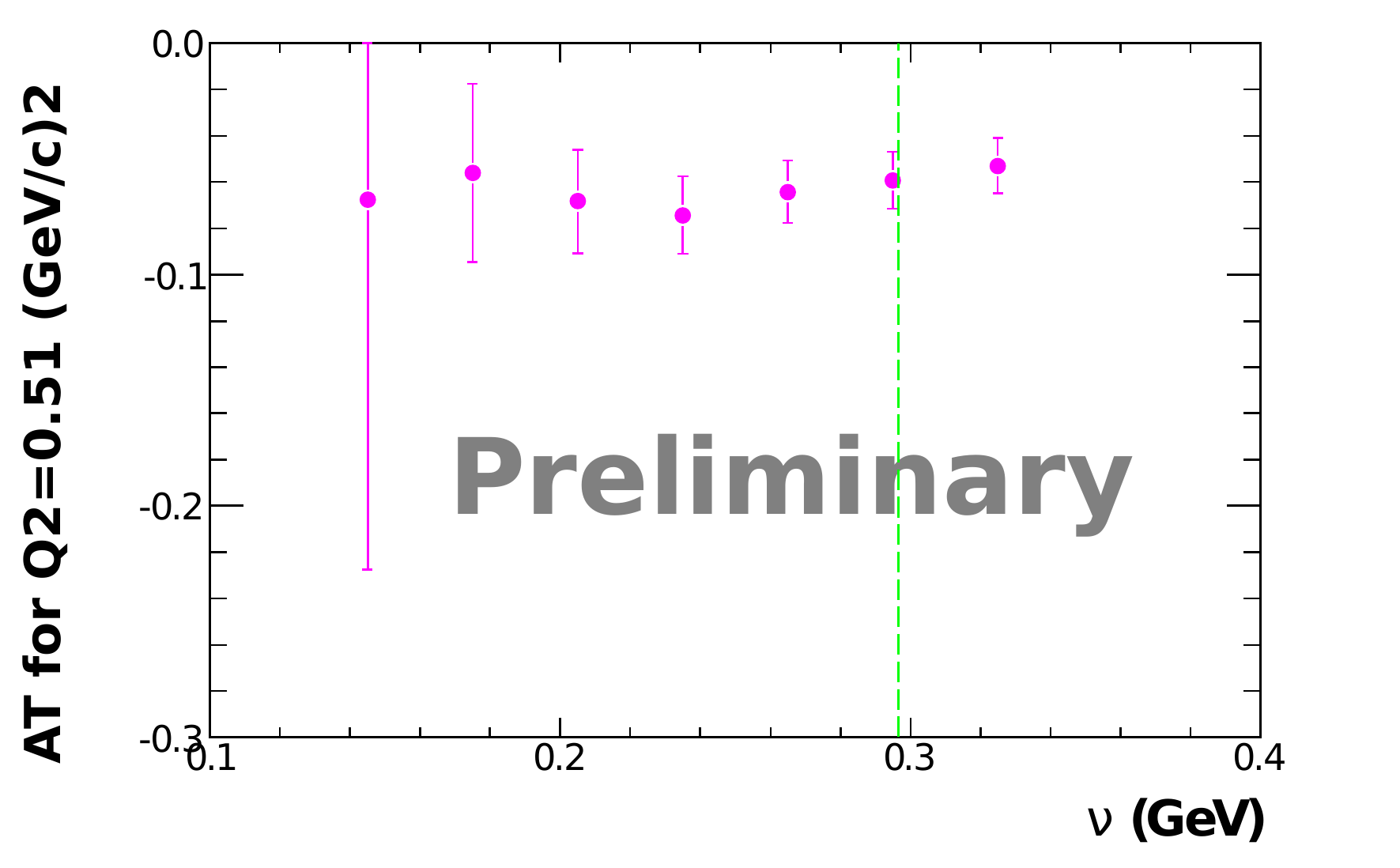}\includegraphics[width=8.75cm]{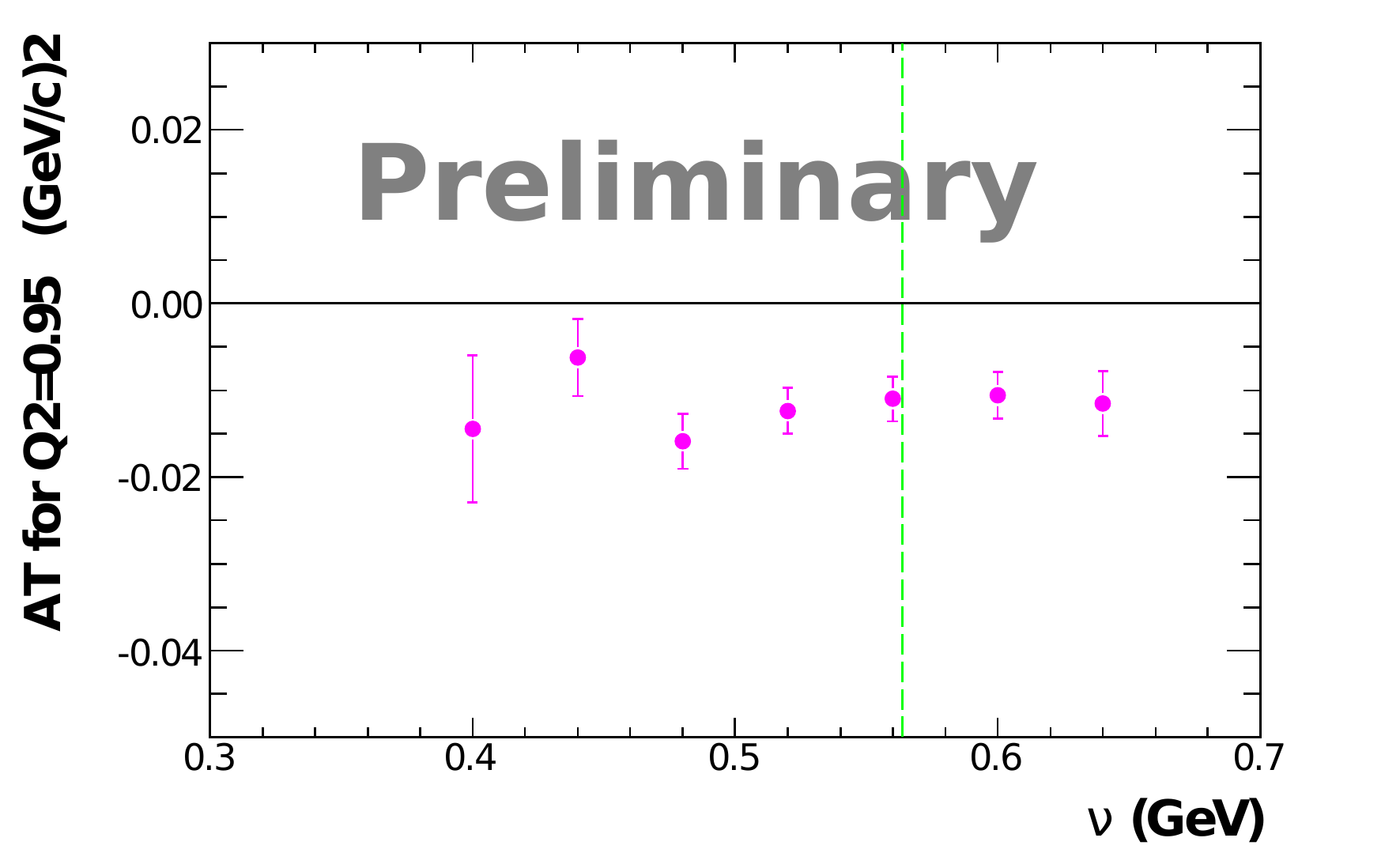}}
\caption{The raw asymmetry for $A_T$ against the energy-transfer $\nu$ is presented at $Q^2=0.505 $(GeV/$c)^2$ on the left and at $Q^2=0.953 $(GeV/$c)^2$ on the right. The green dashed line corresponds to the quasi-elastic peak.}
\label{at-asyms}
\end{figure}

\begin{figure}[hbt]
\center{\includegraphics[width=8.75cm]{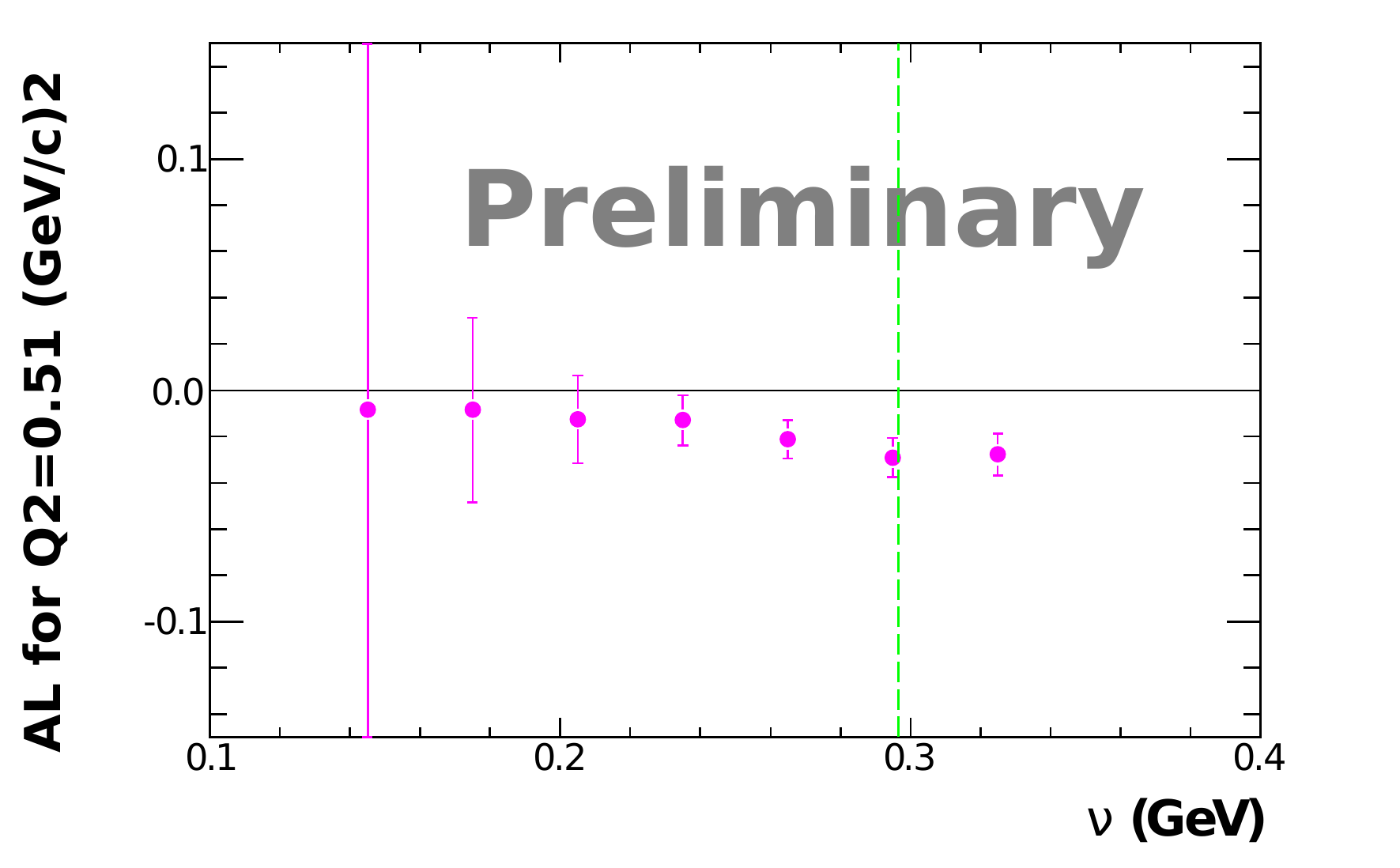}\includegraphics[width=8.75cm]{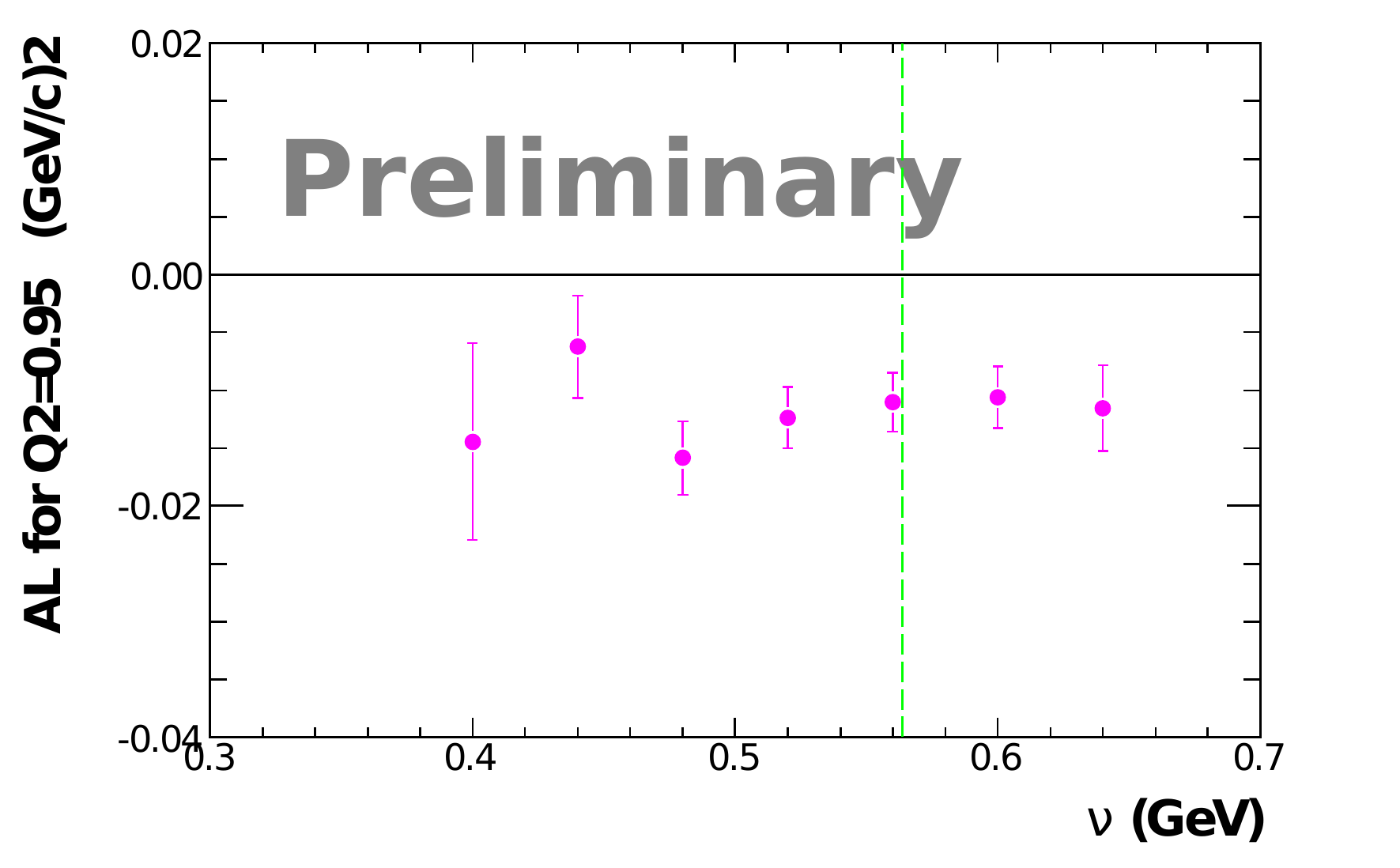}}
\caption{The raw asymmetry for $A_L$ against the energy-transfer $\nu$ is presented at $Q^2=0.505 $(GeV/$c)^2$ on the left and at $Q^2=0.953 $(GeV/$c)^2$ on the right. The green dashed line corresponds to the quasi-elastic peak.}
\label{al-asyms}
\end{figure}

In addition, a similar method was used to determine the target single-spin $^3$He$^{\uparrow}$($e,e'n$) asymmetry $A_y^0$, which ran using the same experimental set-up as E05-102 but with the target polarized in the vertical direction. The A$_y$ observable is sensitive to final state interactions (FSI) and meson exchange currents (MEC). At low Q$^2$, contributions from FSI and MEC are expected to be large and decrease as Q$^2$ increases. Asymmetry measurements on the quasi-elastic peak were taken at $Q^2$ values of 0.13 (GeV/$c)^2$, 0.46 (GeV/$c)^2$, and 0.95 (GeV/$c)^2$. The raw asymmetry measurements are presented in Figure \ref{ay-asym}.

Dilution contributions from target polarization, beam polarization, neutron contamination, and nitrogen admixture in the target are in the process of being finalized for each of the asymmetries mentioned above. These factors will approximately double the magnitude of the raw asymmetries presented.

\begin{figure}[hbt]
\center{\includegraphics[width=8.75cm]{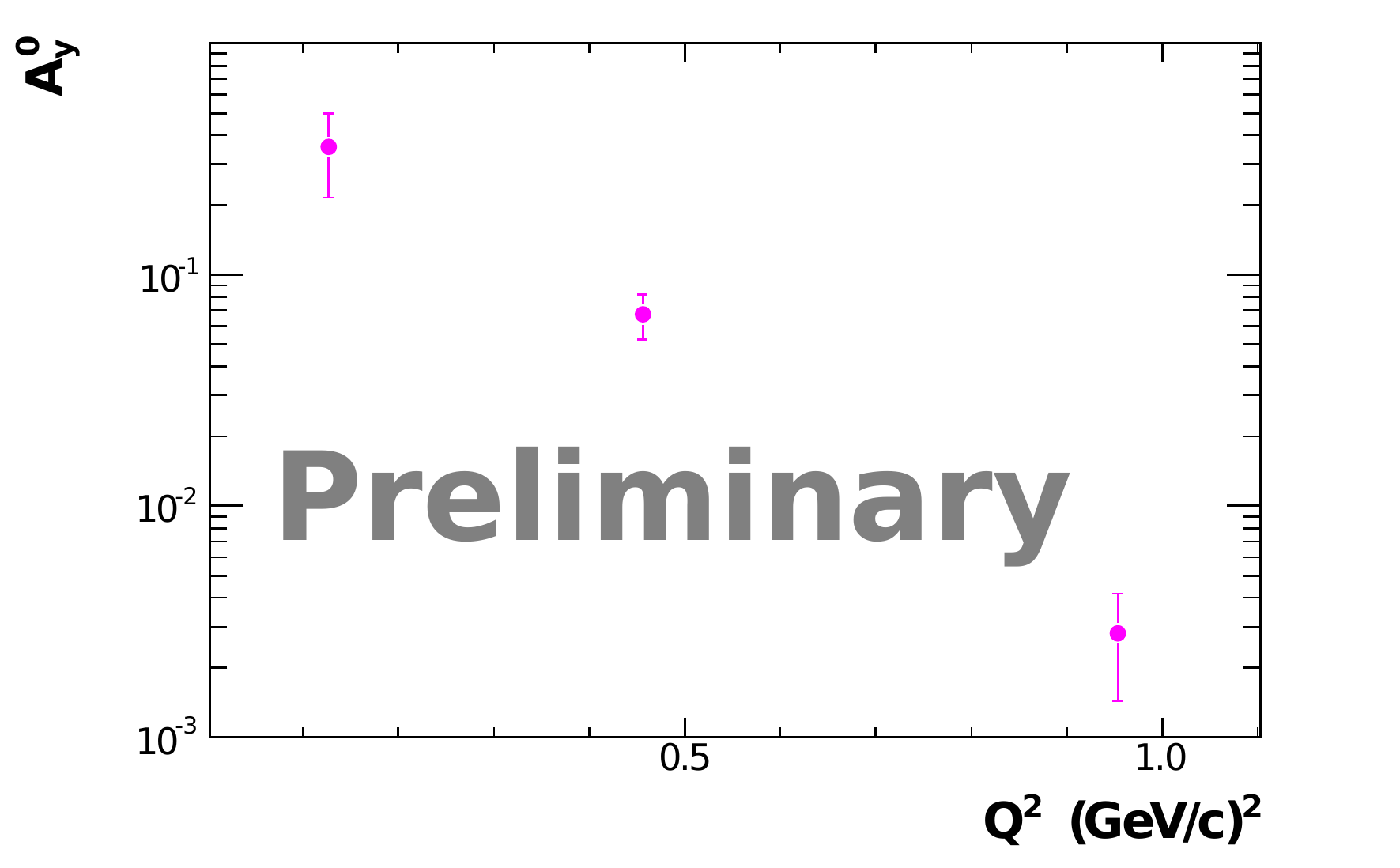}}
\caption{The raw asymmetry for $A_y^0$ against $Q^2$ is presented on a logarithmic scale.}
\label{ay-asym}
\end{figure}

\clearpage \newpage

\subsection{E08-007 - $G_E^p$ at Low $Q^2$}

\begin{center}
\bf Measurement of the Proton Elastic Form Factor Ratio at Low $Q^2$
\end{center}

\begin{center}
J.~Arrington, D.~Day, R.~Gilman, D.W~Higinbotham, G.~Ron, and A.~Sarty, spokespersons,\\
and the E08-007 collaboration.\\
contributed by M. Friedman. 
\end{center}

\subsubsection{Motivation}

The measurement of the proton form factor ratio at low $Q^2$ is important due to several reasons. 
First, the form factors are fundamental properties
of the nucleon that should be measured well to test our understanding of the nucleon. 
Second, although theory generally indicates the
form factors vary smoothly with $Q^2$, there are an unsatisfyingly large number 
of theory calculations, fits, and data points that suggests this might not be the case, and
that there might be narrow structures in the form factors. E08-007-II experiment is
good enough to either confirm or refute existing suggestions of few percent structures
in the form factors, or in the form factor ratio. Third, it has become apparent that the
existing uncertainties in the form factors are among the leading contributions to uncertainties 
in determining other physics quantities, such as the nucleon Zemach radius, the
strange form factors determined in parity violation, and the generalized parton distributions 
determined in deeply virtual Compton scattering. The improvement possible with this 
measurements is substantial. The proton electric and magnetic "radii" are also directly 
related to the form factor slope at $Q^2=0$:
\begin{equation}
\left<r^{2}_{E/M}\right>=-\frac{6}{G_{E/M}(0)}\left(\frac{dG_{E/M}(Q^2)}{dQ^2}\right)_{Q^2=0}.
\end{equation}
Recent results from muonic-hydrogen lamb shift measurements \cite{lamb} suggest a significantly smaller charge radius for the proton
than the established values, and precise measurement of the form factor at very low $Q^2$ may help to resolve this 
discrepancy. 

\subsubsection{The Experiment}

The E08-007-II was run in parallel with experiment E08-027 and the details about the experiment can be found in the experiment section of E08-027.
All E08-007-II runs were done with magnetic field of 5T, and the kinematics are listed in Table \ref{tab:gep_kinematics}. 
\begin{table}
\centering
\begin{tabular}{cccc}
\hline 
$Q^2$ & $E$ & $E^{\prime}$ & $\theta_e$\\
(GeV$^2$) & (GeV) & (GeV) & (deg) \\ \hline
0.013 & 1.157 & 1.150 & 5.7\\ 
0.020 & 1.712 & 1.701 & 4.7\\ 
0.030 & 1.712 & 1.696 & 5.8\\
0.034 & 2.253 & 2.235 & 4.7\\
0.052 & 2.253 & 2.225 & 5.8\\
\hline 
\end{tabular}
\caption{\label{tab:gep_kinematics}Kinematics used for E08-007-II. The $Q^2$ binning is not final}
\end{table}

\subsubsection{Experimental Progress}
The details about optics, helicity and other calibrations are listed in the experimental progress section of E08-027. The target polarization
analysis is completed, with relative uncertainties of 2\%-3\%. These uncertainties are still under investigation. 

A preliminary extraction of the raw data is done, with the available calibrations, for almost all the data. Analysis has begun of the 2.2 GeV data.
\begin{figure}
\centering
\includegraphics[height=80mm,width=100mm]{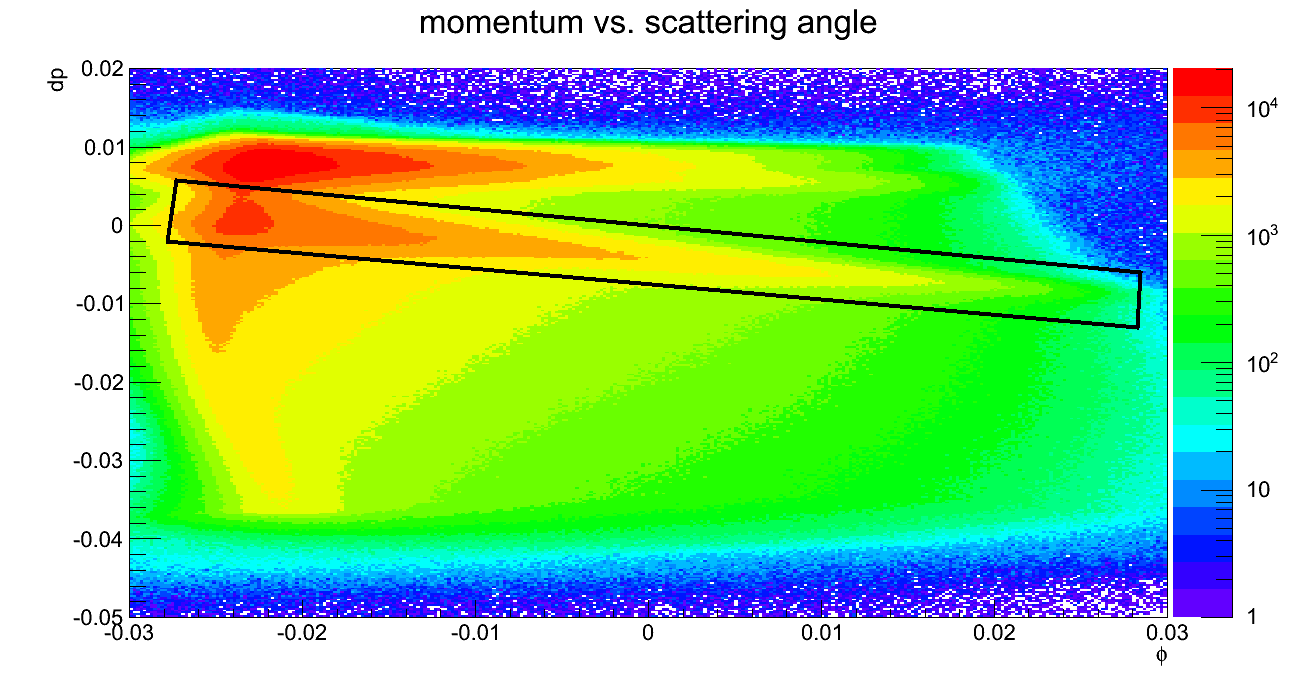}
\caption{\label{fig:gep2D}Scattered electron momentum as function of scattering angle. The black rectangle
identifies the hydrogen elastic stripe. The data was collected by the left HRS in the 2.2 GeV run.}
\end{figure}
Fig. \ref{fig:gep2D} shows 2D data of dp vs $\phi$ for the 2.2 GeV runs. The elastic stripes of the different elements are clearly seen, and the hydrogen
is well separated from the heavy elements.

\begin{figure}
\centering
\includegraphics[height=140mm,width=150mm]{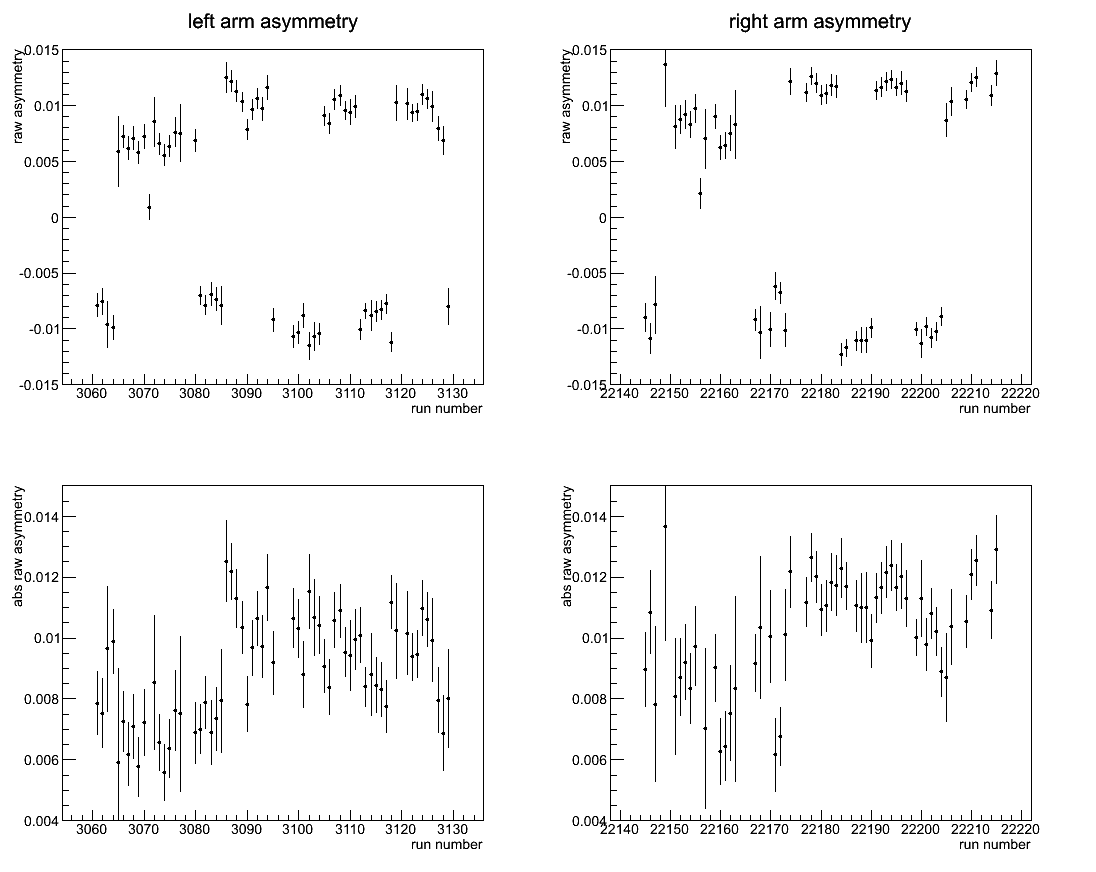}
\caption{\label{fig:raw_asymmetries}Raw asymmetry as function of run number for the 2.2 GeV data. The asymmetry changes sign due
to polarity flipping in the beam and in the target. The bottom plots shows the absolute values of the asymmetries. Note that these are 
raw asymmetries, without corrections for the polarization of the target and the beam, and dilution.}
\end{figure}
Raw asymmetries for the 2.2 GeV data shows consistency for both left and right arms (see Fig. \ref{fig:raw_asymmetries}). The flipping between positive and 
negative asymmetries is due to changes in target field direction and Half-Wave Plate (HWP) flipping during the experiment. The absolute values of the raw asymmetries
does not depend on the target field or the HWP orientation. Based on raw asymmetries for all E08-007-II kinematics, we estimate the statistical uncertainties to be between 
1\%-2\% (depending on bin and bin size).

The dilution factor will be extracted based on experimental data and simulations. 
Experimental dilution factor values are extracted using He and C runs, as illustrated in Fig. \ref{fig:dilution}. 
\begin{figure}
\centering
\includegraphics[height=80mm,width=100mm]{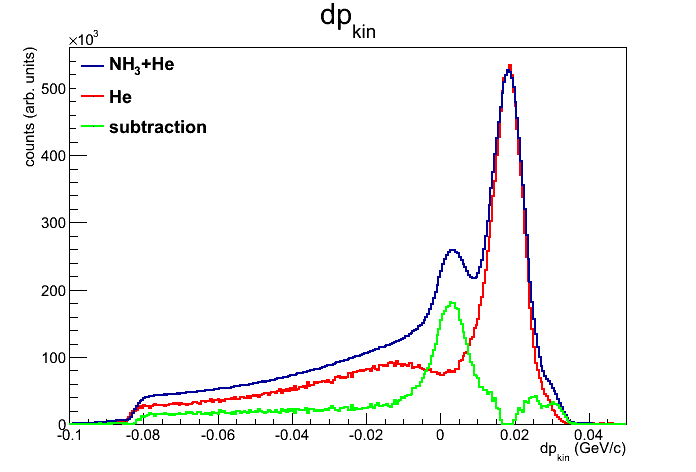}
\caption{\label{fig:dilution}Estimation of the dilution factor using helium data. Subtraction of helium data (red)
from the experimental data (blue) reveals what we interpret as hydrogen elastic events (green).}
\end{figure}
The experimental dilution runs should be corrected for different target thicknesses to be more reliable, and compared to calculations for more precise analysis.
The corrections for the asymmetry contribution by the polarized nitrogen has not been studied.

\clearpage \newpage

\subsection[E08-008 - D Electrodisintegration]{E08-008 - Deuteron Electrodisintegration}
\label{sec:e08008}

\begin{center}
\bf E08-008: Exclusive Study of Deuteron Electrodisintegration near Threshold
\end{center}

\begin{center}
B. Norum, A. Kelleher, S. Gilad, D. Higinbotham, spokespersons, \\
C. Hanretty, Post-Doc, \\
and \\
the Hall A Collaboration.\\
contributed by C. Hanretty.
\end{center}

\subsubsection{The E08-008 Experiment}
Data for the E08-008 experiment were collected from February 17$^{th}$ to 23$^{rd}$, 2011 in the first experiment to run in the Winter 2011 run period of Jefferson Lab's Hall A. In this experiment, polarized electrons with 3.358~GeV of energy were delivered to the Hall via the CEBAF accelerator. These electrons were incident on three different target cells (4~cm LH$_{2}$, 4~cm LD$_{2}$, and 15~cm LD$_{2}$), all with unpolarized target materials. Scattered electrons and recoil protons were detected using the Hall's two High Resolution Spectrometers (HRSs). To reveal the polarization of the recoil proton, a Focal Plane Polarimeter (FPP) was used. This FPP is comprised of an analyzing material placed perpendicular to the proton's trajectory upstream of a set of two straw chambers. Spin-orbit interactions between the polarized proton and the analyzing material result in asymmetries in the $\phi$-distributions seen in the FPP. Data were collected in this manner for $p(\vec{e},e^{\prime}p)$ and $d(\vec{e},e^{\prime}p)n$ reactions for $x_{B}~\epsilon~[1,2]$ as seen in Tab.~\ref{tab:kinSettings}.

\begin{table}[ht]
\centering
\caption[E08-008: Kinematic settings used in analysis of E08-008 data.]{Kinematic settings used in analysis of E08-008 data listed in order of increasing $x_{B}$.}
\begin{tabular}{ | c | c | c | c | c |}
\hline
Target Cell & p$_{LHRS}$ & $\theta_{LHRS}$ & p$_{RHRS}$ & $\theta_{RHRS}$ \\
\hline \hline
4~cm LH$_{2}$ & 0.968~\small{GeV/c} & 57$^{\circ}$ & 2.95~\small{GeV/c} & 16$^{\circ}$ \\ \hline
4~cm LD$_{2}$ & 0.968~\small{GeV/c} & 57$^{\circ}$ & 2.95~\small{GeV/c} & 16$^{\circ}$ \\ \hline
15~cm LD$_{2}$ & 0.625~\small{GeV/c} & 68$^{\circ}$ & 3.14~\small{GeV/c} & 16$^{\circ}$ \\ \hline
15~cm LD$_{2}$ & 0.490~\small{GeV/c} & 68$^{\circ}$ & 3.14~\small{GeV/c} & 16$^{\circ}$ \\ \hline
\end{tabular}
\label{tab:kinSettings}
\end{table}

\subsubsection{Preliminary Analysis}
While the physics goal of E08-008 is to measure the ratio $\mu_{p}(G_{Ep}/G_{Mp})$ at low $Q^{2}$ ($x_{B} \rightarrow 2$, where this quantity is sensitive to N-N interactions), data was also taken in the $x_{B} = 1$ region for calibration purposes. A preliminary analysis of elastic $p(\vec{e},e^{\prime}p)$ and a brief description of the techniques used are presented here.

Events are first subjected to a series of general kinematic and coincidence cuts to ensure the analysis of properly reconstructed $p(\vec{e},e^{\prime}p)$ reactions. Subsequently, events passing these general cuts are then required to pass cuts specific to the focal plane in the HRS. The first of these cuts is termed the ``Conetest'' cut which is used to remove false asymmetries based on the detector geometry. Following this cut is a requirement that the angle $\theta_{fpp}~\epsilon~[5,30)$. The lower limit of this cut removes events that scattered via the Coulomb interaction. Finally a cut is imposed on the secondary scattering vertex which ensures that the analyzed event scattered from the analyzing material. 

Events satisfying the above cuts are used to form $\phi$-distributions for each beam helicity setting (+1 or -1). In order to remove any remaining false asymmetries, the asymmetry between the two helicity states is formed via
\begin{equation}\label{eq:generalHelAsym}
A(\phi) = \left( \frac{N^{+}}{N^{+}_{ave}}\right)-\left(\frac{N^{-}}{N^{-}_{ave}} \right)~,
\end{equation}
where $N^{\pm}$ represents the number of events for a given helicity setting for a bin in $\phi$. These asymmetries are then fit using
\begin{equation}\label{eq:asymFitEqn}
f^{+}(\phi) - f^{-}(\phi) = y_{0} + A_{y}\left[~P^{fpp}_{x}\cos(\phi) - P^{fpp}_{y}\sin(\phi)~\right]
\end{equation}
where $y_{0}$ is a vertical offset term, $A_{y}$ represents the analyzing power, and the expression
\begin{equation}
f^{\pm}(\phi) = \frac{1}{2\pi}\left(1 \pm A_{y}\left[~P^{fpp}_{y}\sin(\phi) - P^{fpp}_{x}\cos(\phi)~\right] \right)
\end{equation}
describes the distribution of events in the FPP as a function of $\phi$. 
A fit of the helicity asymmetry for $\theta_{fpp}~\epsilon~[5,30)$ along with the fit results can be seen in Fig. \ref{fig:phiAsym_rearChambers_kinSetting1_fitResults}.
\begin{figure}[hbt]
\center{\includegraphics[width=9.0cm]{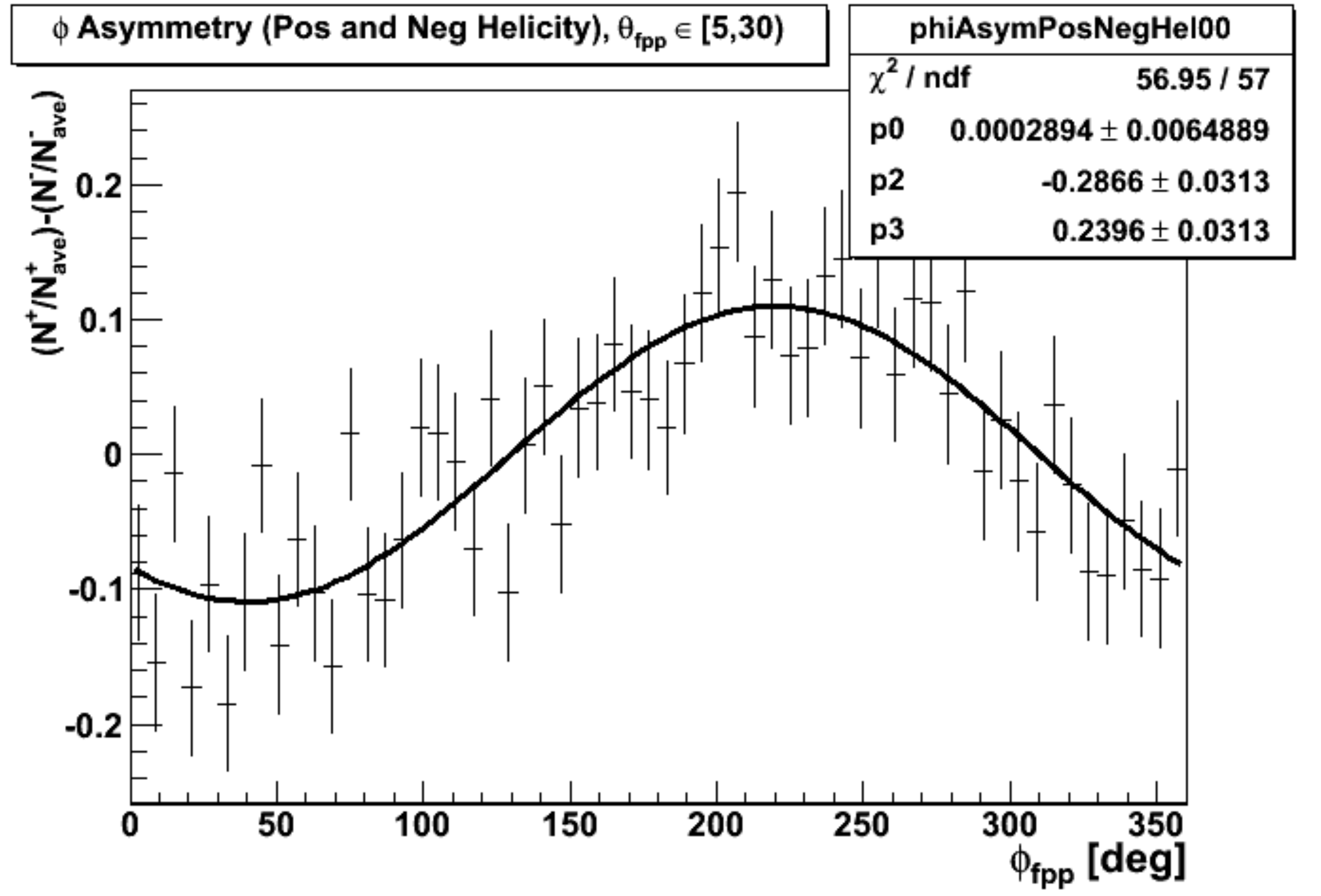}}
\caption[Helicity asymmetry as a function of $\phi$ for $p(\vec{e},e^{\prime}p)$ events.]{The helicity asymmetry $\left(\frac{N^{+}}{N^{+}_{ave}}-\frac{N^{-}}{N^{-}_{ave}}\right)$ as a function of $\phi$ for $p(\vec{e},e^{\prime}p)$ events in the range of $\theta_{fpp}~ \epsilon~[5,30)$. Also shown is the fit to this asymmetry using Eqn. \ref{eq:asymFitEqn} with the fit results. Here, p0 is the vertical offset ($y_{0}$), p2 = $P^{fpp}_{x}$, and p3 = $P^{fpp}_{y}$. The parameter, p1, corresponds to the analyzing power which is determined using model calculations described in \cite{GlisterNIM,ZhanThesis} and set as a fixed parameter.}
\label{fig:phiAsym_rearChambers_kinSetting1_fitResults}
\end{figure}

As the ratio $\mu_{p}(G_{Ep}/G_{Mp})$ is calculated from quantities at the target, measurements in the focal plane of the HRS must be transported ``backwards'' through the spectrometer to the target frame. Two methods are employed to carry out this transport, as described below.

The first of these methods utilizes the ($\phi$) helicity asymmetries seen in the FPP \cite{ZhanThesis}. These asymmetries, instead of being fit to Eq. \ref{eq:asymFitEqn}, are fit to an equivalent expression
\begin{equation}\label{eq:phaseShiftFitEqn}
f^{+}(\phi) - f^{-}(\phi) = C\cos(\phi + \delta),
\end{equation}
where C is a constant in the fit and
\begin{equation}\label{eq:tanDelta}
\tan(\delta) = \frac{P_{y}^{fpp}}{P_{x}^{fpp}}.
\end{equation}
In the dipole approximation, the phase shift $\delta$ as determined from the fit can be used to directly measure the ratio $\mu_{p}(G_{Ep}/G_{Mp})$ via
\begin{equation}\label{eq:phaseShiftFinalEqn}
\mu_{p}\frac{G_{Ep}}{G_{Mp}} = \mu_{p} \cdot K \sin(\chi) \left(\frac{P^{fpp}_{y}}{P^{fpp}_{x}}\right),
\end{equation}
where the kinematic factor, K, is given by
\begin{equation}
K = \frac{E+E^{\prime}}{m_{p}}\tan^{2}\left(\frac{\theta_{e}}{2}\right),
\end{equation}\label{eq:phaseShiftKinFactor}
and $\chi$ represents the precession angle of the proton's spin through the dipole, given by
\begin{equation}\label{eq:phaseShiftChi}
\chi = \gamma(\mu_{p}-1)\Theta_{bend}.
\end{equation}
Seen in Fig. \ref{fig:phiAsym_rearChambers_kinSetting1_deltaFit} is the fit to the helicity asymmetry (Eq. \ref{eq:generalHelAsym}) using Eq. \ref{eq:phaseShiftFitEqn} for $p(\vec{e},e^{\prime}p)$ events. Using the results from the fit and Eqns. \ref{eq:tanDelta}~-~\ref{eq:phaseShiftChi}, $\mu_{p}(G_{Ep}/G_{Mp}) = 0.873801 \pm 0.149866$ for events where $\theta_{fpp}~\epsilon~[5,30)$.
\begin{figure}[hbt]
\center{\includegraphics[width=9.0cm]{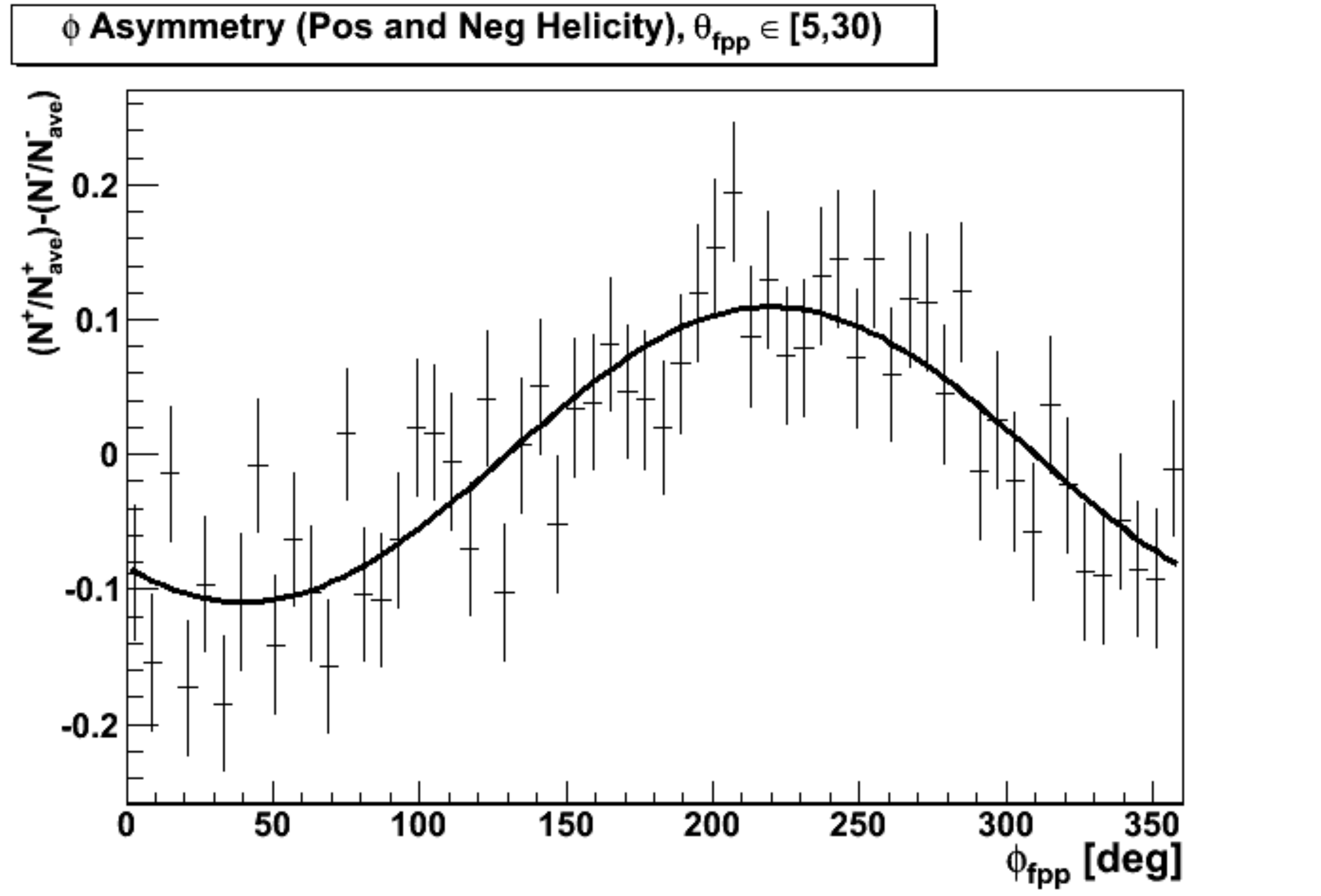}}
\caption[Helicity asymmetry for $p(\vec{e},e^{\prime}p)$ events fit with Eq. \ref{eq:phaseShiftFitEqn}.]{The helicity asymmetry ($\frac{N^{+}}{N^{+}_{ave}}-\frac{N^{-}}{N^{-}_{ave}}$) as a function of $\phi$ for $p(\vec{e},e^{\prime}p)$ events in the range of $\theta_{fpp}~ \epsilon~[5,30)$. Also shown is the fit to this asymmetry using Eqn \ref{eq:phaseShiftFitEqn}. Here, the phase shift $\delta$ is found to be $-0.6962 \pm 0.0838$ rad.}
\label{fig:phiAsym_rearChambers_kinSetting1_deltaFit}
\end{figure}

As important quantities such as analyzing power are not constant over the range of $\theta_{fpp}~\epsilon~[5,30)$, the prudent approach is to bin data in $\theta_{fpp}$ and perform the measurement of $\mu_{p}(G_{Ep}/G_{Mp})$ for each bin. A subsequent fit to these individual measurements can then be performed, allowing for a proper weighting of the data for the entire range of $\theta_{fpp}$. Fig. \ref{fig:phaseShiftResultsInThFpp_kinSetting1} shows the $\mu_{p}(G_{Ep}/G_{Mp})$ measurements using this method binned in $\theta_{fpp}$ as well as a fit to a horizontal line (in the range of $\theta_{fpp}~\epsilon~[5,30)$). Here the cumulative result ($\theta_{fpp}~\epsilon~[5,30)$) is shown as the data point to the far right of the histogram and the horizontal fit result in the stat box. These two quantities, within error, are seen to agree.
\begin{figure}[hbt]
\center{\includegraphics[width=9.0cm]{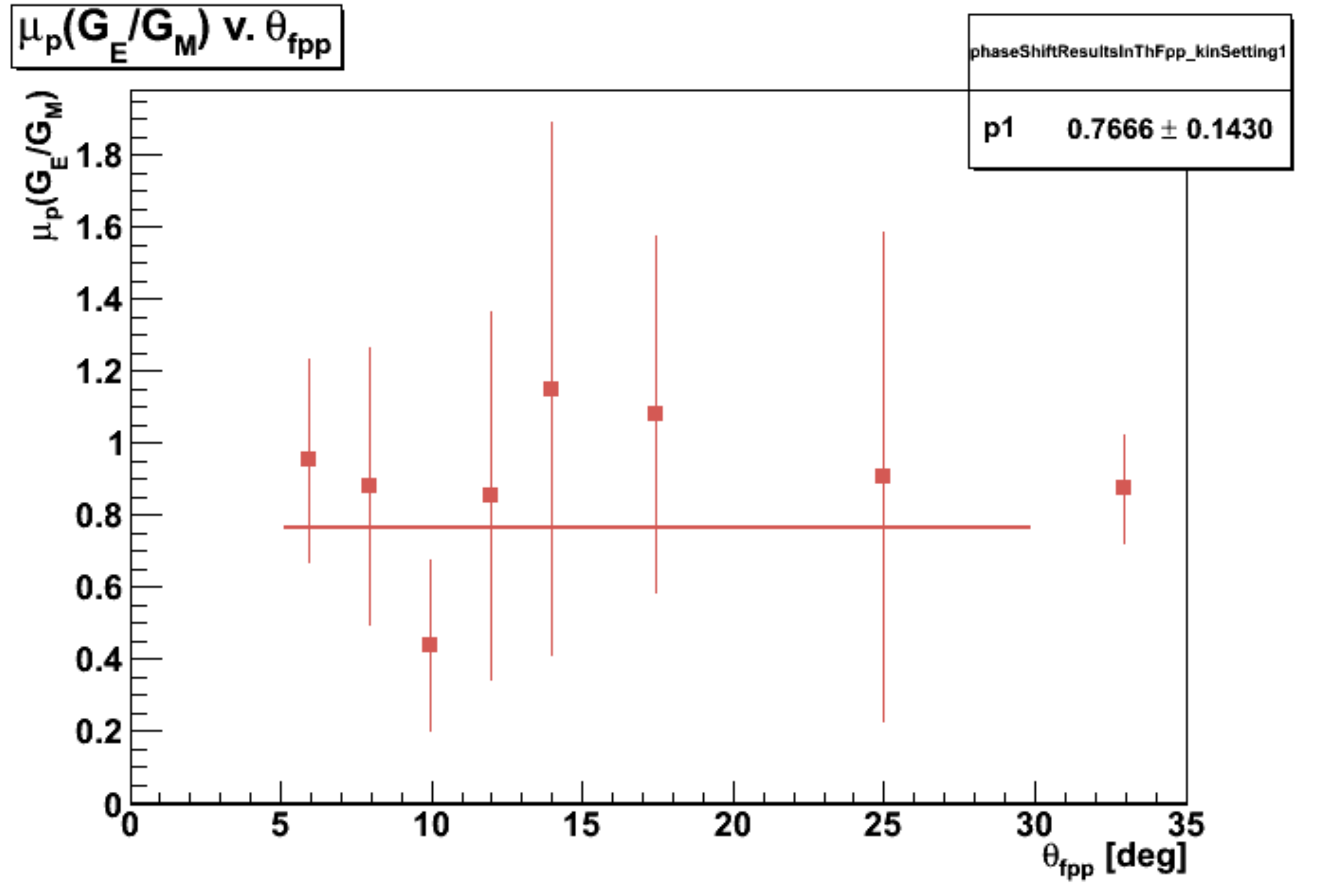}}
\caption[$\mu_{p}(G_{Ep}/G_{Mp})$ measurements (cumulative and binned in $\theta_{fpp}$) for $p(\vec{e},e^{\prime}p)$ events.]{$\mu_{p}(G_{Ep}/G_{Mp})$ measurements using the phase shift method binned in $\theta_{fpp}$. Also shown is a fit to the binned results using a horizontal line (in the range of $\theta_{fpp}~\epsilon~[5,30)$) and the fit result. Here, the cumulative result is shown as the data point to the far right of the histogram.}
\label{fig:phaseShiftResultsInThFpp_kinSetting1}
\end{figure}

The second method used to obtain polarization observables at the target inherits heavily from the Palmetto code used in several Hall A and Hall C analyses. This code, called Sago, uses information from the reconstructed tracks in the FPP and knowledge of the magnetic fields the recoiling proton traverses in the HRS to construct a rotation matrix $\mathbf{S}$. This rotation matrix relates the polarization components in the target ($P_{x}^{tg}, P_{z}^{tg}$) to those seen in the FPP ($P_{x}^{fpp}, P_{y}^{fpp}$):
\begin{equation}\label{basicRotationEqn}
\mathbf{P^{fpp}} = \mathbf{S \cdot P^{tg}}.
\end{equation}
As previously stated, the analyzing power is not constant over the entire range of $\theta_{fpp}$ and events with a higher analyzing power should hold more weight in the analysis. To account for this, Sago uses a weighted sum method when building the rotation matrix ($\mathbf{S}$). The full, explicit form of Eq. \ref{basicRotationEqn} can be written as
\begin{equation}
\begin{pmatrix}
\label{fullRotationEqn}
	\Sigma_{i}\lambda_{x,i} \\
	\Sigma_{i}\lambda_{y,i}
\end{pmatrix}
=
\begin{pmatrix}
    \Sigma_{i}\lambda_{x,i}\lambda_{x,i} & \Sigma_{i}\lambda_{z,i}\lambda_{x,i} \\
    \Sigma_{i}\lambda_{x,i}\lambda_{z,i} & \Sigma_{i}\lambda_{z,i}\lambda_{z,i}		
\end{pmatrix}
\begin{pmatrix}
    P_{x}^{tg} \\
    P_{z}^{tg}
\end{pmatrix}
\end{equation}
where $\Sigma_{i}$ represents a summation of the number of events and
\begin{equation}\label{generalizedLambda}
\lambda_{j} = \eta{h}A_{y}(S_{yj}\sin\phi - S_{xj}\cos\phi)~.
\end{equation}
Here, $\eta$ denotes the beam helicity, $h$ represents the degree of polarization of the incident electrons, $A_{y}$ is the analyzing power, and $S_{yj}$ (and $S_{xj}$) being elements of the rotation matrix $\mathbf{S}$.

Once these matrices are constructed for the data set, Eq. \ref{fullRotationEqn} can be solved for $P_{x}^{tg}$ and $P_{z}^{tg}$ via a matrix inversion. The form factor ratio is then 
\begin{equation}\label{eq:ffRatio}
\mu_{p}\frac{G^{p}_{E}}{G^{p}_{M}} = K\frac{P_{y}^{tg}}{P_{z}^{tg}}~,
\end{equation}\
where
\begin{equation}\label{eq:kinFactor}
K = -\mu_{p}\frac{E_{e} + E_{e^{\prime}}}{2M_{p}}\tan\frac{\theta_{e}}{2}~.
\end{equation}
The described method was used to extract the ratio $\mu_{p}(G_{Ep}/G_{Mp})$ for a subset of the e08008 data. For elastic $p(\vec{e},e^{\prime}p)$ events, this ratio was found to be $0.928243 \pm 0.155129$. As with the method involving the phase shift, the data (and therefore $\mu_{p}(G_{Ep}/G_{Mp})$) were also binned in $\theta_{fpp}$ and the set of measurements fit with a horizontal line from $\theta_{fpp}~\epsilon~[5,30)$ (Fig. \ref{fig:sagoResultsInThFpp_kinSetting1}). Here, the fit of the results from Sago binned in $\theta_{fpp}$ can be seen to match the cumulative result within error.  
\begin{figure}[hbt]
\center{\includegraphics[width=9.0cm]{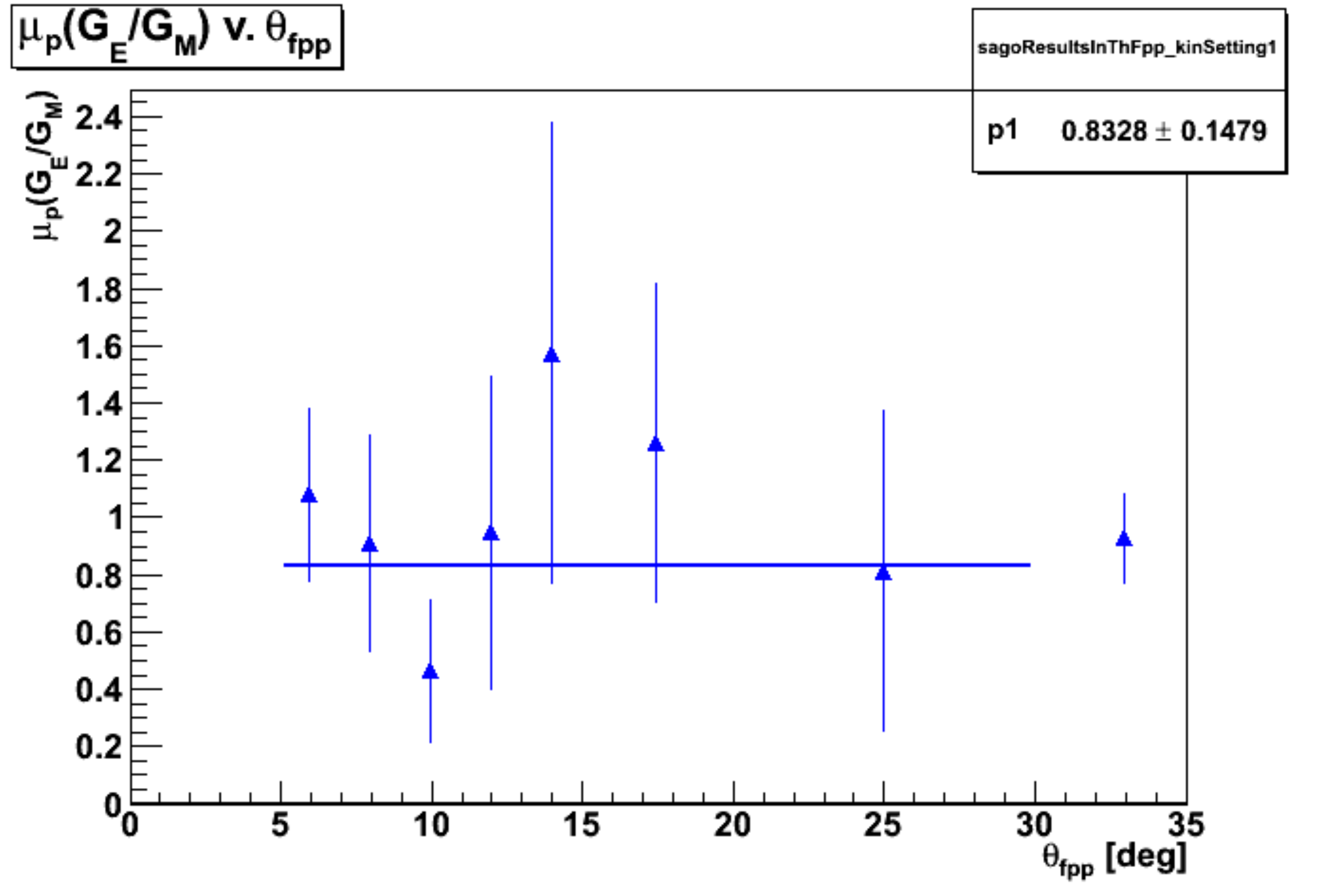}}
\caption[$\mu_{p}(G_{Ep}/G_{Mp})$ measurements (cumulative and binned in $\theta_{fpp}$) for $p(\vec{e},e^{\prime}p)$ events]{$\mu_{p}(G_{Ep}/G_{Mp})$ measurements using Sago binned in $\theta_{fpp}$. Also shown is a fit to the binned results using a horizontal line (in the range of $\theta_{fpp}~\epsilon~[5,30)$) and the fit result. Here, the cumulative result is shown as the data point to the far right of the histogram.}
\label{fig:sagoResultsInThFpp_kinSetting1}
\end{figure}

Finally, a bin-by-bin and cumulative comparison of the results was carried out as seen in Fig. \ref{fig:sagoAndPhaseShiftResultsInThFppBinNum_kinSetting1}. Shown using the red-brown points are the results from the phase shift method while the Sago results are shown in blue. It can be seen that a comparison of the results, both bin-by-bin and cumulative, shows an agreement between these two methods. Furthermore, a comparison of these (preliminary) results to measurements detailed in \cite{ZhanThesis} show that while the error bars of the Zhan results are much smaller than that of the e08008 (preliminary) results, there is an agreement between measurements.
\begin{figure}[hbt]
\center{\includegraphics[width=9.0cm]{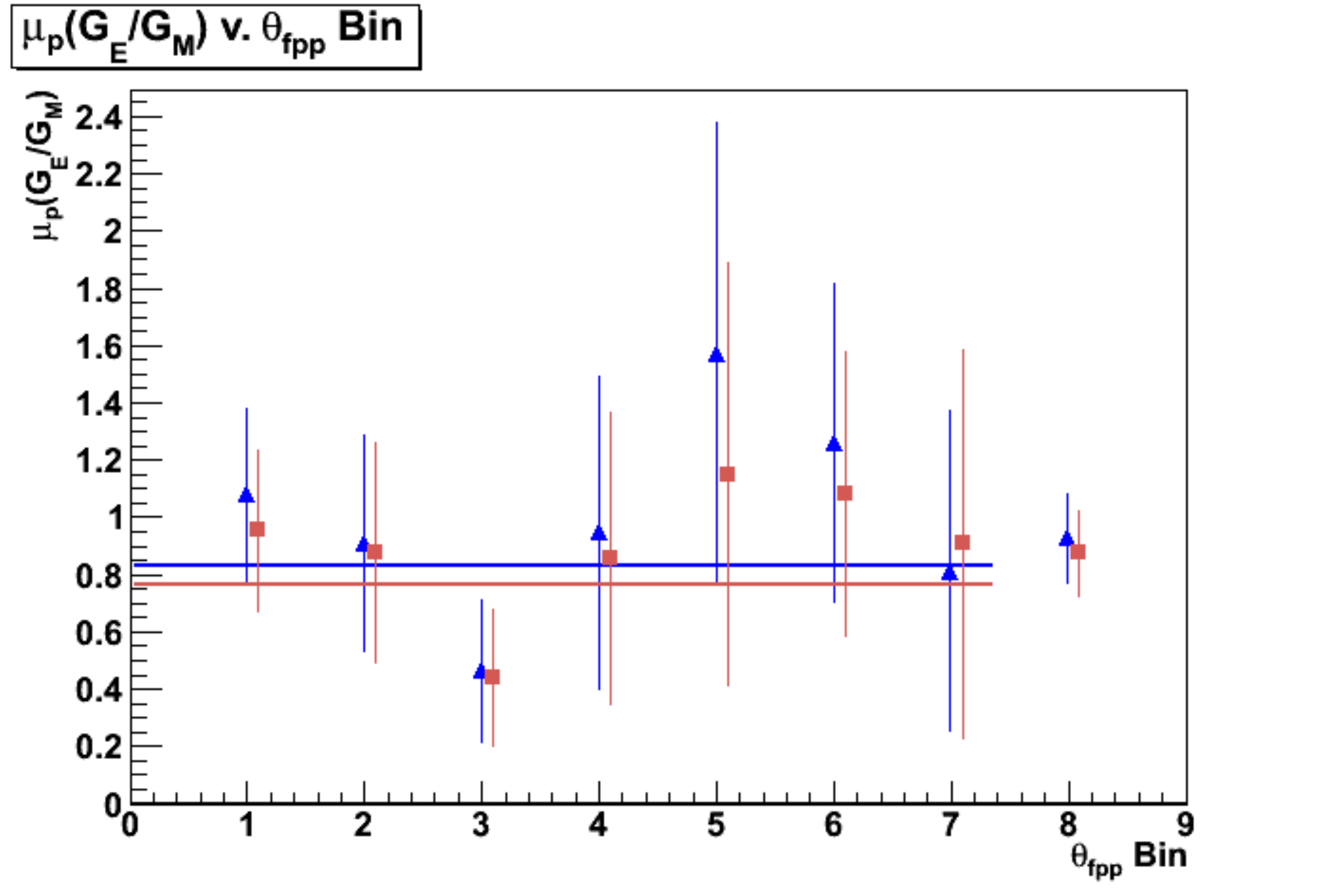}}
\caption[A comparison of $\mu_{p}(G_{Ep}/G_{Mp})$ measurements using the phase shift method and Sago for $p(\vec{e},e^{\prime}p)$ events.]{A comparison of $\mu_{p}(G_{Ep}/G_{Mp})$ measurements using the phase shift method and Sago. Here, the red-brown points are the results from the phase shift method while the Sago results are shown in blue. It can be seen that bin-by-bin and cumulatively, there is agreement between the results from the two methods.}
\label{fig:sagoAndPhaseShiftResultsInThFppBinNum_kinSetting1}
\end{figure}

\subsubsection{Summary}
The e08008 experiment conducted during the Winter 2011 run period in Hall A seeks to shed light on the interactions between the proton and the neutron in the deuterium nucleus through a measurement of the ratio $\mu_{p}(G_{Ep}/G_{Mp})$ for $x_{B}~\epsilon~[1,2]$. Using a beam of polarized electrons and the mechanism of electrodisintegration, data was taken for $p(\vec{e},e^{\prime}p)$ and $d(\vec{e},e^{\prime}p)n$ reactions. Presented here are the results from a preliminary analysis of the elastic $p(\vec{e},e^{\prime}p)$ data using two separate methods which result in a measurement of $\mu_{p}(G_{Ep}/G_{Mp})$ which agree with previous, high-precision results.

%
%

\clearpage \newpage

\subsection[E08-009 - ${}^4$He~$x_b>1$]{E08-009 - $^4\mathrm{He}(e,e'p)^3\mathrm{H}$ at $x_b=1.24$}
\label{sec:e08009}

\begin{center}
\bf Detailed Study of ${}^4\mathrm{He}$ Nuclei through Response Function Separations\\ at High Momentum Transfers
\end{center}

\begin{center}
A.~Saha, D.~Higinbotham, F.~Benmokhtar, S.~Gilad, and K.~Aniol, spokespersons\\
and \\
Students: S.~Iqbal~(CSULA) and N.~McMahon~(CNU)\\
and\\
the Hall A Collaboration.\\
contributed by K.~Aniol.
\end{center}

\subsubsection{Experimental Conditions}\label{sec:conditions}
The data were taken in collaboration with the SRC(E07-006) measurement during April 13 and April 14,
2011 for 16 hours of running. Our measurements provide the low missing momenta data,
0.153 GeV/c and 0.353 GeV/c which complement the high missing momenta data of the
SRC experiment. A 20 cm long cryogenic $^4\mathrm{He}$ target at 20K and 10 atm provided a thickness
of $8\times 10^{22}/\mathrm{cm}^2$. The electron beam energy was 4.454 GeV. This is the first measurement
of $^4\mathrm{He}(e,e'p)X$ at this value of $x_b=1.24$.

\subsubsection{Motivation}\label{sec:motivation}
A theoretical description of $^4\mathrm{He}(e,e'p)X$ is critical for understanding nuclear structure. In particular, one
must be able to include many body forces in the theory. The reaction we measured actually includes
multiple exit channels, that is, X = $^3\mathrm{H}$, $n+^2\mathrm{H}$ and n+n+p nuclear and nucleonic channels. At the
beam energy used here meson production also contributes to X. Our first goal is to compare
the data for $^4\mathrm{He}(e,e'p)^3\mathrm{H}$ to theoretical calculations provided by the Madrid group~\cite{madrid}.   
The missing energy spectra also reveal a broad peak attributed~\cite{fatiha} to the absorption of the virtual photon
on a pair of nucleons.

\subsubsection{Preliminary Results}\label{sec:prelim}
We compare our results to the GEANT simulations. In Fig.~\ref{fig:fig1} we overlay the shapes of the missing energy
spectrum for producing the $^3\mathrm{H}$ ground state, GEANT(blue), with the data for the 0.153 GeV/c kinematic setting.
The simulation in this case only includes geometric, energy loss, multiple scattering and radiative effects. The
simulation predicts a substantially wider distribution in missing energy.\\
In Fig.~\ref{fig:fig2} we see the same spectra except now the GEANT simulation includes weighting by the
Madrid cross section. There is a dramatic dynamical effect on the width of the predicted missing energy width.
In fact, the width of the ground state transition's missing energy spectrum is very well reproduced by the
theory.\\

\begin{figure}[hbt]
\center{\includegraphics[width=9.0cm]{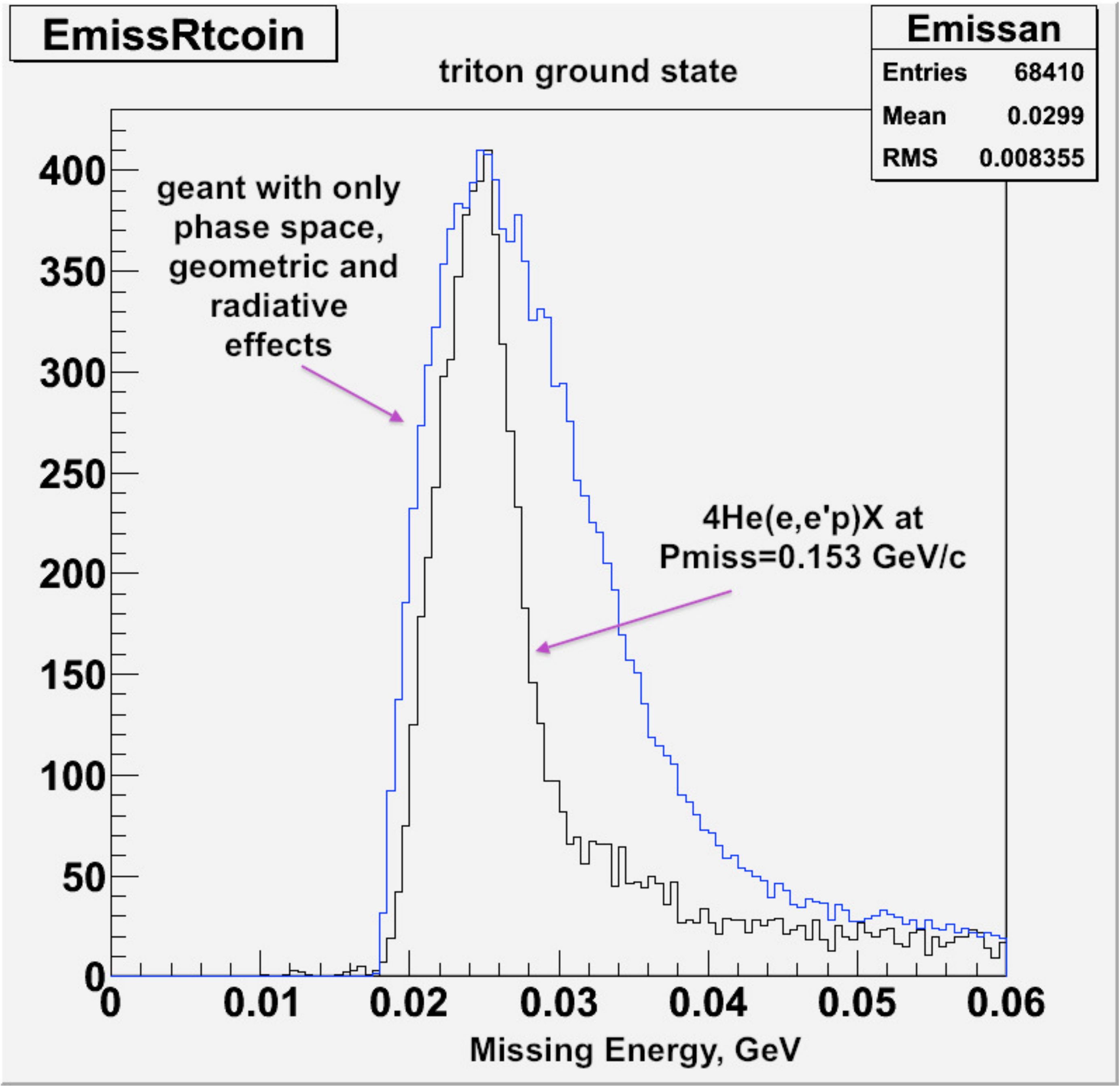}}
\caption{Data(black) compared to GEANT simulation(blue) which does not include 
weighting by the theoretical cross sections. Peaks are normalized to each other.}
\label{fig:fig1}
\end{figure}

\begin{figure}[hbt]
\center{\includegraphics[width=9.0cm]{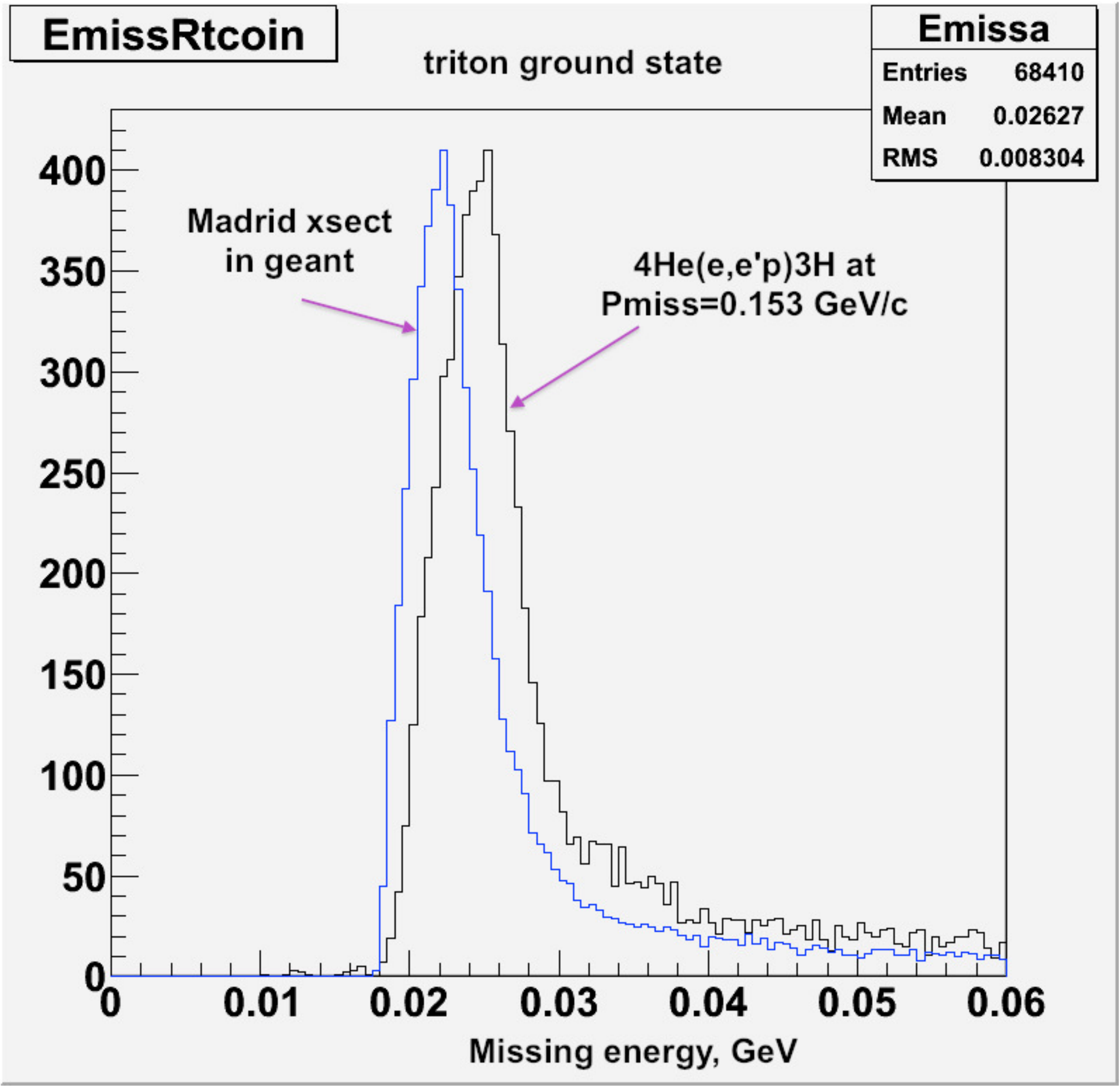}}
\caption{Data(black) compared to GEANT simulation(blue) including weighting
by the theoretical cross sections. Peaks are normalized to each other.}
\label{fig:fig2}
\end{figure}

The large acceptances of the HRSs allows us to bin the data into smaller missing momentum bites. We show
in Figs.~\ref{fig:fig3} and \ref{fig:fig4} the data for two kinematic settings binned into 0.05 GeV/c bins. 
Using the GEANT simulation, as described for Fig.~\ref{fig:fig1}, we can determine the missing momentum
acceptance factor for each bin. These factors enable us to predict a cross section per bin using the
theoretical cross sections folded into the GEANT simulation. Preliminary cross sections from the 0.153 GeV/c
data are in Tab.~\ref{tab:tab1}. Preliminary cross sections from the 0.353 GeV/c
data are in Tab.~\ref{tab:tab2}. Note that these cross sections do not include subtraction of the background
under the triton peak from multi particle break up. An example of the background under the triton peak for
the 0.353 GeV/c kinematic setting and a cut on the missing momentum bin from 0.30 to 0.35 GeV/c is shown
in Fig.~\ref{fig:fig5}.

\begin{figure}[hbt]
\center{\includegraphics[width=9.0cm]{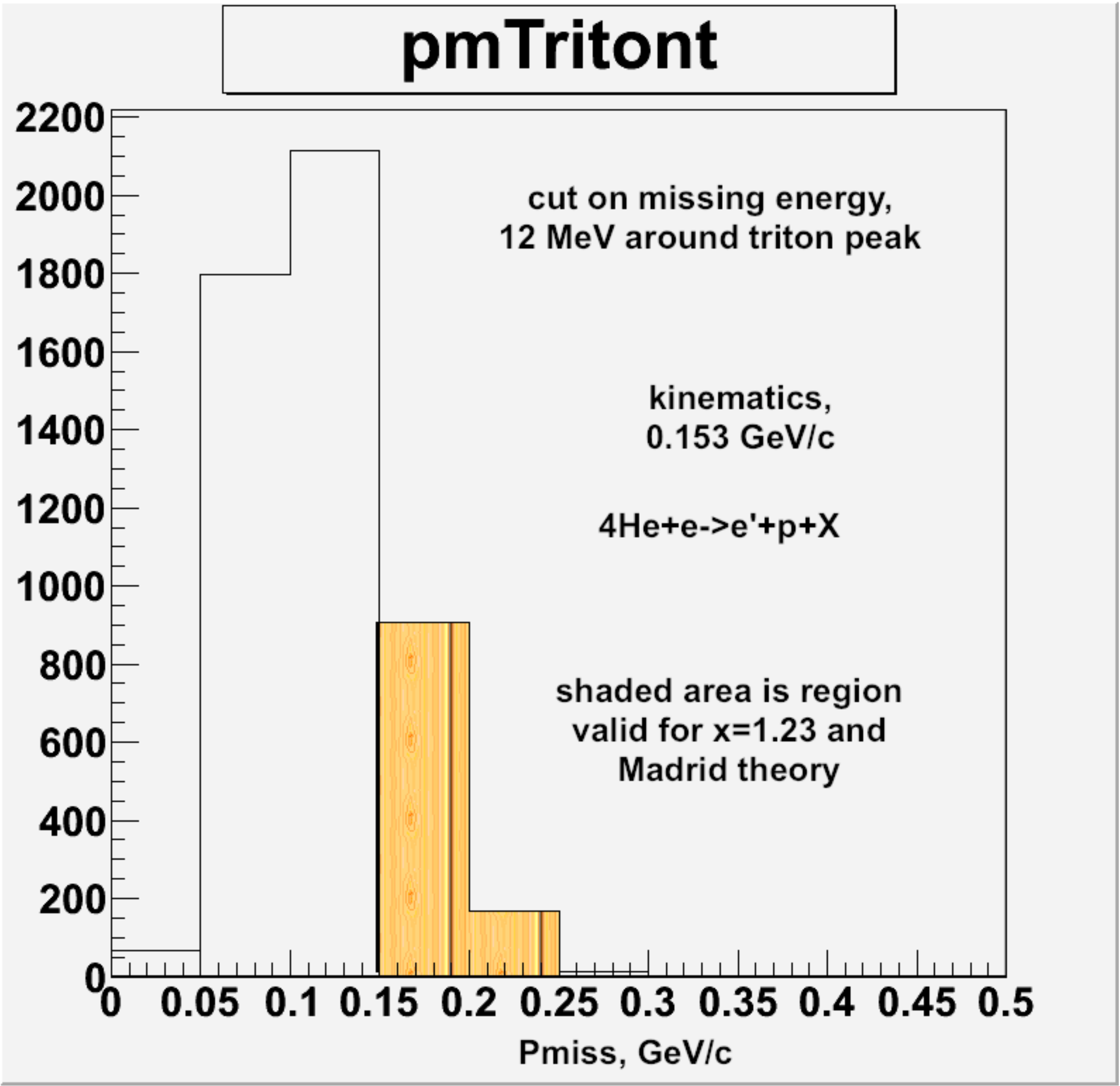}}
\caption{Kinematic setting 0.153 GeV/c. Missing momentum binned in 0.05 GeV/c bins. The shaded area shows which part of the missing momentum is valid for the Madrid theory we have.}
\label{fig:fig3}
\end{figure}

\begin{figure}[hbt]
\center{\includegraphics[width=9.0cm]{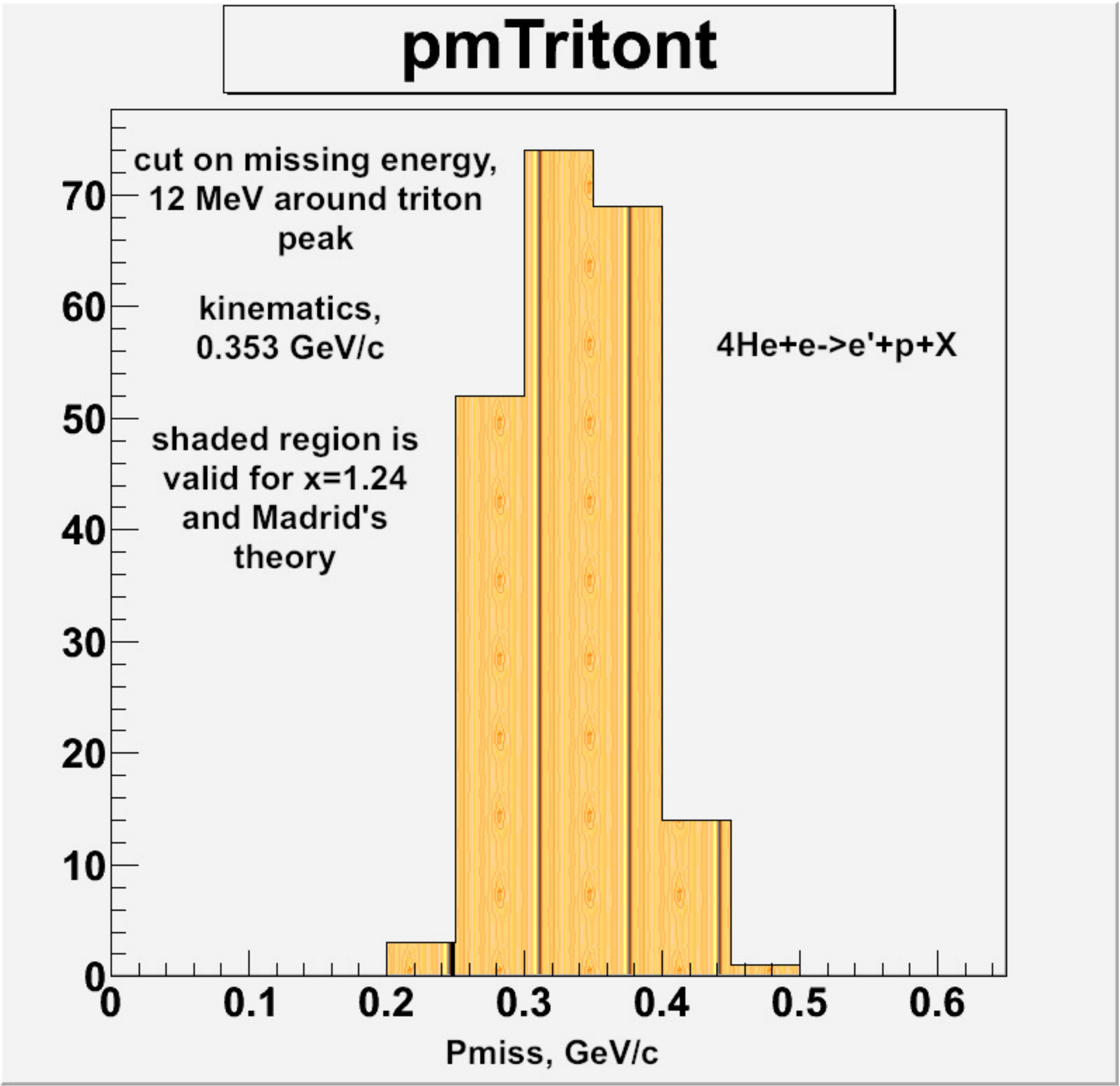}}
\caption{Kinematic setting 0.353 GeV/c. Missing momentum binned in 0.05 GeV/c bins. The Madrid cross sections are valid for all the bins. }
\label{fig:fig4}
\end{figure}

\begin{table}[ht]
\caption{Preliminary cross sections for the 0.153 GeV/c kinematic settings.
Cross sections are in $\mathrm{nb}/\mathrm{MeV}/\mathrm{sR}^2$. The Madrid cross sections are available for $P_{miss}$ greater than
0.15 GeV/c for $x_b = 1.24$.}
\begin{center}
\begin{tabular}{c|c|c|c}
$P_{miss}$(GeV/c) & Cross Sections $\sigma$ & $\delta \sigma$ & Madrid $\sigma$ \\ \hline
0.0 - 0.05  &    1.76     &    0.064  &  N/A       \\
0.05 - 0.10 &    0.69     &    0.009  &  N/A        \\
0.10 - 0.15 &    0.191    &    0.003  &  N/A        \\
0.15 - 0.20 &    0.047    &    0.001  &  0.0334     \\
0.20 - 0.25 &    0.0108   &    0.004  &  0.0078      \\
0.25 - 0.30 &    0.0025   &    0.0003 &  0.0019     \\
0.30 - 0.35 &    0.00043  &    0.0002 &  0.0012     \\
\hline
\end{tabular}
\label{tab:tab1}
\end{center}
\end{table}

\begin{table}[ht]
\caption{Preliminary cross sections for the 0.353 GeV/c kinematic settings.
Cross sections are in $nb/MeV/sR^2$. Background from multi particle breakup have not yet been included
in the cross section extractions.} 
\begin{center}
\begin{tabular}{c|c|c|c}
$P_{miss}$(GeV/c) & Cross Sections $\sigma$ & $\delta \sigma$ & Madrid $\sigma$ \\ \hline
0.20 -0.25  & 0.00084     &  0.0001   & 0.0098     \\
0.25 - 0.30 & 0.00064     &  0.00005  & 0.0015      \\
0.30 - 0.35 & 0.00045     &  0.00003  & 0.0003      \\
0.45 - 0.40 & 0.00051     &  0.0004   & 0.00019     \\
0.40 - 0.45 & 0.00040     &  0.0005   & 0.00015      \\
0.45 - 0.50 & 0.00060     &  0.00015  & 0.00027     \\
\hline
\end{tabular}
\label{tab:tab2}
\end{center}
\end{table}

\begin{figure}[hbt]
\center{\includegraphics[width=9.0cm]{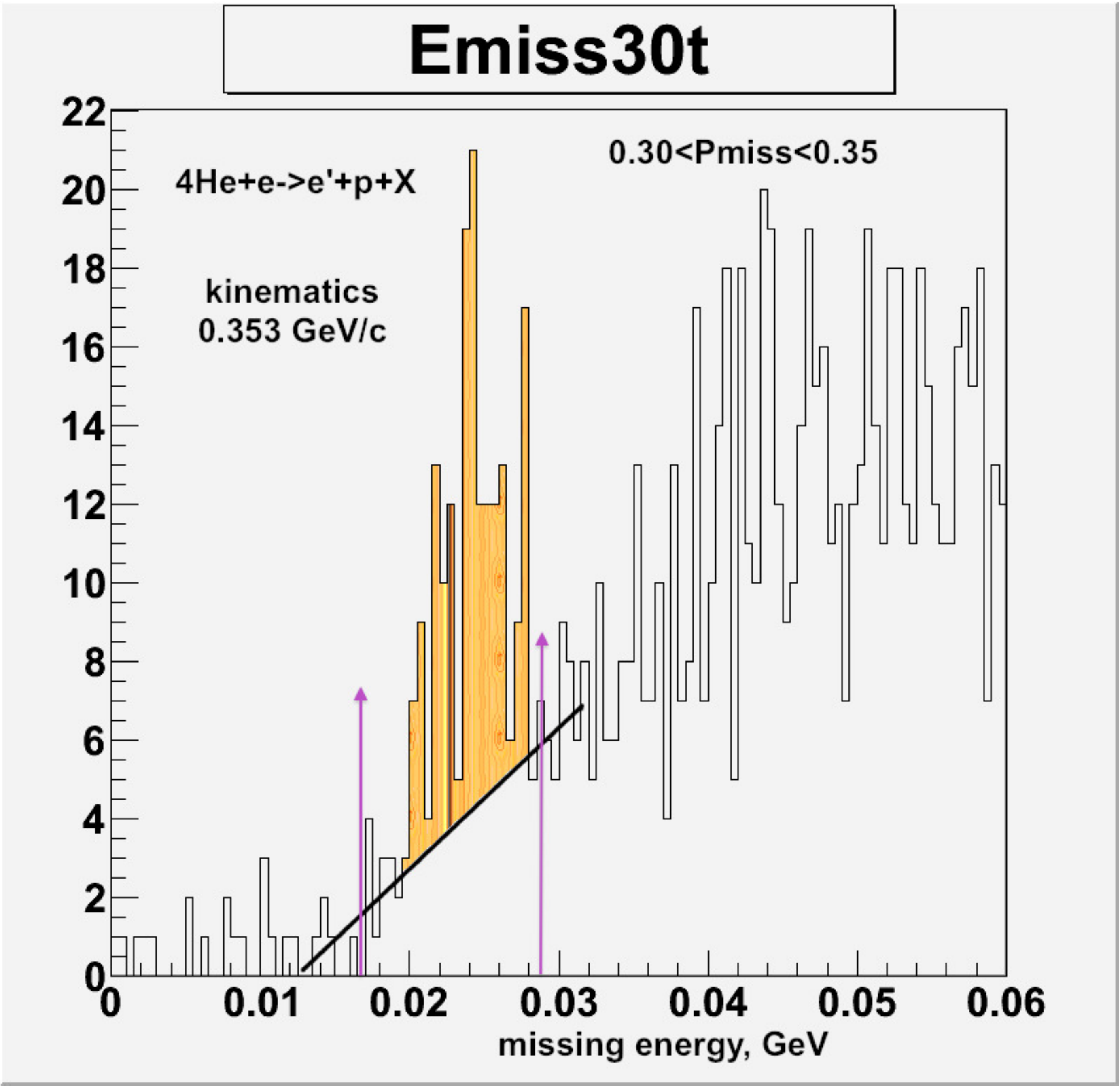}}
\caption{Kinematic setting 0.353 GeV/c. The missing energy spectrum for a cut on
the missing momentum bin from 0.30 to 0.35 GeV/c. The background below the highlighted peak still needs to
be fitted and removed. }
\label{fig:fig5}
\end{figure}

%
%

\clearpage \newpage

\subsection[E08-010 - N - Delta]{E08-010 - N $\rightarrow \Delta$}
\label{sec:e08010}

\begin{center}
{\bf Measurement of the Coulomb quadrupole amplitude at the $\gamma^*p\rightarrow \Delta(1232)$\\
in the low momentum transfer region}
\end{center}

\begin{center}
S.~Gilad, D.~W.~Higinbotham, A.~Sarty and N.~F.~Sparveris, spokespersons, \\
\vskip 0.3cm
Graduate students: D.~Anez (St. Mary's), A.~Blomberg (Temple) \\
\vskip 0.3cm
and the Hall A Collaboration.\\
\vskip 0.3cm
contributed by N.F.~Sparveris.
\end{center}

\subsubsection{Introduction}\label{sec:introduction}

Understanding the origin of possible non-spherical components in the
nucleon wavefunction has been the subject of an extensive
experimental and theoretical effort in recent years
\cite{Ru75,is82,pho2,pho1,frol,pos01,merve,bart,Buuren,spaprl,kelly,spamami,stave,elsner,joo1,St08,ungaro,villano,aznau,dina,sato,dmt00,kama,mai00,multi,said,pv,rev3}.
It is the complex quark-gluon and meson cloud dynamics of hadrons
that give rise to such amplitudes in their wavefunction which in a
classical limit and at large wavelengths will correspond to a
"deformation". The spectroscopic quadrupole moment provides the most
reliable and interpretable measurement of these components; for the
proton, the only stable hadron, it vanishes identically because of
its spin 1/2 nature. As a result, the presence of resonant
quadrupole amplitudes in the $N\rightarrow \Delta$ transition has
emerged as the definitive experimental signature of non spherical
amplitudes. Spin-parity selection rules in the $\gamma^*
N\rightarrow \Delta$ transition allow only magnetic dipole (M1) and
electric quadrupole (E2) or Coulomb quadrupole (C2) photon
absorption multipoles (or the corresponding pion production
multipoles $M^{3/2}_{1+}, E^{3/2}_{1+}$ and $S^{3/2}_{1+}$
($L^{3/2}_{1+}$) respectively) to contribute. The ratios CMR $=
Re(S^{3/2}_{1+}/M^{3/2}_{1+})$ and EMR $=
Re(E^{3/2}_{1+}/M^{3/2}_{1+})$ are routinely used to present the
relative magnitude of the amplitudes of interest. Non-vanishing
resonant quadrupole amplitudes will signify the presence of
non-spherical components in either the proton or in the
$\Delta^{+}(1232)$, or more likely at both; moreover, their $Q^2$
evolution is expected to provide insight into the mechanism that
generate them.

The origin of these components is attributed to a number of
different processes depending on the interpretative framework
adopted. In the quark model, the nonspherical amplitudes in the
nucleon and $\Delta$ are caused by the noncentral, tensor
interaction between quarks \cite{Gl79}. However, the effect for the
predicted E2 and C2 amplitudes \cite{is82} is at least an order of
magnitude too small to explain the experimental results, and even
the dominant M1 matrix element is $\approx$ 30\% low \cite{is82}. A
likely cause of these dynamical shortcomings is that the quark model
does not respect chiral symmetry, whose spontaneous breaking leads
to strong emission of virtual pions (Nambu-Goldstone bosons)
\cite{rev3}. These couple to nucleons as $\vec \sigma \cdot \vec p$,
where $\vec \sigma$ is the nucleon spin, and $\vec p$ is the pion
momentum. The coupling is strong in the p-wave and mixes in nonzero
angular-momentum components. Based on this, it is physically
reasonable to expect that the pionic contributions increase the M1
and dominate the E2 and C2 transition matrix elements in the
low-$Q^2$ (large distance) domain. This was first indicated by
adding pionic effects to quark models \cite{Lu97}, subsequently
shown in pion cloud model calculations \cite{sato,kama}, and
recently demonstrated in chiral effective field theory calculations
\cite{pv,Ga06}. Our current understanding of the nucleon suggests
that at low-$Q^2$ (large distance) the pionic cloud effect dominates
while at high-$Q^2$ (short distance) intra-quark forces dominate.

Recent high precision experimental results
\cite{pho2,pho1,frol,pos01,merve,bart,Buuren,spaprl,kelly,spamami,stave,joo1,St08,ungaro,villano,aznau}
are in reasonable agreement with predictions of models suggesting
the presence of non-spherical amplitudes and in strong disagreement
with all nucleon models that assume sphericity for the proton and
the $\Delta$. With the existence of these components well
established, recent investigations have focused on understanding the
various mechanisms that could generate them. Dynamical reaction
models with pion cloud effects \cite{sato}, \cite{dmt00} bridge the
constituent quark models gap and are in qualitative agreement with
the $Q^2$ evolution of the experimental data. These models calculate
the resonant channels from dynamical equations; they account for the
virtual pion cloud contribution dynamically but have an empirical
parameterization of the inner (quark) core contribution which gives
them some flexibility in the observables. They find that a large
fraction of the quadrupole multipole strength arises due to the
pionic cloud with the effect reaching a maximum value in the region
$Q^2=0.10~(\mathrm{GeV}/c)^2$ (see Fig.~\ref{fig:pioncloud}). Results from
effective field theoretical (chiral) calculations \cite{pv,Ga06},
solidly based on QCD, can also successfully account for the
magnitude of the effects giving further credence to the dominance of
the meson cloud effect in the low $Q^2$ region. Recent results from
lattice QCD \cite{dina} are also of special interest since they are
for the first time accurate enough to allow a comparison to
experiment. The chirally extrapolated \cite{pv} values of CMR and
EMR are found to be nonzero and negative in the low $Q^2$ region, in
qualitative agreement with the experimental results, thus linking
the experimental evidence for the nonspherical amplitudes directly
to QCD while highlighting the importance of future lattice
calculations using lighter quark masses and further refining the
chiral extrapolation procedure.

\begin{figure}[t]
\begin{center}
\includegraphics[width=7.0cm]{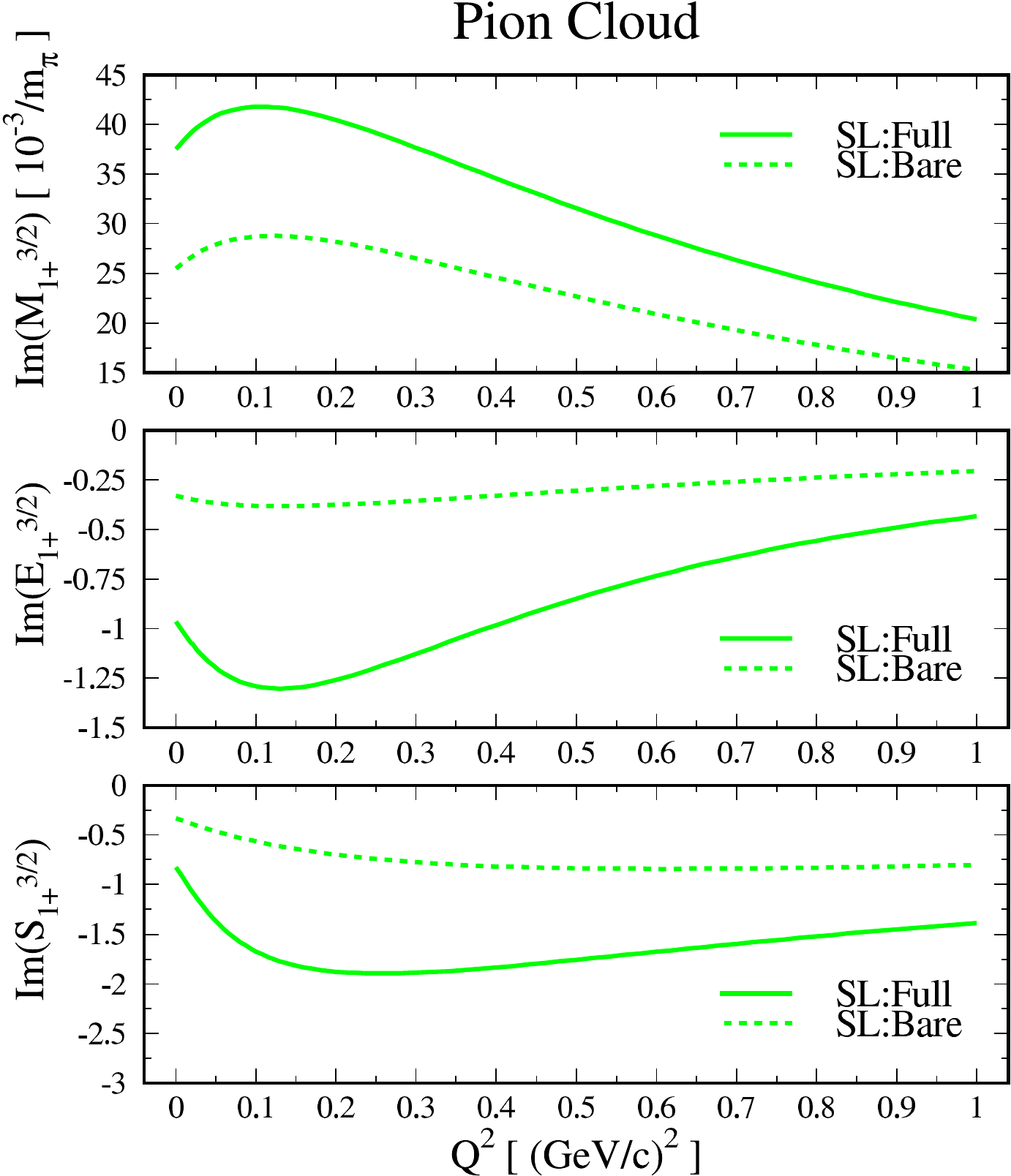}
\end{center}
\caption{\label{fig:pioncloud} The effect of the pionic cloud to the
resonant amplitudes as predicted by the Sato-Lee calculation
\cite{sato}. Solid line includes the pion cloud contribution while
the dashed line neglects the pion cloud effect.}
\end{figure}

\begin{figure}[h]
\vspace*{-0.1in} \centering
\begin{tabular}{cc}
\includegraphics[width=8.0cm]{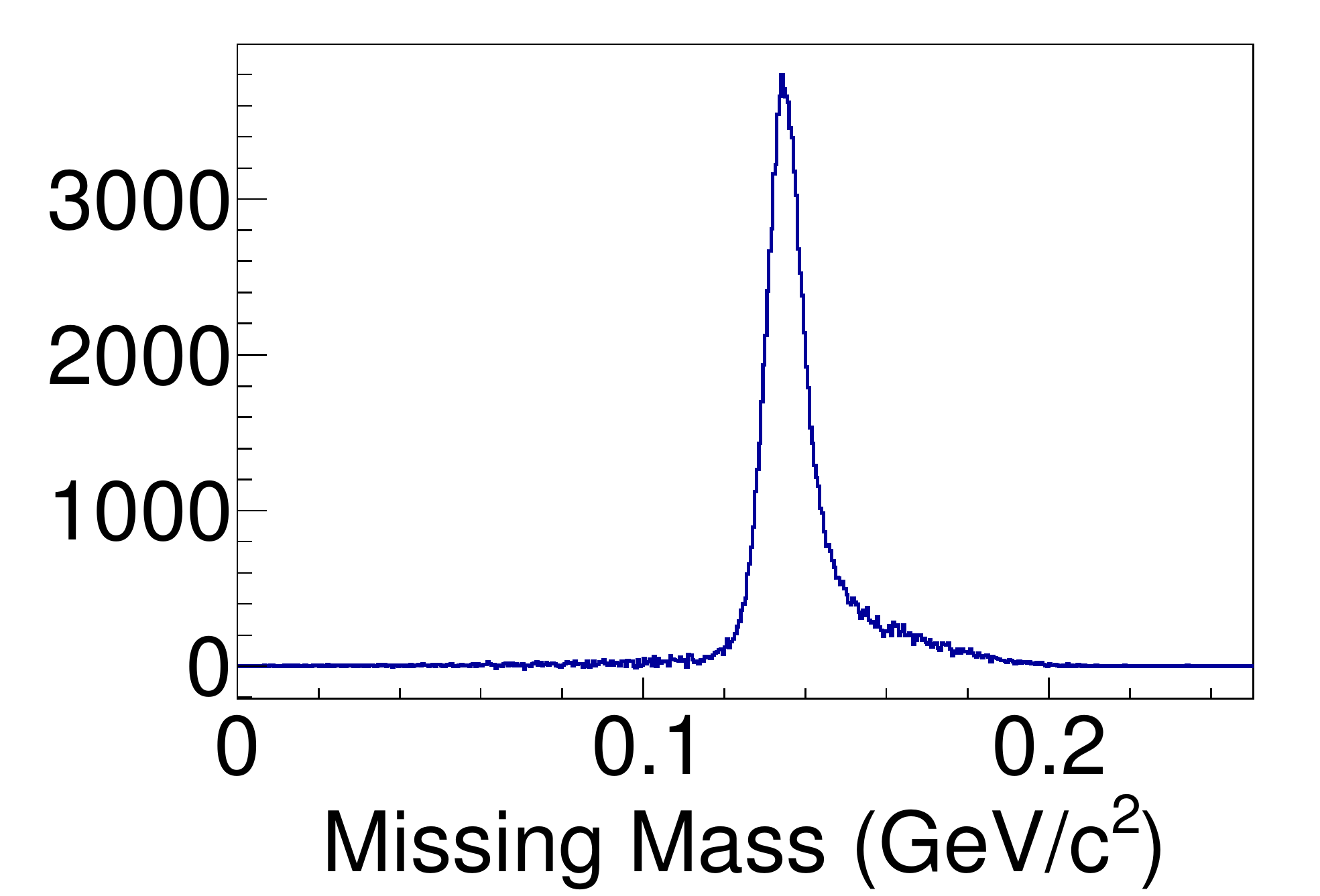}
&
\includegraphics[width=8.0cm,height=5.4cm]{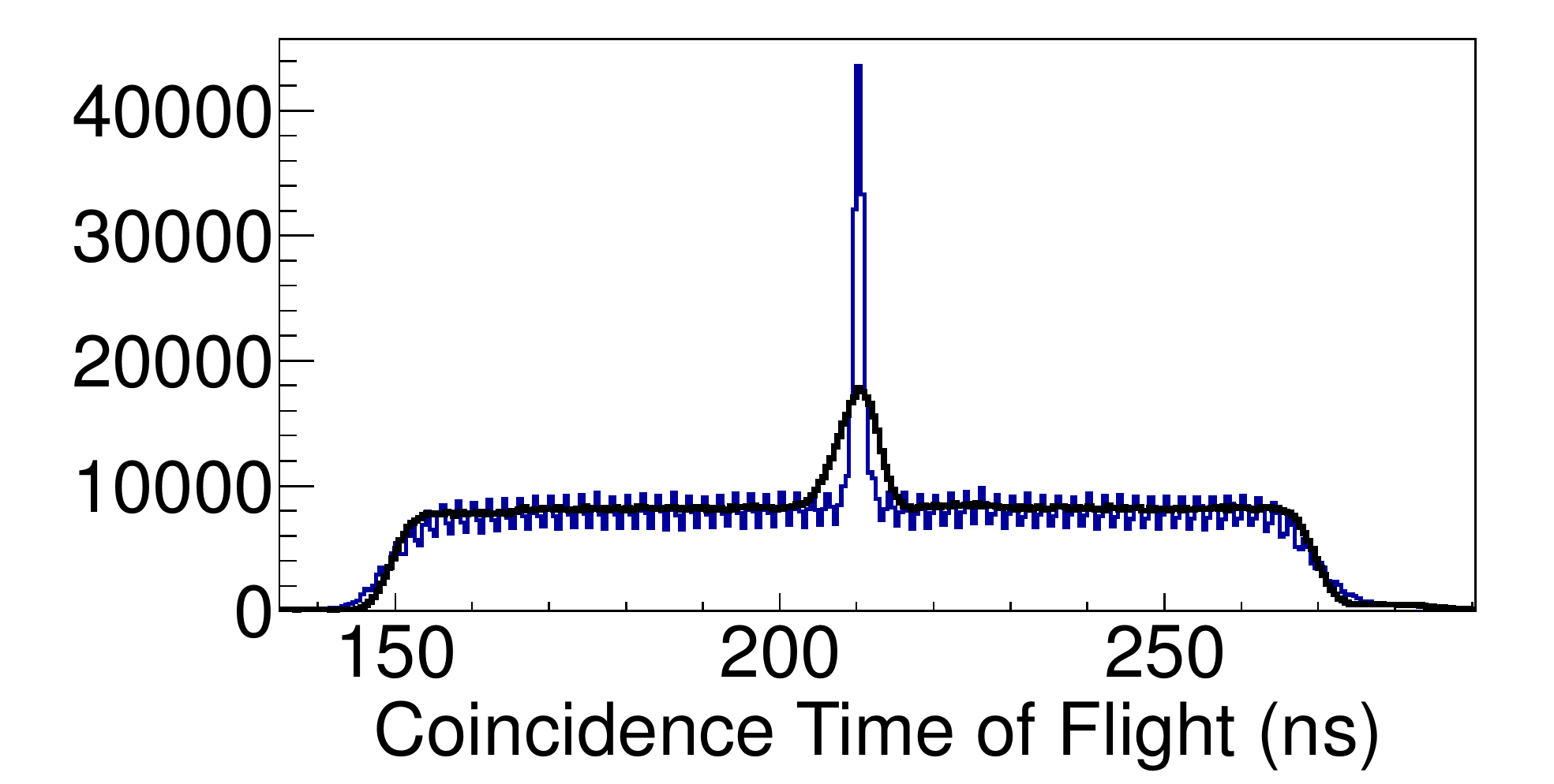}
\end{tabular}
\linespread{0.5} \caption{\label{fig:calib} Left panel: The missing
mass spectrum for the reconstructed (undetected) pion after the
subtraction of accidentals. Right panel: Raw and corrected
coincidence Time of Flight spectrum; an excellent timing resolution
of 1.6~ns has been achieved after the ToF corrections.}
\end{figure}

\subsubsection{The Experiment}\label{sec:experiment}

The E08-010 experiment aim to explore the low momentum transfer
region at the nucleon - $\Delta(1232)$ transition, where the pionic
cloud effects are expected to dominate. The experiment ran in
February and March of 2011 and achieved all the quantitative and
qualitative goals of the experiment proposal. High precision
measurements of the $p(e,e^\prime p)\pi^{\circ}$  excitation channel
were provided. The two High Resolution Spectrometers were utilized
to detect in coincidence electrons and protons respectively while
the 6 cm and 15 cm liquid hydrogen targets and an electron beam of
$E_{o}=~1.15~~\mathrm{GeV}$ at $75~\mu A$ were used throughout the experiment.
High precision measurements were conducted in the
$Q^2=0.04~(\mathrm{GeV}/c)^2$ to $0.13~(\mathrm{GeV}/c)^2$ range. The experiment will
offer results of unprecedented precision in the low momentum
transfer region and will extend the knowledge of the Coulomb
quadrupole amplitude lower in momentum transfer. Furthermore these
measurements will resolve observed discrepancies between
measurements of other labs. Two parallel analysis efforts are
currently in progress, by Temple University and St. Mary's
University, in order to ensure the most efficient outcome and to
provide important cross checks throughout all the steps of the
analysis. At this point the analysis stage involving calibrations
has been completed. In Fig.~\ref{fig:calib} the Missing Mass
spectrum (after background subtraction), corresponding to the
undetected pion, as well as the corrected time of flight spectrum
are presented. An excellent timing resolution of 1.6 ns has been
achieved. Currently the effort has moved on to the kinematical phase
space analysis, the extraction of the spectrometer cross sections
and the extraction of the resonant amplitudes. The projected
uncertainties for the CMR are presented in Fig.~\ref{fig:cmr}. The
new results will allow an in depth exploration of the nucleon
dynamics focusing primarily to the role of the pion cloud. A very
precise signature of the pion cloud will be provided as well as
strong constraints to modern theoretical calculations that will in
turn allow for a more complete understanding of the nucleon
structure.

\begin{figure}[h]
\begin{center}
\includegraphics[width=12.0cm]{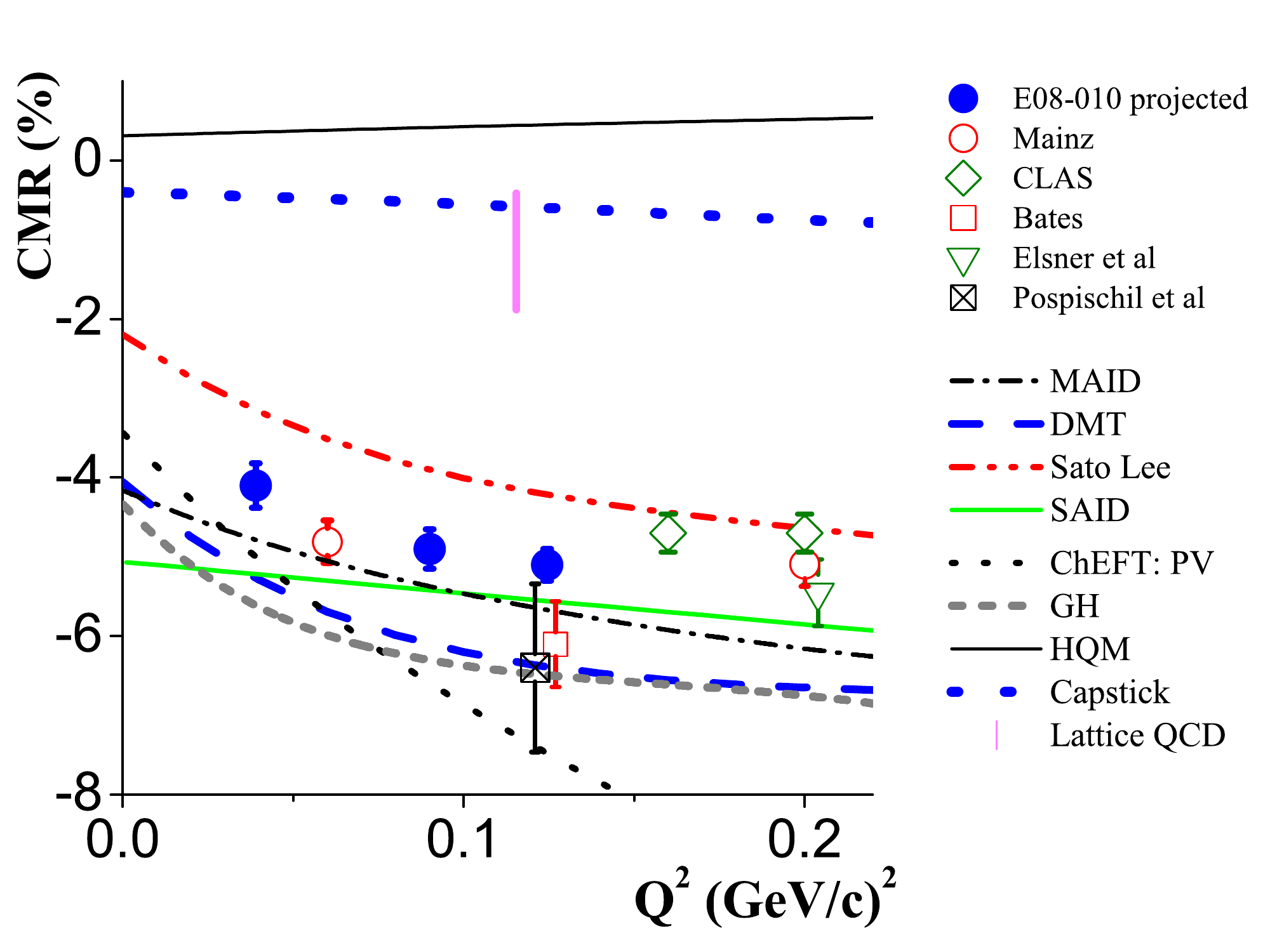}
\end{center}
\caption{The CMR at the low momentum transfer region. The projected
E08-010 uncertainties are presented along with the world data and
the theoretical model predictions} \label{fig:cmr}
\end{figure}

\clearpage \newpage

\subsection{E08-011 - PVDIS}
\label{sec:e08011}

\begin{center}
\bf $\vec e-^2$H Parity Violating Deep Inelastic Scattering (PVDIS) at CEBAF 6 GeV
\end{center}

\begin{center}
R. Michaels, P.E. Reimer, X. Zheng, spokespersons, \\
K. Pan, D. Wang, graduate students, \\
and \\
the Hall A Collaboration.\\
contributed by X. Zheng
\end{center}

The parity violating asymmetry of $\vec e-^2$H deep inelastic scattering (PVDIS)
was measured more than thirty years ago at SLAC~\cite{Prescott:1978tm,Prescott:1979dh},
and was the first experiment that established the value of the Standard Model weak mixing angle
$\sin^2\theta_W$. 
The goal of E08-011 is to provide an up-to-date measurement on the $\vec e-^2$H PVDIS asymmetry. 
A combination of the quark weak axial charges $2C_{2u}-C_{2d}$ will be extracted and will improve
the world knowledge on this value significantly. This experiment also serves as 
as an exploratory step for the future PVDIS program at the 12 GeV Upgrade.

The PV asymmetry of electron deep inelastic scattering (DIS) off a nuclear target
is 
\begin{eqnarray}
 A_{PV}^{DIS} &=& -{{G_FQ^2}\over{4\sqrt{2}\pi\alpha}}
   \left[2g_A^eY_1(y)\frac{F_1^{\gamma Z}}{F_1^Z}+{{g_V^e}}Y_3(y)\frac{F_3^{\gamma Z}}{F_1^Z}\right] \nonumber\\
 &=& -{{G_FQ^2}\over{4\sqrt{2}\pi\alpha}}\left[a_1(x)Y_1(y)+a_3(x)Y_3(y)\right]
\end{eqnarray}
where $G_F$ is the Fermi constant, $\alpha$ is the fine structure constant, 
$x$ is the Bjorken scaling variable, $y=\nu/E$ is the fractional energy loss
of the electron with $E$ the incident electron energy. 
With $r^2=1+{{Q^2}\over{\nu^2}}$ and $R^{\gamma,\gamma Z}$ the ratio of 
the longitudinal and transverse virtual photon electromagnetic 
absorption and the $\gamma-Z^0$ interference cross sections, respectively: 
\begin{eqnarray}
 Y_1&=&\left[\frac{1+R^{\gamma Z}}{1+R^\gamma}\right]
   \frac{1+(1-y)^2-y^2\left[1-\frac{r^2}{1+R^{\gamma Z}}\right]-xy{M\over E}}
   {1+(1-y)^2-y^2\left[1-\frac{r^2}{1+R^{\gamma}}\right]-xy{M\over E}}
\end{eqnarray}
and
\begin{eqnarray}
 Y_3&=&\left[\frac{r^2}{1+R^\gamma}\right]
   \frac{1+(1-y)^2}
   {1+(1-y)^2-y^2\left[1-\frac{r^2}{1+R^{\gamma}}\right]-xy{M\over E}}~.\label{eq:y13}
\end{eqnarray}
In the quark parton model,
\begin{eqnarray}
  a_1(x)=2 g_A^e\frac{F_1^{\gamma Z}}{F_1^Z}=2\frac{\sum{C_{1i}Q_iq_i^+(x)}}{\sum{Q_i^2q_i^+(x)}}
~~&\mathrm{and}&~~
  a_3(x)=g_V^e\frac{F_3^{\gamma Z}}{F_1^Z}=2\frac{\sum{C_{2i}Q_iq_i^-(x)}}{\sum{Q_i^2q_i^+(x)}}~,\label{eq:a3}
\end{eqnarray}
where the summation is over the quark flavor $i=u,d,s\cdots$, $Q_i$ is the corresponding quark 
electric charge, $q_i^\pm(x)$ are defined from the PDF $q_i(x)$ and $\bar q_i(x)$ as 
$q_i^+(x)\equiv q_i(x)+ \bar q_i(x)$ and $q_i^-(x)\equiv q_{i,V}(x)=q_i(x)- \bar q_i(x)$.
For an isoscalar target such as the deuteron, the functions $a_{1,3}(x)$ simplify to
\begin{eqnarray}
  a_1(x)=\frac{6\left[2C_{1u}(1+R_c)-C_{1d}(1+R_s)\right]}{5+R_s+4R_c}~
~~&\mathrm{and}&~~
  a_3(x)=\frac{6\left(2C_{2u}-C_{2d}\right)R_v}{5+R_s+4R_c}~.\label{eq:a1a3}
\end{eqnarray}
Neglecting effects from heavier quark flavors and assuming 
that $u^p=d^n$, $d^p=u^n$ [$u,d^{p(n)}$ are the up and down quark PDF
in the proton (neutron)],  $s=\bar s$, and $c=\bar c$, 
the PDF's give 
\begin{eqnarray}
 R_{c}\equiv\frac{2(c+\bar c)}{u+\bar u+d+\bar d}, ~~ R_{s}\equiv\frac{2(s+\bar s)}{u+\bar u+d+\bar d}, 
 &\mathrm{and}&
 R_V\equiv\frac{u-\bar u+d-\bar d}{u+\bar u+d+\bar d}.
\end{eqnarray}

For E08-011, the central settings of the spectrometer were $Q^2=1.121$ and 
$1.925$~(GeV/$c)^2$, while the acceptance-averaged values from data were $Q^2=1.085$ and
$Q^2=1.901$~(GeV/$c)^2$. These values are comparable to the SLAC experiment. At 
$Q^2=1.901$~(GeV/$c)^2$, the asymmetry
was measured to a $\approx 4\%$ (stat.) level. Not including the uncertainty from 
non-perturbative hadronic effects, 
the electron and quark neutral weak coupling constant combination,
$2C_{2u}-C_{2d}\equiv 2g_V^eg_A^u-g_V^eg_A^d$, can be extracted from this result. 
The asymmetry at $Q^2=1.085$~(GeV/c)$^2$ was measured to a 3\% level (stat.), and a simultaneous
fit to the asymmetries at both $Q^2$ values will set a constraint on the higher twist effect. 

The basic running conditions for E08-011 were reported last year. In the past year, 
our analysis focused on finalizing the systematic uncertainties of the asymmetry measurement.
The majority of this analysis has been completed, including corrections due to beam polarization,
beam asymmetries, counting deadtime of the DAQ, pion and pair-production background 
in the electron trigger, beam transverse asymmetries, aluminum endcaps of the target cell,
and electromagnetic radiative corrections.

As reported last year, the DAQ deadtime has three contributions: the ``path'' 
deadtime caused by summing and discriminating the preshower and shower signals
to form preliminary electron and pion triggers; the ``veto'' deadtime 
caused by combining the preshower/shower triggers with the HRS T1 trigger and 
Cherenkov signals; and the ``final or'' deadtime caused by taking the logical
OR of 6 (8) paths to form the final electron and pion triggers for the left (right)
HRS. A full scale simulation package was developed to study specifically the timing
performance of the DAQ. In the past year, the uncertainty analysis of the deadtime
has been completed and a breakdown into these three components are shown in Table~\ref{tab:deadtime}. 

\begin{table}[!htp]
 \caption{Simulated DAQ deadtime loss in percent for all kinematics and for both narrow (n) 
and wide (w) paths, along with the fractional contributions from group, veto, and OR deadtimes. 
The
fractional deadtime from OR is calculated as one minus those from group and veto, and 
its uncertainty is estimated from the difference between simulation and the analytical results.
The uncertainty of the total deadtime is the uncertainties from group, veto and OR added
in quadrature.
}\label{tab:deadtime}
 \begin{center}
 \begin{tabular}{c|c|c|c|c|c}
  \hline\hline
  HRS, $Q^2$   &  Path  & \multicolumn{3}{c|}{fractional contribution} & Total deadtime \\\cline{3-5}
  (GeV/$c$)$^2$  &        & Group           & Veto   & OR  & loss at 100$\mu$A  \\ \hline
  \multirow{2}{*}{Left, $1.085$}
             & n &$(20.6\pm 2.1)\%$&$(51.3\pm 4.5)\%$ &$(28.1\pm 4.7)\%$& $(1.45\pm 0.10)\%$  \\
             & w &$(29.5\pm 2.4)\%$&$(45.3\pm 4.0)\%$ &$(25.3\pm 4.6)\%$ & $(1.64\pm 0.11)\%$ \\
  \hline
  \multirow{2}{*}{Left, $1.901$} 
             & n &$(5.42\pm 0.8)\%$ &$(81.1\pm 7.1)\%$&$(13.5\pm 7.0)\%$ & $(0.50\pm 0.05)\%$\\
             & w &$(8.39\pm 0.4)\%$ &$(77.3\pm 6.8)\%$&$(14.3\pm 8.0)\%$ & $(0.52\pm 0.06)\%$\\
  \hline
  \multirow{2}{*}{Right, $1.901$} 
             & n &$(2.9\pm 0.2)\%$ &$(80.6\pm 18.5)\%$&$(16.5\pm 12.7)\%$ & $(0.89\pm 0.20)\%$\\
             & w &$(4.3\pm 0.4)\%$ &$(76.6\pm 17.5)\%$&$(19.1\pm 15.5)\%$ & $(0.93\pm 0.22)\%$\\
\hline\hline
 \end{tabular}
 \end{center}
\end{table}

The pion contamination in the electron trigger was found to be below $2\times 10^{-3}$, a major
accomplishment of the DAQ system specifically built for this
experiment, and the effect on the measured asymmetries is negligible.

The blinding factor imposed on the asymmetry analysis was lifted in April 2012. As a result, now we can
use ``pull'' plots to examine the statistical quality of the asymmetry measurement. Here the 
pull value is defined as
\begin{eqnarray}
 p_i &\equiv& (A_i-\langle A\rangle)/{\delta A_i}~,~\label{eq:pull}
\end{eqnarray}
where $A_i$ is the asymmetry extracted from the $i$-th beam helicity pair with the HWP
states already corrected and
$\delta A_i=1/\sqrt{N_i^R+N_i^L}$ its statistical uncertainty with $N_i^{R(L)}$ the
event count from the right (left) helicity pulse of the pair, and $\langle A\rangle$
is the asymmetry averaged over all beam pairs.  From Fig.~\ref{fig:pull} 
one can see that the asymmetry spectrum agrees to 
five orders of magnitude with the Gaussian distribution, as expected from purely statistical
fluctuations.

\begin{figure}[!ht]
 \includegraphics[width=0.5\textwidth,angle=0]{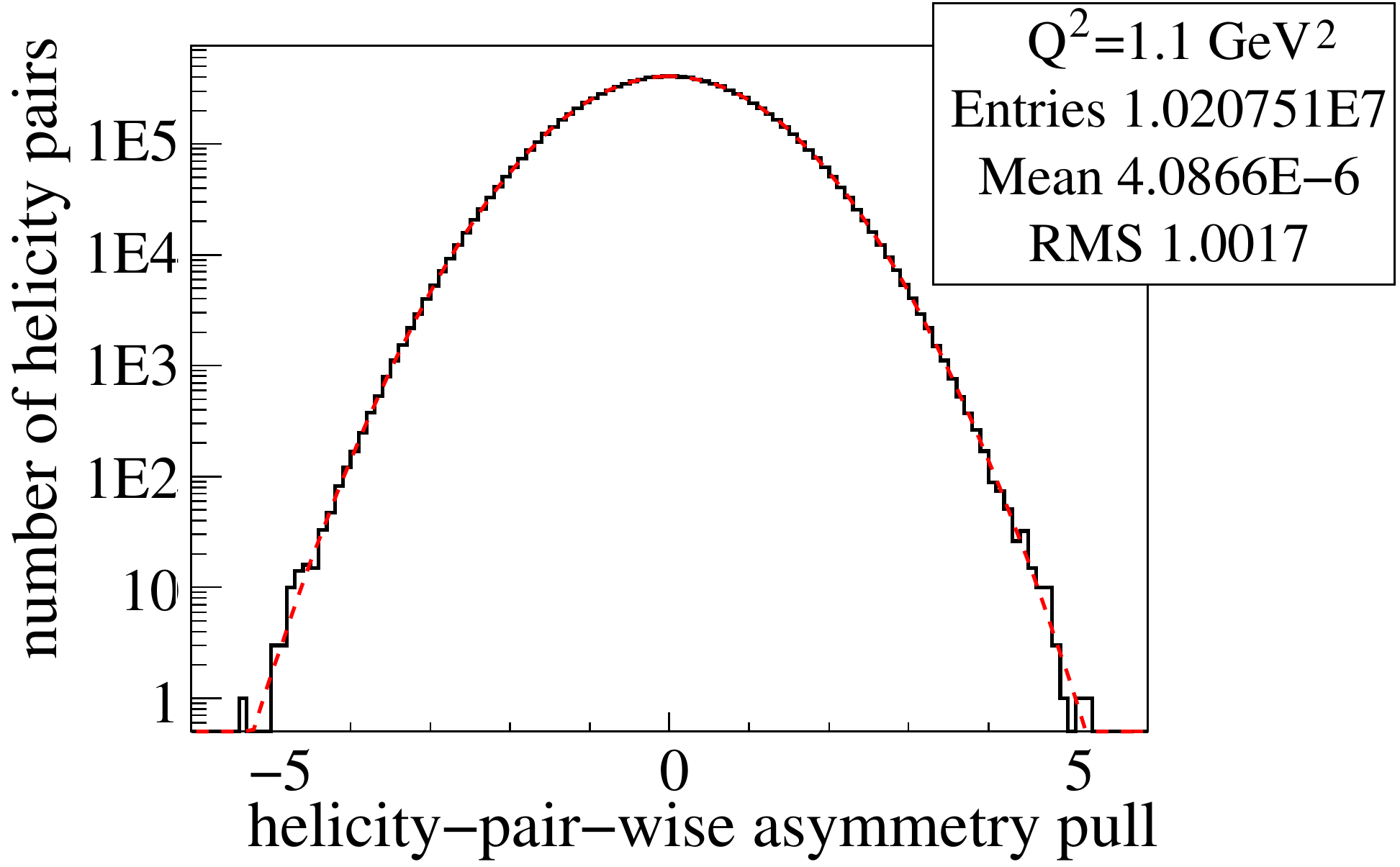}
 \includegraphics[width=0.5\textwidth,angle=0]{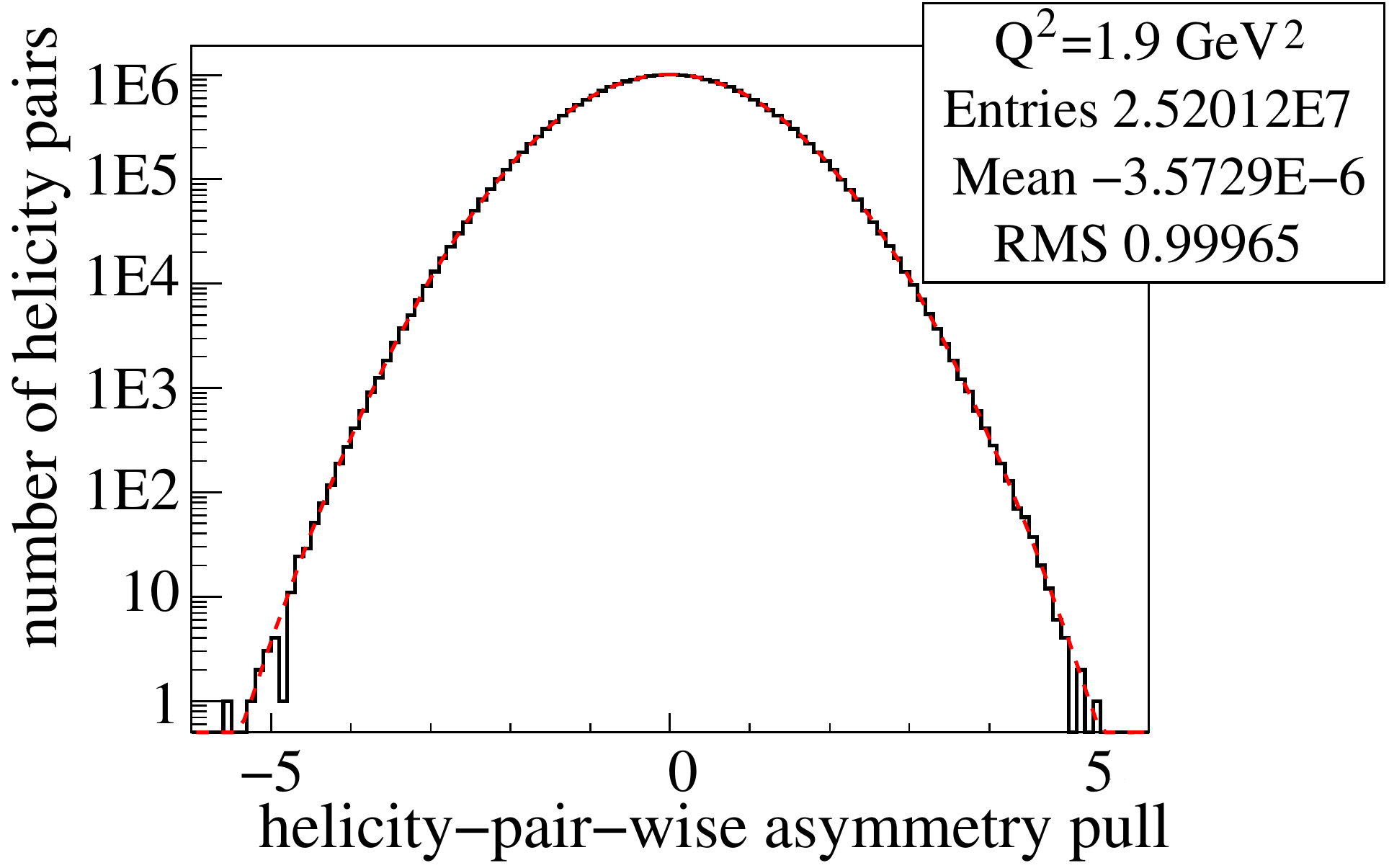}
\caption{[Pull distribution~[Eq.(\ref{eq:pull})] for the global electron narrow trigger 
for $Q^2=1.1$ (top) and $Q^2=1.9$~(GeV/$c$)$^2$ (bottom).
}\label{fig:pull}
\end{figure}

Another major focus of past year's analysis was electromagnetic radiative corrections. 
These corrections account for the energy loss of both incoming and outgoing electrons
due to ionization and bremsstrahlung scattering. Parity violation asymmetries of 
the nucleon resonances were provided by two models~\cite{Matsui:2005ns,Gorchtein:2011mz} 
[with \cite{Matsui:2005ns} for the $\Delta(1232)$ only],
and their uncertainties were determined by how well the model agree with our
measurement of the resonance PV asymmetries, which were performed with a beam energy
of $4.8674$~GeV. Our measured resonance asymmetry for $\Delta(1232)$ is $2-3$ sigma away from
both models, but measurements at the $2^{nd}$ and 
the $3^{rd}$ resonances agreed with the model of Ref.~\cite{Gorchtein:2011mz} very well. 
The uncertainty of the radiative corrections was found to be $2\%$ at $Q^2=1.085$ and
$0.43\%$ at $Q^2=1.901$~(GeV/$c$)$^2$. Our measurement on the nucleon resonance PV asymmetry
will also provide valuable inputs to the box diagram corrections for the Qweak experiment.

In comparing with the asymmetry from the Standard Model, we used the CTEQ/JLab fit, CTEQ6
and CTEQ10~\cite{Pumplin:2002vw,Lai:2010vv}, and 
MRST2008~\cite{Martin:2007bv} and MSTW2010~\cite{Martin:2009iq} 
as PDF inputs to Eq.(\ref{eq:a1a3}). The CTEQ/JLab fit in principle should give the best
values since it is tailored to JLab energies, but the fit does not apply to $Q^2$ below
$1.7$~(GeV/$c$)$^2$. By comparing with other PDF fits, the MSTW2010 provides the closest
values to the CTEQ/JLab fit, and thus was used to fit the measured asymmetries 
at both $Q^2=1.085$ and $1.901$~(GeV/$c$)$^2$. A simultaneous fit of these two measured
asymmetries found the higher twist effect to be very small compared to the statistical
uncertainty of the measurement. 

Currently we are working with theorists to finalize higher-order radiative corrections, 
in particular the effect from the $\gamma\gamma$ and the $\gamma Z$ box diagrams. 
We are also extending our analysis scheme to the resonance measurements as well as extraction
of the pion asymmetries, and we are aiming to
finalize these results within a couple of months. Four draft publications
are been worked on: One targeted for NIM which will report on the construction and the 
performance of the counting DAQ system; one short and one long drafts focusing on the main
PVDIS physics results; and one short draft on the resonance asymmetry measurements.

%
%

\clearpage \newpage

\subsection{E08-014 - $x>2$}
\label{sec:e08014}

\begin{center}
\bf The $x>2$ Experiment
\end{center}

\begin{center}
J. Arrington, D. Day, D. Higinbotham and P. Solvignon, spokespersons, \\
and \\
the Hall A Collaboration.\\
      contributed by J. Arrington, P. Solvignon and Zhihong Ye.
\end{center}

\subsubsection{Motivations}\label{sec:e08014-motivation}
The shell model has been partially successful in describing many features of nuclei such as 
the structure and energies of the nuclear excited states. However, about 30-40\% of the 
nucleonic strength predicted by the shell model to be in shells below the Fermi level is not 
seen in the experimental data~\cite{Lapikas}. This missing strength is thought to be due to 
the nucleon-nucleon (NN) interaction at short distances and the fact that the close packing 
of nucleons in nuclei results in a significant probability of overlapping nucleon wavefunctions. 
These overlapping nucleons belong to a short range correlated cluster and exhibit high momenta, 
well above the Fermi momentum in the nucleus~\cite{Frankfurt1}. 

Short-range correlations (SRC) are now well accepted as a key ingredient in the formulation of 
realistic nuclear wave functions. This means that the experimental characterization of SRC is 
crucial to the development of accurate nuclear structure calculations. Recent results from JLab 
experiment E01-015~\cite{Shneor} confirmed the overwhelming dominance of the proton-neutron pairs 
in two-nucleon SRCs. These two-nucleon knockout experiments are very sensitive to the isospin 
structure as they are able to measure both pp and pn correlations. However, they have also to 
deal with potentially large final state interactions which plague coincidence measurements at 
high missing momentum.

Although inclusive scattering is typically isospin-blind, isospin sensitivity, also 
called ``tensor dominance'', can be identified through a careful choice of complementary targets. 
Isospin-independent and isospin-dependent models predict 25\% differences in the cross-section 
ratios of the two medium-weight nuclei, $^{48}$Ca and $^{40}$Ca. E08-014 complements two-nucleon 
knockout experiments, for which other physical processes make it difficult to extract a 
model-independent and precise quantitative measure of the isospin asymmetry. Further insights 
will be obtained after the JLab 12~GeV upgrade from the use of two light mirror nuclei $^3$He and 
$^3$H~\cite{prop2}; in addition to further enhanced isospin sensitivity in these light nuclei, 
realistic theoretical calculations can be performed where the nucleon-nucleon potential components 
and their amplitudes can be separated. 
\begin{figure}[hbt]
\center{\includegraphics[width=9.0cm]{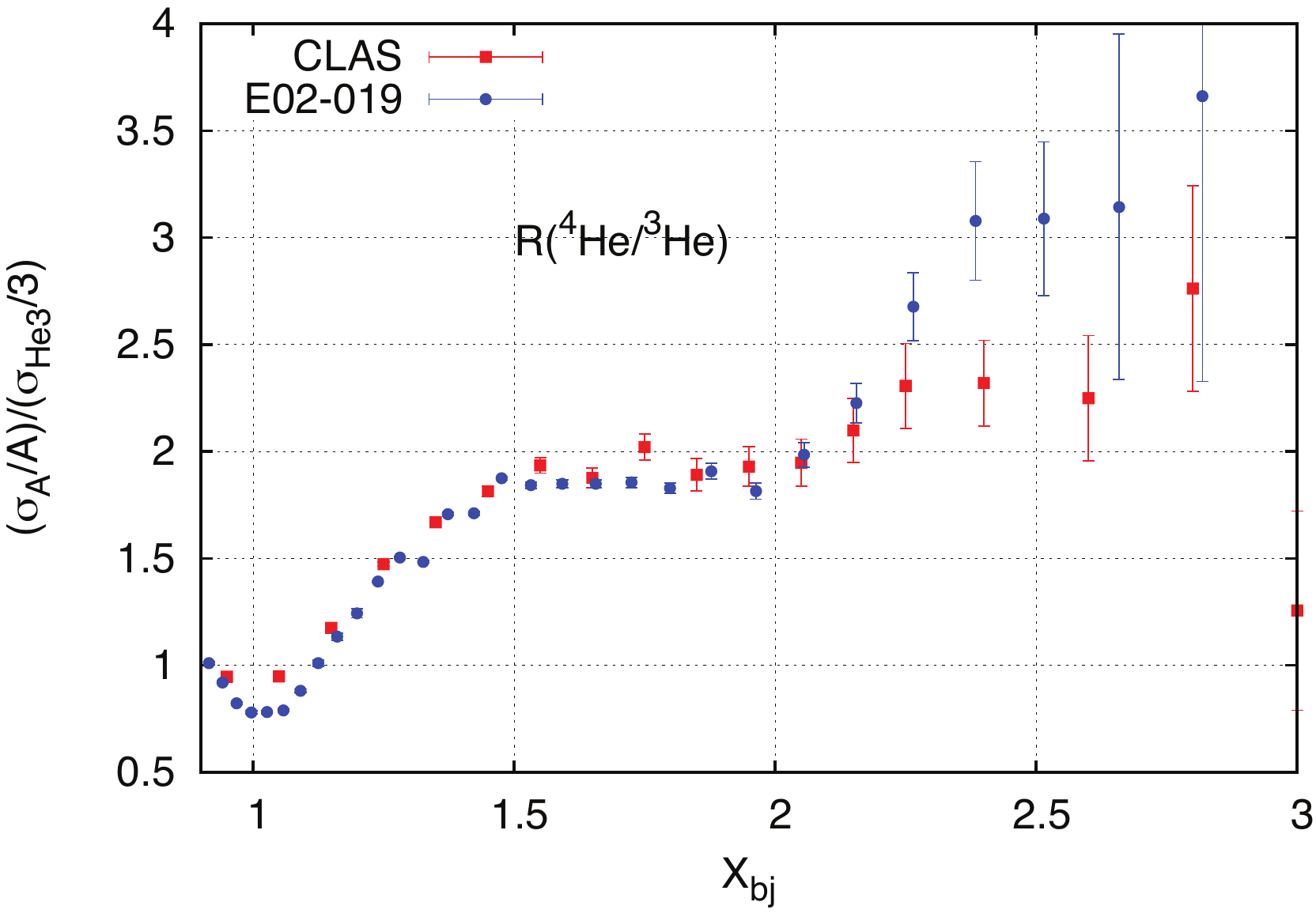}}
\caption[E08014\_mot: Results from Hall C E02-019.]{Results from Hall C experiment E02-019~\cite{Nadia}}
\label{fig:e02019}
\end{figure}

At $x>2$, the cross-sections from nuclei heavier than deuterium are expected to be dominated by 
three-nucleon short-range correlations (3N-SRCs). Results from Hall C experiment 
E02-019~\cite{Nadia} show a discrepancy with the CLAS results~\cite{egiyan05} in the $x>2$ region, 
while being in very good agreement the 2N-SRC region (see Fig.~\ref{fig:e02019}). E02-019 is at 
higher $Q^2$ than CLAS, and this is consistent with the hint of possible $Q^2$ dependence in the 
CLAS results (see figure 3 of the original proposal~\cite{prop}). These new data and observations 
make our measurement decisive in the effort to map precisely the 3N-SRC region and resolve this 
new issue. E08-014 will also be the first measurement of isospin dependence of 3N-SRC. The amplitude 
and properties of SRCs have important implications not only for the structure of the neutron stars 
and their cooling process~\cite{Frankfurt2} but also in the search for neutrino oscillation~\cite{Martini}.

\subsubsection{Analysis status}\label{sec:e08014-details}
JLab experiment E08-014 ran in April-May 2011. This experiment aims at mapping the 2N and 3N-SRC 
scaling behaviors. It should also provide the first test of the SRC isospin dependence in inclusive 
electron-nucleus scattering by using two Calcium isotopes. This experiment used the standard Hall A 
high resolution spectrometers configured for electron detection.

The calibrations of all beam diagnostic elements, spectrometer optics and detectors are done and 
almost all efficiencies have been evaluated.
Recently the analysis efforts were directed on the target density study and the observation of a 
``bump'' in the $^2$H, $^3$He and $^4$He target length spectra as shown in Fig.~\ref{fig:bump} for 
$^2$H and $^3$He. It has been determined that this feature was caused by the density fluctuation 
inside the 20cm long cell. The cooling flow was covering only the upstream half of the cell and 
therefore the gas/liquid density in this part of the cell was higher than in the downstream half. 
Also, as expected, this density gradient between the upstream and downstream parts of the cell 
increases with the beam current. This is mostly an issue for the determination of the target 
luminosity and the radiative corrections. Data were taken at a range of currents and extrapolated 
to zero current to determine the overall density change due to beam heating, while the z-dependence 
of the density as observed from the data is replicated in the simulation to account for the varying 
density. The cross section are generated versus the 
variable $x_{bj}$ with each bin being an average over the target length, i.e. $x_{bj}$ does
not dependent on the vertex position. However the radiative corrections depend on the location 
of the reaction. The density distribution along the cell was extracted by fitting the vertex spectra 
of each cell and then fed into the Monte Carlo simulation to determine the radiative correction 
factor related to the vertex position.
				   
\begin{figure}[hbt]
\center{\includegraphics[width=8.0cm]{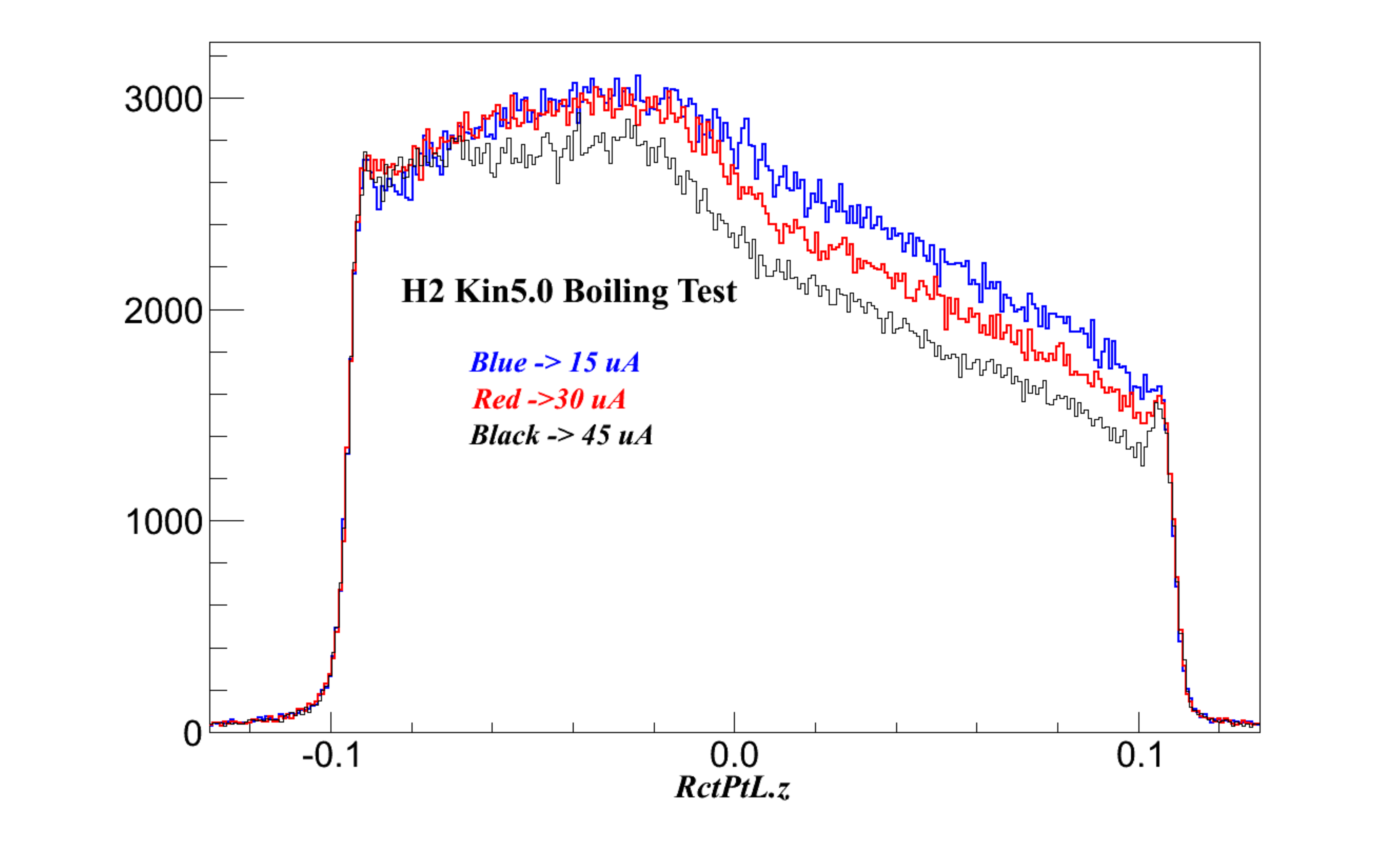}
\includegraphics[width=8.0cm]{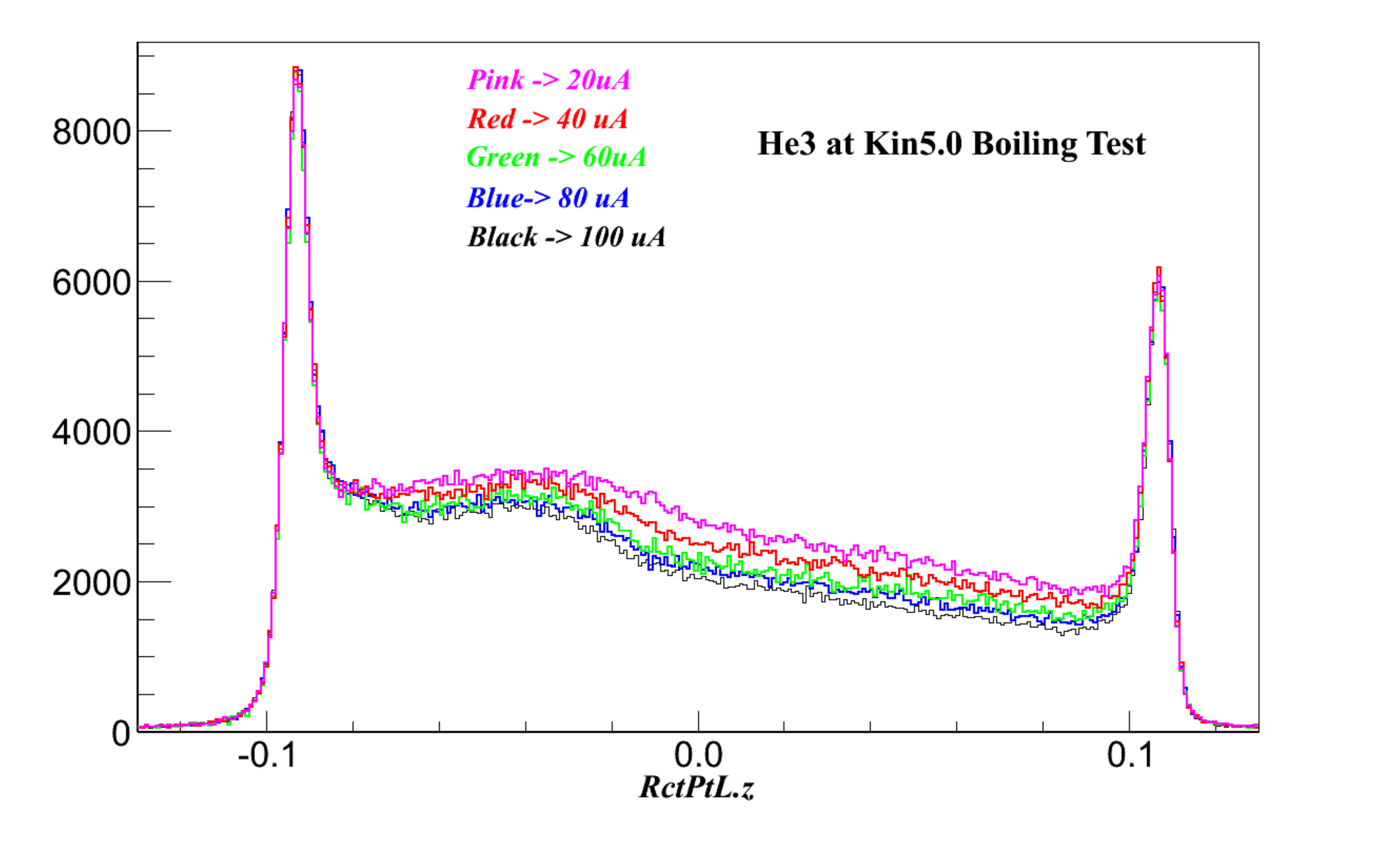}}
\caption{Target spectra along the beam direction for $^2$H (left plot) and for $^3$He (right plot).}
\label{fig:bump}
\end{figure}

At present, the main activity is the iteration of the cross section model (Hall C XEM model) in order 
to fit our kinematical region and then finalize the radiative corrections. Preliminary comparisons of 
this model to the data are shown in Fig.~\ref{fig:xs}.

\begin{figure}[hbt]
\center{\includegraphics[width=8.0cm]{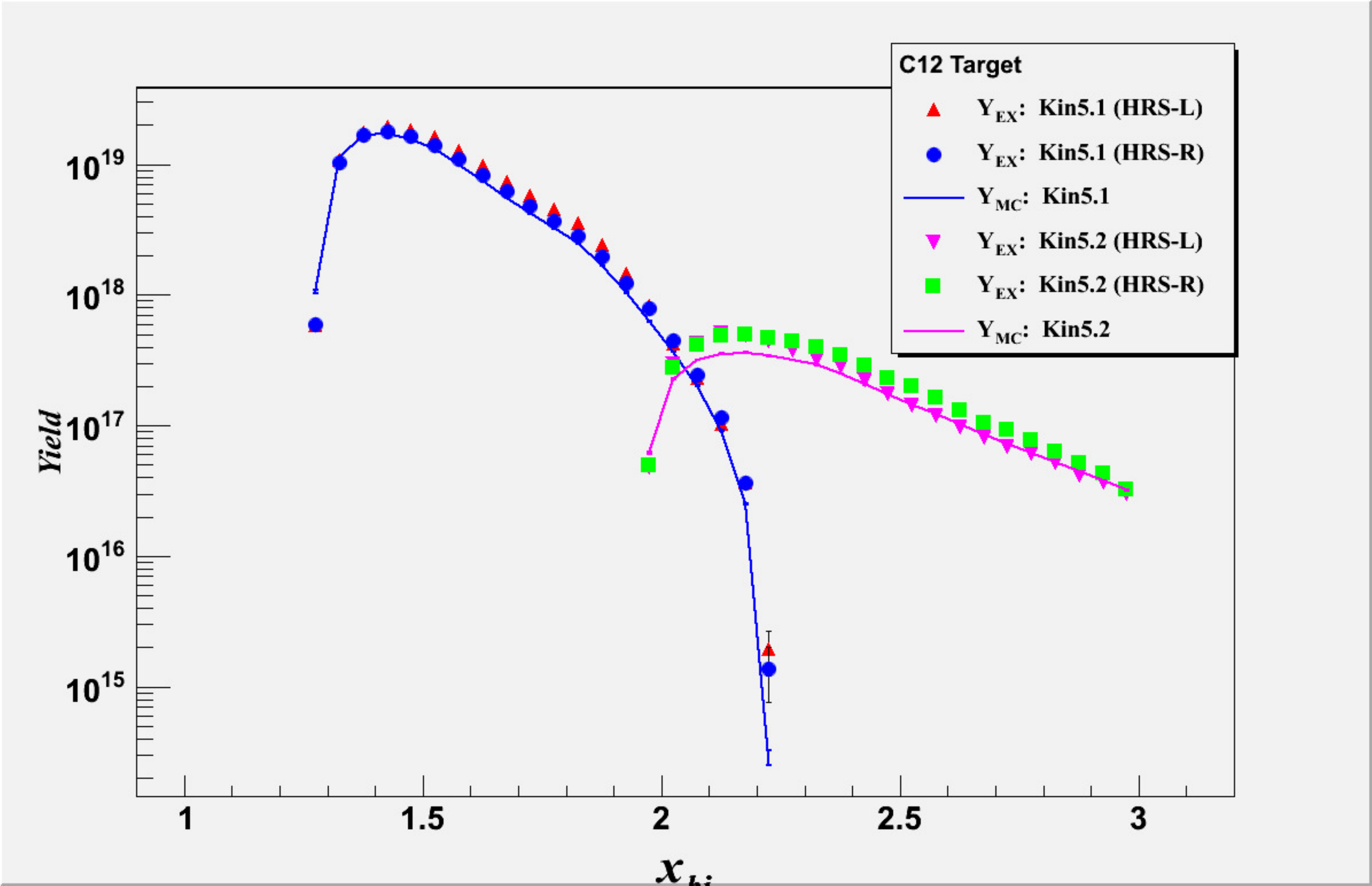}
\includegraphics[width=8.0cm]{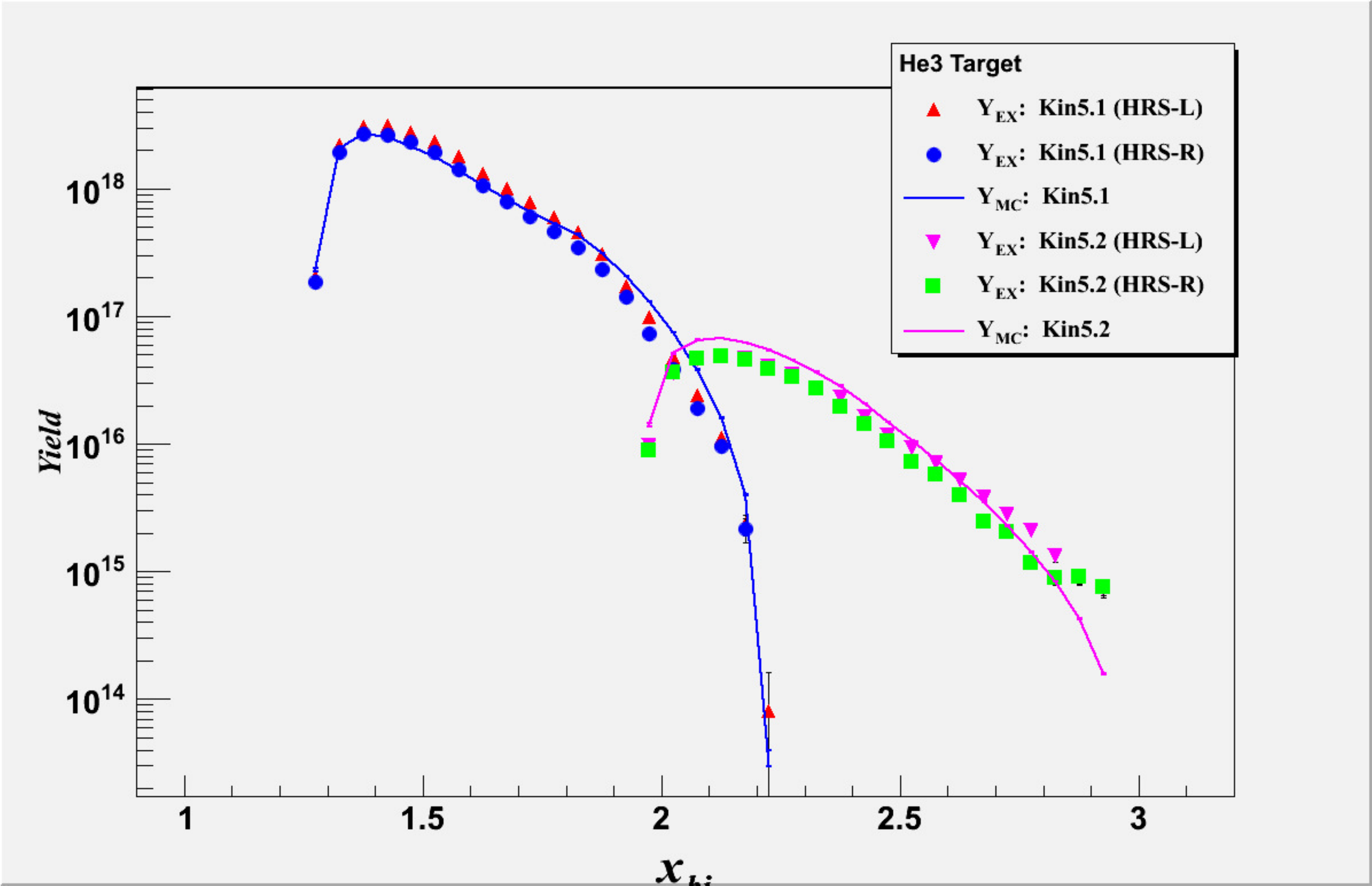}}
\caption{Preliminary carbon (left plot) and $^3$He (right plot) yields for averaged $Q^2$ of 1.8 GeV$^2$.}
\label{fig:xs}
\end{figure}

The analysis of E08-014 is in its final stage for the 2N-SRC and 3N-SRC measurements. The isospin study 
with the data on $^{48}$Ca and $^{40}$Ca has been postponed until the cross section model is optimized. 
Special care needs to be taken for these data because of the non-uniformity of the foils. During E08-014, 
data were taken to look at the thickness and uniformity of the foils. Even so, the systematic 
uncertainty of the isospin ratio is anticipated to be higher than previously estimated in the proposal.

%
%

\clearpage \newpage

\subsection{E08-027 - $g_2^p$}
\label{sec:e08027}

\begin{center}
\bf A Measurement of $g_2^p$ and the Longitudinal-Transverse Spin Polarizability
\end{center}

\begin{center}
A. Camsonne, J.P. Chen, D. Crabb, K. Slifer, spokespersons, \\
and \\
the Hall A Collaboration.\\
contributed by R. Zielinski.
\end{center}

\subsubsection{Motivation}

The inclusive scattering spin structure function (SSF) $g_2^p$ is largely unmeasured at low and moderate $Q^2$ values; the lowest momentum transfer that has been investigated is $1.3$ GeV$^2$ by the RSS collaboration~\cite{Wesselmann:2006mw}. Poor knowledge of $g_2^p$, along with the other SSFs, in the low $Q^2$ region is now a limiting factor in the precision of bound-state QED calculations. The finite size of the nucleon, as characterized by the structure functions has become the leading uncertainty. Furthermore, researchers from PSI~\cite{Pohl:2010zz} have obtained a value for the proton charge radius $\langle R_p\rangle$ via measurements of the Lamb shift in muonic hydrogen, which differs significantly from the value from elastic electron proton scattering. The main uncertainties in the PSI results originate from the proton polarizability and from different values of the Zemach radius. These quantities are determined from integrals of the SSF and elastic form factors, which due to kinematic weighting, are dominated by the low $Q^2$ region.

The existing data has also revealed a striking discrepancy~\cite{Amarian:2004yf}  of $\chi$PT calculations with the longitudinal-transverse polarizability
$\delta_{LT}^n$. 
This is  surprising since
$\delta_{LT}$
was expected to be a good testing ground for the chiral dynamics of QCD~\cite{Bernard:2002bs,Kao:2002cp} 
due to it's relative insensitivity to Delta resonance contributions.
Measurement of $g_2^p$ at low $Q^2$ will give access to $\delta_{LT}^p$ and 
allow an isospin examination of this `$\delta_{LT}$ puzzle'.

\begin{figure}[htdp]
\begin{center}
\includegraphics[angle=0,width=0.45\textwidth]{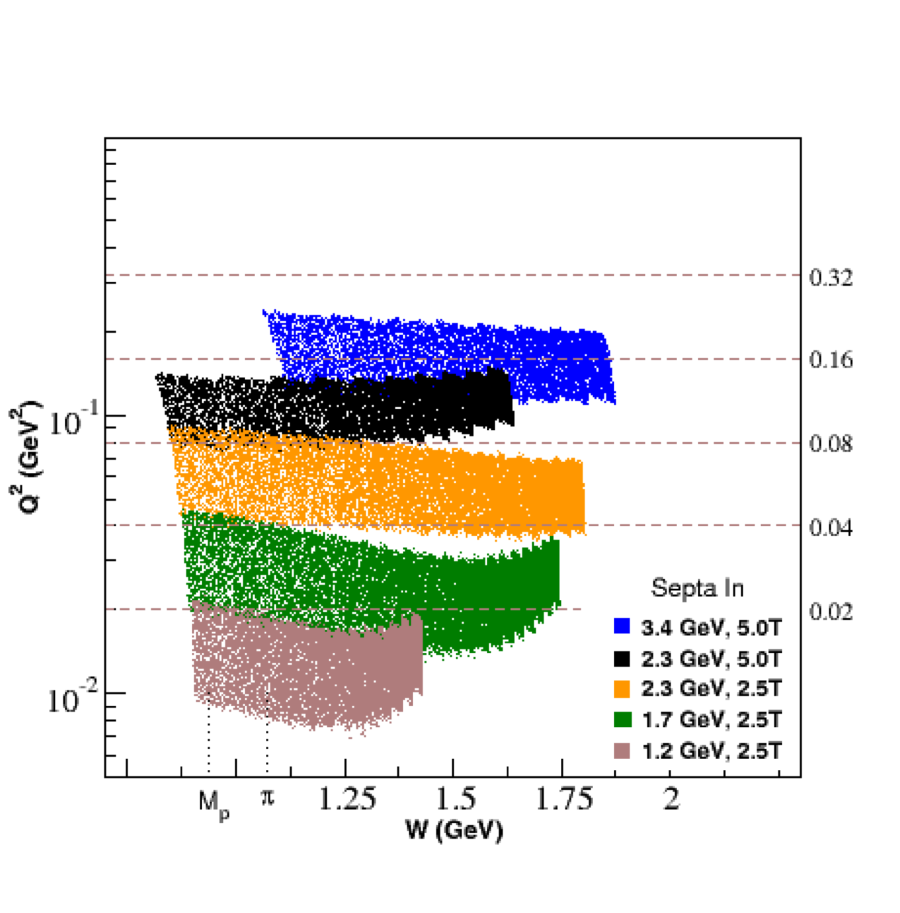}
\caption{\label{KIN}Kinematics covered during experimental run period. The right-hand side vertical axis is the extrapolation to constant $Q^2$. As W increases (and momenta decreases) $Q^2$ rises due to the target field creating a larger scattering angle.  }
\end{center}

\end{figure}

\begin{figure}[]
\begin{center}
\includegraphics[angle=0,width=0.60\textwidth]{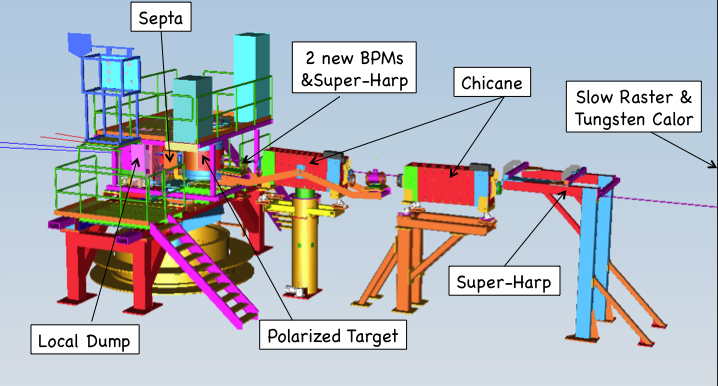}
\caption{\label{PLAT} E08-027 installation in Hall A. The third arm detector was located on the left hand side of the bottom target-platform.}
\end{center}
\end{figure}

\subsubsection{The Experiment}
The experiment successfully ran March - May 2012 (see Table~\ref{STAT}). We performed an inclusive measurement at forward angles of the proton spin-dependent cross sections in order to determine the $g_2^p$ structure function and the longitudinal-transverse spin polarizability $\delta_{LT}$ in the resonance region for 0.02 $< Q^2 < $ 0.20 GeV$^2$ (See Figure~\ref{KIN}). To reach the lowest possible momentum transfer, a pair of room temperature septa magnets were installed to allow detection of scattered electrons at 5.69$^\circ$. During the run period, the right spectrometer (RHRS) septa magnet was damaged twice. After each incident, magnet coils were bypassed, which limited its field strength and momenta range. Dynamical Nuclear Polarization (DNP) was used to polarize the solid ammonia target.  The original target's superconducting magnet coil was damaged during testing, but a similar target was successfully transplanted from Hall B and used during the experiment.

\begin{table}[htdp]

\caption{Statistics of kinematic settings taken during the run period. Running was split between 2.5 T and 5.0 T target field strength configurations. At lower beam energies (and lower momenta) the 5.0 T target field effectively increased the scattering angle of the detected electrons. This would have limited our low $Q^2$ range. To fix this, 2.254 GeV, 1.704 GeV and 1.158 GeV were all run with the target field at half-strength, 2.5 T.}
\begin{center}
\begin{tabular}{|c|c|c|}
\hline
E$_0$(GeV) & Target (T) & Recorded Triggers \\ \hline
2.254 & 2.5 & 3.80E+09 \\ \hline
1.706 & 2.5 & 3.20E+09 \\ \hline
1.158 & 2.5 & 4.00E+09 \\ \hline
2.254 & 5.0 & 7.00E+08 \\ \hline
3.352 & 5.0 & 4.00E+08 \\ \hline
\end{tabular}
\end{center}
\label{STAT}
\end{table}%

The DNP target's strong magnetic field required the installation of two large dipole magnets upstream of the target to provide chicaning of the beam (see Figure~\ref{PLAT}). In order to limit depolarization of the polarized ammonia target, the experiment ran low currents ( $\sim$ 50 - 100 nA), which required the installation of two new super-harps and two new M15 antennae style beam position monitors (BPMs) to fully characterize the beam profile.  To further minimize depolarization, a slow raster was installed to raster the beam over the entire $\sim$ 2 cm diameter target cup. The lower beam currents also led to the installation of a low current tungsten calorimeter to calibrate the beam current monitors (BCMs).

A new third arm detector was designed for the experiment to give a relative measurement of the combination of beam and target polarization at a 10$\%$ level. The detector counted the elastic recoil protons at large scattering angles in order to measure the elastic proton asymmetry. This asymmetry is related to the beam and target polarization. It was installed on the left-side of the beamline at an angle of approximately 70$^\circ$.

\subsubsection{Experimental Progress}


A Geant4 based program was developed to simulate the the physics for the high resolution spectrometers (HRS) together with the target and septum fields.
The full geometries of the experimental setup have been built in (see Figure~\ref{PLAT}). This includes the chicane fields for each kinematic setting as well as the two target field configurations and several versions of the septum fields, which are varied in the number of active coils in the right septum. The optics run plan was determined based on the simulation; we will use it to improve optics calibration and determine the spectrometer acceptance.  

Currently, spectrometer optics data have been optimized for no target field, chicane set to straight-through running on both the left and right HRS.  Straight-though optimization removes the added complexity of the target field and allows for an optics calibration focusing on the septa and HRS magnets. The angle calibration results of the left-HRS straight-through setting are shown in Figure~\ref{Fig:Opt}. Momentum matrix elements have also been calibrated. Work has now shifted to repeating the same analysis but with a longitudinal target field. The next step is to then determine the optics for the production settings with a transverse target field at both 2.5 T and 5.0 T. The effect of the different right-HRS septa coil packages on the optics for both left and right spectrometer optics is also being analyzed. 
 
Beam position information is also needed for the optics optimization. The straight-through calibration of the BPMs is completed; a new method was created to calculate the beam position from the four-antennae BPMs. The BPM analysis package also includes a transfer function from the BPMs to the target, taking into account the 2.5 T and 5.0 T target fields. The current analysis effort is to create this transfer function. We are also looking into the non-linearity of the slow raster and the issue of increased noise in the BPM signal for low-current running. 

\begin{figure}[htdp]
\begin{center}
\includegraphics[angle=0,width=1.0\textwidth]{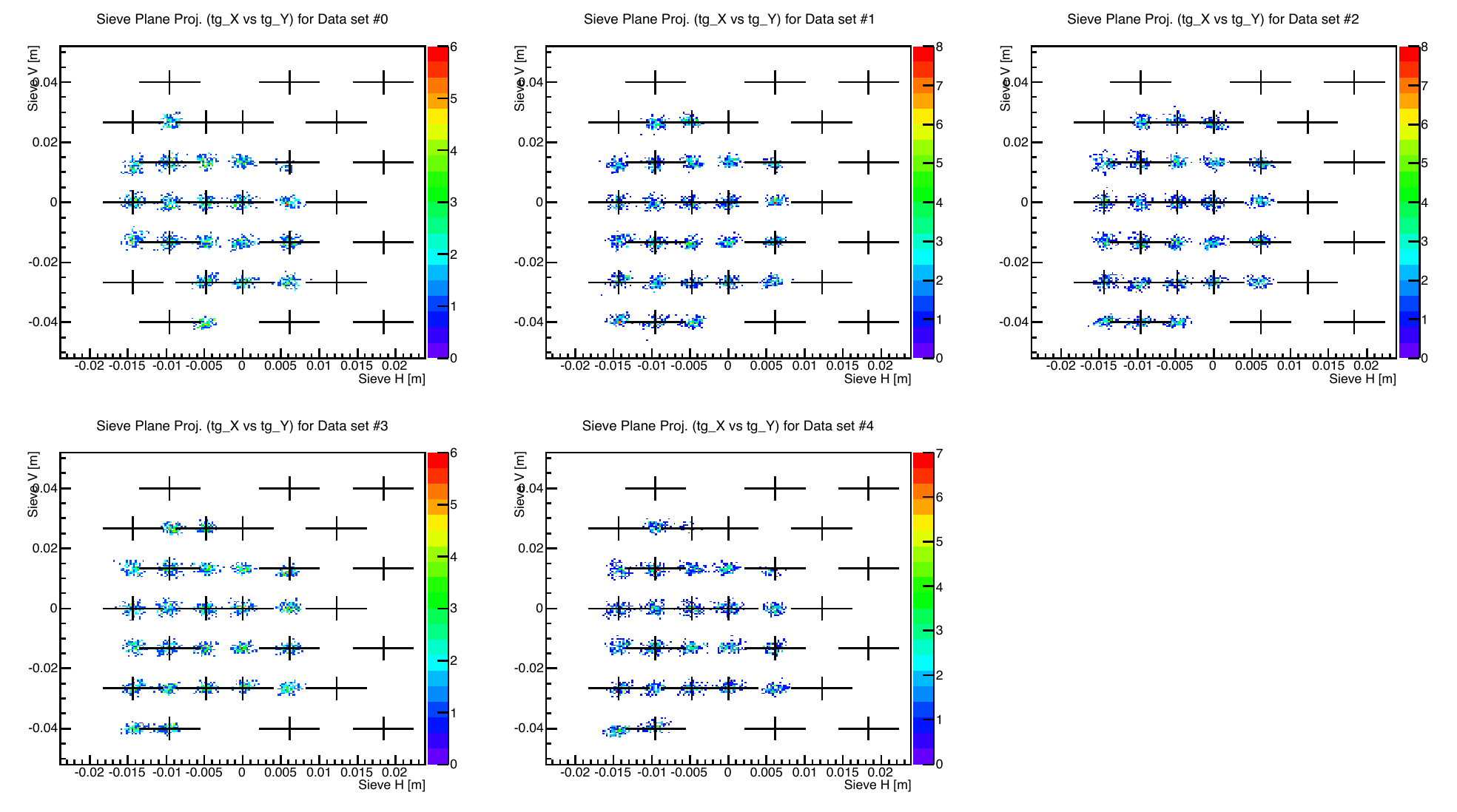}
\caption{\label{Fig:Opt}Sieve slit pattern: X vs. Y. Each cross represents the position of the sieve hole used to calibrate the theta and phi angles. The events from each sieve hole have been aligned well as a sieve pattern is clearly visible. Five panels represent different delta scan runs, which together represent the focal plane momentum coverage, -3.5$\%$, -2$\%$, 0, 2$\%$, 3.5$\%$, respectively.}
\end{center}
\end{figure}

Spectrometer detector efficiencies are needed as a correction to the measured cross section. A first round calibration of the HRS detectors is now completed. This includes particle identification (PID) calibrations on the gas ~\v{C}erenkov and lead glass on both spectrometer arms. Detector and cut efficiency  analysis on the PID detectors is currently underway. The S1 and S2m trigger scintillator efficiency is also finished for both arms. The results  for the left-HRS are shown in Figure~\ref{Fig:Eff}. The trigger efficiency for all good production runs is above 99$\%$. 

An electron asymmetry measurement is needed along with the absolute cross-section to extract $\delta_{LT}^p$. For the asymmetry measurement, beam helicity information is needed. The experiment used the helicity scheme set by Qweak, which was also used in the DVCS experiment. The g$_2^p$ helicity decoder package is a combination of the DVCS and normal Hall A decoders. The package has already been tested during the production running and worked well with very preliminary asymmetries calculated during the production run period.

Measurements of the target polarization are also needed to determine the asymmetries. The calibration constants, used to convert the NMR signal into a polarization, for all target materials and settings have been calculated. They can now be applied to the polarization signals on a run by run basis to obtain averaged target polarizations with relative uncertainties in the 1-4$\%$ range. Still to be done is a quality check of the target NMR signal throughout the experiment, along with some polarization signal fit adjustments to reduce uncertainties.

\begin{figure}[htdp]
\begin{center}
\includegraphics[angle=0,width=0.50\textwidth]{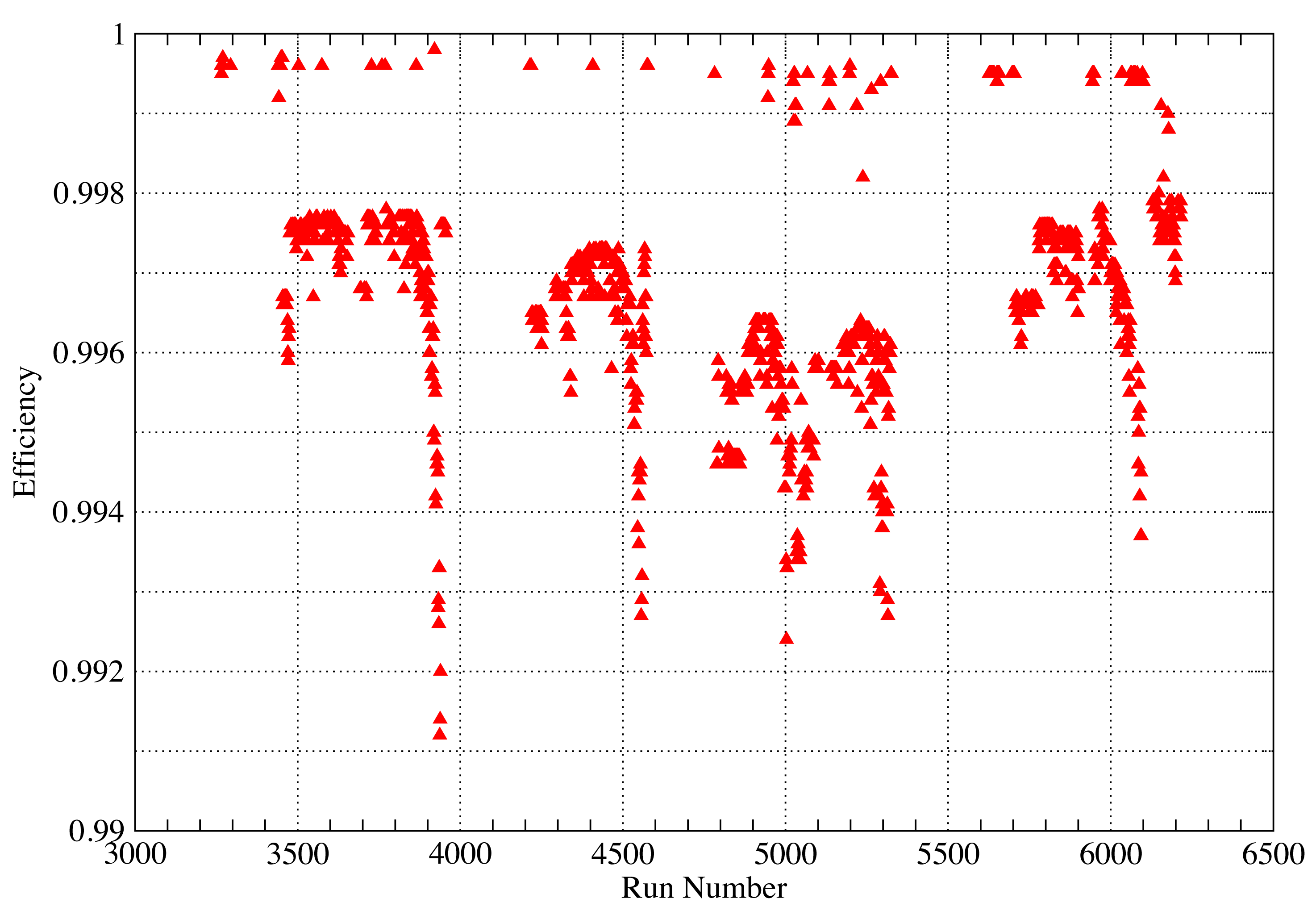}
\caption{\label{Fig:Eff}Left-HRS trigger efficiency results for all production runs.}
\end{center}
\end{figure}

Following the completion of the target analysis, base optics and BPM calibrations, and first round detector calibrations, the experiment is beginning to move into first-pass farm production. Optics will continue with the effort moving to optimization of the production settings and the BPM calibrations for the same settings. The GEANT simulation will also be used to begin work on determining the experimental acceptance needed for a cross section measurement. The detector efficiencies will be finalized, and we will start the asymmetry-dilution analysis, which determines what percentage of the electrons detected by the HRS were not scattered off of a proton from the ammonia target.




%
%

\clearpage \newpage

\subsection{E12-06-122 - $A_1^n$}
\label{sec:e12-06-122}

\begin{center} 
\bf Measurement of the Neutron Asymmetry A1n in the Valence Quark Region
Using 8.8 and 6.6~GeV Beam Energies and the BigBite Spectrometer in Hall A
\end{center}

\begin{center}
contributed by J.~Annand,  T.~Averett, G.D.~Cates, N.~Liyanage, G.~Rosner, B.~Wojtsekhowski, and X.~Zheng.
\end{center}

The PAC36 report said the following about the Hall A A1n experiment: \\
{\it ``These experiments exploit the Hall A and C spectrometers to measure
the down quark polarization $\Delta$d(x) at the highest possible value of x
and level of statistical precision (i.e., beyond that accessible at CLAS).
Their use of a $^3$He target also provides an important systematic check
on possible nuclear effects in the extraction of $\Delta$d.
PAC30 approved the Hall A experiment, but conditionally approved the
Hall C experiment as it was not convinced that Hall A could not be used for the entire program.
The case made at PAC30 was that the Hall A Bigbite spectrometer with
an added Cherenkov detector would not provide
sufficient pion rejection to reach the highest x and Q$^2$ points of the Hall C proposal.
As the Hall A experiment would likely run near the start of 12 GeV
data taking, it seemed sensible to simply wait and see how Bigbite
performed before making a final decision on the need for further data from Hall C.

The updated proposals reviewed by PAC36 included numerous changes that
made the two experiments much more similar.
Perhaps most importantly, the beam time request for the Hall C was
greatly reduced: from 79 to 36 days, which is commensurate with the 23
days requested at Hall A.
Both experiments propose to take two stripes in (x,Q$^2$) that give each
a Q$^2$ lever arm up to similar values of x (0.6 for Hall A and 0.55 for Hall C).
Both have hardware issues.
At Hall C, the dramatic reduction in beam time was accomplished by
proposing a major upgrade
of the target that would permit it to withstand a 60~$\mu$A beam current.
Significant R\&D will be required to accomplish this technical goal.
At Hall A, a more modest target upgrade was invoked
(for 30~$\mu$A maximum beam current),
but the new Cherenkov detector proposed for Bigbite was revealed to be
a RICH with a finely segmented array of over 1,000 phototubes.
The PAC was pleased to see that manpower (the Averett group) was
clearly identified for this task, but even with the recycling of
parts from the decommissioned HERMES RICH, this is an ambitious undertaking.

With the Hall A and Hall C A1n proposals now on a near-equal footing in terms of beam time
and experimental complexity, the PAC concluded that they should have equal approval status.
The difference in letter grade arose from the experiments’ kinematic reach.
The Hall C experiment has two small but distinct advantages in this
area, thanks to its use of the full 11 GeV beam energy.
First, the top x value of 0.77 is the highest of all the g1 experiments.
Though this value is only slightly above the 0.70 accessible at Hall
A, the PAC found Wally Melnitchouk’s
arguments persuasive: the non-linear behavior of higher-twist, target
mass, and other corrections in the quest for the x$\rightarrow$1 limit of the
PDFs makes a strong case for pursuing even slight improvements in x range.
Second, the Hall C experiment has the broader Q$^2$ range (2-8~GeV$^2$
as compared to 2-5~GeV$^2$ at Hall A),
which is also important for constraining higher twist contributions.''} \\

Since the approval of the proposal, the collaboration has put all efforts
 into development of the instrumentation which is briefly outlined below.

\subsubsection{Configuration of the BigBite spectrometer}
\label{sec:BigBite_A1n}

The detector configuration was outlined in the presentation to PAC36.
It includes the front tracker, the Gas Cherenkov counter, the
rear chamber, the two-layer shower calorimeter, and the timing
hodoscope.
Following the PAC advice, we have reconsidered the source of the PMTs for
the Gas Cherenkov counter and decided to use the 29~mm PMTs from
the DIRC detector of BABAR instead of the HERMES RICH PMT array.
The detector configuration was presented to the DOE review in October
2011, which reviewed the SBS projects, as well as ``the dependencies'' such
as the Bigbite, which will be used as an electron arm
in two form factor experiments and in the SBS SIDIS
experiment.  These detectors are under construction by the collaboration.
The INFN collaborators are in charge of the front tracker.
UVa is providing the larger rear chamber.
Glasgow University is leading construction of the timing hodoscope.
W\&M is leading development of the Gas Cherenkov counter.
Most of the funding required for such an upgrade of the Bigbite detector
is being provided by users.   However, about \$70k is requested from Hall A
for the construction of a vessel of the Gas Cherenkov counter and its
front-end electronics.
The following sections present the status and configuration of those detectors.

\paragraph{Gas Cherenkov}
\label{sec:GasCher}

\begin{enumerate}
\item The focus of the  experiment is to measure $A_1^n$ with a statistical precision on the level of 0.003 and similar systematics.
{\bf{Thus we will require the Cherenkov detector to have $\eta_{\pi}^{GC} \leq0.1$}}
while assuming modest performance of the lead-glass system: $\eta_{\pi}^{LG} = 0.1$,
where $\eta_{\pi}^{GC}$ and $\eta_{\pi}^{LG}$ are the pion detection efficiencies for gas Cherenkov and lead-glass respectively.

\item The detector must fit within a 90~cm keep-out zone between the GEM chambers in the new BB detector frame.
This requires the maximum length of GRINCH is $\leq$ 85 cm.

\item In the $A_1^n$ experiment, the Bigbite spectrometer is set at a scattering angle of $30^{\circ}$.
The target length will be 60 cm,
compared with 40~cm used previously which brings the total material in the beam from 0.31~g/cm$^2$ to 0.36~g/cm$^2$.
The beam current will be increased from $12.5\,\mu$A to $40\,\mu$A.
These factors will increase the total luminosity by a factor of 4,
assuming no other changes to the material in the beamline.
Assuming the worst case scenario, with no improvement to the beamline, the background rates are also expected to increase by a factor of 4.
To separate the signals from the large background, few improvements have to be applied.
        \begin{enumerate}
        \item The Cherenkov ring will be detected by an array of 29mm diameter Electron Tubes 9125B PMTs with 25mm active diameter.
        The surface area of these PMTs compared with those used in $d_2^n$ (5") is 0.052.
        In addition, the face of the PMTs is made from glass that is 3x thinner.
        A (8)9x30 array of tubes,
        where 8 for even rows and 9 for odd rows will be instrumented with custom NINO-based front end electronics that include an amplifier,
        discriminator and pulse height sensitivity to provide a crude measurement of the pulse height.
        The NINO boards will be connected to FASTBUS multi-hit TDCs.
        We may also have FASTBUS ADCs that could be used for low rate running and calibration.

        \item The primary information from the detector will be timing-based.
        Spatially localized clusters of PMTs that detect a signal within a 10 ns timing window will be identified in the offline analysis.
        Based on simulation, the average number of photoelectrons produced per tube is expected to be around 2.5,
        and on average 9 PMTs are expected to ``fire'' ($\ge 1$~p.e.) for an electron event.
        The discriminator thresholds will be set to approximately 20-30\% of the single ph.e. peak to provide good detection efficiency for single p.e. hits.
        Offline analysis will be even less sensitive to background as the timing cut is reduced to $<10$~ns.

        \item The PMT array will be located on the large angle side of Bigbite where the rates are 10x smaller as demonstrated in the Hall A technical note for the Bigbite Cherenkov Detector\cite{d2n_note}.
        \item The heavy gas radiator $\mathrm{C}_4\mathrm{F}_8\mathrm{O}$ with index $n=1.00135$ at 1~atm and as long active length of gas as possible will be used to generate enough photons to identify the electrons. And the index gives a pion threshold of 2.7 GeV/c.
        \end{enumerate}

\item Because of the open geometry of the Bigbite magnet, there is a significant magnetic field (of 15 Gauss) at the PMT location.
        This requires the shielding should be added in the final design.

\item Experimental studies~\cite{neutron_data} have demonstrated that neutron induced $\alpha$ particles have a very short path in the glass
and a negligible probability of inducing a signal in the PMT.
The probability was found of $3 \times 10^{-4}$ for the borosilicate window R7525-HA and
$3 \times 10^{-6}$ for the quartz window R7600U-200-M4 PMTs, respectively.
Again note that the thickness of the glass window of the 29 mm PMTs is 3x less than the 5" tubes.
The contribution of neutrons to the background by conversion in the PMT glass is expected to
be very small relative to the electromagnetic background discussed next.

The Monte Carlo simulations~\cite{MC}  have shown that a significant background of low-energy electrons is present
due to production from material near the beam line and in the target scattering chamber which was not under vacuum.
Bench tests~\cite{BW} have shown that $\sim1$ MeV electrons from a $^{106}$Ru source will produce Cherenkov radiation
in the PMT glass which produces a single-photoelectron hit 30\% of the time.
Reasonable agreement between MC and the measured rates in $d_2^n$ was achieved.
Using this information, we can estimate the background contribution from this process,
which again will be reduced by a factor of 3 due to smaller glass window thickness.
Another source of background is photons that are produced in Bigbite or via line of sight to the target or beam line.

\end{enumerate}

To satisfy these requirements,
simulation and optimization of the proposed design has been completed using  Geant4 to optimize the geometry of the vessel and the detector array.
The optimization was focused on producing the largest possible electron efficiency and pion efficiency.
This translates to generating the largest number of photo-electrons per tube per event with the smallest possible Cherenkov ring diameter.

In the Geant4 simulation,
the detector vessel with an average active path length of 66 cm are installed with entrance (40cm(W)x150cm(H)x0.01cm(T)) and exit windows (50cm(W)x200cm(H)x0.01cm(T)) shown in Figure~\ref{fig:grinch_tank}.
It will be attached to the gas vessel such that 4 cylindrical mirrors with radius of curvature 1.3 meters can be used to transport all Cherenkov light onto the array.
As mentioned previously, the PMTs with glass on the front surface, light-catchers and 61 iron/mu-metal magnetic shielding bars
will be arranged in a tall, narrow array located on the large angle side of the BB spectrometer.
The distance from mirror center to the front surface of PMT is 65 cm. Angle between beam and PMT array is $55^{\circ}$.
The detector will be filled with 1~atm (absolute) of the ``heavy gas" C$_4$F$_8$O.
The above selected geometry will provide an average of 2.5 photo-electrons per tube, and an average of 9 tubes will fire for an event.
\begin{figure}[h]
        \centering
        \includegraphics[angle=0,scale=0.50]{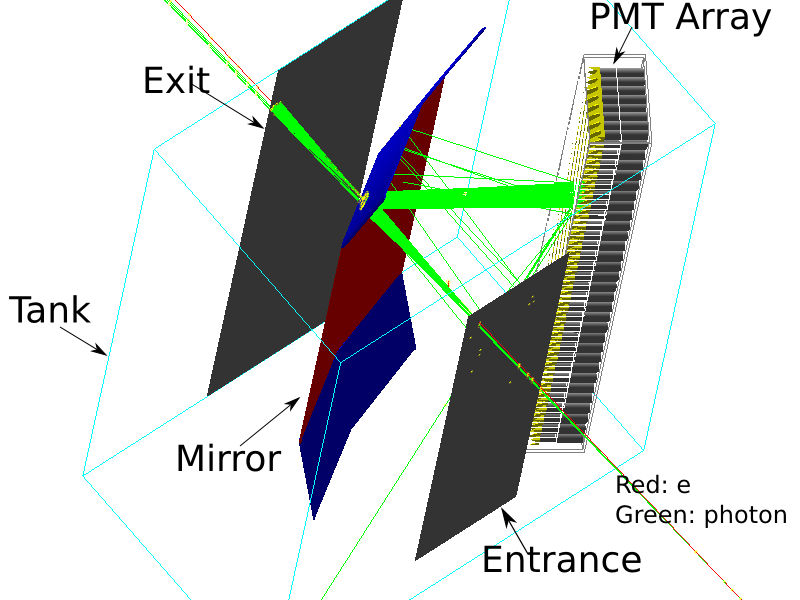}
        \caption{\tiny{GRINCH Tank.}}
        \label{fig:grinch_tank}
\end{figure}
\begin{itemize}
        \item Four mirrors are stacked as shown in Figure~\ref{fig:grinch_tank}.
        The size of middle two mirrors (red and blue) is 70cm(W)x60cm(H),
        for the top and bottom ones (gray and green) is 70cm(W)x40cm(H).
        Mirrors could be made using a technique that has been proven in the HRS Cherenkov detectors~\cite{HRS_mirror} and in the CLAS6 spectrometer.
        A thin Lexan sheet, which is a polycarbonate material because it has less trouble with stress at the surface than acrylic, will be formed using pressure and heat to achieve the desired curvature.
        This sheet will be coated with reflective aluminum and the back will be glued to honeycomb or low-density foam, followed by an outer shell.
        Because this detector is not based on the concept of focusing a given ring onto a single PMT, it is much less sensitive to mirror geometry.
        Cylindrical mirrors instead of the usual spherical mirrors will be used, making fabrication easier and potentially cheaper.
        Cylindrical mirrors are sufficient for transporting the light to a PMT array.

        \item
        Recently, several thousand PMTs were obtained by Hall A from the decommissioned BaBar DIRC detector~\cite{DIRC},
        and some PMT bases and HV cables were included.
        There they were used in a ring-imaging Cherenkov detector that transported light from quartz crystal radiators to the  using water.
        The tubes were removed after 10 years of service, and the faces of the tubes showed signs of surface damage due to immersion in ultra-pure water.
        The quantum efficiency of the tubes was found to have decreased by approximately 20-25\%~\cite{DIRC_info}.
        A thin disk of fused silica was glued onto the window at Jefferson Lab and  70 of these tubes were tested for dark noise/current,
        gain and relative quantum efficiency~\cite{PMT_test}.
        Data from before and after adding the glass disks were compared.
        The average increase in quantum efficiency was ~10\% after adding the glass disks.
        The PMTs will be mounted in a hexagonal close packed configuration with a spacing of $31$~mm,
        which means the sensitive area of the PMTs will cover $\sim60\%$ of the area of the array (see Figure~\ref{fig:pmt_array}).
     Each tube will be outfitted with a reflective cone attached to the perimeter of the photocathode as the yellow component shown in Figure~\ref{fig:pmt_array_3d_view}.
        The cone will be constructed from thermoplastic with a reflective coating and will be used to collect light  that
        is not incident on the active area of a given tube.
        Based on simulation, we expect a 20\% increase in the number of detected photons.
        \begin{figure}
                \centering
		\begin{subfigure}{0.49\textwidth}
		\includegraphics[angle=0,width=1.0\textwidth]{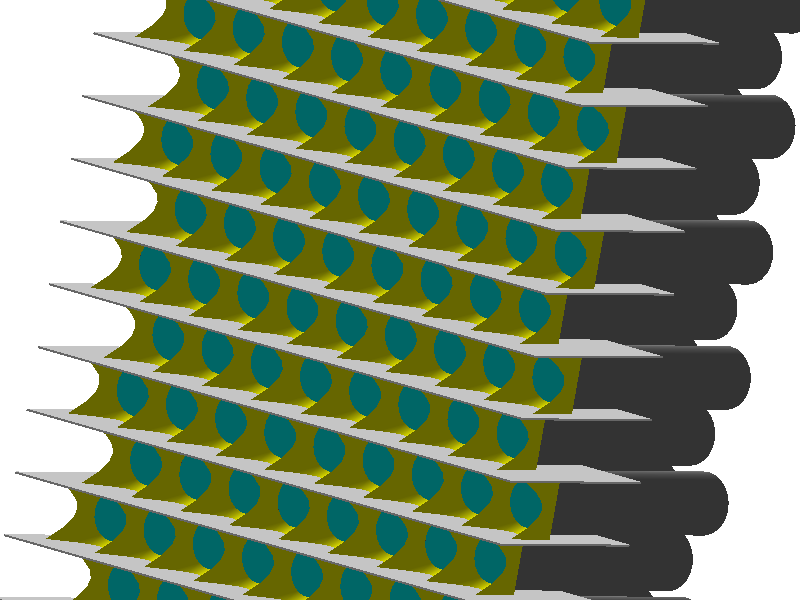} 
                \caption{Array 3D view.} \label{fig:pmt_array_3d_view}
		\end{subfigure} \begin{subfigure}{0.49\textwidth}
		\includegraphics[angle=0,width=1.0\textwidth]{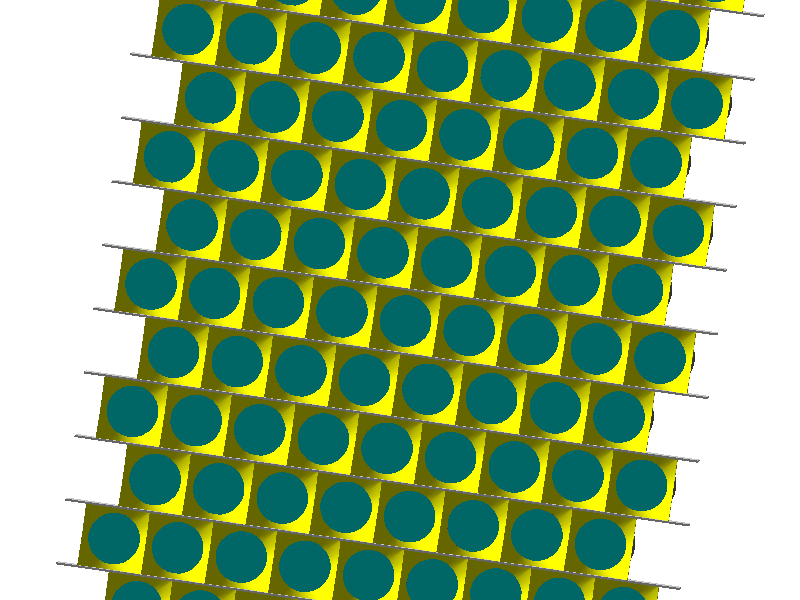}  
                \caption{Array Front view.}\label{fig:pmt_array_front_view}
		\end{subfigure}\\
                \begin{subfigure}{0.49\textwidth}
		\includegraphics[angle=0,width=1.0\textwidth]{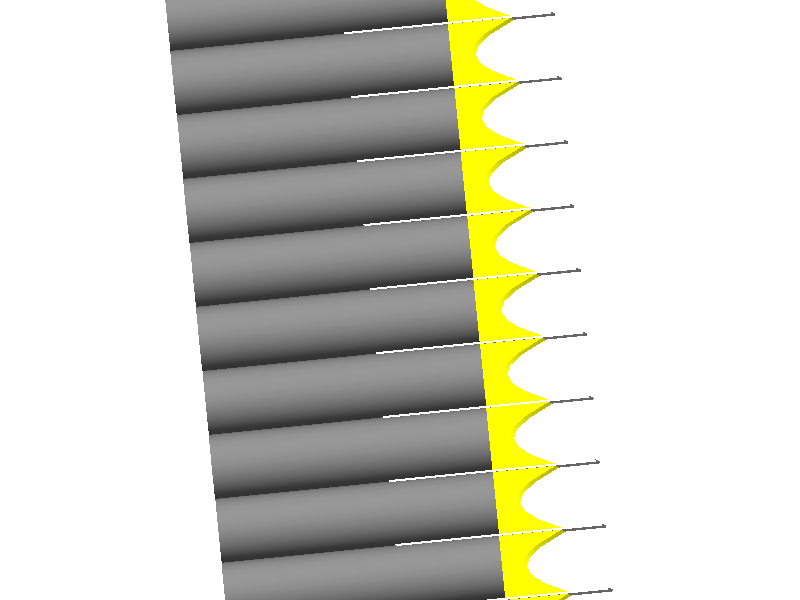} 
		\caption{Array Side view.}\label{fig:pmt_array_side_view}
		\end{subfigure} \begin{subfigure}{0.49\textwidth}
		\includegraphics[angle=0,width=1.0\textwidth]{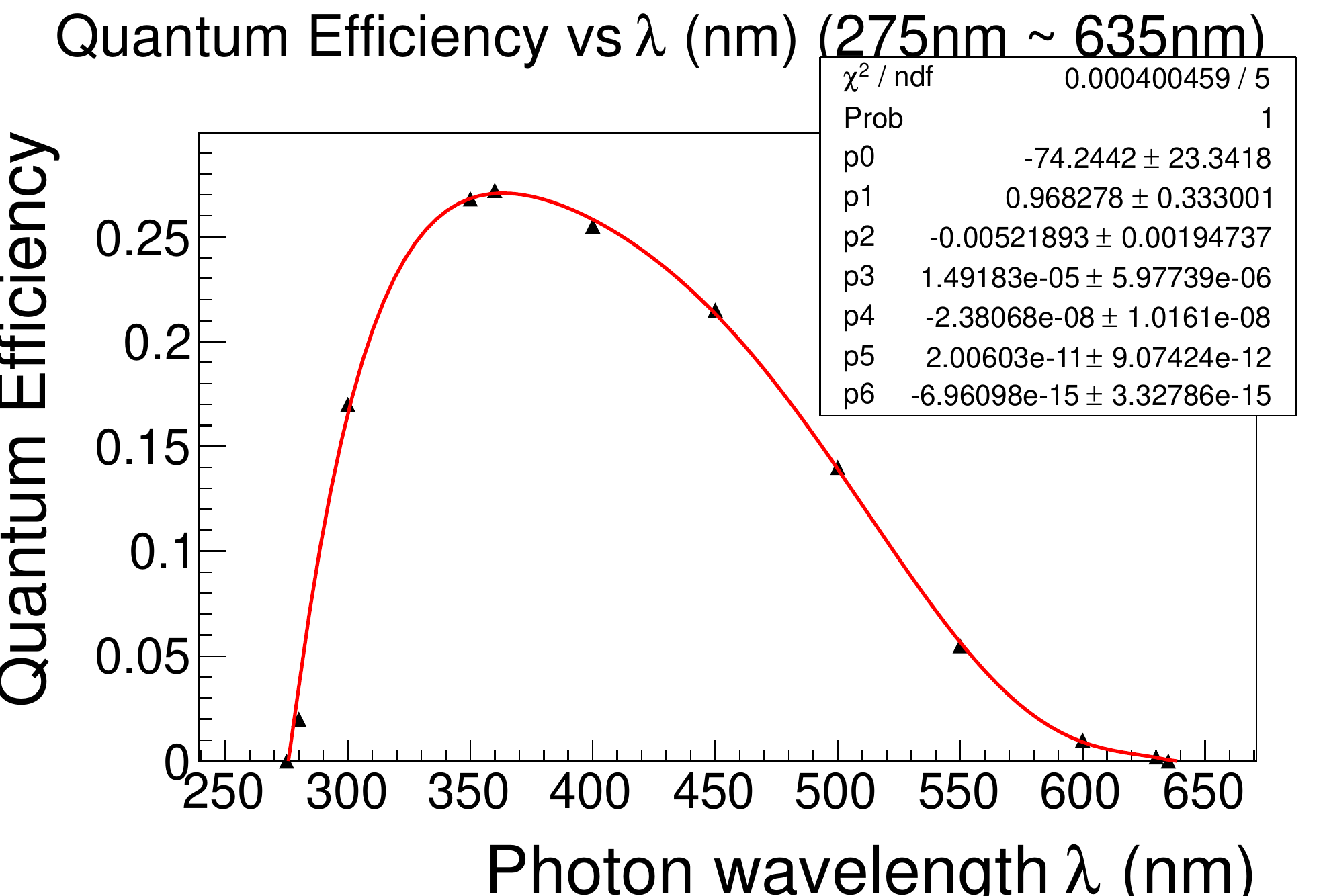} 
                \caption{PMT quantum efficiency.}\label{fig:pmt_qe}
		\end{subfigure}

                \caption{PMT array. White is shielding bars (29.9cm(W)x10cm(H)x0.1cm(T)),half of height is below PMT surface. Cyan is glass layer with 3mm thickness on PMT. Yellow is light-catcher. Grey is insensitive area of PMT and blue is sensitive area (25mm diameter).}
                \label{fig:pmt_array}
        \end{figure}

        \item
        The array will be surrounded with a rectangular steel (1010) box with five sides, the sixth side being open for the faces of the PMTs, which are located approximately 5~cm down inside the box.
        The detector and shielding geometry is being studied using TOSCA.
        The properties of the BigBite field at the PMT array are well understood from TOSCA simulation
        and measurements.
        PMT dynodes will be oriented horizontally to minimize the effect of the residual magnetic field.

\end{itemize}

In the MC simulation, the electron detection efficiency $\eta_e$  is defined as the probability
that an incident electron will be detected as a good electron  event.
The pion  detection efficiency  $\eta_{\pi}$ is defined as the probability that an incident pion
will be detected as a good electron event.

At pion momenta below the Cherenkov threshold, $\eta_\pi$ is limited by the delta
electron knockout process in the gas, which has probability of 2\%.
The total number of p.e.'s in that process has a wide distribution due to variation
of the origin of the delta-electron along the pion track through the gas.

Figure~\ref{fig:eta_pi} shows the pion detection efficiency $\eta_{\pi}$ for the different amplitude thresholds, averaged over momentum 1.6-3.3 GeV/c.
The $\pi/e$ ratio increases as momentum decreases.
We will require $\eta_{\pi}\leq 0.1$.
Using Figure~\ref{fig:eta_pi}  with 1~atm of gas, we require $\geq 11$ photoelectrons.
Using this cut, Figure~\ref{fig:eta_e} gives electron efficiencies of $\eta_e\sim 0.98$ (est.) for 1 ~atm.

The electronics for the GRINCH may provide some pulse-height information which would allow us to make a crude measurement and cut on the number of photoelectrons, but the detector is designed to work using timing clusters.
Therefore, if we are to make a cut at 9 p.e.'s, what we really need is a cut on the number of PMTs fired that will be efficient for requiring 9 or more p.e.'s per event.
\begin{figure}
        \centering
	\begin{subfigure}{0.5\textwidth}
        \caption{$\eta_e$ for different $N_{p.e}$ cut.} \label{fig:eta_e}\includegraphics[angle=0,width=1.0\textwidth]{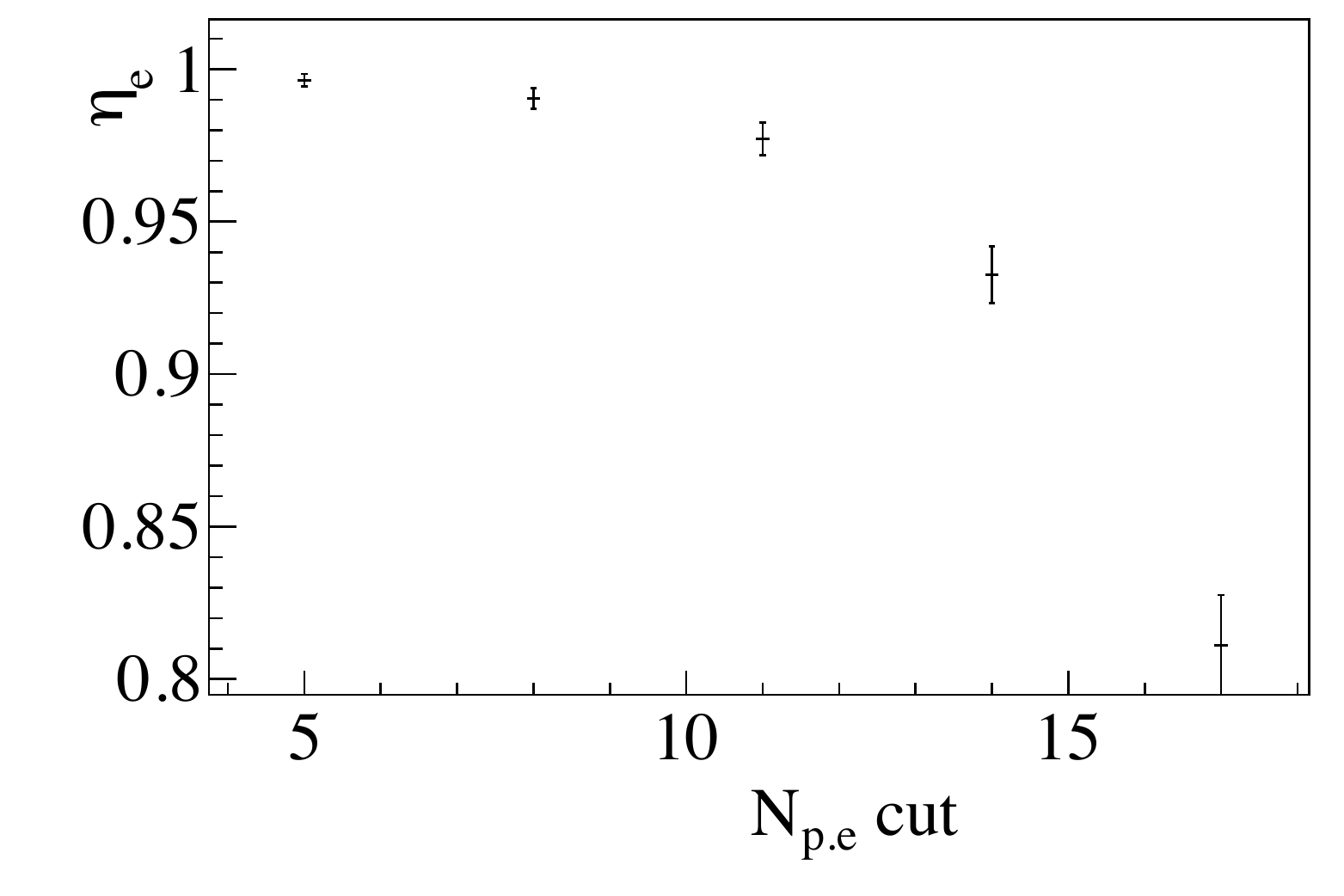}  
	\end{subfigure}\begin{subfigure}{0.5\textwidth}

        \caption{$\eta_\pi$ for $P$ for different $N_{p.e}$ cut.}\label{fig:eta_pi}\includegraphics[angle=0,width=1.0\textwidth]{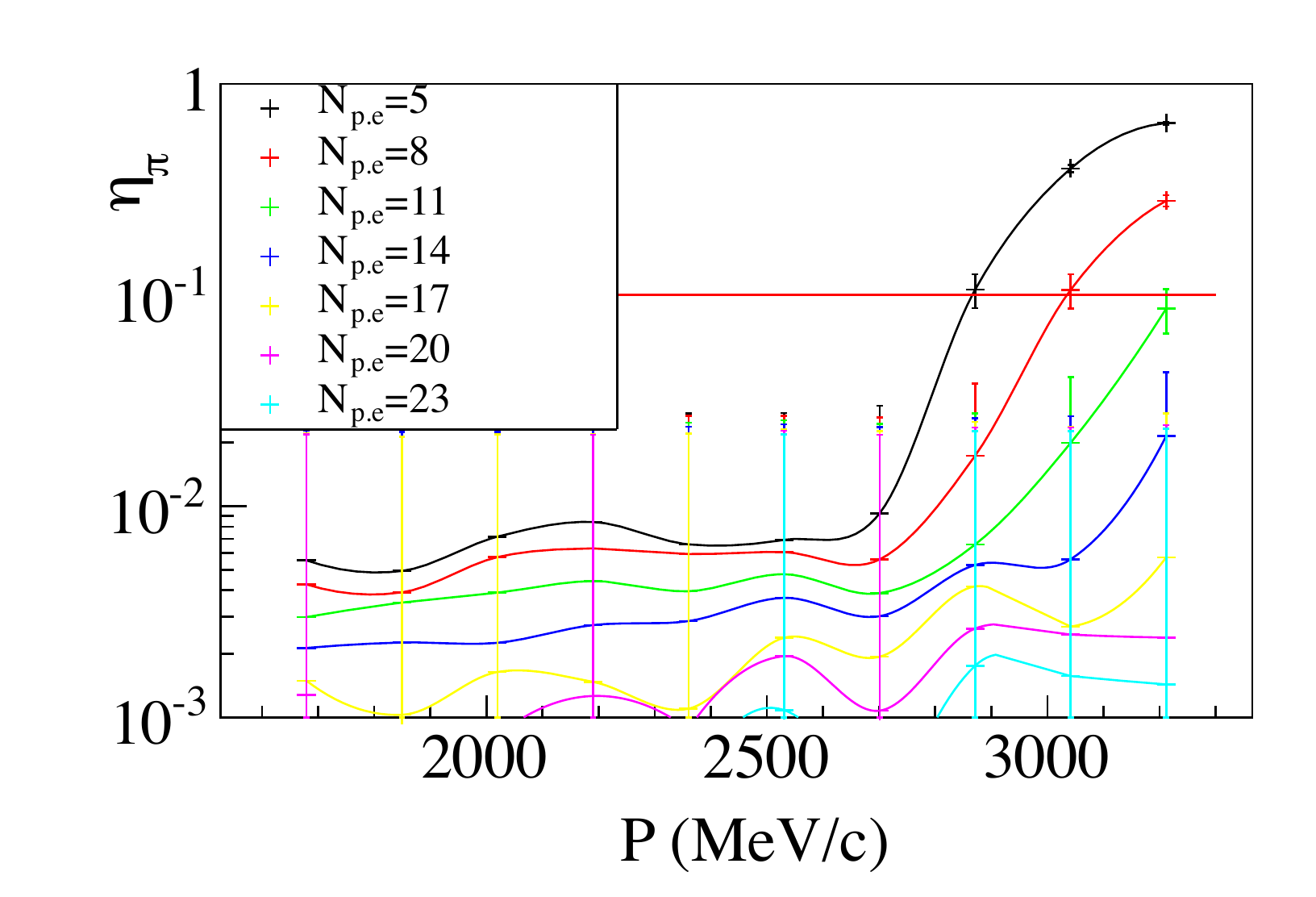} 
	\end{subfigure}
        \caption{Efficiency}
        \label{fig:eta}
\end{figure}

\paragraph{Timing Hodoscope}
\label{sec:TimingHodoscope}
The A1n experiment, in common with many new JLab experiments, plans
to run the BigBite spectrometer at higher luminosity than previously
achieved in Hall A. To accomplish this, the detector stack of BigBite
is being upgraded. Part of this upgrade involves an increase in the
segmentation of the plastic scintillator hodoscope, used to determine
the time of an electron hit precisely. The new hodoscope will consist
of 90 elements of EJ200 plastic (Fig.\ref{fig:BB-t-hodo}), each of
size $25\times25\times600$~mm and each read out at either end of
the bar via 24~mm diameter lucite light guides. The PMT is clamped
in position at the end of the light guide by a housing, which excludes
both stray light and any He in the Hall A atmosphere. Since the cross
section dimensions of a bar are less than the diameter of the PMT
housing, each alternate element has curved light guides to allow the
bars to be packed closely. The PMTs are type ELT9125 units, which
have been obtained from the BaBar spectrometer. They were originally
used on a DIRC system, and when coupled to a scintillator, the output
pulse was found to saturate. This has been cured by a redesign of
the PMT voltage divider base.

\begin{figure}[th]
\begin{center}\includegraphics[width=0.9\columnwidth]{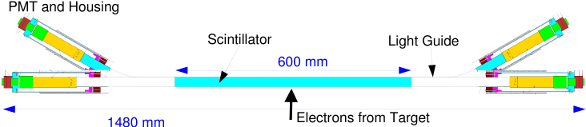}\end{center}

\caption{\label{fig:BB-t-hodo}Top view of the BigBite timing hodoscope elements.}
\end{figure}

The hodoscope is under construction in Glasgow and one bar has been
tested using cosmic rays. The interaction position of the cosmic rays
was localized using 2 trigger plastic scintillators of size $40\times40\times10$~mm
placed above and below the bar. Fig.\ref{fig:Pulses-PMT.1} shows
the variation in the pulse form from a single PMT (1) attached to
the hodoscope bar, as the trigger counter is moved along the length
of the bar. The rise time from the PMT is $\sim4$~ns, which is relatively
slow for a 28~mm diameter linear focused PMT, and the total pulse
length from a hodoscope bar is $\sim25$~ns.

\begin{figure}[th]
\begin{center}\includegraphics[width=0.9\columnwidth]{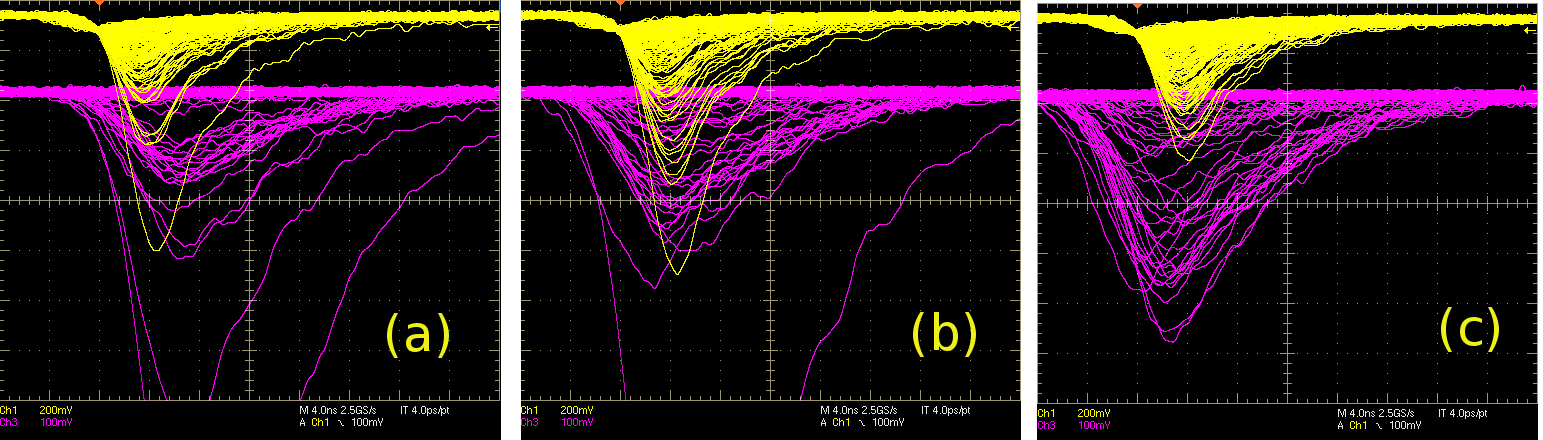}\end{center}

\caption{\label{fig:Pulses-PMT.1}Pulses from PMT.1 of the hodoscope bar under
test (magenta) and the upper trigger counter (yellow). a) Trigger
the far end of the hodoscope bar. b) Trigger at the middle of the
hodoscope bar. c) Trigger at near end of the hodoscope bar.}
\end{figure}

Both trigger and hodoscope signals were fed into a NINO card (Sec.
\ref{sec:NINO}) and in addition the trigger signals
were split and fed to 2, type 934 Ortec Constant Fraction Discriminators
(CFD). The latter provided reference times and also a physical trigger
signal for the DAQ system. Pulse amplitudes from the NINO card were
recorded by a CAEN V792 QDC and times recorded by a V1190A multi-hit
TDC, which accepts LVDS signals and has 100~ps resolution. The TDC
was programed to record both the leading and trailing edges of the
logic pulses, so that time over threshold could be measured. This
is correlated to the pulse amplitude, and is intended for time walk
correction. The CFD NIM-logic outputs were converted to LVDS and also
input to the V1190A.

\begin{figure}[th]
\begin{center}\includegraphics[width=0.9\columnwidth]{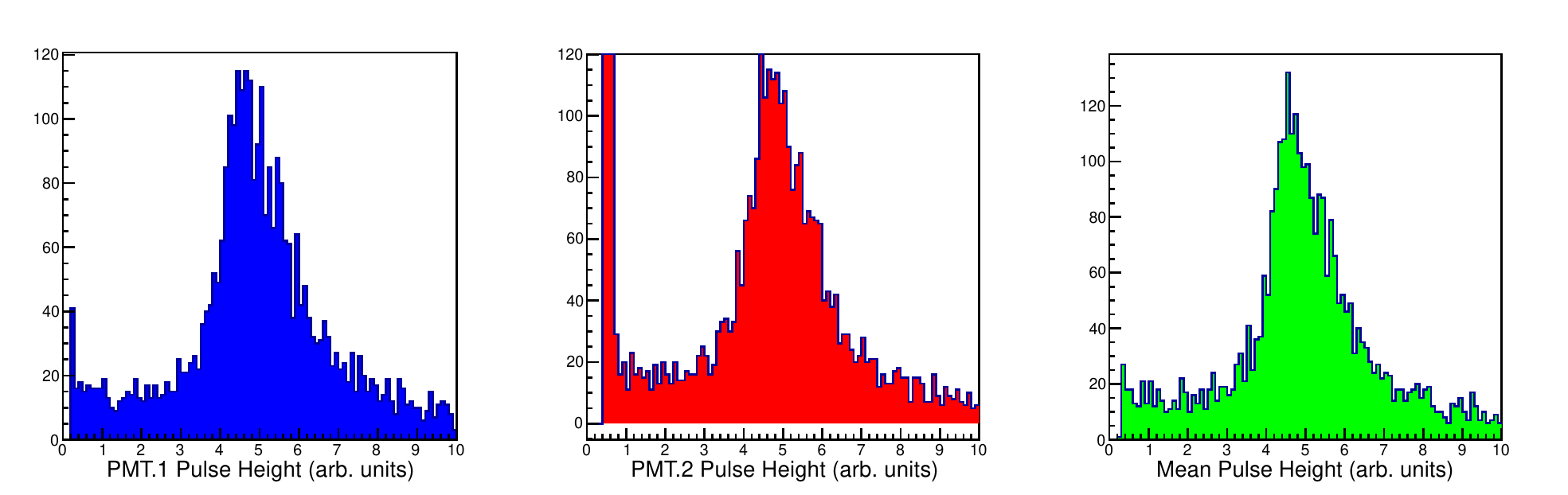}\end{center}

\caption{\label{fig:QDC}Pulse amplitude spectra from the hodoscope bar (blue:
PMT.1, red: PMT.2, green: geometric mean pulse height of PMT 1 and 2).}
\end{figure}

\begin{figure}[th]
\begin{center}\includegraphics[width=0.9\columnwidth]{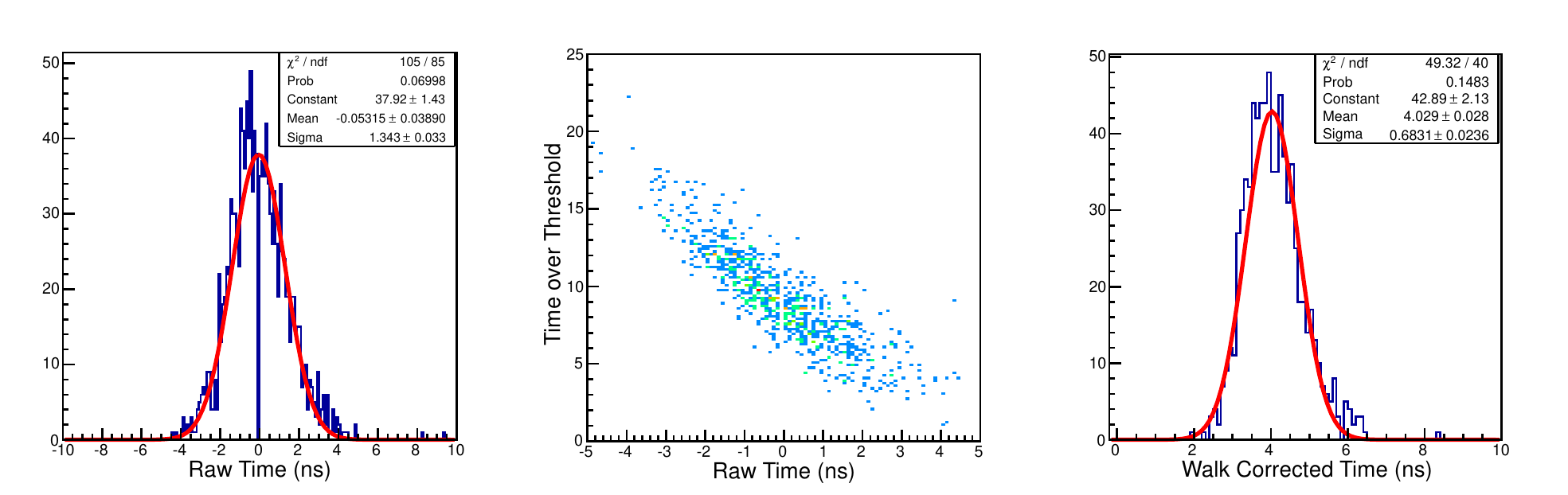}\end{center}

\caption{\label{fig:TDC}TDC spectra for PMT.2. The left panel shows the uncorrected
coincidence time distribution. The middle panel shows the same distribution
plotted against the pulse time over threshold (walk effects). The
right panel shows the coincidence time distribution after correction
for walk.}
\end{figure}

Data were recorded with the trigger scintillators placed at various
points along a hodoscope bar. Fig.\ref{fig:QDC} shows pulse height
spectra from the two PMTs attached to the bar and the geometric mean
pulse height, given by $Q_{mean}=\sqrt{Q_{1}Q_{2}}$. The TDC spectrum
from PMT.2 is shown in Fig.\ref{fig:TDC} (left). When plotted against
the time over threshold (middle) the effect of time walk is evident.
The walk effect was corrected, assuming a linear dependence of walk
on time-over-threshold, producing the distribution on the right panel.
The hit time in a hodoscope bar is obtained from the mean time of
the 2 PMTs $T_{mean}=(T_{1}+T_{2})/2$ and the hit position along
the bar from the time difference $T_{diff}=T_{1}-T_{2}$. Typical
distributions are displayed in Fig.\ref{fig:Time-distributions}.
The mean-time resolution of the hodoscope bar is $\sim0.4$~ns while
the time-difference resolution is $\sim0.7$~ns. By comparison the
coincidence time resolution of the top/bottom trigger detectors using
a CFD is $\sim0.4$~ns. 

\begin{figure}[th]
\begin{center}\includegraphics[width=0.9\columnwidth]{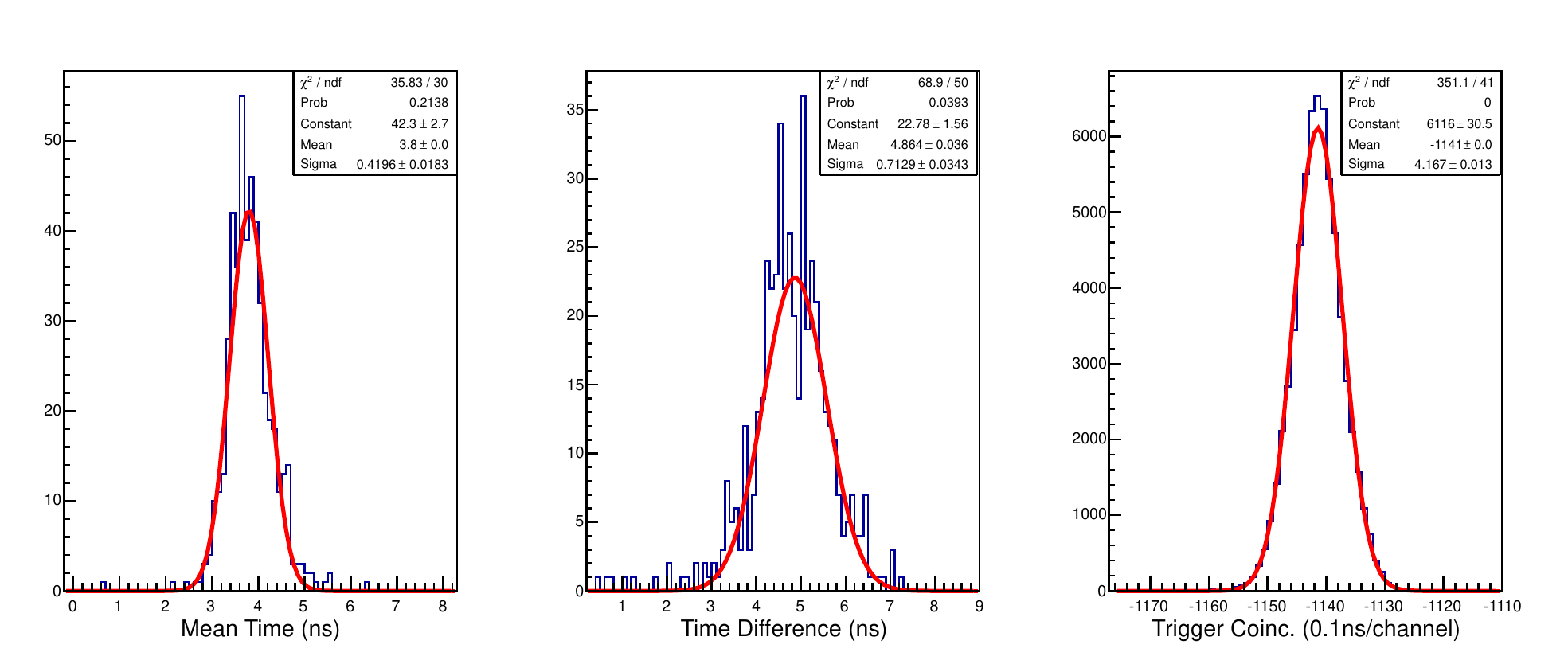}\end{center}

\caption{\label{fig:Time-distributions}Coincidence time distributions. Left:
mean time $T_{mean}$, center: time difference $T_{diff}$, right:
top-bottom trigger coincidence.}
\end{figure}

\begin{figure}[th]
\begin{center}\includegraphics[width=0.9\columnwidth]{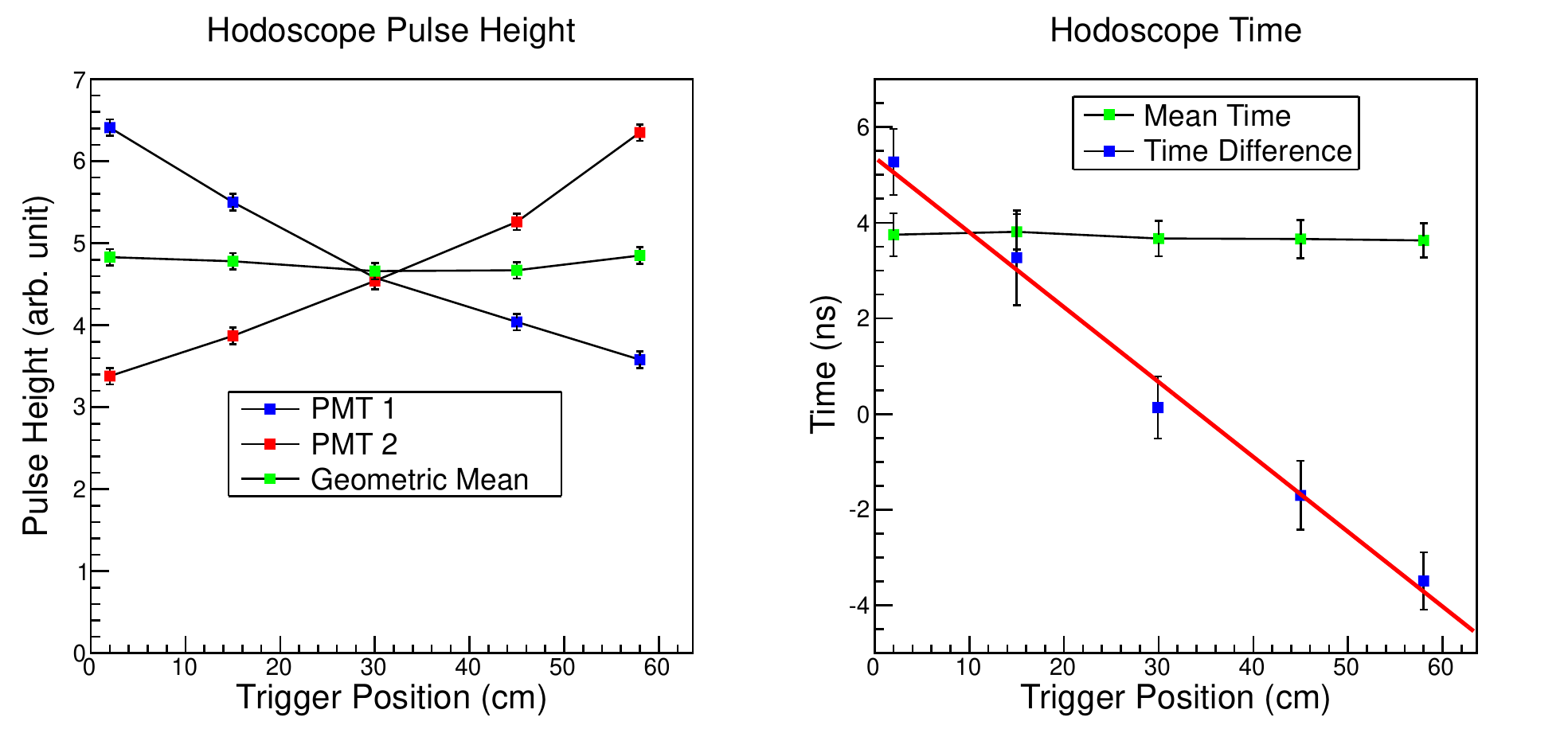}\end{center} 

\caption{\label{fig:Variation}Variation of pulse height (left) and hit time
(right) as the trigger position is varied. The pulse height is obtained
from the minimum-ionizing peak position. The error bars on the time
plot are the widths obtained from Gaussian fits. The red line is a
linear fit to the time-difference dependence.}
\end{figure}

The variation in pulse height and timing distributions for different
positions of the trigger counters is shown in Fig.\ref{fig:Variation}.
While the pulse height from individual PMTs varies significantly over
the length of the bar, the geometric mean is almost position independent.
Similarly the mean time is position independent while the time difference
dependence on position is approximately linear. A linear fit to the
time difference dependence produces a slope of -0.16 ns/cm which translates
to a position resolution along the bar of $\sim4.5$~cm.

\paragraph{Front-end NINO card}
\label{sec:NINO}
The GRINCH gas Cerenkov and Timing Hodoscope components of the BigBite
detector stack will employ the best part of 1000 photomultipliers
(PMT) for readout and will use PMTs, originally from the DIRC system
of the BaBar spectrometer. The electronics will be mounted close to
the PMTs to avoid long runs of coaxial cable, and must therefore be
compact and economical on power consumption. A low cost implementation,
based on the NINO ASIC \cite{NINO} from CERN, has been designed.
NINO was designed originally for the Multigap, Resistive Plate Chambers
(MRPC) which form the time of flight systems at the ALICE experiment
at LHC. It expects to receive a differential input signal and thus,
for the present PMT application, some additional circuitry is required.
A diagram of the NINO circuit for BigBite and general Hall-A use is
given in Fig. \ref{fig:NINO}.

\begin{figure}[th]
\begin{center}\includegraphics[width=0.55\columnwidth]{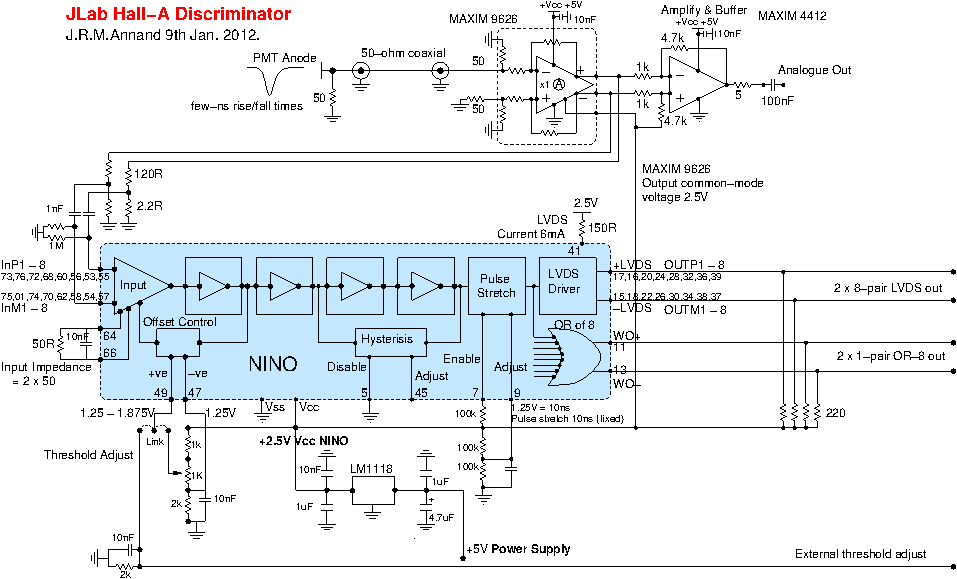}\hspace{0.5cm}
\includegraphics[width=0.2\textwidth]{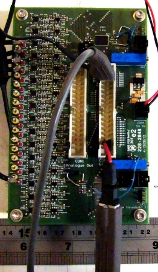}\end{center}

\caption{\label{fig:NINO}Left: circuit diagram of the NINO based amplifier/discriminator
card. Right: photograph of the prototype NINO card under test at Glasgow.}
\end{figure}

Each NINO ASIC has 8 discriminator channels and the Hall-A card (Fig.
\ref{fig:NINO} right panel) will mount 2 of these, giving 16 channels
in total. On each channel the signal from the PMT anode is input,
via coaxial cable, to a fast, fully-differential amplifier (MAXIM
9626). The differential output from this unity-gain amplifier is then
attenuated and fed to the NINO input. The MAXIM 9626 output also feeds
to a buffer stage (MAXIM 4412), providing an output for pulse amplitude
measurement.

The 16 NINO logic outputs conform to LVDS standard and in addition
2 OR-of-8 LVDS outputs are provided. The LVDS outputs operate in time-over-threshold
mode, where the width of the logic pulse is the duration that the
input analogue pulse exceeds the discriminator threshold plus 10~ns.
Each NINO ASIC has a common threshold circuit for the 8 channels,
and the threshold level may be adjusted via an on-board potentiometer
or external voltage level. A voltage difference of 650~mV applied
to the threshold circuit (Fig. \ref{fig:NINO}) produces a threshold
voltage of 180~mV on the input signal, equivalent to a charge of
$\sim45$~pC for the hodoscope signal. The relation of voltage difference
on the threshold circuit to PMT output voltage is given in Fig. \ref{fig:NINO-thresh}.
The overall range may be shifted by changing or removing the attenuation
factor for pulses input into the NINO chip. 

The card is powered by a single +5~V supply and draws 1.76~A. The
+2.5~V necessary for the NINO chips is derived from +5~V using an
on-board voltage regulator. 

Tests of the performance of the discriminator card in conjunction
with elements of the timing hodoscope is described in Sec.\ref{sec:TimingHodoscope}. 

\begin{figure}[th]
\begin{center}\includegraphics[width=0.5\columnwidth]{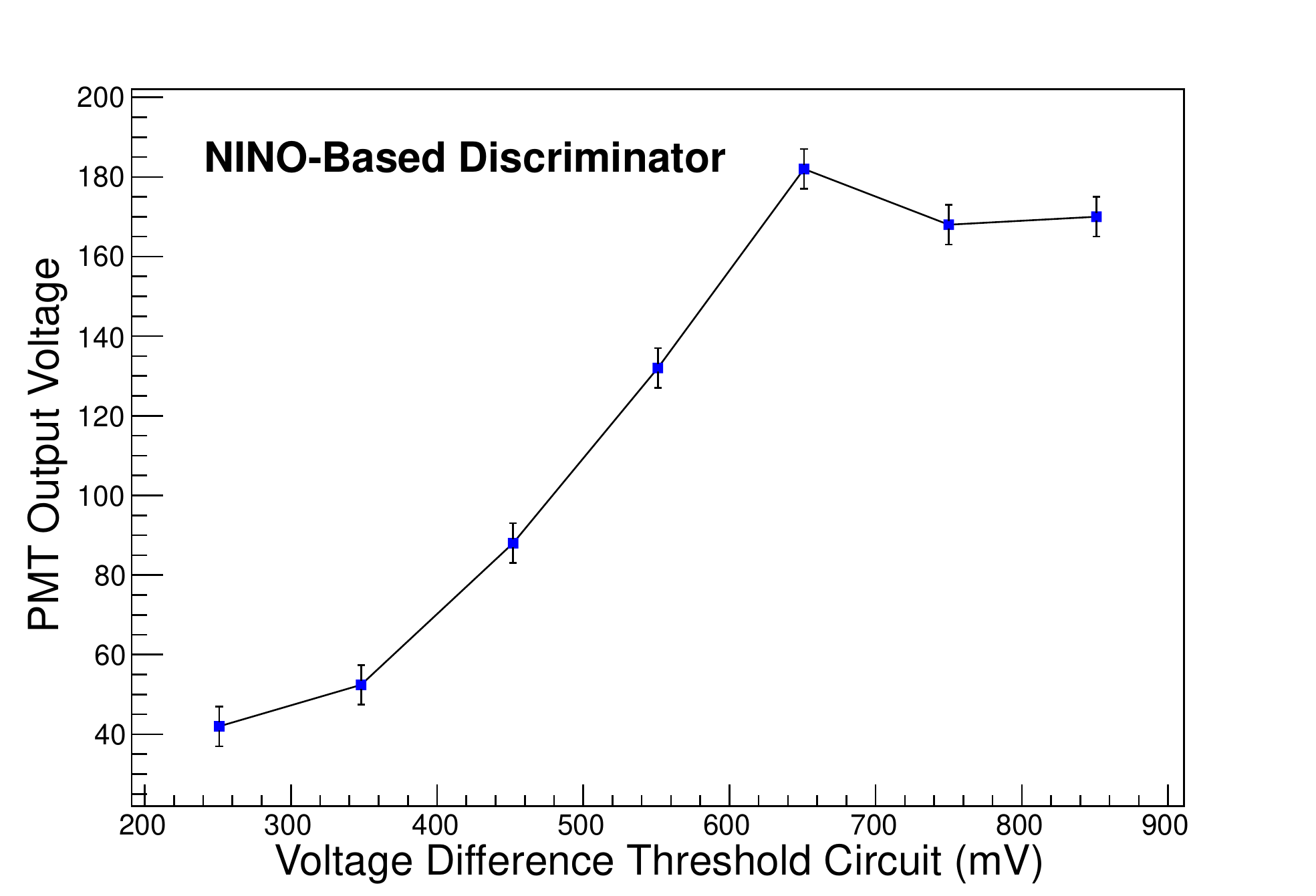}\end{center}

\caption{\label{fig:NINO-thresh}Relation of voltage difference on the threshold
circuit with PMT output voltage.}
\end{figure}

\subsubsection{High Luminosity polarized $^3$He target}
\label{sec:He3_A1n}

The most important requirement for the polarized $^3$He target for the Hall A $A_1^n$ experiment
is the ability to maintain high polarization despite being run at high luminosity.  Indeed, this 
requirement is shared by most of the other 12~GeV-era experiments that will use polarized 
$^3$He targets.  Another key feature is cost.  Here, it is important to consider both the cost of the target that
is being built specifically for $A_1^n$, as well as the polarized-target costs that will be incurred as the 12~GeV program progresses.  

The $A_1^n$ polarized $^3$He target will be based on alkali-hybrid
spin-exchange optical pumping~\cite{hybrid_cells}, a variant of
spin-exchange optical pumping that has resulted in significant
improvements in target performance~\cite{hybrid_targets}.  While the
design of the $A_1^n$ polarized target is still in progress, one
feature that is settled is that it will have ``convection-based gas
flow".  The polarized $^3$He targets are sealed glass cells, and they
have historically included two chambers, a pumping chamber in which
the $^3$He is polarized, and a target chamber through which the
electron beam passes.  With convection-based gas flow the transfer of
gas between the two chambers, proposed back in 2002 by
B.~Wojtsekhowski~\cite{loop-cell}, is much more rapid than was the case in earlier designs.  
The convection-based cells, the basic geometry of which is illustrated in Fig.~\ref{fig:convection}a, are discussed in the next section, and are described in detail in paper by P.A.M. Dolph {\it et al.}~\cite{dol11}.  

Another issue that has received considerable attention in formulating a conceptual design is whether the target cells should have one pumping chamber, as they have in the past, or two pumping chambers, as is likely to be the case for some of the highest-luminosity upcoming polarized $^3$He experiments.  With two pumping chambers, there would be a fairly large margin of safety in delivering the desired performance for $A_1^n$.  Two pumping chambers, however, would also require significantly more development than a single pumping chamber, and would thus be more expensive.  In what follows, we focus particularly on the question of the performance that can be expected when using a single-pumping-chamber design.  The discussion includes the results of simulated-beam tests with a prototype single-pumping-chamber cell.    We next describe calculations for the target's magnetic holding field. Up to this point, these calculations have been oriented more toward a two-pumping-chamber system, and will need to be extended to fully understand the implications of going to a single-pumping chamber system.  Collectively, the information presented here provides most of what is needed for a conceptual design, thus enabling work on a more detailed design to proceed.

\begin{figure}[th]
\begin{center}\includegraphics[width=1.0\columnwidth]{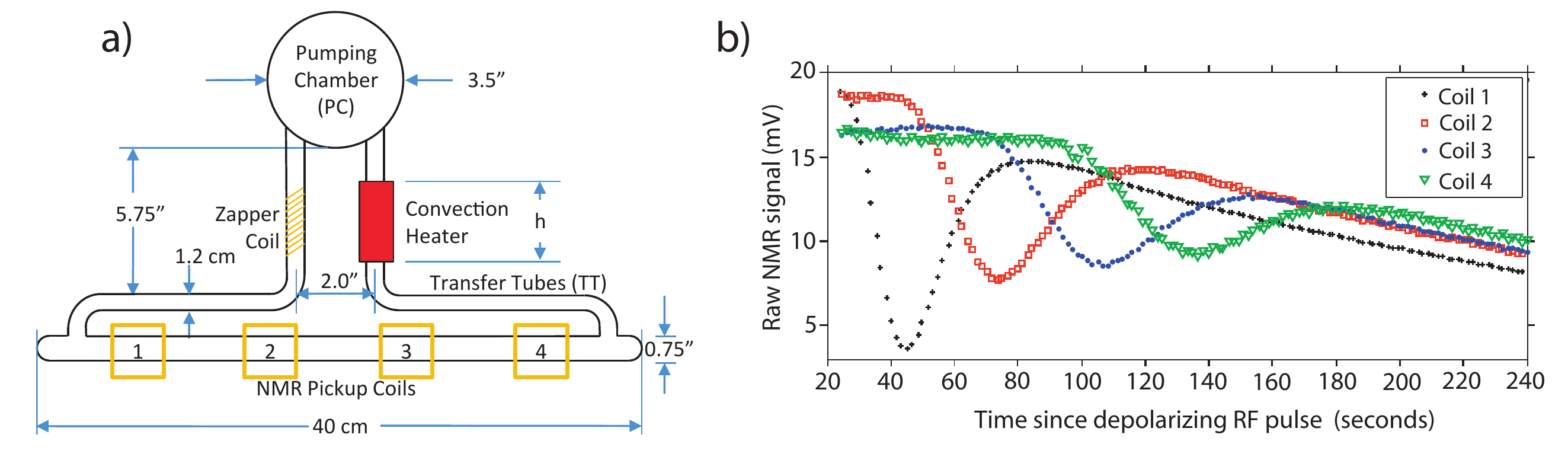}\end{center}
\caption{\label{fig:convection}Shown in a) is the basic design for convection-based target cells, together with the instrumentation used to visualize gas flow.  In b) NMR signals from the four pick-up coils are shown as a function of the time following the application of a depolarizing pulse of RF by the ``zapper coil".  The dip in the signal from each coil marks the passage of a depolarized slug of gas.}
\end{figure}
 
\paragraph{Convection-based target cells and polarization gradients}
\label{sec:Cell}

The polarized $^3$He targets that have been used at JLab thus far are comprised of two chambers: 1) a pumping chamber in which the spin-exchange optical pumping takes place and 2) a target chamber through which the electron beam passes.  Historically, these two chambers have been connected by a single ``transfer tube", and gas moved between the two chambers by diffusion. The polarization of the $^3$He in the target chamber, however, is determined by a balance between the rate at which it is being depolarized, and the rate at which it is being replenished with gas from the pumping chamber. At high luminosities, spin-relaxation due to the electron beam is quite significant, and diffusion between the two chambers becomes a bottleneck resulting in a significant ``polarization gradient" between the pumping and target chambers. This problem would become quite severe at the higher luminosities planned for future experiments if no provision were made to address the issue.

The ratio of the equilibrium polarization in the target chamber, $P_\mathrm{tc}$ to the equilibrium polarization in the pumping chamber, $P_\mathrm{pc}$, is given by~\cite{dol11}
\begin{equation}
{{P_\mathrm{tc}}\over{P_\mathrm{pc}}} = {{1}\over{1+\Gamma_\mathrm{tc}/d_\mathrm{tc}}}
\label{eq:pol_grad}
\end{equation}
where $\Gamma_\mathrm{tc}$ is the spin-relaxation rate specific to the target chamber and $d_\mathrm{tc}$ is the rate at which atoms are leaving the target chamber.  Clearly, it is important to keep the ratio $\Gamma_\mathrm{tc}/d_\mathrm{tc} \ll 1$, something that becomes more difficult as the beam current is increased.

Depolarization of polarized $^3$He in spin-exchange-based targets is well understood both experimentally~\cite{cou1989} and theoretically~\cite{bon1988}.  For our conditions, the formula for the beam-induced spin-relaxation rate of $^3$He nuclei in a cylindrical target cell of cross-sectional 
area $A_{tc}$ is well approximated by:
\begin{equation}
\Gamma_\mathrm{beam} = \left(5\times10^{-3}{{\rm cm^2}\over{\rm\mu A\,hr}}\right) I / A_{tc}\ \ ,
\label{eq:beam_depol}
\end{equation}
where $I$ is the current of the beam.  Multiple comparisons with experimental data from both SLAC and JLab have provided confidence that this formula can be trusted at the level of 10--20\%.  A good example involves a target cell we will refer to as Brady, that was used during the most recent set of polarized $^3$He experiments at JLab.  More specifically, Brady was to study both single~\cite{qia2011} and double-spin~\cite{huang2012} asymmetries in semi-inclusive deep inelastic scattering.  Because these experiments used a transversely polarized target, they are referred to collectively as the Transversity experiments.  Using dimensions and operational parameters discussed in \cite{hybrid_targets} and \cite{dol11}, we can use Eq.~\ref{eq:beam_depol} to compute the beam-induced relaxation rate in the target chamber ({\it not} averaged over the whole cell) to be $\rm\Gamma_\mathrm{beam} = 1/32\,hrs$ at $\rm 12\mu A$, which incidentally, is in good agreement with what was deduced from the experimental data.  Other contributions to relaxation in the target chamber during Transversity came from frequent AFP flips ($\rm\sim 1/ 60\,hrs$) and the cell's intrinsic relaxation rate ($\rm\sim 1/20\,hrs$).  Collectively, this resulted in a net relaxation rate in the target chamber, $\rm\Gamma_\mathrm{tc} \sim1/10\,hrs$.  For Brady, $d_\mathrm{tc} = 0.72\,hr^{-1}$, and using Eq.~\ref{eq:pol_grad}, we find $P_\mathrm{tc}/P_\mathrm{pc} = 0.88$.  For $A_1^n$, if the cell Brady were used, a beam current of $45\,\mu A$ would be required to achieve the proposed luminosity.  This would result in $\rm\Gamma_\mathrm{beam} = 1/8.6\,hrs$, and a net $\Gamma_\mathrm{tc} =\rm 1/6\,hrs$, and an expected ratio $P_\mathrm{tc}/P_\mathrm{pc} = 0.81$.  The average polarization in the target chamber during Transversity was 55.4\%~\cite{qia2011}.  Using this as a starting point, the polarization that one would expect at $45\,\mu A$ would be 48.4\%.  At the highest luminosities currently projected for approved experiments, the maximum in-beam polarization would be more like 38.4\%.

We have developed a new target-cell design that addresses the problem of polarization gradients.  The target cells still have two chambers as before, but rather than connecting these two chambers by a single ``transfer tube", the chambers are connected by two transfer tubes.  With a single transfer tube, gas moves between the two chambers by diffusion.  With two transfer tubes, one tube can be heated while leaving the other tube at room temperature, a condition that leads to convective flow.  In fact, we have demonstrated that we can vary the flow rate of gas moving through the target chamber from just a few cm/min up to around 80 cm/min in a very reproducible manner~\cite{dol11}.  The geometry of the cell is illustrated in Fig.~\ref{fig:convection}a. In Fig.~\ref{fig:convection}b, NMR signals are plotted as a function of time for the four pickup coils shown in Fig.~\ref{fig:convection}a.  The dips associated with the data from each coil are due to the passage of a slug of gas that was depolarized by the ``zapper coil" at time zero.

When using a convection-based cell, even a modest flow rate of $\rm6\,cm/s$ results in a value of $d_\mathrm{tc} =\rm 9\,hr^{-1}$ for a target chamber with a length of $\rm 40\,cm$.  If a convection-based cell were used under the conditions during which the target cell Brady was run, all other parameters being equal, we would expect the ratio $P_\mathrm{tc}/P_\mathrm{pc} > 0.98$ while in beam, and $P_\mathrm{tc}/P_\mathrm{pc} > 0.99$ with no beam.  Not only would this significantly increase the polarization of the gas in the target chamber, it would also have significant advantages for polarimetry, as will be discussed in the section on polarimetry.  This is because an absolute calibration of NMR signals in the pumping chamber could be easily translated into an absolute calibration of NMR signals in the target chamber since the polarization in the two chambers would be nearly the same.  Indeed, during calibrations,  the flow rate could be substantially increased.  At $\rm60\,cm/s$, for example, under calibration conditions, we would expect $P_\mathrm{tc}/P_\mathrm{pc} > 0.999$, which would greatly aid cross calibrations between the pumping and target chambers.  

\paragraph{High luminosity and target size}
\label{sec:Size}

Totally apart from the question of polarization gradients, there is also the question of how to maintain high cell-averaged polarization in the face of the depolarizing effects of the electron beam at high currents.  The solution to this issue is that it is necessary to polarize $^3$He at a rate that is rapid compared to the rate at which it is being depolarized.  Eq.~\ref{eq:beam_depol} gives the beam-induced spin-relaxation rate in the target chamber.  If instead we want to know the beam-induced spin-relaxation rate {\it averaged over the entire cell}, it can be written as
\begin{equation}
\rm\Gamma_\mathrm{beam}^\mathrm{av} = f_\mathrm{tc}\,\Gamma_\mathrm{beam} = f_\mathrm{tc}\, \left(5\times10^{-3}{{\rm cm^2}\over{\rm\mu A\,hr}}\right) I / A_{tc}  
\end{equation}
where $f_\mathrm{tc}$ is the fraction of $^3$He atoms in the target chamber.  If we want to make $\Gamma_\mathrm{beam}^\mathrm{av}$ small, we need to make $f_\mathrm{tc}$ small, which in turn means increasing the volume of the pumping chamber.  For the Transversity target cell Brady considered earlier, $f_\mathrm{tc} = 0.43$ under operating conditions, yielding  $\rm\Gamma_\mathrm{beam}^\mathrm{av} = 1/75\,hrs$.  For $A_1^n$, however, where the beam current would need to be $\rm45\,\mu A$ if we were to use Brady, $\rm\Gamma_\mathrm{beam} = 1/8.6\,hrs$, and $\rm\Gamma_\mathrm{beam}^\mathrm{av} = 1/20\,hrs$. To avoid having a lower cell-averaged polarization, we would need to use a pumping chamber with a larger volume.

In Fig.~\ref{fig:cell_options}a, we show a target-cell geometry with
two pumping chambers and a $\rm60\,cm$ long target chamber for which
we developed a relatively detailed oven design as well as
magnetic-field calculations.  Under operating conditions,
$f_\mathrm{tc} = 0.23$ for both this cell and the geometry shown in
Fig.~\ref{fig:cell_options}b.  
Using Transversity as a benchmark, these two double-pumping-chamber
geometries can be expected to deliver over 60\% polarization with a
$\rm60\,\mu A$ electron beam.  
Originally, we had hoped to use one of these two geometries for the
Hall A $A_1^n$ experiment, which was proposed to run with a beam
current of $\rm30\,\mu A$.  
This plan would provide a considerable margin of safety, and provide
valuable data so that the target design could be tweaked for
subsequent higher-luminosity experiments.  
In fact, however, again using Transversity as a benchmark, the
single-pumping-chamber geometry shown in Fig.~\ref{fig:cell_options}c,
which has a target-chamber length of $\rm40\,cm$,  should in principle
be nearly adequate for Hall A $A_1^n$ if the beam current is increased
to $\rm45\,\mu A$ (to compensate for the shorter target chamber).  
This statement is relatively consistent with an optimistic
interpretation of a simulated beam test we performed with Protovec-I,
a cell whose geometry is quite close to that of Fig.~\ref{fig:cell_options}c.  
As will be discussed shortly, a pessimistic interpretation of this same simulated beam test suggests that a target with the geometry of Fig.~\ref{fig:cell_options}c would achieve 50\% polarization, a performance level with an effective figure of merit around 2/3 of the proposal value.

\begin{figure}[th]
\begin{center}\includegraphics[width=0.9\columnwidth]{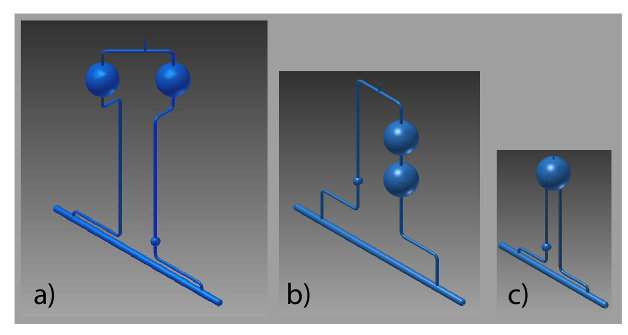}\end{center}
\caption{\label{fig:cell_options}Shown are 3D renderings of three cell geometries that have been considered.  In a) and b) there are two pumping chambers and the target chambers are $\rm60\,cm$ in length.  In c) there is one pumping chamber and the target chamber is $\rm40\,cm$ in length.  The geometry shown in c) is nearly identical to Protovec-I, bench tests of which are described in the text.}
\end{figure}

We describe next how to estimate the expected performance of the geometry shown in Fig.~\ref{fig:cell_options}c using Transversity as a benchmark.  The cell-averaged equilibrium polarization, $P_\mathrm{He}^\mathrm{eq}$,  that any given target can achieve, can be written:
\begin{equation}
P_\mathrm{He}^\mathrm{eq} = \langle P_\mathrm{alk} \rangle {{\langle \gamma_\mathrm{se}\rangle}\over{\langle \gamma_\mathrm{se}\rangle(1+X) + \Gamma_\mathrm{nx}}}
\end{equation}
where $\langle P_\mathrm{alk} \rangle$ is the average alkali polarization in the pumping chamber, $\langle \gamma_\mathrm{se}\rangle$ is the spin-exchange rate averaged over the cell, $X$ quantifies a spin-relaxation rate that is roughly proportional to $\langle \gamma_\mathrm{se}\rangle$, and $\Gamma_\mathrm{nx}$ is the spin relaxation due to everything other than spin exchange and the relaxation rate described by the parameter $X$.  The ultimate polarization of a particular target is complicated by the fact that the parameter $X$ varies from cell to cell in a manner that is not understood, as is also the case with the wall-induced spin relaxation. We can proceed, however, by assuming that both $X$ and the wall relaxation are identical to the case of Brady for which $\rm\Gamma_\mathrm{nx} \sim 1/20\,hrs + 1/75\,hrs + 1/60\,hrs = 1/12.5\,hrs$, where the first term was due to the ``intrinsic" relaxation rate (a combination of wall relaxation and bulk relaxation due to $^3$He-$^3$He collisions), the second term was due to the beam and the third term was due to the frequency of AFP spin flips that were used during Transversity.  During $A_1^n$, it is reasonable to expect that the effective relaxation rate due to AFP measurements will be negligible.  If we demand for some new target cell that $\rm\Gamma_\mathrm{nx} = 1/12.5\,hrs$, as was the case for Brady during Transversity, and further assume a beam current of $\rm45\,\mu A$, we are essentially demanding that 
\begin{equation}
\rm\Gamma_\mathrm{nx} = 1/12.5\,hrs = 1/20\,hrs + f_\mathrm{tc}(1/8.6\,hrs)
\end{equation}
which is satisfied if $f_\mathrm{tc}= 0.26$.  This value of $f_\mathrm{tc}$ is quite close to the value of 0.29 associated with the geometry of Fig.~\ref{fig:cell_options}c.  Indeed, we can probably adjust the design shown in Fig.~\ref{fig:cell_options}c to make up the small difference.  

The preceding argument suggests that the single-pumping-chamber geometry shown in Fig.~\ref{fig:cell_options}c, which is extremely close to the geometry of the test cell Protovec-I,  is cable of delivering 60\% polarization at full luminosity.  The argument is based on an extrapolation based on the proven performance achieved during Transversity.  With that said, the argument also leaves no margin of safety, something that is of particular concern given that the $A_1^n$ target will be breaking new ground in several ways.  We describe in the next section bench tests of Protovec-I which suggest that a target cell with the geometry shown in Fig.~\ref{fig:cell_options}c is capable of achieving at least 50\% polarization at full luminosity, and perhaps even higher polarization.  

We return briefly to the cell geometries shown in Fig.~\ref{fig:cell_options}a and b, both of which (due to their size and an extrapolation from Transversity such as that described above), can be expected to deliver polarization in excess of 60\% at luminosities twice that of the Hall A  $A_1^n$ experiment .  We have developed in some detail certain aspects of a target-system design that could accommodate the geometry shown in Fig.~\ref{fig:cell_options}a.  These studies include magnetic-field calculations that are presented in their own section.  More recently we have considered the advantages of the geometry shown in Fig.~\ref{fig:cell_options}b, which due to the vertical stacking of the pumping chambers, can be pumped from almost any direction in the horizontal plane.  For this reason,  a vertically stacked geometry is practical for at least four of the approved polarized $^3$He experiments.  Also, a vertically stacked geometry might reduce the magnetic field inhomogeneities experienced by the target for several of the magnet systems under consideration.  The great success of the Hall A polarized $^3$He program has been due in part to the fact that the target designs, on paper, have always included a conservative margin of safety.  The cell geometry shown in Fig.~\ref{fig:cell_options}b would provide such a margin of safety for $A_1^n$, and would also be suitable for multiple polarized $^3$He experiments.  All other things being equal, we view the vertically-stacked geometry of Fig.~\ref{fig:cell_options}b as our preferred choice, and believe that it would be worthwhile to explore the question of whether pursuing its development sooner rather than later might ultimately prove to be the wisest choice, both in terms of cost and the success of the program.

\paragraph{Tests of Protovec-I}
\label{sec:protovec_tests}

We have constructed two prototype target cells with geometries nearly identical to that shown in Fig.~\ref{fig:cell_options}c and will refer to them as Protovec-I and Protovec-II.  For Protovec-I, pictured in Fig.~\ref{fig:protovec_i}b, we have performed polarization tests in which we simulated the spin relaxation due to the electron beam by subjecting the cell to periodic pulses of RF.  These RF pulses were applied to a small spherical volume that was situated along one of the two transfer tubes.  The RF pulses tipped the nuclear spins of the $^3$He contained within that small volume so that they were at an angle of 90$^{\circ}$ with respect to the magnetic holding field, causing them to precess and decay away.  The same coil used to apply the pulses of RF was also used to detect the precessing spins, thus providing a measure of the polarization.  The polarization as a function of time is shown in Fig.~\ref{fig:protovec_i}a, and is seen to saturate at 49\%.  We note further that the convection heater was set during these measurements to produce a gas velocity in the target chamber of $\rm6\,cm/min$, which is the gas velocity at which we plan to run.

\begin{figure}[th]
\begin{center}\includegraphics[width=0.95\columnwidth]{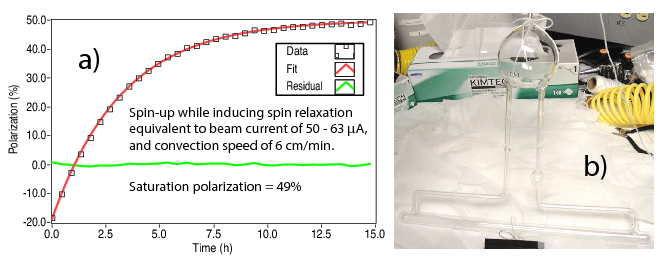}\end{center}
\caption{\label{fig:protovec_i} In a) we show a ``spin-up",  polarization as a function of time while a target is being polarized. In b) we show a photograph of the cell being tested, Protovec-I.  The polarization measurements shown in a) utilized pulse NMR, and the frequency of the RF pulses was chosen to simulate the depolarization that would result from a beam with a current in the range of $\rm50-63\,\mu A$.}
\end{figure}

To the extent that we understand the polarization losses associated with the RF pulses, the data shown in Fig.~\ref{fig:protovec_i}a provide a measure of the expected performance of Protovec-I under running conditions.  The uncertainty in the induced polarization losses are such that we believe we were simulating a beam current in the range of $\rm50-63\,\mu A$.  This was the first such test we have performed, and we expect to greatly reduce the uncertainties in the induced polarization losses in the future.  We view these tests as a valuable tool for evaluating the expected performance of individual target cells under running conditions.  For the test shown in Fig.~\ref{fig:protovec_i}a, a pessimistic interpretation suggests that Protovec-I is capable of achieving a polarization of 49\% with roughly the beam current  required to achieve full luminosity ($\rm45\,\mu A$). 

The simulated beam test shown in Fig.~\ref{fig:protovec_i}a is quite encouraging.  The test was performed using only three lasers, whereas we expect to use four or five lasers when actually running.  Furthermore, the measured properties of Protovec-I were not very good when compared with our better target cells.  Brady, for example, was initially measured at UVa as having a spin-relaxation rate of less than $\rm1/30\,hrs$, compared to just under $\rm1/20\,hrs$ for Protovec-I.  Also, the value of the $X$ parameter, while only measured roughly for Protovec-I, appeared significantly larger than that measured for Brady.  In short, there are good reasons to expect that future simulated beam tests with new target cells will establish performance levels that exceed those achieved with Protovec-I.  Even with no improvements, however, our tests suggest that a target such as Protovec-I would deliver an effective (polarization adjusted) luminosity of at least 2/3 of that which appears in the Hall A $A_1^n$ proposal.

\paragraph{Magnetic Field}
\label{sec:field}

The polarized $^3$He target uses a magnetic holding field that is typically on the order of $\rm20\,G$. The current plan for the source of this magnetic field is to use the existing coils that were used for the Transversity experiment, although the question of whether these coils are adequate for the larger target cells that will be used for  $A_1^n$ is not yet definitively answered.  One design constraint is that magnetic field inhomogeneities need to be held to an acceptable level, something that is more challenging with a larger target volume.  A second design issue is minimizing the deviation of the holding field direction from that which is desired, and in particular, controlling the vertical component of the magnetic field due to the fringe field of the BigBite dipole.  

We present below the results of calculations performed using the program Tosca for two configurations, both of which are illustrated in Fig.~\ref{fig:helmholtz_configs}.  In both cases we assume that the field is produced using the two largest coils that were used during Transversity, and that both coils are oriented to produce a field whose direction is in the horizontal  plane.  In fact, the largest coil used during Transversity was used to produce field in the vertical direction, and orienting that coil to produce field in the horizontal direction would require new mounting hardware.  At the time of this writing, the target system design being explored for $A_1^n$ would {\it not} use the largest of the three Transversity coils.  For this and other reasons, the field studies presented here cannot be considered as being final.  Nonetheless, they provide considerable insight as we proceed toward a final design.

In the first of the two configurations presented, shown in Fig.~\ref{fig:helmholtz_configs}a, a single field clamp is used.  In the second configuration presented, shown in Fig.~\ref{fig:helmholtz_configs}b, two symmetric field clamps are used.  Also, in this second configuration, two coils are wrapped around each of the two field clamps in order to energize them and produce a small vertical field to offset that due to the BigBite dipole.
\begin{figure}[th]
\begin{center}\includegraphics[width=0.85\columnwidth]{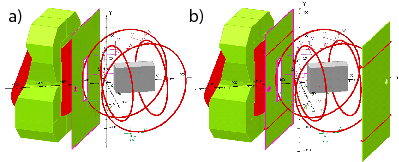}\end{center}
\caption{\label{fig:helmholtz_configs} Shown are two configurations whose magnetic field properties have been studied.  In a), a single field clamp is used.  In b), two symmetric field clamps are used, each of which is energized with two coils in order to produce a small vertical field to offset that due to the BigBite dipole.}
\end{figure}

One of the design criteria for the magnetic holding field is the longitudinal spin-relaxation rate, $1/T_1$, that is caused by magnetic field inhomogeneities. Since the magnetic field can have three components, and each component can have a non-zero derivative in three directions, the magnetic field inhomogeneities are described by a nine-component tensor.  
The longitudinal spin relaxation caused by magnetic field inhomogeneities under static conditions (as opposed to magnetic resonance conditions) is given by:
\begin{equation}
{1\over{T_1}} = {{|\vec\nabla B_x|^2 + |\vec\nabla B_y|^2}\over{B_z^2}}D
\label{eq:long_t1}
\end{equation}
where $\vec B$ is the static holding field, which is presumed here to be in the $z$ direction, and $D$ is the diffusion coefficient of the $^3$He~\cite{cat1988}.  For the configuration shown in Fig.~\ref{fig:helmholtz_configs}a (single field clamp), when the field is along the beamline (the $z$ direction), we have used the program Tosca to compute the resulting magnetic field and show the magnitude of $B_z$ as a function of $z$ in Fig.~\ref{fig:z_direction}a. Each point corresponds to values of $x$ and $y$ that are (for appropriate values of $z$) within one of the pumping chambers of the double-pumping-chamber cell geometry shown in Fig.~\ref{fig:cell_options}a.  Also shown, in Fig.~\ref{fig:z_direction}b, is the quantity $(|\vec\nabla B_x|^2 + |\vec\nabla B_y|^2)^{1/2}$, the square of which appears in the numerator in Eq.~\ref{eq:long_t1}.
\begin{figure}[th]
\begin{center}\includegraphics[width=0.75\columnwidth]{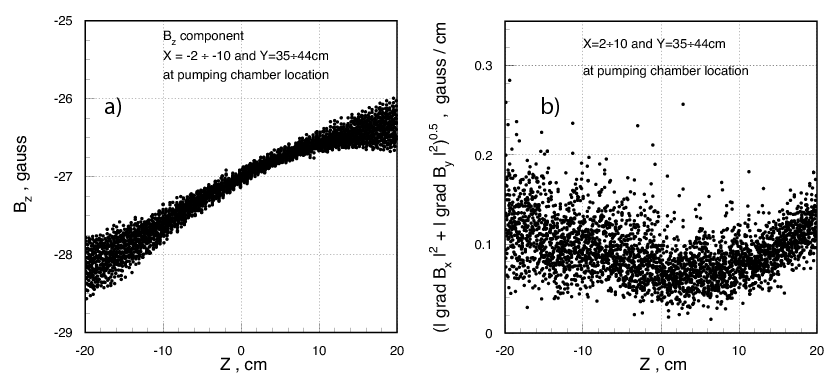}\end{center}
\caption{\label{fig:z_direction} For the vicinity of one of the pumping chambers of the cell depicted in Fig.~\ref{fig:cell_options}a, and the single-field-clamp magnetic field configuration, we plot, as a function of $z$, the magnitude of the $z$ component of static holding field (a) and the indicated magnetic field inhomogeneities (b).  The field direction is nominally along the beam-line direction, which is the $z$ axis.}
\end{figure}
In round numbers, we see that the inhomogeneities are of the order of $\rm0.1\,G/cm$.  For a 10 amagat sample of $^3$He, the diffusion constant $D\rm\sim 0.2\,cm^2/s$.  This would result in $1/T_1 \sim\rm 1/56\,hrs$, which in principle is tolerable, but would definitely cause some limitation to the polarization that would be achievable. 

Another design criterion when considering the magnetic holding field are the polarization losses caused by magnetic field inhomogeneities during the NMR technique of adiabatic fast passage, or AFP, that we use for polarimetry.  In this case, the losses are caused largely by $1/T_{1\rho}$, the longitudinal spin-relaxation rate in a frame that is rotating about the $z$ axis at the Larmor frequency.  On resonance, this quantity is given by~\cite{cat1988b}
\begin{equation}
{1\over{T_{1\rho}}} = {{|\vec\nabla B_z|^2}\over{B_1^2}}D
\label{eq:t1rho}
\end{equation}
where $B_1$ is the magnitude of the RF field, which is typically significantly smaller than the static holding field.  The magnitude of the quantity ${|\vec\nabla B_z|}$ for the configuration shown in Fig.~\ref{fig:helmholtz_configs}a is shown in Fig.~\ref{fig:t1rho} for a collection of points, again located inside one of the pumping chambers of the cell shown in Fig.~\ref{fig:cell_options}a. It can be seen that the numbers span a considerable range, with an average that is, perhaps, somewhere around $\rm0.075\,G/cm$.  Using $B_1=\rm0.05\,G$, a fairly typical value, we find the relaxation rate $1/T_{1\rho} =\rm 0.45\,s^{-1}$, much faster than before, but it only occurs during magnetic resonance conditions.  Using the results of ref.~\cite{cat1988b}, it can be shown that the fractional relaxation that occurs during AFP is well approximated by
\begin{equation}
\mathrm{fractional\ relaxation} = {1\over{T_{1\rho}}}\,{{\pi\,B_1}\over{2(\partial B_z/\partial t)}}
\label{eq:afp_loss}
\end{equation}
where $\partial B_z/\partial t$ is the sweep rate of the magnetic holding field.  Using $\partial B_z/\partial t=1.2\,\rm G/s$, we find a fractional loss of 0.029, or about 3\%.  In contrast, the same calculation for the case of $|\vec\nabla B_z|=20\,\rm mG/cm$ gives a loss of 0.21\% (or 0.42\% per polarization measurement), and for $|\vec\nabla B_z|=10\,\rm mG/cm$, a loss of 0.05\%.  In general, with the Hall A polarized $^3$He target, efforts have been made to keep inhomogeneities in the range of $\rm10-20\,mG$, and the losses per polarization measurement (which involves two AFP spin flips), have generally been 0.5\% or less,  consistent with the predictions from Eq.~\ref{eq:afp_loss}.

If the AFP losses are truly 3\% per spin flip, or 6\% per polarization measurement, AFP cannot be used as the primary on-line monitor of polarization.  It should still be possible to use pulse NMR polarimetry, however, by arranging to have the portion of the target cell being probed in a ``sweet spot" with low field inhomogeneities.

\begin{figure}[th]
\begin{center}\includegraphics[width=0.5\columnwidth]{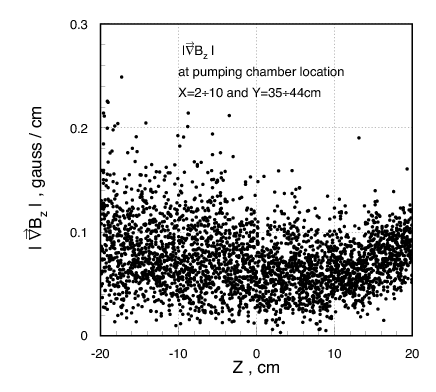}\end{center}
\caption{\label{fig:t1rho} For the configuration shown in Fig.~\ref{fig:helmholtz_configs}a, with a single field clamp, we plot the magnitude of $\vec\nabla B_z$ in the vicinity of one of the pumping chambers of the cell depicted in Fig.~\ref{fig:cell_options}a.  The quantity $\vec\nabla B_z$ is a determining factor in the polarization losses that occur during the NMR technique of AFP.  The inhomogeneities are seen to be significantly larger than the range of $\rm10-20\,mG/cm$ that has typically been the goal when operating the Hall A polarized $^3$He target system.}
\end{figure}

As mentioned earlier, another design criterion that needs to be considered is the {\it direction} of the magnetic field.  For $A_1^n$ this is not quite as critical as for some measurements (such as $G_E^n$), but it is still important.  The fringe fields from the BigBite dipole are large, and in particular, there tends to be a significant unwanted vertical component in the field.  Even when using a field clamp, as is the case in the configuration of Fig.~\ref{fig:helmholtz_configs}a, there is still a significant vertical component, as is shown in Fig.~\ref{fig:vertical_comp}a.  By using two symmetrical field clamps, and energizing each with two coils (the configuration shown in Fig.~\ref{fig:helmholtz_configs}b), the vertical component can be greatly reduced, as is shown in Fig.~\ref{fig:vertical_comp}b.  At the time of this writing, we are still studying the two-field-clamp configuration, but it certainly seems to have some advantages. 

\begin{figure}[th]
\begin{center}\includegraphics[width=0.75\columnwidth]{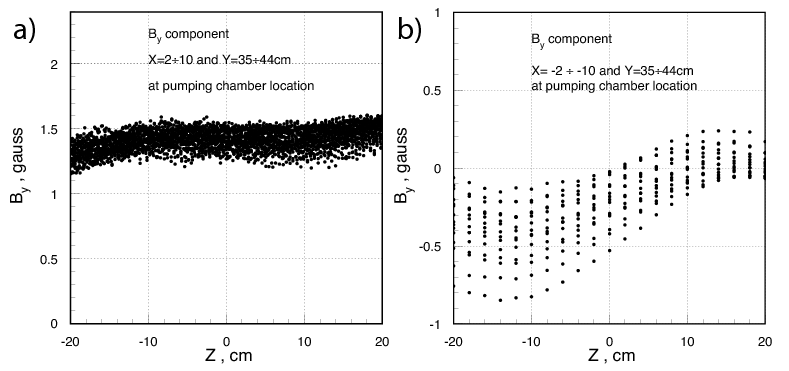}\end{center}
\caption{\label{fig:vertical_comp} Shown is the vertical magnetic-field component, $B_y$, for points in the vicinity of one of the pumping chambers of the cell depicted in Fig.~\ref{fig:cell_options}a.  The plots a) and b) correspond to the single- and double-field-clamp configurations, respectively, depicted in Figs.~\ref{fig:helmholtz_configs}a and b.  The nominal magnetic-field direction is along the $z$ axis, which is parallel to the beam line.}
\end{figure}

The magnetic-field calculations presented above focused on locations that are relevant for a two-pumping-chamber design, and assumed that the magnetic field was being produced using the largest two of the three coils used during Transversity.  In fact, however, the configuration that would be easiest and cheapest to implement would use the two smallest of the three coils used during Transversity.  Using smaller coils almost certainly means larger gradients at equivalent locations.  On the other hand, a single-pumping-chamber geometry allows for a more symmetric placement of the target cell, which may well mean that the target cell would experience smaller gradients, even when using the smaller coils.  This question can only be answered by performing additional Tosca calculations.

In summary, the calculations presented above seem to suggest that a magnetic field generated using coils from the Transversity setup will suffice for $A_1^n$, although more calculations are needed to confirm this statement.  The calculations also seem to indicate that we are likely to have gradients a bit larger than would be ideal.  It is thus essential that we complete our calculations using Tosca as soon as possible, since the Hall A designers are already beginning to work on a detailed design for the target system.  In parallel, we are also undertaking additional studies, both theoretical and experimental, on the effects of inhomogeneities on polarization losses while performing AFP.  These studies are already underway, and will better define the levels of inhomogeneities that are tolerable.

\paragraph{Polarimetry}
\label{sec:polarimetry}

The online monitoring of polarization for the $A_1^n$ polarized $^3$He target will be accomplished using one or both of two NMR techniques.  Historically, in the Hall A polarized $^3$He target system, the NMR technique of AFP has been used, and provides very precise and reproducible measurements.  One drawback of AFP is that it requires reversing the orientation of {\it all} of the spins within the target, something that is problematic if some portions of the target are made out of metal (which blocks the RF).  This would be the case, for instance, if we used metal end windows on the target chamber, something that we are actively developing.  AFP would also be problematic if large parts of the target experienced excessive magnetic field inhomogeneities.  The second polarimetry technique being considered is based on pulse NMR.  In this technique the spins within a small region of the target cell are tipped from the direction of the holding field, and subsequently precess, resulting in what is often called a free induction decay, or FID.  We routinely use pulse NMR for polarimetry at UVa, and it was this technique that was used to measure polarization during the simulated beam test that is presented in Fig.~\ref{fig:protovec_i}.  Pulse NMR polarimetry has the advantage that it can be performed on a small well-localized region of the cell, which can be chosen to be well away from any parts of the cell that are made of metal, an in a region of the magnetic field that is sufficiently homogeneous.  Pulse NMR polarimetry has the disadvantage, however, that we currently only trust it at the 5--10\% level, which is not yet sufficient for the $A_1^n$ experiment.  There are no obvious obstacles, however, to refining pulse NMR polarimetry so that it can be trusted at the level of several percent.   

In discussing polarimetry, it is important to distinguish between the measurements that will monitor the target polarization on an ongoing basis, and the measurements that will calibrate those measurements.  One technique for determining the absolute $^3$He polarization is the observation of polarization-induced frequency shifts in the electron paramagnetic resonances associated with the alkali vapors~\cite{rom1998}.  This can only be performed in the pumping chamber, however, even though the quantity of interest is the polarization in the target chamber.  It is thus important to understand the ratio of the polarization in the target chamber to the polarization in the pumping chamber, $P_\mathrm{tc}/P_\mathrm{pc}$, which is given by Eq.~\ref{eq:pol_grad}.  In the past, the uncertainty in $P_\mathrm{tc}/P_\mathrm{pc}$ has led to a 1-2\% contribution to the relative uncertainty in polarimetry.   In convection-based targets, however, this uncertainty can be reduced to a fraction of a percent, well below the level of most other sources of error in the polarimetry system.

To better understand the source of polarimetry errors associated with polarization gradients, we consider again the cell Brady used during the Transversity experiment.  To compute $P_\mathrm{tc}/P_\mathrm{pc}$ using Eq.~\ref{eq:pol_grad}, we need to know both $d_\mathrm{tc}$, which for Brady was $\rm (0.72 \pm 0.10) hr^{-1}$ and $\Gamma_\mathrm{tc}$. Ideally, the calibration of the NMR system would be done in the absence of beam, which would also typically mean the absence of the rapid polarization reversals that were used during Transversity running.  The room-temperature spin-relaxation rate for Brady was measured at JLab to be $\rm 1/20\,hr$, and if we take this as the value for $\Gamma_\mathrm{tc}$, we find $P_\mathrm{tc}/P_\mathrm{pc} = 0.935$. The quantity $\Gamma_\mathrm{tc}$ is not measured directly, however.  It is really only the {\it cell-averaged} relaxation rate that is known with accuracy.   The uncertainty in $\Gamma_\mathrm{tc}$ thus leads to something like a 1--2\% uncertainty in $P_\mathrm{tc}/P_\mathrm{pc}$, and hence in the target polarization.

The situation is dramatically different in a convection based cell.  We are planning to run our convection-based cells at a gas velocity of $\rm6\,cm/hr$, which as discussed earlier, results in a value of $d_\mathrm{tc} =\rm 9\,hr^{-1}$ for a target chamber with a length of $\rm 40\,cm$.  If we again take $\Gamma_\mathrm{tc} =\rm 1/20\,hr$, we have $P_\mathrm{tc}^\infty/P_\mathrm{pc}^\infty = 0.994$.  Even large changes in the assumed value of $\Gamma_\mathrm{tc}$ would change this number by at most a few tenths of a percent.  Furthermore, when actually performing a calibration, faster gas velocities would be used, resulting in uncertainties of well under $0.1\%$.

The above discussion has important implications for the strategy we will employ for polarimetry.  Historically, AFP signals in the target chamber have been compared to AFP signals from thermally polarized water contained in a glass cell with a geometry that is similar, but not identical, to that of the target cells.  To compare the water signals to the $^3$He signals, one must also take into account geometric factors, such as the slightly different size of the two chambers, and their placement in the NMR pickup coils.  One of the most thorough such comparisons was done by Michael Romalis~\cite{rom97}, during the experiment E154 at SLAC~\cite{abe97}.  The error in this comparison was quoted as being 1.6\%.  Even if one did twice this well, it would still dwarf the error with which the quantity $P_\mathrm{tc}/P_\mathrm{pc}$ is known.  Thus, during $A_1^n$, there is no advantage in obtaining water signals from the location of the target chamber, and this will not be attempted.  Calibration of $^3$He signals with thermally polarized water will still be made, either directly or indirectly, but these comparisons will be obtained under more optimal conditions.

\paragraph{Mechanical aspects of the $A_1^n$ target}
\label{sec:Mechanics}

At the time of this writing, the design of the mechanical aspects of the Hall A $A_1^n$ target system is only just beginning.  The strategy that is currently being explored involves using as much of the hardware developed for Transversity as possible.  The oven, which will accommodate a convection-based cell, will necessarily be different, as will the placement of the coils, which must accommodate the planned kinematics.  Many other aspects of the hardware, however, will be be either partially or fully reused.  Examples include the support and drive mechanisms for the oven/target ladder, the laser and optics, the support structure for the Helmholtz coils, and many aspects of the polarimetry system. The primary reason for reusing as much as possible the Transversity hardware is to keep costs down.

A considerable amount of effort has gone into developing a conceptual design for the $A_1^n$ target that goes beyond what might best be described as an adaptation of the Transversity target.  The primary reason for this is to accommodate a double-pumping-chamber cell of the sort shown in Figs.~\ref{fig:cell_options}a and b.  The cell pictured in Fig.~\ref{fig:cell_options}b, with vertically stacked pumping chambers, has the distinct advantage that it can be pumped from almost any direction that falls in the horizontal plane, making it practical for at least four of the planned polarized $^3$He experiments.  It could even be fitted with an oven that could rotate about a vertical axis, thus obviating the need to build a custom oven for each experiment.  We have spent some time considering the advantages of mounting the cell and oven assembly on a frame that would be supported from the bottom, thus making weight less of an issue.  There is little doubt that such a system would be more expensive than our current strategy in the short term.  When considering all the experiments that such a system could serve, however, it may actually be cheaper than adapting the existing Transversity hardware. 

Regardless of whether we use a single or double pumping chamber, and regardless of many other details of the target system, an important innovation in the $A_1^n$ polarized $^3$He target will be the inclusion of radiation shielding between the pumping chamber and the target chamber.  When target cells have exploded during running, it is usually after prolonged exposure to the beam, and a reasonable assumption is that the target has suffered weakening due to radiation damage.  It is usually the pumping chamber that ruptures, and this is presumably because the stresses in the relatively large pumping chamber are larger than those associated with the smaller-diameter target chamber.  It is thus reasonable to expect that providing radiation shielding for the pumping chamber would make it possible to substantially increase the integrated luminosity over which the targets can be used.  Radiation shielding is not practical for non-convecting target cells because the pumping and target chambers must be kept relatively close to one another.  In a convection-driven cell, however, the distance between the pumping and target chambers can be quite large, making it practical for the first time to shield the pumping chamber.  

An important question that is currently open regarding mechanical aspects of the target is whether there will be metal end windows on the target chamber.  This is particularly relevant because the current design being explored would utilize a single-pumping-chamber cell with a target chamber that is $\rm40\,cm$ in length instead of $\rm60\,cm$, as in the original proposal.  With the shorter target cell, full luminosity would only be achieved by using a $\rm45\mu A$ beam instead of the proposed value of $\rm30\mu A$.  While we are fairly confident that glass end windows will tolerate a $\rm30\mu A$ beam (this has actually been briefly tested), it is less clear that glass end windows will tolerate a $\rm45\mu A$ beam.  We are currently developing a design for beryllium end windows, but the approach has not yet been fully validated.  Thus, achieving full luminosity with a $\rm40\,cm$ long target chamber may be contingent on the success of our metal-end-window development work.

\paragraph{Target Design Summary}
\label{sec:target_summary}

It is useful to list a few selected conclusions from our conceptual design for the Hall A $A_1^n$ target.
\begin{itemize} 
\item The use of a convection-style target cell will be essential to achieving high polarization.
\item A double-pumping chamber target cell can be expected to comfortably achieve the performance assumed in the proposal.
\item A single-pumping-chamber cell may not achieve the desired polarization of over 60\% at the proposed high luminosity.  A polarization over 50\% seems fairly likely.
\item A single-pumping chamber cell can be mechanically accommodated by modifying the existing Transversity target hardware.
\item Achieving full luminosity with a target cell that has a single pumping chamber and a $\rm40\,cm$ target chamber may be contingent on the successful development of suitable metal end windows.
\item The magnetic-field calculations that have been performed thus far show magnetic field inhomogeneities that would make online AFP polarimetry impossible.  The configurations considered, however, do not correspond to the design currently being considered, so more calculations are needed.  
\item Pulse NMR polarimetry may be necessary if either 1) parts of the target cell are made of metal, or 2) is the magnetic field inhomogeneities are too large for performing AFP frequently.
\item The absolute calibration of the polarimetry system can take advantage of the fact that the polarizations in the pumping and  target chambers are nearly equal in a convection-driven cell.  One consequence of this is that there is no advantage to studying signals from thermally polarized water at the location of the target chamber.

 \end{itemize}
 
From the bullet points listed above, we conclude that a credible path forward exists that would make maximal use of existing hardware from Transversity, and that would achieve a figure of merit at least 2/3 of what was assumed in our proposal.  Risks associated with this plan include some uncertainty as to whether  suitable metal end windows can be constructed, and the possible need to use pulse NMR polarimetry for online monitoring of the polarization.  The target assumed in the original proposal, which is larger and would probably have two pumping chambers, would almost certainly deliver the full figure of merit assumed in the proposal.  It would share both of the risks mentioned above, and it would be more expensive.  Such a target, however, might need only limited modification to be used for at least three experiments other than the Hall A $A_1^n$ experiment.

\subsubsection{Summary}
\label{sec:a1n:Summary}
The A1n collaboration is working in close contact with the SBS
collaboration, which will use the BigBite for several experiments.

The new timing hodoscope for BigBite consists of 90 25$\times25\times600$~mm
bars of plastic scintillator. It is currently under construction.
Tests of an element of the hodoscope, in conjunction with the NINO
card, have been made using cosmic rays. The mean pulse height and
mean time from readout of the 2 PMT attached to a hodoscope bar is
close to position independent. The time difference of the PMTs gives
a horizontal position resolution of $\sim4.5$~cm. The vertical position
resolution is set by the 25~mm lateral dimension of the bar. The
time resolution is $\sim0.4$~ns. This is possibly limited by the
type ELT9125 PMTs which will be used for the A1n experiment. For an
inclusive $(e,e')$ measurement such as A1n, this level of time resolution
is entirely adequate. However for a coincidence $(e,e'n)$ measurement,
such as in the $G_{M}^{n}$ experiment, improved resolution may be
desirable. The hodoscope bars will also be tested with alternative
PMTs with faster rise time. We also envisage testing of the NINO card
and hodoscope elements under high intensity conditions at Mainz.

The NINO chip gives good timing performance with photomultiplier signals.
Its time over threshold logic signal is an effective alternative to
a pulse amplitude measurement in correcting for time walk in the discriminator.
The tests described here relate to a scintillator hodoscope, but the
NINO card will also be suitable for the BigBite gas Cerenkov counter
GRINCH.

\clearpage \newpage

\subsection{E12-07-108 - $G_M^p$}
\label{sec:e12-07-108}

\begin{center} 
\bf Precision Measurement of the Proton Elastic Cross Section at High $Q^2$
\end{center}

\begin{center}
J. Arrington, M. Christy, S. Gilad, B. Moffit, V. Sulkosky and B. Wojtsekhowski, spokespersons, \\
and \\
the Hall A collaboration.  \\
contributed by V.~Sulkosky and B.~Wojtsekhowski, for the GMp collaboration.
\end{center}

The GMp experiment was proposed to measure the elastic electron-proton cross-section 
from 7 to 17~GeV$^{2}$ in $Q^2$ with an unprecedented precision of less than 
2\%.  The experiment has two main goals.  The first goal is a precise measurement 
of the proton elastic cross-section at the kinematics of the upgraded JLab 12~GeV  
facility for accurate normalization of other experiments.  The second objective is   
to determine with improved precision the proton magnetic form-factor $G_{M}^{p}$ at 
high $Q^2$.  The JLab PAC recommended a beam time assignment of 24 days by 
asking the experimentalists not to pursue measurements at the two highest Q$^2$ points, 
which limits the highest planned $Q^2$ to 14~GeV$^{2}$.

The GMp and DVCS~\cite{DVCS_12GeV} experiments are scheduled to be the first two experiments 
to run after the 12~GeV upgrade in Hall A.  The tentative plan is to concurrently run the 
two experiments to the greatest extent possible.  The GMp collaboration is working in close 
contact with the DVCS collaboration, which will use the left high resolution 
spectrometer (HRS).  Since the approval of the proposal, the collaboration has 
made a significant step in the instrumentation preparation which is briefly presented 
below.


\subsubsection{Electron beam quality requirements}
\label{sec:Beam_GMp}
To achieve the goals outlined above, the GMp experiment requires the highest beam 
energy possible after the upgrade, $\sim$ 11~GeV into Hall A.  The $Q^{2}$ range is 
achieved by using three standard beam energies (6.6~GeV, 8.8~GeV and 11~GeV) and two 
non-standard energies at 4.8~GeV and 5.8~GeV.  These later two energies are necessary 
to study the $\epsilon$ dependence in order to constrain two photon exchange (TPE) 
corrections.  

Of concern for the experiment is the accuracy of the beam energy.  The current plan 
is to use the ARC energy method, which has provided a precision of (3--4)$\times$10$^{-4}$.  
The ARC energy system is in the process of being upgraded for the 12~GeV beam.  After 
the upgrade, the system will be recalibrated using elastic scattering from hydrogen 
and tantalum targets with 2.2~GeV incident electrons.   The beam stability and energy 
spread will be monitored by using the Optical Transition Radiation viewer, Synchrotron 
Light Interferometer and Tiefenback energy readback, the latter which is based on the 
beam positions from the beam position monitors (BPMs) located in the Hall~A arc.

The incident beam angle can be determined by using the BPMs located near the target in 
Hall A with an uncertainty of 0.1~mrad.  The beam charge is typically known to $\pm$0.5\%
with careful monitoring and calibration of the beam current monitors. 

\subsubsection{Detector configuration of the HRS spectrometers}
\label{sec:HRS_GMp}
The detector configuration, shown in Fig.~\ref{fig:HRS_config}, includes the 
vertical drift chambers (VDCs), the S0 and S2m counters, the Gas Cherenkov, the lead 
glass calorimeter, and one of the front FPP chambers.  The last is an existing detector 
used many times in the left HRS.  It is the only non-standard item in an otherwise 
traditional electron arm configuration.  By using this chamber, we plan to resolve a 
long-standing problem of HRS tracking analysis whose efficiency of track reconstruction 
is less than 95\% in spite of a high chamber efficiency of 99.5\%.  For the VDCs, the 
amplifier/discriminator (A/D) cards are in the process of being replaced with those developed
for the BigBite wire chambers~\cite{BB_MWDC}.  These new cards use MAD chips~\cite{MAD}, whose 
output stage works with LVDS signals, which produce significantly less feedback than ECL signals 
used in A/D cards by LeCroy Research Systems.  This allows the VDCs to be operated at lower 
discriminator thresholds and high voltages.  The status of these detectors is presented in 
Section~\ref{sec:HRS_GMp}. 

\begin{figure}[ht]
\begin{center}
\includegraphics[width=0.5\textwidth, angle=0.]{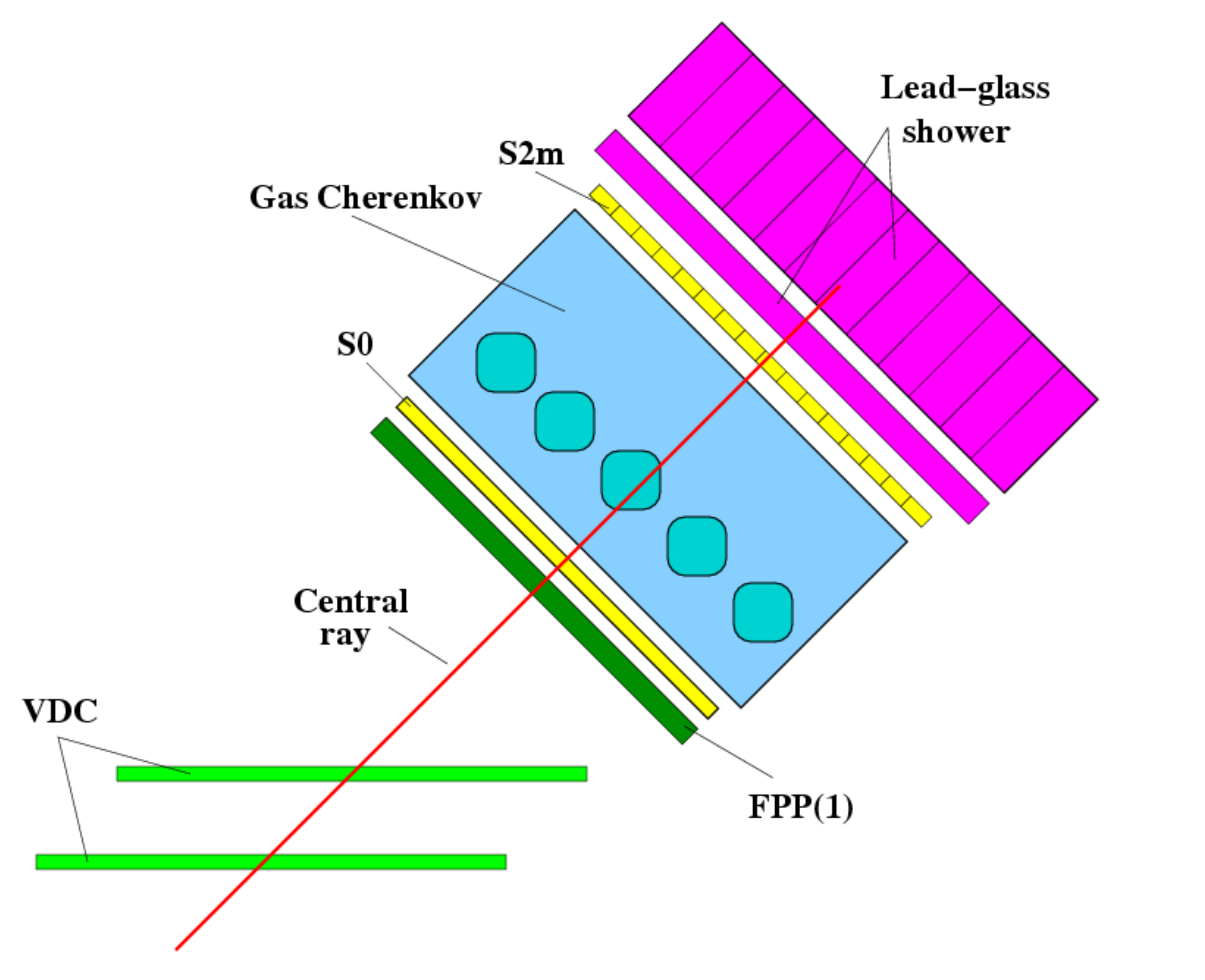}
\caption{The planned detector configuration for the left and right HRSs for GMp with three
wire chambers.}
\label{fig:HRS_config}
\end{center}
\end{figure}

For the spectrometer optics calibration, a sieve-slit placed at the entrance to the 
spectrometers will be used to calibrate the solid angle acceptance, and carbon foils
will be used along the beam line to calibrate the vertex position.

\subsubsection{Configuration of the target}
\label{sec:Cryotarget_GMp}
The standard Hall A cryogenic target ladder will be used with
a liquid hydrogen target and carbon and aluminum solid targets
for spectrometer optics calibrations and target background 
measurements, respectively.  The thickness of the hydrogen target 
needs to be well defined.  We plan to use 15 cm LH$_{2}$ ``race-track'' 
style cells that feature vertical flow of cryogenic fluid to reduce 
the effects of target density fluctuations.  Earlier experiments used
20-cm long race-track cells.

\paragraph{Vertical-flow liquid hydrogen cell and performance\\} 
\label{sec:VertFlow} 
In 2002, tests had shown that the standard cigar-shaped cells displayed 
large density fluctuations with liquid hydrogen even with moderate beam
currents at high fan speeds and large raster sizes.  For this reason, the race-track 
cell~\cite{Race_track} with vertical fluid flow was designed and built by the 
\mbox{Cal State LA} group in coordination with the JLab target group for the HAPPEx-II 
and HAPPEx-He experiments~\cite{Happex2}.  With the new race-track cells, the density fluctuations 
for an LH2 target were an order of magnitude smaller under similar raster sizes and 
fan speeds compared to the standard cells.  The collaboration determined that with a 
5 mm $\times$ 5 mm raster size that density fluctuations would contribute negligibly 
for a hydrogen target at 100~$\mu$A.  However, during the 2005 run period, the fan 
speed had to be increased from the nominal 60~Hz to 91~Hz to minimize the density 
fluctuations.  We plan to perform a detailed study of the density fluctuations from
the race-track cell during commissioning of the experiment.

\subsubsection{Swing arm wire target}
\label{sec:Pointing}
Accurate knowledge of the scattering angle is very important in the GMp experiment.  
However, the mechanical stability of the Hall A spectrometers has a known problem, which 
could lead to a several mm displacement of the vertex.  Regular checks of the spectrometer 
pointing by using the optics target mounted on the long ladder of the cryo-target provide 
an indication of stability but are insufficient for the level of accuracy needed in the
GMp experiment.  We have proposed a solution to this problem by using a tungsten wire target
mounted on the scattering chamber via a short arm, see Fig.~\ref{fig:Swing_Arm}.

\begin{figure}[ht]
\begin{tabular}{ll}
\includegraphics[width=0.475\textwidth, angle=0.]{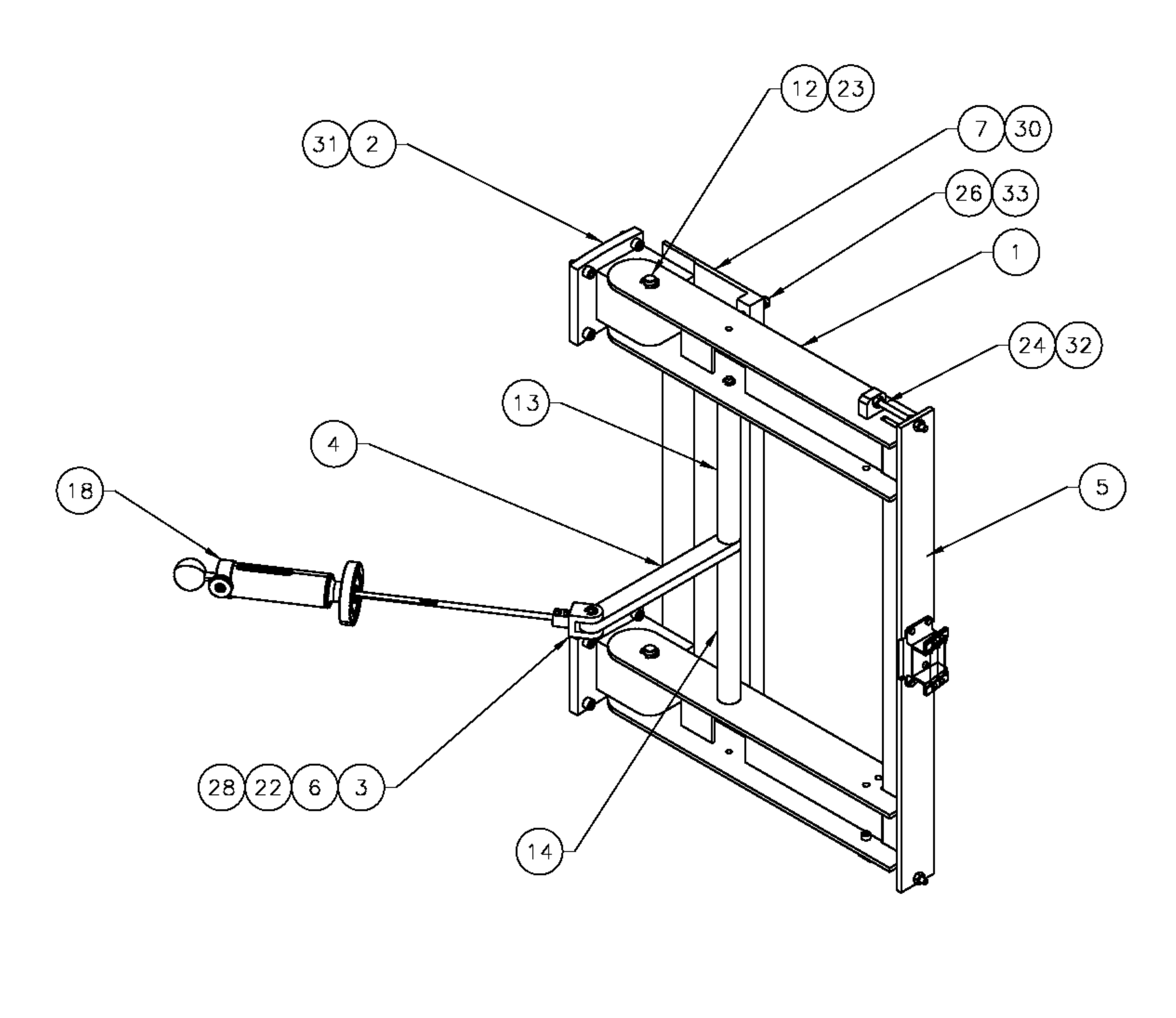}
&\includegraphics[width=0.47\textwidth, angle=0.]{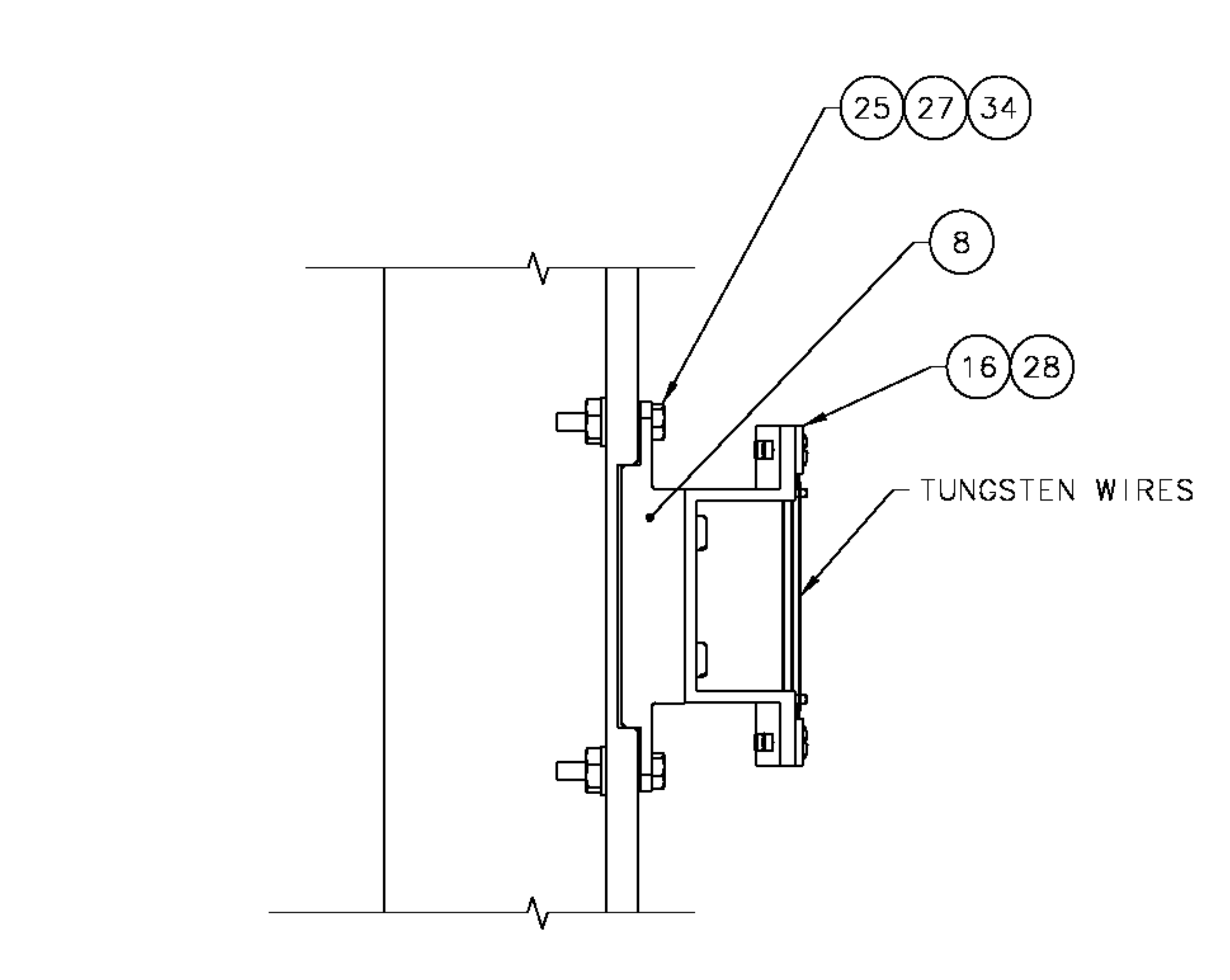}
\end{tabular}
\caption{Shown is a schematic view of the swing arm target.  The right-hand side figure illustrates
the tungsten wires that will be placed in the beam.}
\label{fig:Swing_Arm}
\end{figure}

\subsubsection{Systematic Uncertainty and Expected Results}
\label{sec:GMpexpected}
Achieving the total expected precision on the cross-section measurement requires 
serious work to minimize the systematic uncertainties on several parameters.  Some
of the most critical include the scattering angle, detector efficiencies, deadtime, 
spectrometer acceptance, target density and beam charge.  A few of these items were
discussed in the previous sections.  The contributions to the total uncertainty 
expected on the differential cross section are summarized in Table~\ref{tab:GMp_systs}.

\begin{table}[htb]
\begin{center}
\begin{tabular}{lr} \hline \hline
Source                                             &$\Delta\sigma/\sigma$ (\%) \\ \hline
{\bf Point to Point uncertainties}                 & \\ 
\hspace{0.35cm}Incident Energy                     &$<$0.3 \\
\hspace{0.35cm}Scattering Angle                    &0.1--0.3 \\
\hspace{0.35cm}Incident Beam Angle                 &0.1--0.2 \\
\hspace{0.35cm}Radiative Corrections*               &0.3 \\
\hspace{0.35cm}Beam Charge                         &0.3 \\
\hspace{0.35cm}Target Density Fluctuations         &0.2 \\
\hspace{0.35cm}Spectrometer Acceptance             &0.4--0.8 \\
\hspace{0.35cm}Al Endcap Subtraction               &0.1 \\
\hspace{0.35cm}Detector efficiencies and dead time &0.3 \\ \hline
\hspace{0.75cm}{\it quadratic sum}                 &0.8--1.1 \\ \hline \hline
{\bf Normalization uncertainties}                  & \\ 
\hspace{0.35cm}Beam Charge                         &0.4 \\
\hspace{0.35cm}Target Thickness/Density             &0.5 \\
\hspace{0.35cm}Radiative Corrections*              &0.4 \\
\hspace{0.35cm}Spectrometer Acceptance             &0.6--1.0 \\
\hspace{0.35cm}Al Endcap Subtraction               &0.1 \\
\hspace{0.35cm}Detector efficiencies and dead time &0.4 \\ \hline
\hspace{0.75cm}{\it quadratic sum}                 &1.0--1.3 \\ \hline \hline
\hspace{0.75cm}{\it Statistics}                    &0.5--0.8 \\ \hline \hline
{\bf Total} (Scale+Rand.+Stat.)                    &{\bf 1.2--1.7} \\ \hline \hline
* Not including two photon exchange                & \\
\end{tabular}
\caption[GMp expected systematic uncertainties]{GMp expected systematic uncertainties on the cross section.}\label{tab:GMp_systs}
\end{center}
\end{table}

In Figure~\ref{fig:gmp_expected}, the published world data on the form factor $G^{p}_{M}$ 
versus $Q^2$ is shown along with the range and uncertainties for the expected results 
of the planned experiment.
\begin{figure}[htb]
\begin{center}
\includegraphics[width=0.9\textwidth, angle=0.]{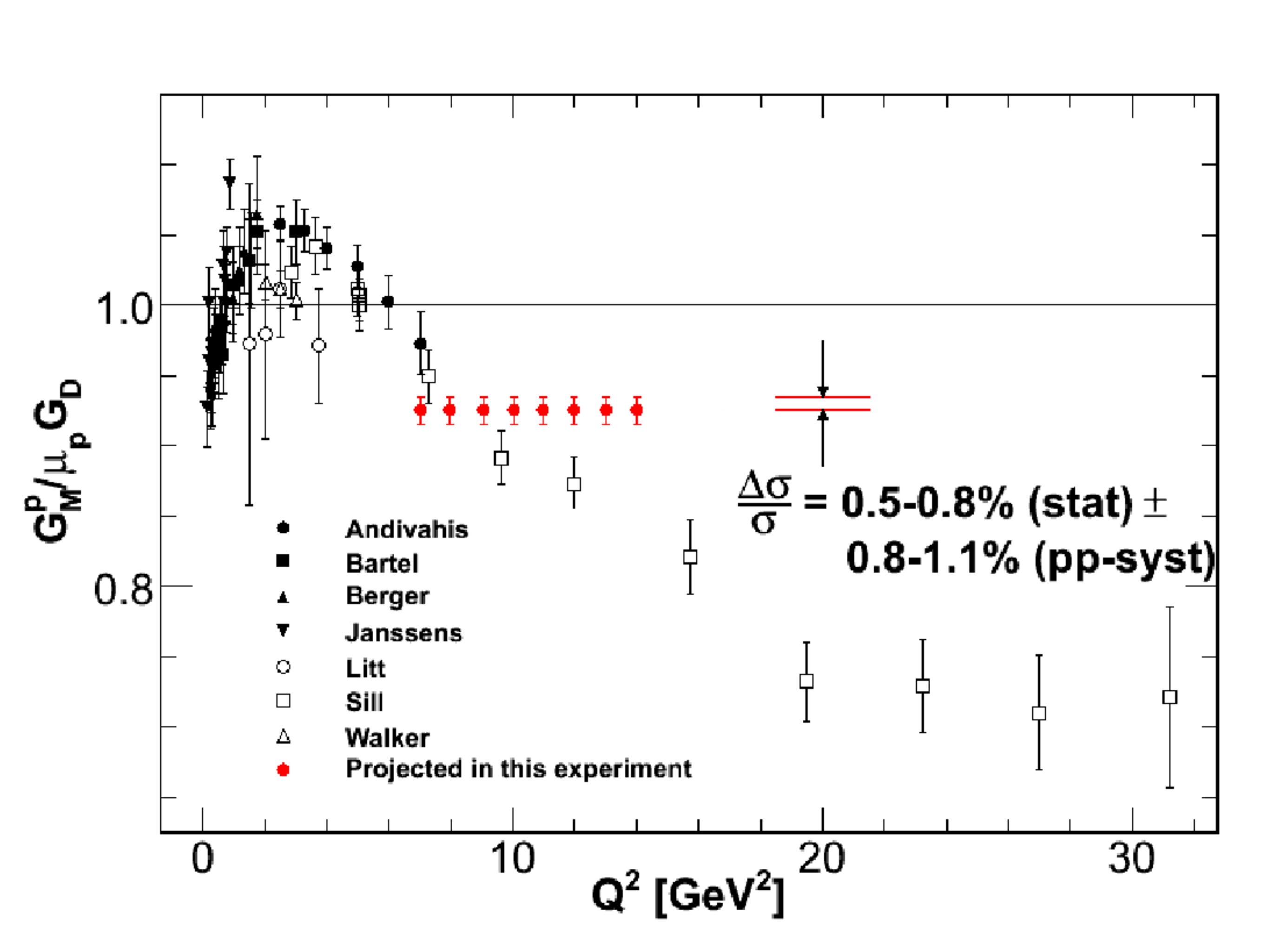}
\caption{Published world data (Refs.~\cite{Andivahis:1994,Bartel:1973,Berger:1971,Janssens:1966,Litt:1970, Walker:1994,Still:1993}) for $G^{p}_{M}/\mu_{p}G_{D}$ as a function of $Q^2$, and the expected 
results for the proposed measurements.  The uncertainties do not include the normalization 
uncertainty.}
\label{fig:gmp_expected}
\end{center}
\end{figure}

\subsubsection{Summary}
\label{sec:GMpSummary}
For the GMp experiment, we aim to measure the proton elastic cross section for $Q^2$ = 
7--14~GeV$^2$ with a total precision of less than 2\%.  A few additional general purpose 
devices and improved analysis techniques will be used to aid our understanding of the 
systematic uncertainties.  The precision of these measurements will greatly benefit
other measurements of the proton form factors that rely on the proton cross section, and
this measurement will also provide an extraction of $G_{M}^{p}$, which is needed for accurate 
determination of $G_{E}^{p}$ from future polarization transfer measurements.

\clearpage \newpage

\section{Publications}
\newcommand{\etal}{{\em et~al.}}
\def\NCA{\rm Nuovo~Cimento}
\def\NIM{\rm Nuclear~Instruments~\&~Methods }
\def\NIMA{{\rm Nucl.~Instrum.~Methods}~A}
\def\NPA{{\rm Nucl.~Phys.}~A\,}
\def\NPB{{\rm Nucl.~Phys.}~B\,}
\def\PLB{{\rm Phys.~Lett.}~B\, }
\def\PRL{\rm Phys.~Rev.~Lett.\,}
\def\PR{{\rm Phys.~Rev.}}
\def\PRD{{\rm Phys.~Rev.}~D\,}
\def\PRC{{\rm Phys.~Rev.}~C\,}
\def\ZPA{{\rm Z.~Phys.}~A\,}
\def\ZPC{{\rm Z.~Phys.}~C\,}
\def\EJP{\rm Eur.~Phys.~J.}
\def\RMP{\rm Rev.~Mod.~Phys.\,}

\begin{enumerate}
\item{O. Glamazdin,  Moeller (iron foils) existing techniques, \NCA~C035N04, 176(2012)}

\item{H.~Fonvieille \etal, Virtual Compton Scattering and the Generalized Polarizabilities of the Proton at $Q^2=0.92$ and 1.76 GeV$^2$, {\rm Phys.~Rev.}~{\bf C86}, 015210 (2012)}

\item{S.~Abrahamyan \etal, Measurement of the Neutron Radius of ${}^{208}$Pb Through Parity-Violation in Electron Scattering, \PRL {\bf 108}, 112502 (2012)}

\item{Z.~Ahmed \etal, New Precision Limit on the Strange Vector Form Factors of the Proton, \PRL {\bf 108}, 102001 (2012)}

\item{J.~Huang \etal, Beam-Target Double Spin Asymmetry $A_{LT}$ in Charged Pion Production from Deep Inelastic Scattering on a Transversely Polarized He-3 Target at $1.4 < Q^2 < 2.7$~GeV$^2$, \PRL {\bf 108}, 052001 (2012) }

\item{A.~Puckett \etal, Final Analysis of Proton Form Factor Ratio Data at $Q^2 =$ 4.0, 4.8 and 5.6~GeV${}^2$, {\rm Phys.~Rev.}~{\bf C85}, 045203 (2012)  }

\item{M.~Mihovilovic \etal, Methods for Optical Calibration of the BigBite Hadron Spectrometer, \NIM {\bf 686}, 20 (2012)}
\item{S. Abrahamyan \etal, New Measurements of the Transverse Beam Asymmetry for Elastic Electron Scattering from Selected Nuclei, Phys.~Rev.~Lett.~{\bf 109}, 192501 (2012)}

\item{M.~Friend, \etal, Upgraded photon calorimeter with integrating readout for the Hall A Compton polarimeter at Jefferson Lab. \NIM {\bf A 676}, 96 (2012)}

\item{D.S.~Parno, \etal, Comparison of Modeled and Measured Performance of GSO Crystal as Gamma Detector. arXiv:1211.3710 [physics.ins-det]}

\end{enumerate}

\clearpage
\newpage

\section{Theses}
\begin{enumerate}
\item{\emph{Polarized ${}^3\mathrm{He}(e,e'n)$ Asymmetries in Three Orthogonal Measurements} \\ Elena Long, Kent State\\ \url{http://arxiv.org/abs/1209.2739}}
\item{\emph{Measurement of double polarized asymmetries in quasi-elastic processes ${}^3\vec{He}(\vec{e},e' d)$ and ${}^3\vec{He}(\vec{e},e' p)$} \\Miha Mihovilovic, University of Ljublijana\\ \url{http://arxiv.org/abs/1208.0748}}
\item{\emph{Precision Measurement of Electroproduction of $\pi^0$ near Threshold} \\Khem Chirapatpimol, University of Virginia\\  \url{https://userweb.jlab.org/~khem/thesis/thesis.pdf}}

\item{\emph{Probing Novel Properties of Nucleons and Nuclei via Parity Violating Electron Scattering} \\ Luis Mercado, University of Massachusetts, Amherst \\ \url{http://scholarworks.umass.edu/cgi/viewcontent.cgi?article=1588&context=open_access_dissertations}}

\item{\emph{Probing the Strangeness Content of the Proton and the Neutron Radius of ${}^{208}$Pb using Parity-Violating Electron Scattering} \\ Rupesh Silwal, University of Virginia\\ \url{http://hallaweb.jlab.org/experiment/HAPPEX/silwal_thesis.pdf} }

\item{\emph{A Precision Measurement of the Proton Strange-Quark Form Factors at $Q^2 = 0.624$~GeV$^2$} \\ Megan Friend, Carnegie Mellon University \\ \url{http://www-meg.phys.cmu.edu/~mfriend/onlinethesis.pdf}}

\item{\emph{Measurement Of Neutron Radius In Lead By Parity Violating Scattering Flash ADC DAQ} \\ Zafar Ahmed, Syracuse University\\ \url{http://surface.syr.edu/phy_etd/120/}}

\end{enumerate}

\clearpage
\newpage

\section{Hall A Collaboration Member List}
{\parindent 0cm 
{\bf Argonne National Lab}\\
John Arrington\\
Frank Dohrmann\\
Paul Reimer\\
Jaideep Singh\\
Xiaohui Zhan\\

{\bf Budker Institute of Nuclear Physics}\\
Dima Nikolenko\\
Igor Rachek\\

{\bf Cairo University}\\
Hassan Ibrahim\\

{\bf California Institute of Technology}\\
Xin Qian\\

{\bf California State University}\\
Konrad A. Aniol\\
Martin B. Epstein\\
Dimitri Margaziotis\\

{\bf Carnegie Mellon University}\\
Gregg Franklin\\
Vahe Mamyan\\
Brian Quinn\\

{\bf The Catholic University of America}\\
Tanja Horn\\

{\bf Commissariat a l’Energie Atomique - Saclay}\\
Maxime Defurne \\

{\bf China Institute of Atomic Energy (CIAE)}\\
Xiaomei Li\\
Shuhua Zhou\\

{\bf Christopher Newport University}\\
Ed Brash\\

{\bf College of William and Mary}\\
David S. Armstrong\\
Todd Averett\\
Juan Carlos Cornejo\\
Melissa Cummings\\
Wouter Deconinck\\
Keith Griffioen\\
Joe Katich\\
Lubomir Pentchev\\
Charles Perdrisat\\
Huan Yao\\
Bo Zhao\\

{\bf Dapnia/SphN}\\
Nicole d'Hose\\
Franck Sabatie\\

{\bf Duquesne University}\\
Fatiha Benmokhtar\\

{\bf Duke University}\\
Steve Churchwell\\
Haiyan Gao\\
Calvin Howell\\
Min Huang\\
Simona Malace\\
Richard Walter\\
Qiujian Ye\\

{\bf Faculte des Sciences de Monastir (Tunisia)}\\
Malek Mazouz\\

{\bf Florida International University}\\
Armando Acha\\
Werner Boeglin\\
Luminiya Coman\\
Marius Coman\\
Lei Guo\\
Hari Khanal\\
Laird Kramer\\
Pete Markowitz \\
Brian Raue\\
Jeorg Reinhold\\

{\bf The George Washington University}\\
Ramesh Subedi\\

{\bf Gesellschaft fur Schwerionenforschung (GSI)}\\
Javier Rodriguez Vignote\\

{\bf Hampton University}\\
Eric Christy\\
Leon Cole\\
Ashot Gasparian\\
Peter Monaghan\\

{\bf Harvard University}\\
Richard Wilson\\

{\bf Hebrew University of Jerusalem}\\
Moshe Friedman\\
Aidan Kelleher\\
Guy Ron \\

{\bf Huangshan University}\\
Hai-jiang Lu\\
XinHu Yan\\

{\bf Idaho State University}\\
Mahbub Khandaker\\
Dustin McNulty\\

{\bf INFN/Bari}\\
Raffaele de Leo\\

{\bf INFN/Catania}\\
Vincenzo Bellini\\
Sutera Concetta Maria\\
Antonio Guisa\\
Francesco Mammoliti\\
Giuseppe Russo\\
Maria Leda Sperduto\\

{\bf INFN/Lecce}\\
Roberto Perrino\\

{\bf INFN/Roma}\\
Marco Capogni\\
Francesco Cusanno\\
Alessio Del Dotto \\
Salvatore Frullani\\
Franco Garibaldi\\
Franco Meddi\\
Guido Maria Urciuoli\\
Evaristo Cisbani\\
Rachel di Salvo\\
Mauro Iodice\\

{\bf Institut de Physique Nucleaire}\\
Alejandro Marti Jimenez-Arguello\\

{\bf Institute of Modern Physics, Chinese Academy of Sciences}\\
Xurong Chen\\

{\bf Institut de Physique Nucleaire - Orsay}\\
Camille Desnault\\
Rafayel Paremuzyan\\

{\bf ISN Grenoble}\\
Eric Voutier\\

{\bf James Madison University}\\
Gabriel Niculescu\\
Ioana Niculescu\\

{\bf Jefferson Lab}\\
Kalyan Allada\\
Alexandre Camsonne\\
Larry Cardman\\
Jian-Ping Chen\\
Eugene Chudakov\\
Kees de Jager\\
Alexandre Deur\\
Ed Folts\\
David Gaskell\\
Javier Gomez\\
Ole Hansen\\
Douglas Higinbotham\\
Mark K. Jones\\
Thia Keppel\\
John Lerose\\
Bert Manzlak\\
David Meekins\\
Robert Michaels\\
Bryan Moffit\\
Sirish Nanda\\
Noel Okay\\
Andrew Puckett\\
Yi Qiang\\
Lester Richardson\\
Yves Roblin\\
Brad Sawatzky\\
Jack Segal\\
Dennis Skopik\\
Patricia Solvignon\\
Mark Stevens\\
Riad Suleiman\\
Stephanie Tysor\\
Bogdan Wojtsekowski\\
Jixie Zhang\\

{\bf Jozef Stefan Institute}\\
Miha Mihovilovic\\
Simon Sirca\\

{\bf Kent State University}\\
Bryon Anderson\\
Mina Katramatou\\
Elena Khrosinkova\\
Elena Long\\
Richard Madey\\
Mark Manley\\
Gerassimos G. Petratos\\
Larry Selvey\\
Andrei Semenov\\
John Watson\\

{\bf Kharkov Institute of Physics and Technology}\\
Oleksandr Glamazdin\\
Viktor Gorbenko\\
Roman Pomatsalyuk\\
Vadym Vereshchaka\\

{\bf Kharkov State University}\\
Pavel Sorokin\\

{\bf Khalifa University}\\
Issam Qattan\\

{\bf Lanzhou University}\\
Bitao Hu\\
Yi Zhang\\

{\bf Longwood University}\\
Tim Holmstrom\\
Keith Rider\\
Jeremy St. John\\
Wolfgang Troth\\

{\bf Los Alamos Laboratory}\\
Jin Huang\\
Xiaodong Jiang\\
Ming Xiong Liu\\

{\bf LPC Clermont-Ferrand France}\\
Pierre Bertin\\
Helene Fonvielle\\
Carlos Munoz Camacho\\

{\bf Mississippi State University}\\
Dipangkar Dutta\\
Mitra Shabestari\\
Amrendra Narayan\\
Nuruzzaman\\

{\bf Massachusetts Institute of Technology}\\
Bill Bertozzi\\
Shalev Gilad\\
Navaphon ``Tai'' Muangma\\
Kai Pan\\
Cesar Fernandez Ramirez\\
Rupesh Silwal\\
Vincent Sulkosky\\

{\bf Negev Nuclear Research Center}\\
Arie Beck\\
Sharon Beck\\

{\bf NIKHEF}\\
Jeff Templon\\

{\bf Norfolk State University}\\
Vina Punjabi\\

{\bf North Carolina Central University}\\
Benjamin Crowe\\
Branislav (Branko) Vlahovic\\

{\bf Northwestern University}\\
Ralph Segel\\

{\bf Ohio University}\\
Julie Roche\\

{\bf Old Dominion University}\\
Gagik Gavalian\\
Mohamed Hafez\\
Wendy Hinton\\
Charles Hyde\\
Hashir Rashad \\
Larry Weinstein\\

{\bf Peterburg Nuclear Physics Institute}\\
Viacheslav Kuznetsov \\

{\bf Regina University}\\
Alexander Kozlov\\

{\bf Rutgers University}\\
Lamiaa El Fassi\\
Ron Gilman\\
Gerfried Kumbartzki\\
Katherine Myers\\
Ronald Ransome\\
Yawei Zhang\\

{\bf Saint Norbert College}\\
Michael Olson\\

{\bf Seoul National University}\\
Seonho Choi\\
Byungwuek Lee\\

{\bf Smith College}\\
Piotr Decowski\\

{\bf St Mary's University}\\
Davis Anez
Adam Sarty\\

{\bf Stanford Linear Accelerator}\\
Rouven Essig\\

{\bf Syracuse University}\\
Zafar Ahmed\\
Richard Holmes\\
Paul A. Souder\\

{\bf Tel Aviv University}\\
Nathaniel Bubis\\
Or Chen\\
Igor Korover\\
Jechiel Lichtenstadt\\
Eli Piasetzky\\
Ishay Pomerantz\\
Ran Shneor\\
Israel Yaron\\

{\bf Temple University}\\
David Flay\\
Eric Fuchey\\
Zein-Eddine Meziani\\
Michael Paolone\\
Matthew Posik\\
Nikos Sparveris\\

{\bf Tohoku University}\\
Kouichi Kino\\
Kazushige Maeda\\
Teijiro Saito\\
Tatsuo Terasawa\\
H. Tsubota\\

{\bf Tsinghua University}\\
Zhigang Xiao\\

{\bf Universidad Complutense de Madrid (UCM)}\\
Joaquin Lopez Herraiz\\
Luis Mario Fraile\\
Maria Christina Martinez Perez\\
Jose  Udias Moinelo\\

{\bf Universitat Pavia}\\
Sigfrido Boffi\\

{\bf University ``La Sapienza'' of Rome}\\
Cristiano Fanelli \\
Fulvio De Persio\\

{\bf University of Glasgow}\\
John Annand\\
David Hamilton\\
Dave Ireland\\
Ken Livingston\\
Dan Protopopescu\\
Guenther Rosner\\
Johan Sjoegren\\

{\bf University of Illinois}\\
Ting Chang\\
Areg Danagoulian\\
J.C. Peng\\
Mike Roedelbronn\\
Youcai Wang\\
Lindgyan Zhu\\

{\bf University of Kentucky}\\
Dan Dale\\
Tim Gorringe\\
Wolfgang Korsch\\

{\bf University of Lund}\\
Kevin Fissum\\

{\bf University of Manitoba}\\
Juliette Mammei\\

{\bf University of Maryland}\\
Elizabeth Beise\\

{\bf University of Massachusetts, Amherst}\\
Krishna S. Kumar\\
Seamus Riordan\\
Jon Wexler\\

{\bf University of New Hampshire}\\
Toby Badman\\
Trevor Bielarski\\
John Calarco\\
Greg Hadcock\\
Bill Hersman\\
Maurik Holtrop\\
Donahy John\\
Mark   Leuschner\\
James Maxwell\\
Sarah Phillips\\
Karl Slifer\\
Timothy Smith\\
Ryan Zielinski\\

{\bf University of Regina}\\
Garth Huber\\
George Lolos\\
Zisia Papandreou \\

{\bf University of Saskatchewan}\\
Ru Igarashi\\

{\bf University of Science and Technology of China (USTC)}\\
Yi Jiang\\
Wenbiao Yan \\
Yunxiu Ye\\
Zhengguo Zhao\\
Yuxian Zhao \\
Pengjia Zhu\\

{\bf University of South Carolina}\\
Steffen Strauch\\

{\bf University of Virginia}\\
Khem Chirapatpimol\\
Mark Dalton\\
Donal Day\\
Xiaoyan Deng\\
Gordon D. Gates\\
Gu Chao\\
Charles Hanretty\\
Ge  Jin\\
Sudirukkuge Tharanga Jinasundera\\
Richard Lindgren\\
Jie Liu\\
Nilanga Liyanage\\
Vladimir Nelyubin\\
Blaine Norum\\
Kent Paschke\\
Peng Chao\\
Oscar Rondon\\
Kiadtisak Saenboonruang\\
William ``Al'' Tobias
Diancheng Wang\\
Kebin Wang\\
Zhihong Yi\\
Zhiwen Zhao\\
Xiaochao Zheng\\
Jiayao Zhou\\

{\bf University of Washington}\\
Diana Parno\\

{\bf Yamagata University}\\
Seigo Kato\\
Hiroaki Ueno\\

{\bf Yerevan Physics Institute}\\
Sergey Abrahamyan\\
Nerses Gevorgyan\\
Edik Hovhannisyan\\
Armen Ketikyan\\
Samvel Mayilyan\\
Artush Petrosyan\\
Galust Sargsyan \\
Albert Shahinyan\\
Hakob Voskanian\\

{\bf Past Members}\\
Mattias Anderson\\
Maud Baylac\\
Hachemi Benaoum\\
J. Berthot\\
Michel Bernard \\
Louis Bimbot\\
Tim Black\\
Alexander Borissov\\
Vincent Breton\\
Herbert Breuer\\
Etienne Burtin\\
Christian Cavata\\
George Chang\\
Nicholas Chant\\
Jean-Eric Ducret\\
Zhengwei Chai\\
Brandon Craver \\
Natalie Degrande\\
Pibero Djawotho\\
Chiranjib Dutta\\
Kim Egiyan\\
Stephanie Escoffier\\
Catherine Ferdi\\
Megan Friend \\
Robert Feuerbach\\
Mike Finn\\
Bernard Frois\\
Oliver Gayou\\
Charles Glashausser\\
Jackie Glister\\
Brian Hahn\\
Harry Holmgren\\
Sebastian Incerti\\
Riccardo Iommi\\
Florian Itard\\
Stephanie Jaminion\\
Steffen Jensen\\
Cathleen Jones\\
Lisa Kaufman\\
James D. Kellie\\
Sophie Kerhoas\\
Ameya Kolarkar\\
Norm Kolb\\
Ioannis Kominis\\
Serge Kox\\
Kevin Kramer\\
Elena Kuchina\\
Serguei Kuleshov\\
Jeff Lachniet\\
Geraud Lavessiere\\
Antonio Leone\\
David Lhuillier\\
Meihua Liang\\
Han Liu\\
Robert Lourie\\
Jacques Marroncle\\
Jacques Martino\\
Kathy McCormick\\
Justin McIntyre\\
Luis Mercado\\
Brian Milbrath\\
Wilson Miller\\
Joseph Mitchell\\
Jean  Mougey\\
Pierre Moussiegt\\
Alan Nathan\\
Damien Neyret\\
Stephane Platchkov\\
Thierry Pussieux\\
Gilles Quemener\\
Abdurahim Rakhman\\
Bodo Reitz\\
Rikki Roche\\
Philip Roos\\
David Rowntree\\
Gary Rutledge\\
Marat Rvachev\\
Arun Saha\\
Neil Thompson\\
Luminita Todor\\
Paul   Ulmer\\
Antonin Vacheret\\
Luc Van de Hoorebeke\\
Robert Van de Vyver\\
Pascal Vernin\\
Dan Watts\\
Krishni Wijesooriya\\
Hong Xiang\\
Wang Xu\\
Jingdong Yuan\\
Jianguo Zhao\\
Jingdong Zhou\\
Xiaofeng Zhu\\
Piotr Zolnierczuk\\
}

\end{document}